\documentclass[a4paper,twoside]{article}
\usepackage{ulem}
\usepackage{graphicx}
\usepackage[varg]{txfonts}
\usepackage{amssymb}
\usepackage[english]{babel}
\usepackage[utf8]{inputenc}
\usepackage{array}
\usepackage{float}
\usepackage[colorinlistoftodos]{todonotes}
\usepackage[]{titlesec} 
\titleformat{\subsubsection}{\normalfont\large\bfseries}{OP \thesubsubsection}{1em}{}
\usepackage{natbib}
\bibpunct{(}{)}{;}{a}{}{,} %% natbib format for A&A and ApJ

\def\farcs{\hbox{$.\!\!^{\prime\prime}$}}
\def\arcsec{\hbox{$^{\prime\prime}$}}

\newcommand{\kms}{\mbox{km~s$^{-1}$}} 
\newcommand{\ion}[2]{{#1}\,{\sc #2}}

\newcommand{\CaI}{\ion{Ca}{I} }
\newcommand{\CaII}{\ion{Ca}{II} }
\newcommand{\CaIIK}{\ion{Ca}{II}~K }
\newcommand{\CaIIH}{\ion{Ca}{II}~H }
\newcommand{\FeI}{\ion{Fe}{I} }
\newcommand{\FeII}{\ion{Fe}{II} }

\newcommand{\NaI}{\ion{Na}{I} }
\newcommand{\MgI}{\ion{Mg}{I} }
\newcommand{\HeI}{\ion{He}{I} }
\newcommand{\HI}{\ion{H}{I} }
\newcommand{\SiI}{\ion{Si}{I} }
\newcommand{\SrI}{\ion{Sr}{I} }
\newcommand{\TiI}{\ion{Ti}{I} }
\newcommand{\Halpha}{\ion{H}{$\!\alpha$} }

\usepackage[pdftex, hyperfootnotes=false]{hyperref}

\usepackage{fancyvrb} % allows to use verbatim text in footnotes  % Alex Feller
\VerbatimFootnotes  % Alex Feller

\usepackage{adjustbox}
\usepackage{lscape}      % For landscape tables
\usepackage{graphicx}    % Figures if needed

\usepackage{caption}
\usepackage{booktabs}    % Better tables
\usepackage{longtable}   % For multi-page tables
\usepackage{siunitx}     % For units
\usepackage{tabularx}
\usepackage{placeins}
\usepackage{float} % in preamble
\usepackage{makecell}
\usepackage{booktabs}
\usepackage{tcolorbox}
\newcommand{\OPtabletitleCustom}[4]{\bigskip{\bfseries \large OP Table~\hyperref[#4]{#1}: #2}\phantomsection\label{#3}\par}

\newlength{\MySpanTwoColumn}
\begin{document}
%------------------------------------------------------------------------
\begin{center}
\thispagestyle{empty}
\begin{minipage}{19cm}{
\includegraphics[height=4cm]{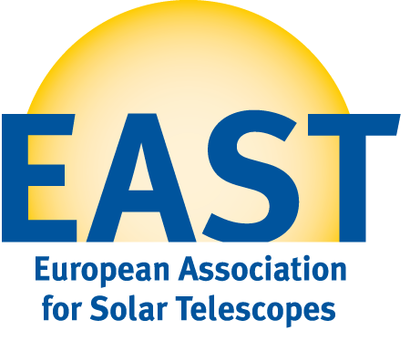} \hspace{3.5cm}
\includegraphics[height=4cm]{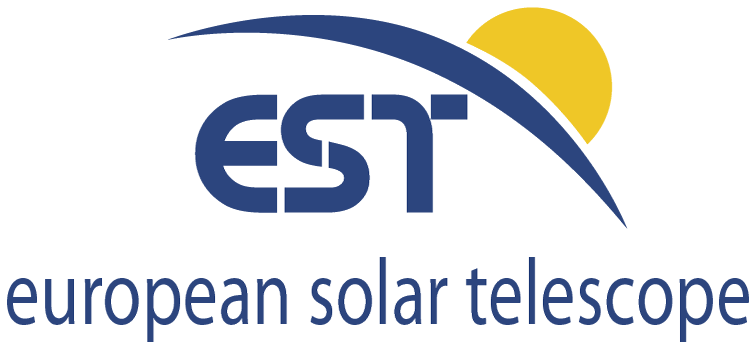} 
}\end{minipage}

\vspace*{2cm} 
\LARGE

{Science Requirement Document (SRD) \\ for the \\European Solar Telescope (EST)\\[4ex]  

%{\Large 2nd edition, December 2019} \\[0ex]
{\Large 3rd edition, December 2025} \\[6ex]

%{\large with updates in October 2025 by chair of SAG,  Rolf Schlichenmaier} \\[6ex]

{\large Schlichenmaier, R.$^{1}$; Bellot Rubio, L.R.$^{2}$; Collados, M.$^{3,4}$; Erdelyi, R.$^{5,6,27}$; Feller, A.$^{7}$; Fletcher, L.$^{8,13}$; Jur\v{c}\'{a}k, J.$^{9}$; Khomenko, E.$^{3}$; Leenaarts, J.$^{10}$; Matthews, S.$^{11}$; Belluzzi, L.$^{12,1}$; Carlsson, M.$^{13,14}$; 
%Dalmasse, K.$^{15}$; 
Danilovic, S.$^{10}$; Gömöry, P.$^{16}$; Kuckein, C.$^{3}$; Manso Sainz, R.$^{17}$; Martínez González, M.$^{3}$; Mathioudakis, M.$^{18}$; Ortiz, A.$^{24,25}$; Riethm\"uller, T.L.$^{7}$; Rouppe van der Voort, L.$^{13,14}$; Simoes, P.J.A.$^{19}$; Trujillo Bueno, J.$^{3,20}$; Utz, D.$^{21}$; Zuccarello, F.$^{22}$; de la Cruz Rodriguez, J.$^{10}$; Giovanelli, L.$^{23}$, Jafarzadeh, S.$^{18}$, Jess, D.B.$^{18}$; Milic, I.$^{1}$; Nelson, C.$^{26}$; van Noort, M.$^{7}$; Ruiz de Galarreta, C.${^3}$; Zeuner, F.$^{12}$ \\
(Author affiliations given on page \pageref{affiliations}.)}\\[3ex]
 
}

\vfill

\includegraphics[height=2cm]{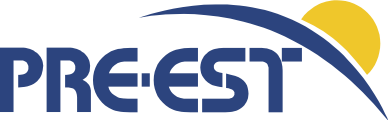}

\vspace{0.5cm}

\includegraphics[height=2cm]{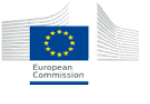}
\parbox[b]{13cm}{\large
“This project has received funding from the European Union’s Horizon 2020 research and innovation programme under grant agreement No 739\,500”
\vspace*{4ex}
}

\end{center}

%------------------------------------------------------------------------
\newpage
\phantomsection
\label{affiliations}
{\Large Author affiliations:}
\bigskip\bigskip

\begin{itemize}
\item[$^{1}$:] Institut für Sonnenphysik (KIS), Schöneckstr. 6, 79104 Freiburg, Germany \\[-4ex]
\item[$^{2}$:] Instituto de Astrof\'isica de Andaluc\'ia (IAA), Glorieta de la Astronomía s/n, Granada, Spain \\[-4ex]
\item[$^{3}$:] Instituto de Astrof\'isica de Canarias (IAC), V\'ia L\'actea s/n, 38205 La Laguna, Tenerife, Spain \\[-4ex]
\item[$^{4}$:] Departamento de Astrofísica, Universidad de La Laguna, 38206 La Laguna, Tenerife, Spain \\[-4ex]
\item[$^{5}$:] SP2RC, University of Sheffield, Sheffield, UK \\[-4ex]
\item[$^{6}$:] Dept of Astronomy, E\"otv\"os University, Budapest, Hungary \\[-4ex]
\item[$^{7}$:] Max-Planck-Institut für Sonnensystemforschung (MPS), Justus-von-Liebig-Weg 3, 37077 Göttingen, Germany \\[-4ex]
\item[$^{8}$:] SUPA School of Physics and Astronomy, University of Glasgow, Glasgow G12 8QQ, UK \\[-4ex]
\item[$^{9}$:] Astronomical Institute of the Czech Academy of Sciences, Fri\v{c}ova  298, 25165 Ond\v{r}ejov, Czech Republic \\[-4ex]
\item[$^{10}$:] Institute for Solar Physics, Department of Astronomy, Stockholm University, Albanova University Center, 10691 Stockholm, Sweden \\[-4ex]
\item[$^{11}$:] UCL Mullard Space Science Laboratory, Holmbury St Mary, Dorking RH5 6NT, UK \\[-4ex]
\item[$^{12}$:] Istituto Ricerche Solari Locarno (IRSOL), Via Patocchi 57, CH-6605 Locarno Monti, Switzerland \\[-4ex]
\item[$^{13}$:] Rosseland Centre for Solar Physics, University of Oslo, PO Box 1029, Blindern 0315, Oslo, Norway \\[-4ex]
\item[$^{14}$:] Institute of Theoretical Astrophysics, University of Oslo, PO Box 1029, Blindern 0315, Oslo, Norway \\[-4ex]
%\item[$^{15}$:] IRAP, Université de Toulouse, CNRS, CNES, UPS, F-31028 Toulouse, France \\[-4ex]
\item[$^{16}$:] Astronomical Institute of the Slovak Academy of Sciences, 05960 Tatransk\'{a} Lomnica, Slovakia \\[-4ex]
\item[$^{17}$:] Third Institute of Physics, University of G{\"o}ttingen, Friedrich-Hund-Platz 1, 37077 G{\"o}ttingen, Germany \\[-4ex]
\item[$^{18}$:] Astrophysics Research Centre, School of Mathematics and Physics, Queen’s University Belfast, Belfast, BT7 1NN, Northern Ireland, U.K. \\[-4ex]
\item[$^{19}$:] Centro de Rádio Astronomia e Astrofísica Mackenzie, Escola de Engenharia, Universidade Presbiteriana Mackenzie, São Paulo, Brazil \\[-4ex]
\item[$^{20}$:] Consejo Superior de Investigaciones Cient\'ificas, Spain \\[-4ex]
\item[$^{21}$:] IGAM/Institute of Physics, Karl-Franzens University Graz, Austria \\[-4ex]
\item[$^{22}$:] Dipartimento di Fisica e Astronomia “Ettore Majorana”, University of Catania, Via S. Sofia 78, 95123 Catania, Italy \\[-4ex]
\item[$^{23}$:] Department of Physics, University of Rome Tor Vergata, 00133, Rome, Italy.\\[-4ex]
\item[$^{24}$:] Expert Analytics AS, Oslo, Norway\\[-4ex]
\item[$^{25}$:] European Solar Telescope Canarian Foundation (EST-CF)\\[-4ex]
\item[$^{26}$:] European Space Agency (ESA), European Space Research and Technology Centre (ESTEC), Keplerlaan 1, 2201 AZ Noordwijk, The Netherlands  \\[-4ex]
\item[$^{27}$:] Gyula Bay Zoltan Solar Observatory (GSO), Hungarian Solar Physics Foundation (HSPF), Pet\H{o}fi t\'er 3, Gyula, H-5700, Hungary. \\[-4ex]
%\item[$^{25}$:] ...\\[-4ex]
\end{itemize}
%------------------------------------------------------------------------
\newpage
\bigskip
\begin{minipage}{15cm}
\vspace*{2cm}
\huge{\bf Preface to 3rd edition}
\vspace{3cm}
\end{minipage}

The European Strategy Forum on Research Infrastructures (ESFRI) included the European Solar Telescope (EST) as an “ESFRI Project” in its 2016 Roadmap and confirmed this status during the preparatory phase in 2021. The Preparatory Phase, partially funded as PRE-EST under the H2020 Framework Programme, ran from April 2017 to 2022 and was one of three ESFRI projects in the field of Physical Sciences and Engineering.

During this phase, the EST Science Advisory Group (SAG) was established in 2017. Its first task was to revise the Science Requirements Document (SRD), originally formulated during the Preliminary Design Phase of EST between 2008 and 2011. This second edition of the SRD was published in December 2019 \citep{2019arXiv191208650S}.

Since 2019, the EST Project Office has advanced the telescope design and developed a highly sophisticated Science Instrumentation Suite (SIS). All telescope subsystems have now passed their Preliminary Design Review, and the SIS has successfully completed its Conceptual Design Review. Technical and design developments were iterated in close consultation with the SAG. As a result, the SIS has reached a high degree of maturity, following several years of coordinated scientific consolidation, structured community engagement, and the application of systems engineering methodologies.

Aligned with the discussions and resolutions of the SAG, the SIS comprises three categories of first-generation instruments:
\begin{enumerate}
\item Tunable Imaging Spectropolarimeters coupled with Fixed Band Imagers (TIS/FBIs), employing large-aperture Fabry-Pérot etalons;
\item Integral Field Spectropolarimeters based on microlens arrays (IFS-M);
\item The near-infrared spectropolarimeter EMBER (spectropolariMeter Based on slicEr-mirrors for the near-infraRed), using image-slicing technology.
\end{enumerate}

Together, these instruments are distributed across seven spectral channels spanning the Blue, Visible, Red, and Infrared optical arms of the EST Coudé light-distribution system. This configuration enables simultaneous observations of the same two-dimensional solar region across a broad wavelength range (380–2200 nm), giving EST a uniquely powerful multi-wavelength observing capability.

The significant evolution of the telescope and instrumentation design has made it necessary to revise the Observing Programmes so that they accurately reflect the capabilities of the SIS and ensure that all science objectives can be met. This third edition of the EST SRD integrates the outcomes of numerous SAG meetings held between 2020 and 2025 and incorporates all design developments up to 2025. It updates the Observing Programmes accordingly and assesses their alignment with the scientific objectives defined by the SAG.

\bigskip
Freiburg, December 10th, 2025 \hfill Rolf Schlichenmaier

%------------------------------------------------------------------------
\newpage

\setcounter{tocdepth}{2}
\tableofcontents
%------------------------------------------------------------------------

\part{Introduction}

The European Solar Telescope (EST)%
\footnote{EST web link: \verb+http://www.est-east.eu+ } 
is a research infrastructure for solar physics. It is planned to be an on-axis solar telescope with an aperture of 4\,m and equipped with an innovative suite of spectro-polarimetric and imaging post-focus instrumentation. The EST project was initiated and is driven by EAST%
\footnote{EAST web link: \verb+ http://www.est-east.eu/est/index.php/people/ + }%
, the European Association for Solar Telescopes. EAST was founded in 2006 as an association of 14 European countries. Today, as of December 2019, EAST consists of 26 European research institutes from 18 European countries. 

The Preliminary Design Phase of EST was accomplished between 2008 and 2011. During this phase, in 2010, the first version of the EST Science Requirement Document (SRD)
%\footnote{SRD: Science Requirement Document} 
was published. After EST became a project on the ESFRI%
\footnote{ESFRI: European Strategy Forum on Research Infrastructures. Web link to ESFRI roadmap in March 2016: \verb+ http://www.esfri.eu/roadmap-2016 + } 
roadmap 2016, the preparatory phase started. This phase is partially supported by EU funding through the PRE-EST H2020 project%
\footnote{\label{footnote:preest}PRE-EST: The EST Preparatory Phase has received funding from the European Union's Horizon 2020 research and innovation programme under grant agreement Nr. 739 \,500.}%
. The goal of the preparatory phase is to accomplish a final design for the telescope and the legal governance structure of EST. A major milestone on this path is to revisit and update the Science Requirement Document (SRD).
 
The EST Science Advisory Group (SAG) has been constituted by EAST and the Board  of the PRE-EST$^{\begin{NoHyper} {\ref{footnote:preest}} \end{NoHyper}}$
EU project in November 2017 and has been charged with the task of providing with a final statement on the science requirements for EST. Based on the conceptual design, the SRD update takes into account technical and scientific developments, to ensure that EST provides significant advancement beyond the current state-of-the-art.

The present update of the EST SRD has been developed and discussed during a series of EST SAG meetings:
\begin{enumerate}
\item[] 1st telecon meeting on Nov 5th, 2017
\item[] 2nd meeting in Freiburg, Nov 24, 2017 
\item[] 3rd telecon meeting, Dec 15, 2017 
\item[] 4th telecon meeting, March 26, 2018
\item[] 5th meeting in Belfast, April 16 \& 17, 2018 
\item[] 6th meeting in Naxos, June 16, 2018
\item[] 7th telecon meeting, January 14, 2019
\item[] 8th telecon meeting, October 11, 2019
\item[] 9th telecon meeting, October 22, 2019
\item[] 10th telecon meeting, December 3, 2019
\end{enumerate}

The SRD develops the top-level science objectives of EST into individual science cases. Identifying critical science requirements is one of its main goals. Those requirements will define the capabilities of EST and the post-focus instrument suite. The technical requirements for the final design of EST will be derived from the SRD. 

The science cases presented in Part II (Sects. 1 to 8) are not intended to cover all the science questions to be addressed with EST, but rather to provide a precise overview of the capabilities that will make of EST a competitive state-of-the-art telescope to push the boundaries of our knowledge over the next few decades. The science cases contain detailed observing programmes specifying the type of observations needed to solve specific science problems. An effort is being made to define the parameters of the required observations as accurately as possible, taking into account both present capabilities and technological developments expected in the near future. The tables of the observing programmes corresponding to the science cases are compiled in Sect. 10.
The EST science cases represent challenging observations that put strong constraints on the telescope and its instrument suite.  Ultimately, they will be translated into Technical Requirement Document (TRD) leading to the final EST design to be implemented during the construction phase.

The unique design advantages of the EST concept is presented in Section 11. The effect of the science cases on the EST design are discussed in Section 12 and summarized in Section 13.

\part{Top-level science goals}
\label{partII}

%%%%%%%%%%%%%%%%%%%%%%%%%%%%%%%%%%%%%%%%%%%%%%%%%%%%%%%%%%%%%%%%%%%%%%%%%%%%
%From SG4: Many structures and phenomena observed on the Sun are a result of the interaction between moving plasma and large-scale magnetic fields. The magnetic field gives rise to a variety of physical effects:
%\begin{itemize}
%\itemsep 0pt
% \item[--] it exerts a force, which may accelerate plasma or create structures
% \item[--] it stores energy, which may later be released as an eruption or solar flare
% \item[--] it acts as a thermal blanket, giving rise to sunspots or stabilising a prominence
% \item[--] it channels fast particles and plasma
% \item[--] it drives instabilities and supports waves.
%\end{itemize}
%Each of these effects is crucial for solar activity related phenomena. Novel high-resolution and high-polarimetric-sensitivity observations are necessary to understand the interaction between plasma motions and magnetic fields both in the solar photosphere and chromosphere. And since photospheric and chromospheric processes are the fundamental driver of the dynamic and hotter outer solar atmosphere, these processes are essential for our understanding of the solar corona.\\
%%%%%%%%%%%%%%%%%%%%%%%%%%%%%%%%%%%%%%%%%%%%%%%%%%%%%%%%%%%%%%%%%%%%%%%%%%%%%
%
%-----------------------------------------------------------
\section{Structure and evolution of magnetic flux} 
\label{sec_qs}
{Authors: Luis Bellot Rubio, Dominik Utz, Sanja Danilovic} %\footnote{({version: 2018-Dec-07})}

%==========================================

Magneto-convection is an ubiquitous process in the solar surface. The
interaction of magnetic fields and granular convection leads to the
formation of magnetic features and a broad range of flows and waves on
very small spatial and temporal scales. This makes the photosphere highly
dynamic and the place where heating of the upper solar atmosphere could 
be initiated via small-scale flux emergence, flux cancellation, braiding 
of magnetic field lines, or generation and upward propagation of waves.

These processes need to be studied at their intrinsic spatial and 
temporal scales, of order 100 km and 10 s, respectively. For the most
part, small-scale magnetic fields in the photosphere and chromosphere
are weak, producing very small polarisation signals. To detect them,
high polarimetric sensitivity is needed. The three requirements have
been difficult to meet with existing telescopes and, as a result,
temporal resolution has usually been sacrificed. Thus, our knowledge
of the temporal evolution of quiet sun magnetic fields is very
limited. With its large aperture primary mirror and dedicated suite of
instruments, EST will for the first time provide the necessary
sensitivity and spatial resolution to study the structure, dynamics,
and evolution of the weak fields pervading the quiet solar surface,
which according to estimates harbor orders of magnitude more
magnetic flux than active regions and sunspots.

\subsection{Formation and disappearance of kG flux concentrations in 
the solar photosphere}

\label{sec:conv_coll}

Outside of active regions, the solar magnetic field is organized in
small-scale ($\sim$100 km) magnetic flux tubes and flux sheets
\citep{1972SoPh...22..402H, 1973SoPh...32...41S,1984A&A...131..333S,
2004A&A...428..613B}. These features are observed in intergranular
lanes and form the photospheric network, located at the borders of
supergranular cells. Small-scale flux tubes are believed to play an
important role in the irradiance variations of the Sun
\citep{2014A&A...568A..13R} and the heating of the chromosphere 
and corona through a variety of dynamical processes, including the
channeling of waves, shock fronts, and solar tornadoes
\citep[see][]{2012Natur.486..505W}.

The magnetic field of network flux tubes is as strong as 1500~G 
\citep[e.g.,][]{2000ApJ...535..489B, 2013A&A...554A..65U}, i.e., well 
above the equipartition value of a few hundred G. Therefore, in
addition to the kinematic concentration of flux by horizontal
convective motions, another mechanism capable of enhancing the field
up to kG strengths is required. In the late 1970s, it was
theoretically proposed that the field is amplified by a thermal
instability known as convective collapse \citep{1978ApJ...221..368P, 
1979SoPh...61..363S}. This process leads to rapid downflows in
the tube's interior, causing a strong evacuation of the magnetic
element and a concentration of the field. The spatial and temporal
scales associated with the formation of intense flux tubes are on the
order of 100~km and 30~s, according to numerical simulations by
\cite{2010A&A...509A..76D}.

Unfortunately, convective collapse has proven very difficult to
detect. Only a limited number of single case studies and statistical
analyses have been presented in the literature 
\citep[e.g.,][]{2001ApJ...560.1010B, 2008ApJ...677L.145N, 
2009A&A...504..583F, 2011A&A...529A..79N} Thus, we still lack a clear,
convincing observational picture of the formation of kG flux
concentrations on the solar surface. One of the main problems is that
the intensification events recorded so far do not produce stable kG
features. In most cases, rapid upflows develop while the field is
amplified, leading to the disappearance of the magnetic features
\citep{2001ApJ...560.1010B, 2005ApJ...620L..71S}. Some
 studies suggest that the newly created elements undergo
oscillations in field strength and size \citep{2014ApJ...789....6R,
2014ApJ...796...79U}. This may partly explain the observed lack of
stability: the features would seem to disappear if the polarisation
signals go below the detection threshold during specific phases of the
oscillation. Better polarimetric sensitivity than currently available
is needed to verify this conjecture.

Interestingly, the rapid photospheric downflows generated by
convective collapse should produce even faster flows in the
chromosphere, due to its lower density. This provides another 
means to detect convective collapse events. However, only
\cite{2009A&A...504..583F} have studied the response of the
chromosphere to convective collapse. Using Dopplergrams in the
\ion{Mg}{i} b$_2$ line taken by the Hinode NFI, they concluded that a
significant fraction of the field amplification events observed in the
photosphere show rapid downflows near the temperature minimum
region. Clearly, a full description of the formation process of kG
features calls for simultaneous spectropolarimetric measurements in
the photosphere, temperature minimum region, and chromosphere, at very
high cadence ($\sim 10$~s). The required multi-line observations are
currently out of reach for existing telescopes, due to the lack of
proper instrumentation (integral field spectropolarimeters or a suite
of imaging polarimeters working in parallel).

The evolution and physical processes that lead to the disappearance of
kG flux tubes are not well understood either. On large scales, network
regions do not change significantly in the course of hours/days, but
on the smallest scales the flux evolves rapidly
\citep[e.g.,][]{2016ApJ...820...35G}. Small-scale magnetic elements
change shape, move around pushed by granular convection, and interact
with other flux concentrations, merging or cancelling with them
\citep{2014ApJ...789....6R}. During their lifetime,
they may brigthen considerably and be detected as magnetic bright
points (MBPs). It is possible to study these features using broad-band
imaging. However, the brightness enhancement is transitory and MBPs
usually disappear much earlier than the magnetic field itself. We
still do not know why. To understand these processes, we need
ultra-high resolution imaging and spectropolarimetry, in order to
determine the full vector magnetic field. Unfortunately,
spectropolarimetric measurements are scarce, due to the very small
spatial and temporal scales involved. The evolution and eventual
disappearance of flux tubes would be a natural consequence of their
interaction with convective motions if they are a surface phenomenon,
as indicated by numerical simulations. On the other hand, the
persistence of some network features suggests that they are rooted
below the solar surface. In that case, other mechanisms might be
responsible for the disappearance of the flux.

\subsubsection{Formation and evolution of intense flux tubes in the 
solar atmosphere}\label{OP1.1.1}

\vspace{-2ex}
See table on page \pageref{OP1.1.1table}.

% -----------------------------------------------------------------------

\subsection{Internal structure of small-scale flux concentrations}

Our understanding of small-scale magnetic flux concentrations 
heavily relies upon magneto-convection simulations
\citep[e.g.,][]{1998ApJ...495..468S, 2005A&A...429..335V, 
2006ApJ...642.1246S, 2009A&A...504..595Y,
2014ApJ...789..132R}. Conversely, an accurate observational
characterisation of their internal structure may help validate and
improve current computer models. Particularly interesting are the
interfaces between magnetic and field-free regions. One example of
such interfaces are the canopies of flux tubes expanding with height
in the solar atmosphere. These are regions where one can expect
electric currents, Joule dissipation, and a variety of dynamical
flows.

Magnetic canopies and the flows associated with them were soon
recognized as essential ingredients for explaining the asymmetries of
the polarisation profiles that emerge from flux concentrations at a
resolution of 1\arcsec\/. However, a direct observational detection 
of such interfaces was not possible at the time. Only with Hinode and
SUNRISE have we started to resolve the internal structure of magnetic
elements, including their canopies.

\cite{2007A&A...476L..33R} presented evidence of canopies in magnetic
elements from the spatial variation of the Stokes $V$ area asymmetry
of the \ion{Fe}{i} 630~nm lines recorded by Hinode. The observed
variations appear to agree well with those predicted by radiative
magneto-hydrodynamic simulations. \cite{2012ApJ...758L..40M}
characterised the canopy of a large network flux concentration from an
analysis of \ion{Fe}{i} 525.02 measurements taken by SUNRISE. They too
found a spatial distribution of Stokes $V$ area asymmetries that
agrees well with the results of MHD models. In particular, they
located the height of the canopy as a function of radial distance from
the center of the flux patch, finding an expanding structure, and
determined the jumps in field strength and velocity across the
discontinuity. \cite{2015A&A...576A..27B} derived the
3D structure of plage flux concentrations using a spatially coupled
inversion of the \ion{Fe}{i} lines recorded by Hinode. For the first
time, these authors were able to map the height variation of the
vector magnetic field and flow field across the magnetic element. They
verified the existence of a ring of strong downflows surrounding the
flux concentration in deep photospheric layers and found hints of
opposite magnetic polarities associated with the downflows. These
features go well beyond the simple picture of magnetic flux tubes.

The next step is to determine the temporal evolution of flows and
fields across magnetic elements, relating the observed changes to
interactions with the surrounding granular convection
\citep{2014ApJ...789....6R,2015ApJ...810...79R}. This demands time
series with cadences of less than 1 minute that cannot be obtained
with existing telescopes due to insufficient photon flux.

It is also important to examine the differences between magnetic
canopies in internetwork, network and plage regions through
statistical analyses. How high are magnetic canopies located in the
quietest and more active regions? What fraction of the surface is
covered by strongly inclined fields in the chromosphere? What are
their average lifetimes? To answer these questions, a good height
coverage of the chromosphere and high polarimetric sensitivity are
required.

Magnetic interfaces do not exist only as canopies in flux
concentrations, but also as true discontinuities in other solar
structures, e.g., magnetic field lines emerging in the photosphere or
horizontal chromospheric fields \citep{2006ASPC..354..345S}. Hinode
observations show a non-negligible fraction of Stokes V profiles with
extreme asymmetries. Such asymmetries can only be produced by sharp
magnetic field discontinuities in structures that do not seem to fit
the picture of classical flux tubes
\citep{2012ApJ...748...38S}. The observed profiles change in the
course of minutes, indicating a rapid evolution of the field and/or
associated flows. EST, with its ultra-high resolution capabilities,
will contribute to the characterisation of those fields and their
temporal evolution in the photosphere and the chromosphere.

\subsubsection{Internal structure and evolution of magnetic elements}\label{OP1.2.1}

\vspace{-2ex}
See table on page \pageref{OP1.2.1table}.

%-------------------------------------------------
\subsection{Magnetic bright points}

Interestingly, the cross-section of small-scale kG flux tubes 
and flux sheets can be identified as bright points when observed 
in Fraunhofer's G band region of the solar spectrum
\citep[e.g.,][]{2001A&A...372L..13S, 2013SoPh..tmp..276B}. These
features are called G-band bright points or more generally magnetic
bright points (MBPs). MBPs are important because they represent the
photospheric part of flux tubes reaching up to the upper
atmosphere. Their buffeting and violent displacement by the
surrounding granulation \citep{1989SoPh..119..229M} create
magnetohydrodynamic waves \citep[e.g.,][]{2013SSRv..175....1M}
traveling up into the upper atmosphere (see Sect. 2). Moreover, they
have different total and spectral irradiances compared with their
surroundings.  Thus, changes in the appearance of MBPs over the solar
cycle have consequences for the total solar irradiance variation and
therefore for the climate on Earth.

Although MBPs are fundamental building blocks of the solar magnetism
\citep{1987ARA&A..25...83Z}, many basic questions are still
unanswered. For example, it is not known if these flux elements can be
created with arbitrarily small sizes. All studies trying to reveal the
size distribution of MBPs so far had the problem that MBPs can be
found all the way down to the diffraction limit of the telescope
\citep[e.g.,][]{2010ApJ...725L.101A}. Thus we need larger telescopes
to detect the lower size limit of these features. Besides, even the
shape of the size distribution itself is under discussion. Clearly,
the physics creating an unlimited exponential distribution toward the
smallest scales would be different from the physics creating a
lognormal or Gaussian distribution of sizes.

Similarly, there have been problems in measuring a fundamental
property of MBPs - their magnetic field strength. Due to their small
size and highly dynamic evolution, the inferred field strength
distributions contain a lot of noise and uncertainty
\citep[see, for example,][]{2007A&A...472..607B, 2013A&A...554A..65U}. 
Therefore, it is impossible to say if MBPs always appear with field strengths of
roughly 1.3 kG as predicted by theory \citep{1979SoPh...61..363S} and
supported by MHD simulations \citep{2014A&A...568A..13R}, or if they
indeed possess a significant weak field component as deduced from real
observations. To answer this question we need higher spatial and
temporal resolution spectro-polarimetric data with increased
sensitivity.

Another fundamental property of MBPs is their movement on the solar
surface. We do not know if they follow random and erratic paths or
more deterministic, regular trajectories. This is important in at
least two ways. Firstly, as mentioned before, the driving of MBPs will
create MHD waves which travel into the upper atmosphere and deposit
energy there, thus contributing to chromospheric/coronal heating
\citep[e.g.,][]{2004A&A...422.1085H}. Clearly, the kind of motion
experienced by MBPs will strongly influence the type and amplitude of
the resulting MHD waves \citep{1994A&A...283..232M}. Secondly, the way
MBPs move tells us about the diffusion of magnetic fields
\citep[][]{2010A&A...511A..39U,2011ApJ...743..133A}, hence by studying 
them we gain knowledge also on how large-scale structures such as
sunspots disappear.

Besides of these basic properties we need to study the evolution of
MBPs in more detail \citep[e.g.,][]{2014ApJ...796...79U}. While we
believe they are created via the convective collapse process (see
Sect. \ref{sec:conv_coll}), we do not know much about their
dissolution or about the processes they experience between creation
and dissolution. This is due to their very short lifetimes (of the
order of 2.5 min). With the cadences delivered by existing
spectroscopic instruments, generally in the order of 30~s
\citep{2010ApJ...725L.101A, 2010A&A...511A..39U, 2018ApJ...856...17L},
we only have about 5 measurements of the properties of MBPs from
creation to dissolution, which prohibits detailed studies of the
involved physical processes. With EST it will be possible to increase
the spatial resolution and, more importantly, the temporal
cadence. This will enable us to study not only the basic properties
of MBPs but also their whole evolution from birth to death, teaching us
in that way fundamental plasma physics.

\subsubsection{Magnetic bright points}\label{OP1.3.1}

\vspace{-2ex}
See table on page \pageref{OP1.3.1table}.

\subsubsection{Dynamic parameters and evolution of MBPs}\label{OP1.3.2}

\vspace{-2ex}
See table on page \pageref{OP1.3.2table}.

%-----------------------------------------------------------------------
\subsection{Small-scale flux emergence in the quiet Sun}

In the last two decades, high-resolution observations by ground-based
and space-borne telescopes have led to the discovery of different
modes of magnetic flux emergence in the quiet Sun. These events occur
at the spatial scale of granular convection and seem to be very
frequent.

\cite{2007ApJ...666L.137C}  and \cite{2008A&A...481L..25I} described the
appearance of low-lying magnetic loops in granules. The loops are
characterised by footpoints located in intergranular lanes and
horizontal fields in between, above granular cells. It has been
suggested that they emerge from subsurface layers, carried by the
upflows of granular convection. After emergence, the distance between
the footpoints increases rapidly, and the linear polarisation signals
indicating horizontal fields disappear in only a few minutes. About
25\% of the loops seem to rise into the chromosphere, producing
measurable polarisation signals and brightnenings on their way up
\citep{2009ApJ...700.1391M}.  The loops have a range of sizes and 
are observed to emerge down to the smallest spatial scales of 
$\sim$$0\farcs2$ achievable nowadays \citep{2012ApJ...755..175M}.

\cite{2008A&A...481L..33O} reported on the emergence of apparently
vertical fields in quiet Sun granules. The typical lifetimes of those
events are of order 10-20 minutes. The flux concentrations appear at
the center of granules and grow with time, associated with upward
motions. They fade gradually until they become undetectable below the noise
level. Sometimes, the magnetic flux is observed to disappear with the
hosting granule. Throughout the process, no opposite polarities are
detected in the field of view. Such events may represent the emergence
of fields from subsurface layers. An alternative explanation is that
the fields do not come from subsurface layers, but are present in the
photosphere all the time until they show up above the noise level by
the action of some unknown mechanism.

The events described by \cite{2008A&A...481L..33O} may just be
individual examples of the unipolar flux appearances observed in the
solar internetwork by Hinode and SUNRISE
\citep[e.g.,][]{2010ApJ...720.1405L, 2016ApJ...820...35G,
2017ApJS..229...17S}. Unipolar appearances are very common in the
quiet Sun and may contribute a significant fraction of the total
internetwork flux, but their origin is a mystery.

An observational characterisation of the magnetic properties
(strengths, inclinations) and evolution of both small-scale loops and
unipolar flux patches in the quiet Sun still needs to be performed. To
that end, multi-line spectropolarimetry at high cadence is required,
given the short lifetimes of the features.

Furthermore, the total amount of flux carried to the solar surface by
emerging loops and unipolar features has not been evaluated properly
so far. Due to the limited sensitivity of the measurements, a
significant fraction of the flux emerging on the surface may have gone
unnoticed, although estimates from Hinode and SUNRISE suggest that the
amount of flux carried by internetwork flux concentrations is much
larger than that of active regions \citep{2011SoPh..269...13T, 
2016ApJ...820...35G, 2017ApJS..229...17S}. To solve this problem, 
long time sequences of multi-line spectropolarimetric observations 
covering a full supergranule are needed, at the highest sensitivity possible.

Finally, it is important to clarify the role of emerging internetwork
loops and unipolar patches in heating the solar atmosphere. These
features are known to interact with pre-existing fields, merging and
cancelling with them. Cancellations of opposite-polarity flux patches
are likely to result in magnetic reconnection and energy release at
chromospheric heights. This process needs to be quantified in the
solar internetwork.

\subsubsection{Emergence and evolution of magnetic fields in granular convection}\label{OP1.4.1}

\vspace{-2ex}
See table on page \pageref{OP1.4.1table}.

%----------------------------------------------------------------

\subsubsection{Properties of magnetic fields emerging in the quiet Sun}\label{OP1.4.2}

\vspace{-2ex}
See table on page \pageref{OP1.4.2table}.

%-----------------------------------------------------------------------

\subsection{Magnetic flux cancellations in the quiet Sun}

Flux cancellation occurs when magnetic features of opposite polarity
come into close contact and disappear partially or totally as a
consequence of the interaction. In that way, magnetic flux is removed
from the solar surface.

Cancellation events are very common, particularly among the small flux
features of the quiet Sun internetwork. However, we still do not know
if they are the result of the submergence of $\Omega$-loops or the
rise of U-loops, with or without previous reconnection of unrelated
magnetic flux systems. All these mechanisms may be a source of
chromospheric/coronal heating and transient events, due to the release
of magnetic energy. To distinguish among the different possibilities,
simultaneous observations of the photosphere and chromosphere need to
be carried out at the highest angular resolution possible. By
analyzing the photospheric and chromospheric flow fields at the
cancellation sites, and the timing difference between the events
occurring in the different layers, it will be possible to draw
definite conclusions about the mechanisms responsible for the removal
of magnetic flux from the solar surface.

If magnetic reconnection is the main driver behind cancellations, then
the goal is to determine how the magnetic topology of the cancelling
features changes, the height at which the reconnection takes place,
and whether current MHD simulations are able to quantitatively
reproduce observations as we go down to extremely small spatial
scales. We also need to determine how much these processes contribute
to the overall heating of the solar atmosphere.

Observations show that current instruments are unable to
resolve these processes and reliably detect pre- and post-reconnection
loops \citep{2010ApJ...713..325I, 2010ApJ...712.1321K,
2014ApJ...793L...9K}. Furthermore, sampling of the layers where the
magnetic energy gets deposited is difficult because of the rapid
temperature and density increase which can result in dramatic changes
of the formation heights of the observables \citep{2017ApJS..229....5D}.

This observational program requires the highest spatial resolution
possible combined with high spectropolarimetric sensitivity so that
the magnetic field vector can be constrained reliably. Additionally,
simultaneous observations in multiple spectral lines is mandatory in
order to retrieve the 3D configuration of these events. This is
especially true for determining the location and length of the 
current sheet and tracing the path of the outflowing material.

\subsubsection{Magnetic field topology, dynamics and energy release at flux cancellation sites}\label{OP1.5.1}

\vspace{-2ex}
See table on page \pageref{OP1.5.1table}.

%-----------------------------------------------------------------------

\subsection{Quiet Sun internetwork fields}

A substantial (perhaps dominant) fraction of the total magnetic flux
of the solar photosphere is stored in internetwork regions, i.e., the
interior of supergranular cells
\citep{2014ApJ...797...49G}. Internetwork fields produce very small
polarisation signals either because they are intrinsically weak, or
because the magnetic filling factor is small, or both. Thus, their
observational characterisation represents a significant
challenge. Prior to Hinode, our understanding of these fields was
mainly based on long integration (1 min) snapshots with spatial
resolutions not better than 1\arcsec\/ \citep{2000ApJ...532.1215S,
2003A&A...408.1115K, 2004ApJ...613..600L, 2007A&A...464..351L,
2008A&A...477..953M}

Measurements with the Hinode spectropolarimeter at a resolution of
0\farcs3 revealed that internetwork fields are weak and highly
inclined. The field strength distribution peaks at hG values,
monotonically decreasing toward stronger fields. The field inclination
distribution shows a maximum at 90 degrees, which corresponds to
purely horizontal fields. However, the exact shape of the distribution
is still under debate, with some authors favoring a quasi-isotropic
distribution \citep{2009ApJ...701.1032A, 2014A&A...572A..98A} and
others pointing out that the distribution corresponds to very inclined
but not isotropic fields \citep{2012ApJ...751....2O,
2016A&A...593A..93D}. Furthermore, the magnetic filling factor (the
fractional area of the resolution element covered by fields) amounts
to 0.2-0.3, even at the resolution of Hinode. Thus, there is still
room for subpixel structuring of the field on very small scales. To
investigate this possibility, a resolution of 0\farcs1 or better is
required.

While circular polarisation signals are seen in almost 100\% of the
internetwork surface area \citep{2008ApJ...672.1237L}, linear polarisation is
much more difficult to detect due to the weakness of the field. This
represents a serious drawback for the accurate determination of the
magnetic properties of the internetwork, in particular the field
inclination. Linear signals exist, but they are buried in the noise
\citep{2012ApJ...757...19B, 2016A&A...596A...5M}. To improve the 
field inclination estimates and resolve the controversy about the
shape of the distribution, it is necessary to detect those signals and
invert them together with Stokes V. This requires much more sensitive
observations than are feasible with any existing instrument, by at
least one order of magnitude.

The height variation of internetwork fields is an important aspect 
of the quiet Sun magnetism. It has been addressed by
\cite{2016A&A...593A..93D}, using Hinode measurements, that internetwork fields do not show
linear polarisation signals except in a tiny fraction of the field of
view. Here too the analysis would benefit from increased sensitivity
leading to the detection of linear polarisation all over the field of
view. Another way to improve the accuracy of the results is to observe
as many spectral lines as possible simultaneously, in order to extract
information on the different atmospheric layers in a more direct way
and not through extrapolation.

Another aspect that needs to be studied is the temporal evolution of
internetwork fields. The Stokes profiles emerging from internetwork
flux patches are often asymmetric and show multiple lobes, indicating
a very dynamic nature. It is therefore important to resolve their
evolution on short time scales. The large photon collecting power of
EST will make it possible to obtain the time series required for that.

Observing the temporal evolution of internetwork fields is also
important to understand their nature. A fundamental question is
whether internetwork fields are the result of a global or a local
dynamo. Magneto-convection simulations of local dynamo action
seem to explain the predominance of inclined fields deduced from
Hinode observations \citep{2008A&A...481L...5S, 2010A&A...513A...1D,
2014ApJ...789..132R}. However, other mechanisms can also produce
inclined fields in the quiet Sun internetwork, such as the emergence
of horizontal fields into the solar surface
\citep{2008ApJ...680L..85S}. One way to distinguish between scenarios
is to compare the short-term evolution of internetwork fields with
existing MHD simulations, which will be possible for the first time
with EST.

Lastly, it is necessary to study the interaction of internetwork
fields with other internetwork and network fields. These processes
release energy and can be expected to contribute to chromospheric
heating in the quiet Sun. First indications that this is the case have
been gathered by \cite{2018ApJ...857...48G}, but the observations lack
the sensitivity required to detect the weakest internetwork fields,
and therefore a large fraction of the quiet Sun surface remains
unexplored.

\subsubsection{Physical properties of internetwork magnetic fields}\label{OP1.6.1}

\vspace{-2ex}
See table on page \pageref{OP1.6.1table}.

\subsubsection{Short-term evolution of internetwork fields}\label{OP1.6.2}

\vspace{-2ex}
See table on page \pageref{OP1.6.2table}.

%-----------------------------------------------------------------------

\subsection{Polar magnetic fields}

The magnetic fields in the poles of the Sun are believed to be an
essential ingredient of the solar dynamo. As the activity cycle
proceeds, flux from mid-latitude decaying active regions is
transported to the poles by supergranular diffusion and meridional
circulation. This flux accumulates in the polar caps and eventually
reverses the magnetic polarity of the poles. Among other reasons, the
polar caps are important because the fast solar wind emanates from
them and because they host a variety of magnetic structures such as
faculae (one of the main sources of solar irradiance variations),
polar plumes (observed in the EUV), and small-scale X-ray polar jets
\citep{2007PASJ...59S.745S, 2007Sci...318.1580C}.

Far from the direct influence of the lower-latitude activity belt, the
polar region is also the obvious place to investigate whether or not
small-scale photospheric flux concentrations are the result of local
dynamo action. Observations with Hinode suggest that the only
difference between the magnetism of the quiet Sun at the disk center
and the polar region is the existence of strong, relatively large
unipolar flux patches, with the small-scale fields showing nearly the
same distribution of parameters \citep{2008ApJ...688.1374T,
2010ApJ...719..131I}. This seems to favor a turbulent dynamo. However,
most of the solar surface in those regions remains unexplored as a
consequence of the strong projection effects, which reduce both the
amplitude of the polarisation signals and the effective spatial
resolution of the observations. EST, with its enormous light
collecting power, will give access to very weak polarisation signals
that current instruments cannot detect. Polar field studies will also
benefit from the high spatial resolution capabilities of EST, which
permit a significant decrease in the solar surface area sampled by the
resolution element.

Of particular interest are the properties of polar faculae in the
lower atmospheric layers and how they appear and disappear. The
short-term evolution of polar faculae is essentially unknown. However,
the results of \cite{2015ApJ...799..139K} based on Hinode measurements
suggest that these features are formed by advection of flux by
supergranular flows rather than by flux emergence. Eventually polar
faculae fade and disappear on the spot, with or without previous
fragmentation. This mode of disappearance is not well understood and
requires further investigation. One possibility is that the magnetic
patches fragment in elements too small to be detected. The fragments
may later cancel out with opposite polarity elements that reach the
polar region around the maximum of the solar cycle. EST will confirm
or refute this scenario.

Other scientific questions that need to be addressed include how the
flux from the old cycle is replaced in the polar caps (either by
reconnection, field submergence, or field expulsion), whether or not the
properties of strong flux concentrations are the same in the polar
regions and the equatorial belt, and how the magnetic field drives the
various phenomena observed at high latitudes, in particular X-ray
jets. These problems should be investigated at their relevant spatial
scales (100 km).

Further progress in our understanding of polar fields will be allowed
by magnetic stereoscopy from coordinated observations of EST and Solar
Orbiter. Magnetic stereoscopy is the only reliable way to solve the
180$^{\circ}$ azimuth ambiguity of Zeeman measurements, which is
particularly problematic near the poles as it renders the orientation
and even the polarity of the field unknown
\citep[e.g.,][]{2010ApJ...719..131I}. For meaningful comparisons of 
the same structure from the two vantage points, the high spatial
resolution of EST is essential, as other ground-based telescopes will
not be able to isolate the small surface area contained in the Solar
Orbiter resolution element. Thus, coordination of EST and Solar
Orbiter will likely result in new ground-breaking science.

\subsubsection{Structure of polar faculae}\label{OP1.7.1}

\vspace{-2ex}
See table on page \pageref{OP1.7.1table}.

\subsubsection{Properties, distribution and evolution of polar magnetic fields}\label{OP1.7.2}

\vspace{-2ex}
See table on page \pageref{OP1.7.2table}.

%-----------------------------------------------------------------------------------------------------------------------------------------------
\subsection{Network dynamics}

The magnetic network is a well-known large-scale feature of the 
solar photosphere that outlines the borders of supergranular cells
\citep[e.g.,][]{2012ApJ...758L..38O}. It contains a significant fraction
of the total magnetic flux of the quiet Sun, and is relatively stable
on supergranular time scales ($\sim$~days). The magnetic network also
leaves clear imprints on the chromosphere, where it can be easily
identified in \ion{Ca}{ii} H filtergrams as a bright network structure
indicating excess temperatures but also shock fronts and waves caused
by the dynamics of the underlying photospheric network
\citep[e.g.,][]{2008ApJ...680.1542H}.

Open questions concerning the magnetic network include its obvious
unipolar character on large scales (compared to the smaller bipolar
internetwork fields populating the interior of supergranules), its
flux balance \citep{1997ApJ...487..424S}, and its evolution over the
solar cycle \citep{2014ApJ...796...19T}.

Moreover, strong network elements have lifetimes of several minutes to
hours only \citep[e.g.,][]{2005AA...441.1183D}, while the network
itself persists for days. Thus, while appearing stable, the network is
fully dynamic: strong magnetic elements are continually formed,
reshuffled, and dissolved there. During those processes, magnetic flux
can be lost due to submergence, reconnection, cancelation of opposite
polarity elements, and various other mechanisms
\citep{1997ApJ...487..424S}. As a result, the network flux needs 
to be replenished by new flux elements appearing on the solar
surface. The source of such elements is still unknown. Some papers
suggest that quite a large fraction can be provided by internetwork
magnetic features, which are swept to the network via supergranular
flows \citep[see][]{2014ApJ...797...49G}. A contradicting older
hypothesis would state that most of the network flux is actually
coming from decaying ephemeral regions.

For a more thorough understanding of the flux balance of the magnetic
network and its dynamics, it is necessary to lower the detection
threshold of internetwork magnetic features in terms of both field
strength and size, to capture the more abundant lower end of the
distributions. Only then the contribution of internetwork fields to
the maintenance of the network will be fully understood. On the other
hand, a better characterisation of the interaction between
internetwork and network fields is necessary to understand how
internetwork fields get converted into network fields and what this
means for the dynamics of the upper atmosphere.

\subsubsection{Network evolution and dynamics}\label{OP1.8.1}

\vspace{-2ex}
See table on page \pageref{OP1.8.1table}.

 %\newpage \input{SG1_tables}
%-----------------------------------------------------------
\newpage\section{Wave coupling throughout solar atmosphere} 
{Authors: Robertus Erdelyi, Elena Khomenko, Mihalis Mathioudakis} %\footnote{({version: 2018-Dec-14})}
\label{sec_waves}

%\section{SG2 ``Wave coupling throughout solar atmosphere''}

The Sun generates a wealth of waves, mostly acoustic in their generated character, by turbulent convection beneath the photosphere \citep{1952RSPSA.211..564L, 1967SoPh....2..385S, 1990ApJ...363..694G, 1994ApJ...424..466G, 1998A&A...335..673B, 1994ApJ...423..474M, 2001ApJ...546..576N, 2001ApJ...546..585S}. The energy of these waves in the photosphere is believed to be sufficiently large to account for the heating of the upper layers \citep{1989A&A...222..171U, 1994ApJ...423..474M, 2002A&A...386..971F, 2006ApJ...646..579F, 2007SoPh..246....3B, 2007AN....328..726E, 2009SSRv..149..229T, 2009Sci...323.1582J, 2015RSPTA.37340261A}. For this to manifest, however, during their propagation from the (sub)photosphere to higher, magnetically dominated regions of the solar atmosphere, the waves may need to transform their acoustically dominated sub-photospheric nature to waves that account for the presence of the atmospheric magnetic fields. Once they reach the mid to upper atmosphere, they often dissipate there, at least in part, and release their non-thermal energy in the form of heat. Studying the wave coupling through the solar atmosphere consists of studying how the waves propagate and evolve, interact, transform and dissipate during their passage through the photosphere, chromosphere, transition region and corona of the Sun \citep{2006RSPTA.364..351E}. Such research includes studying the waves in the presence of atmospheric stratification, the wave interactions with the often dynamic magnetic structures present everywhere over the surface of the Sun. Wave interaction with these structured and stratified magnetic fields results in the generation of a large variety of wave modes. At the same time structuring, whether magnetic or thermodynamic, also serves as guide for the progression of wave energy and momentum, and connect all the layers up to the corona and interplanetary space. The common line for all observational studies of the wave coupling is the need of simultaneous data at several layers of the solar atmosphere, which EST will routinely provide in unprecedented spatial, temporal and spectral resolution. 

Traditionally, many wave studies focus on a single layer of the solar atmosphere \citep[e.g.,][]{2000ApJ...534..989B, 2003ApJ...595L..63D, 2007Sci...317.1192T, 2008A&A...489L..49E, 2009A&A...494..269V, 2009A&A...500.1239R, 2009ApJ...702.1443F, 2009A&A...503..577C, 2011ApJ...730L..37M,2012SoPh..tmp..303R}.  However, there is increasing evidence that observing a single layer does not give us sufficient information to understand plasma heating or to carry out plasma diagnostics by means of solar magneto-seismology (SMS) in the required details. The state-of-the-art in wave studies consists of using multi-instrument multi-layer data combined simultaneously to follow the wave propagation over several layers. Examples of observations of such kind can be found elsewhere \citep{2005ApJ...624L..61D, 2005A&A...438..713T, 2006ApJ...648L.151J, 2006ApJ...647L..77M, 2006ApJ...640.1153C,  2007Sci...318.1574D, 2010ApJ...722..131F, 2010A&A...524A..12K, 2012ApJ...746..119R, 2013ApJ...779..168J, 2014A&A...561A..19Y, 2014ApJ...791...61F, 2017ApJ...847....5K}. For a  review of multi-wavelength studies of waves in the chromosphere see \citet{2015SSRv..190..103J}. 
 
Observational studies are frequently accompanied by realistic or semi-realistic simulations which have been specifically designed to study wave dynamics \citep{2007A&A...467.1299E, 2007SoPh..246...41M, 2012A&A...542L..30N, 2009A&A...505..763K, 2009A&A...508..951V, 2010ApJ...722..131F, 2011ApJ...730L..24K, 2011ApJ...727...17F, 2012ApJ...755...18V, 2015ApJ...799....6M, 2015MNRAS.449.1679M, 2016ApJ...819L..11S, 2016ApJ...817...45R}. Some of the simulations include heights up to the corona \citep{2011ApJ...731...73P, 2011ApJ...743..142H, 2016A&A...590L...3S, 2015ApJ...812L..15K}. Yet another promising direction focuses on data-driven simulations \citep{2007ApJ...655..624D, 1997ApJ...481..500C,  2011ApJ...735...65F, 2015ApJ...812L..15K, 2017ApJ...845L..18D, 2018ApJ...857..125S}. 

The wave studies can be sub-divided into two large categories: (1) when the waves are studied in order to measure the energy being carried and dissipated to heat the atmosphere,  (2) when the observed wave properties are used to infer the properties of the waveguides \citep[i.e. solar magneto-seismology, see reviews in][]{2005LRSP....2....6G, 2007AN....328..305E,2007AN....328..286C}. In each of these two cases, a particular manifestation of the wave phenomena depends on the magnetic structure they interact with: e.g., sunspots, pores, a range of magnetic elements in the quiet Sun, prominences or filaments.  

Regarding sunspot waves, our current understanding is summarized in \citet{2010SoPh..267....1M, 2015LRSP...12....6K}. The global picture of the wave propagation in these structures is addressed in details successfully and it is understood that waves observed in these structures are different manifestations of the same phenomena. While there is still no agreement weather sunspot waves are internally or externally excited, there is an agreement that the umbral chromosphere contains slow magneto-acoustic waves, propagating along the field lines and forming shocks \citep{2006ApJ...640.1153C}. In the penumbra, a mixture of fast and slow waves are observed, the periods of the slow waves following the rule of the ramp effect, affecting the cutoff frequency for inclined fields \citep{2013ApJ...779..168J, 2014A&A...561A..19Y}. Clear detections of Alfv\'en waves are still elusive \citep{2007Sci...317.1192T, 2007Sci...318.1572E, 2008ApJ...676L..73V, 2009Sci...323.1582J}. Studies using DST, ROSA and HMI/AIA data by  \citet{2015ApJ...812L..15K} demonstrate how the particular wave fronts can be traced propagating from the photospheric source to the corona. These kind of observations are, however, sparse. It remains to determine how sunspot waves interact with fine structure of the magnetic field in the umbra (umbral dots) and penumbra, how they propagate through the transition region, and their energy content and dissipation mechanisms in the chromosphere and corona. For that, one would need high resolution and high polarimetric precision data to finally measure elusive oscillations of the magnetic field. The question is even more challenging to be answered for smaller magnetic structures, e.g. from pores to tiny inter-granular magnetic flux concentrations.

The coupling between the different solar layers by means of waves, propagating in quiet Sun structures has received a considerable attention.  High-frequency acoustic halos surrounding active regions have been attempted to be explained in terms of the wave mode conversion physics \citep{2009A&A...506L...5K, 2016ApJ...817...45R}.  Waves with photospheric periodicities are systematically observed overall in the mid to upper atmosphere, even in the solar corona. Nevertheless, it is still an open question how the $p$-mode energy would propagate efficiently up there because of a set of obstacles that waves encounter on their way up, such as: cut-off layer, wave speed gradients, transition region, multi-component plasma, etc. \citep{2004Natur.430..536D, 2005ApJ...624L..61D, 2006ApJ...647L..77M, 2009ApJ...692.1211C}. Vortex motions (magnetic tornadoes), observed connecting all layers are another potential source of coupling and energy transport to the upper atmosphere \citep{2012Natur.486..505W, 2013ApJ...776L...4S, 2017ApJ...848...38I,2018ApJ...852...79Y, 2018arXiv180402931L}. 

Studying the processes related to coupling observationally requires resolving and measuring the magnetic field strength and topology at smallest scales. These kind of acquisitions are already at the limit of the current instrumentation, especially when it comes to measuring the magnetic field in the chromosphere and higher up.  Photospheric flows may alter the magnetic field topology not only in the photosphere but also in higher layers of the solar atmosphere. Therefore, the chromospheric magnetic fields may become non-potential and electric currents are generated. The currents can then be dissipated through various dissipative processes, e.g.,  ambipolar diffusion neutral-related mechanisms, yielding important highly localized energy deposits and allowing the dissipation of even magnetic non-compressible waves in an efficient manner. Dissipation of the Alfv\'en waves seems to be extremely difficult to test observationally with the existing instrumentation, because these waves have velocity and magnetic field perturbations perpendicular to the magnetic field lines (i.e. frequently perpendicular to the line of sight) within constant magnetic surfaces. Each such surface may have its own frequency-dependent transversal magnetic oscillation. It is a real observational challenge to determine unambiguously these constant magnetic surfaces \citep{2007Sci...318.1572E, 2009Sci...323.1582J}.
The wave physics must also be explored in the range of very high frequencies. It is not yet established how the high-frequency and small-wavelength waves contribute to the energetic connectivity between the various layers of the solar atmosphere \citep{2006ApJ...646..579F, 2010MmSAI..81..777F}. Simulations and observations have been controversial in this respect, mainly because observational detections are extremely hard and challenging to be made.

Wave diagnostics based on polarimetry or filter imaging have their advantages and shortcomings, and the information is often masked by radiative transfer effects and limited spatial and temporal resolution. The combination of different techniques with the high-resolution instrumental capabilities of EST will provide the necessary tool to clarify the connectivity by means of waves in the Sun.

%%%%%%%%%%%%%%%%%%%%%%%%%%%%%%%%%%%%%%%%%%%%%%%%%%
\subsection{MHD waves in localized quiet Sun structures}
%%%%%%%%%%%%%%%%%%%%%%%%%%%%%%%%%%%%%%%%%%%%%%%%%%

Energy propagation is intimately linked to the magnetic field strength and topology. Magnetic fields in the quiet Sun are swept to the boundaries of intergranular lanes, and to the boundaries of larger network cells, forming kG magnetic field concentrations. These flux tube-like structures expand with height due to the pressure fall-off, producing a carpet of strongly inclined fields at chromospheric heights. The plasma-$\beta$ changes with height in the solar atmosphere, from pressure dominated regions in the photosphere (including sunspots) to the magnetically dominated regions in the chromosphere and above. Measuring the strength and topology of the fields is crucial for studies of the wave propagation. For a review on measurements of photospheric and chromospheric magnetic fields see \citet{2017SSRv..210...37L}.

\paragraph{MHD wave propagation in MBPs}
Magnetic Bright Points (MBPs) are small-scale kG field magnetic flux concentrations found in the dark intergranular lanes \citep{2013MNRAS.428.3220K, 2014A&A...566A..99K, 2018ApJ...856...17L}. Formed by the process of convective collapse, these highly dynamic features are ubiquitous across the solar surface. They exhibit transverse motions of a few km sec$^{-1}$ caused by granular buffeting and can lead to the excitation of MHD waves. Due to their strong localised field, MBPs can act as wave guides and transfer energy into the upper parts of the solar atmosphere \citep{2009ApJ...702L.168D, 2013SSRv..175....1M}. Their ubiquitous nature means that they can be a significant contributor to the non-thermal energy budget of the solar atmosphere. \citet{2012ApJ...744L...5J} and \citet{2013A&A...554A.115S} showed that MBPs display upward propagating waves and that longitudinal MBP oscillations in the photosphere can drive H-alpha spicule oscillations in the chromosphere. EST observations will allow us to observe variations in the Stokes profiles in the presence of MHD waves at the location of MBPs in the presence of MHD waves. We will be able to {\it determine the phase relations} between the Stokes profiles, Doppler shifts and intensity perturbations of MHD waves at these locations. Therby we can determine uniquely the energy associated with these wave modes and their contribution to localised heating. 

\subsubsection{Sausage and kink oscillations in MBPs}\label{OP2.1.1}

\vspace{-2ex}
A range of magneto-hydrodynamic waves have now been observed in pores and sunspots (including MHD sausage and kink modes). Whether similar waves could be present in smaller concentrations of strong magnetic field, such as MBPs is, however, currently unknown. The abundant presence of such waves, if they are observed to propagate through the solar atmosphere, could contribute significant amounts of energy into the solar corona where it can be dissipated. Using extremely high-resolution imaging data from EST, it will be possible to identify whether area perturbations (consistent with the presence of sausage mode oscillations) can be identified in MBPs.

See table on page \pageref{OP2.1.1table}.

\subsubsection{Magneto-acoustic waves in spicules on disk}\label{OP2.1.2}

\vspace{-2ex}
RBEs and RREs are the disk counterparts of the H-alpha spicules \citep{2012ApJ...750...51K, 2013ApJ...779...82K, 2015ApJ...802...26K}. It is known that spicules are excellent waveguides for a wide range of MHD waves whether slow or fast, whether sausage or kink. It is even claimed that torsional Alfv\'en waves are detected in spicules at limb. However, limb observations are hard, it requires a detection of higher photon count than at disk. Therefore, the hunt for the various MHD waves  present in concentrations of strong magnetic field, such as RBEs/RREs is, more feasible. Similar to MBPs, pores or sunspots, the propagation of such waves, could carry significant amounts of energy flux contributing into the non-thermal energy of the upper solar atmosphere where it can be dissipated. Using the combination of extremely high-resolution spectro-polarimeric imaging data from EST, it will be possible to identify whether these perturbations (consistent with the presence of waves in spicules) can be identified and their roles assessed for energising the local plasma in RBEs/RREs.

See table on page \pageref{OP2.1.2table}.

\subsubsection{Magneto-acoustic waves in spicules at limb}\label{OP2.1.3}

\vspace{-2ex}
A key challenge with spicule observations off-limb require to disentangle the superposition of structures along the line of sight.  One may employ \FeI\ 630.2, \Halpha, \CaII\ and He I 1083.0 to measure magnetic field across the lower solar atmosphere. Point the telescope at the limb or close to limb harbouring the spicules. Sit and stare may be used for 2D spectroscopy. Full-Stokes polarimetry of \FeI\ 630.2, Ca II 854.2 and He I 1083.0 is suggested to measure the magnetic field in the chromosphere, spectroscopy in H$\alpha$ to track the spicules to larger heights, spectroscopy in Ca II K to have the highest spatial resolution if possible. Context imaging using FP or BBF is desirable. Coordination with space missions is highly desirable to trace spicule evolution in the (E)UV to greater (e.g. transition region or low coronal) heights and temperatures. Science questions can be addressed such as: How do the fine structure of spicules manifest and evolve; What type of wave motions do spicules exhibit (MA longitudinal, kink, sausage, torsional AW, etc.);  What is the energy and momentum flux and power spectrum of the waves? What is the magnetic field structure in spicules?

See table on page \pageref{OP2.1.3table}.

\subsubsection{Torsional Alfv\'en wave (TAW) propagation in spicules}\label{OP2.1.4}

\vspace{-2ex}
Here, the main focus is to find conclusive evidence for the most elusive waves in solar magnetic structures, the Alfv\'en wave. Given the likely axially symmetric structure of spicules, the most expected form of these magnetic waves would be the torsional Alfv\'en waves (TAWs). 

See table on page \pageref{OP2.1.4table}.

\subsection{Magnetic twist and torsion}

Given their size, pores (or sunspots) are ideal magnetic structures for studying MHD waves in spite of their complexity. In 1942, Hannes Alfv\'en theoretically predicted the existence of a very special type of MHD wave in which the sole restoring force is provided by the tension of the magnetic field lines. In 1970 he won the Nobel prize in Physics for this contribution to the field and for the numerous applications in plasma physics that derive from it. Indeed, Alfv\'en waves are expected to play a key role in heating and energy dissipation within both laboratory and space plasmas including the solar atmosphere. Despite the many laboratory proofs, and several claims in the astrophysical literature focusing on the upper layers of the Sun's atmosphere, Alfv\'en waves have not been {\it directly} observed in the lower solar atmosphere, where torsional oscillations of magnetic flux iso-contours can now be measured directly. Thus, Alfv\'en waves represent the most elusive and physically intriguing class of MHD waves. 

The aim is to find not only the $m=0$ torsional Alfv\'en wave but also its higher harmonic, the $m=\pm1$. Discovering the existence of the latter could be carried out by identifying (a)symmetric torsional motions e.g. in the lobs of a pore (or sunspot) separated by e.g. its light bridge. Directly proving the existence of these torsional Alf\'en modes would be a significant step to observationally prove the completeness of eigenmodes of the MHD Hilbert operator.

\subsubsection{Magnetic torsion and torsional oscillations of pores or sunspots}\label{OP2.2.1}

\vspace{-2ex}
See table on page \pageref{OP2.2.1table}.

\subsubsection{Observation of magnetic vortices and tornadoes}\label{OP2.2.2}

\vspace{-2ex}
{\ref{OP2.2.2}.1:} Swirl penetration into the chromosphere

Intensity (density) swirls have been widely observed from the photosphere up to the corona in both quiet-Sun region and coronal holes \citep[e.g.,][]{2012Natur.486..505W}. It is then natural to infer the presence of {\it magnetic} swirls and tornadoes in the solar atmosphere, having considered the magnetic nature of the solar plasma. However, high-spatial and temporal observations of magnetic swirls and tornadoes are still urgently needed to answer the following scientific questions: 1) Are intensity and magnetic swirls coupled with each other in space and time? 2) How are they coupled? 3) What is the role of MHD waves (especially Alfv{\'e}n waves) in the propagation of intensity and magnetic swirls from the photosphere into the chromosphere?  \cite{2018ApJ...tmp...ToL} developed an automated swirl detection algorithm (ASDA) aiming to automatically detect vortex motions at different layers of the solar atmosphere. This algorithm will also be adapted to magnetic field observations to perform automated detection of magnetic swirls.

\noindent {\ref{OP2.2.2}.2:} Swirls in very high-cadence observations

Having considered the very short lifetime of previously detected swirls ($> 70\%$ less than 6 sec), it is further proposed to investigate the coupling of intensity and magnetic swirls with very high-cadence observations. The penetration of swirls from the photosphere into the chromosphere and whether tornadoes are formed from swirls will also be further studied into high details using ultra-high cadence observations.

\noindent {\ref{OP2.2.2}.3:} Observation of swirls under 2-fluid condition

At a time scale less than 1 sec, the single-fluid MHD approximation will be broken down and the two-fluid condition will apply. We propose to further investigate photospheric intensity swirls under the two-fluid condition. Extremely high cadence (0.1 sec) will be needed.

See table on page \pageref{OP2.2.2table}.

\subsubsection{Formation of magnetic swirls in intergranular lanes}\label{OP2.2.3}

\vspace{-2ex}
It has been found by their statistical study \citep{2018ApJ...tmp...ToL}, that, more than 70\% of swirls are located in intergranular lanes. The small-scale photospheric magnetic swirls are widely hypothesised to form as a natural consequence to granular flows \citep[e.g.,][]{1995ApJ...447..419W, 2009A&A...493L..13A} in the quiet Sun and could play a key role in the supply of energy to the upper solar atmosphere either, for example, through the build up of magnetic energy in twisted field lines in the corona or through the channeling of MHD waves \citep[e.g.,][]{1999SSRv...87..339V, 2011AnGeo..29..883S}. However, the exact mechanism of how these magnetic swirls are generated in the intergranular lanes remain unclear. High-resolution observations of the velocity, intensity and magnetic field inhomogeneity in the intergranular lanes are needed to perform both case and statistical studies on the role that MHD instabilities resulting from granular flows may play in generating magnetic and intensity swirls, as suggested by Liu et al. (2019).

See table on page \pageref{OP2.2.3table}.

\subsubsection{Vortex flows in the lower solar atmosphere}\label{OP2.2.4}

\vspace{-2ex}
Due to their small-scale in the solar photosphere, not much
insight into vortex flows, especially on small scales such as isolated
flux tubes (about 200 km diameter), has been obtained in this layer from
direct observations so far (Bonet et al.\ 2008; Liu et al.
2018). While the observational evidence is quite scarce, in
simulations, however, vortex flows seem to be abundant both in
hydrodynamical simulations (see, e.g., Calvo et al.\ 2016) as well as
in MHD simulations (Shelyag et al.\ 2011). Thus,
vortex features and the associated swirling vortex motions may play an
important role in the creation and amplification of magnetic fields as
well as in the transportation of energy into the higher layers via the
generation of MHD waves \citep{2011ApJ...727...17F}. Moreover, it seems as if
magnetic fields and vortex motions are to a great extent coupled
phenomena.

For tracking down the vortex motions directly in observations
extremely highly resolved images with a sufficient high cadence are
necessary, as can be seen, e.g., in \cite{2018ApJ...856...17L}, who studied
MBPs, proxies for isolated strong small-scale
magnetic flux tubes, and have shown a case of an apparently turning
bright point. However, due to the still limited observational
capabilities (1m diameter telescope; $\sim$70~km diffraction limit;
feature size $\sim$200~km) they were not able to reinforce this
single case study measurement by statistical data. Clearly, a 4 m class
telescope would be necessary to directly track the rotation of a
feature via its morphologic changes.

On the other hand, indirect methods exist and are applied for instance
in the case of Alfv\`en wave detection \citep{2009Sci...323.1582J} where
rotational motions of small-scale features can be detected as
spectro-polarimetric variations in line widths. To conclude, for
successful revelations of the true nature of vortex flows and their
interaction with the solar magnetic field a 4 m class solar telescope
with imaging and spectro-polarimetric capabilities in the photosphere
and chromosphere is needed.

See table on page \pageref{OP2.2.4table}.

%-----------------------------------------------------------------------------------------------------------------------------------------------

\subsection{Wave propagation in active regions}

Sunspots are structures with strong magnetic fields in active regions which are thought to be ``portals'' through which energy of e.g. acoustic $p$-modes can penetrate into the upper atmosphere via a range of wave coupling processes.

One of such processes is mode conversion. Theoretical mode conversion models suggest it is a two-step process. In the first place, acoustic $p$-modes convert to fast and slow magneto-acoustic waves at heights where the plasma-$\beta$ is equal to unity. The slow mode (acoustic in nature) would then continue along the magnetic field lines to the upper chromosphere. The fast (more magnetic in nature) mode would refract and reflect due to the gradients in the Alfv\'en speed. Around heights of this reflection the Alfv\'en mode itself would be generated through the second mode transformation. All these processes are strongly dependent on the wave parameters (e.g. frequency) and magnetic field inclination and azimuth. This theoretical process could provide a way of transferring the energy to the chromosphere efficiently via the generation of Alfv\'en waves at heights close to the transition region, amplifying their possibility to escape this barrier without reflection. Observational confirmation of this process is still missing. 

Among other wave-coupling theoretical mechanisms acting in sunspots (and pores) are the resonant coupling of $p$-modes and the formation of photosphere-chromosphere cavity. 

\subsubsection{Excitation mechanisms of sunspot waves}\label{OP2.3.1.1}\label{OP2.3.1.2}

\vspace{-2ex}
Several theories exist regarding the mechanisms of excitation of sunspot waves, and regarding the relation between wave phenomena observed in different layers. Using multi-instrument and multi-layer sunspot observations, \citet{2016ApJ...830L..17Z} has shown that  helioseismic $p$-mode waves are able to channel up from the photosphere through the chromosphere and transition region into the corona, and that the magneto-acoustic waves observed in different atmospheric layers are the same waves originating from the photosphere but exhibiting differently under distinct physical conditions. Do this observation suggest that waves are locally excited in sunspots by the weak convection? How do these waves continue traveling up to the corona, and excite oscillations in coronal loops?

Long-time series of data are needed at moderate resolution, bi-dimensional Doppler velocity data at several height. Magnetic field information is required to determine the wave propagation speeds and directions. The observed area should be large to cover the sunspot and the nearby quiet Sun. 

See tables OP2.3.1.1 and OP2.3.1.2 on page \pageref{OP2.3.1.1table}.

\subsubsection{Alfv\'en waves in sunspots}\label{OP2.3.2.1}\label{OP2.3.2.2}

\vspace{-2ex}
This observing program aims at finding observational evidence for Alfv\'en waves in sunspots. It aims answering the following questions: at what height are the Alfv\'en waves generated, depending on magnetic field strength and topology; how large is the energy carried by Alfv\'en waves up to the transition region and corona; are the Alfv\'en waves able to penetrate into the corona without significant reflections; what is the power spectrum of the velocity component transverse to the magnetic field.

This observing campaign would require low-noise polarimetric data to measure oscillations of the magnetic field and velocity, magnetic field topology, and their mutual relation. Data should be taken at different locations across the disc.

See tables OP2.3.2.1 and OP2.3.2.2 on page \pageref{OP2.3.2.1table}.

\subsubsection{Magnetic field oscillations in sunspots}\label{OP2.3.3}

\vspace{-2ex}
Detecting oscillations of magnetic field has been illusive \citep{2000ApJ...534..989B, 2009ApJ...702.1443F}. High-precision polarimetry should make such detection possible. As discussed in the literature, detected oscillations in the magnetic field are not always due to ``real'' oscillations in the magnetic field vector, but can be rather due to opacity effects, shifting the formation heights of spectral line up and down \citep{2003ApJ...588..606K}. In chromospheric lines, such as \CaII\ 854, the effects of shifts of line formation heights on the passage of shocks are able to produce detection of ``false'' magnetic field oscillations with amplitudes as large as $50-100$ G, while intrinsic oscillations do not exceed few G in amplitude \citep{2014ApJ...795....9F}. In order to separate real oscillations from these effects, one has to have information about magnetic field gradient, and also compare data from different spectral lines, providing complimentary information. 

Magnetic field oscillations, together with oscillations in other quantities, and the phase information of propagating waves at several layers would be required to identify the wave modes observed in sunspots, depending on the region (umbra, penumbra) and height.  Several lines with different sensitivity to temperature need to be observed simultaneously. The resolution can be moderate, with the primarily goal to have low-noise signal. 

See table on page \pageref{OP2.3.3table}.

\subsubsection{Sunspot penumbral waves in the photosphere and above}\label{OP2.3.4.1}\label{OP2.3.4.2}

\vspace{-2ex}
Running penumbral waves (RPW) are disturbances, detected mainly in chromospheric lines, that appear to be emitted from the umbra and expand concentrically with constant velocity around $10--15$ km/s \citep{1972ApJ...178L..85Z, 1972SoPh...27...71G}. \citet{2012ApJ...746..119R, 2013ApJ...779..168J, 2013A&A...554A.146K} have investigated the role of the magnetic field topology in the propagation characteristics of umbral and running penumbral waves at chromospheric, transition region and coronal layers. They found an increase of the oscillatory period of brightness oscillations as a function of distance from the umbral center. \citet{2012ApJ...746..119R} and \citet{2013ApJ...779..168J} concluded that running penumbral wave phenomena are the chromospheric signature of upwardly propagating magneto-acoustic waves generated in the photosphere.  \citet{2015A&A...580A..53L} have found signatures of RPWs that are found at photospheric layers, with some evidence for being there a relation between field inclination and observed wave propagation velocities. Possible connections between RPW and umbral flushes has to be explored further, following the waves through several layers of the atmosphere simultaneously. It is not known well what happens to the penumbral waves at layers above the chromosphere, how they penetrate through the transition region to the corona and if the inclination facilitate this process. 
 
The purpose of this particular observing campaign is to track the wavefronts of RPWs from photospheric heights where they presumably are generated, across the penumbra in the chromosphere and higher up. It also aims at finding observational evidences for simultaneous up- and downward wave propagation in the penumbral chromosphere, suggestive for possible reflections from the transition region. The campaign would reveal how RPWs are generated and if they can efficiently penetrate to the corona, depending on their frequency and field inclination. 

\paragraph{Fine structure of the penumbra and its relation to waves}
The photospheric penumbra is very inhomogeneous. The question to be answered here is how does this inhomogeneity affect wave propagation? Why do running penumbral waves apparently disappear at the boundary of the visible penumbra? The observational setup of this campaign is the same as OP2.3.4 above.

\paragraph{MHD wave propagation in pores}
Pores are intermediate-scale kG field magnetic flux concentrations found around sunspots and around/in active regions. They exhibit a range of perturbations, from sausage, transverse kink or Alfv\'en modes. Due to their strong and compact magnetic field, pores can act as wave guides and transfer energy into the upper parts of the solar atmosphere \citep{2009ApJ...702L.168D, 2013SSRv..175....1M, 2014ApJ...791...61F, 2016ApJ...817...44F} . In spite of their less ubiquity when compared to e.g. MBPs, they may still have an important contributor to the non-thermal energy budget of the solar atmosphere. EST observations will allow us to track variations in their area, intensity and Stokes profiles pointing towards the presence of MHD waves in pores. Phase relations between the Stokes profiles, Doppler shifts and intensity perturbations of MHD waves at these locations will enable the unique identification of wave modes in pores, therefore, allowing us to determine the associated energy flux with these wave modes with their contribution to localised heating.\\ 

See tables OP2.3.4.1 and OP2.3.4.2 on page \pageref{OP2.3.4.1table}.

\subsubsection{Sausage and kink oscillations in pores}\label{OP2.3.5}

\vspace{-2ex}
Here, the most important aspect is to have a combined intensity (area) and LoS component magnetic field observations at high cadence. \\

See table on page \pageref{OP2.3.5table}.

\subsubsection{(Torsional) Alfv\'en waves in pores}\label{OP2.3.6}

\vspace{-2ex}
Here, the aim is to find observational evidence for (torsional) Alfv\'en waves in pores, i.e., at what height are the (T)AWs generated, depending on magnetic field strength and topology; how large is the energy carried out by (T)AWs up to the transition region and corona; are (T)AWs able to penetrate into the corona without significant reflections; what is the power spectrum of the velocity component transverse to the magnetic field.

This observing campaign would require low-noise polarimetric data to measure oscillations of the magnetic field and velocity, magnetic field topology, and their mutual relation. Data should be taken at different locations across the disc.

See table on page \pageref{OP2.3.6table}.

%%%%%%%%%%%%%%%%%%%%%%%%%%%%%%%%%%%%%%%%%%%%%%%%%%
\subsection{Wave propagation in the quiet Sun}

Relations between waves and magnetic topology have been hard to determine for waves in the quiet Sun, due to more complex and weaker fields. Unlike in sunspots, magnetic field in the quiet regions is generally weaker, less organized, and, as a consequence, it is more challenging to measure the field components. 

\subsubsection{High-frequency wave propagation and dissipation from the photosphere to the chromosphere}\label{OP2.4.1}

\vspace{-2ex}
The granular buffeting and vortical motions detected at the top of the solar convection zone can lead to the generation of MHD waves either directly or as a result of reconnection in the magnetic canopy region. The photospheric driver of MHD waves is typically of too low a frequency for significant dissipation on its transit of the chromosphere. However, if sufficient wave reflection occurs at the transition region it may be possible for a turbulent cascade to be set up in the chromosphere.  Theoretical predictions suggest that dissipation mechanisms, such as resonant absorption, are very efficient at high frequencies \citep{2010ApJ...718L.102V}. Observations obtained with the ROSA high cadence imager, suggest that there is a frequency-dependent velocity power of transverse waves. \citet{2014ApJ...784...29M} show that the velocity power of transverse waves in the chromosphere as a function of frequency has increased power at higher frequencies. In certain small-scale solar structures higher frequency waves transport more energy through the chromosphere. 

%%%%%%%%%%%%%%%%%%%%%%%%%%%%%%%%%%%%%%%%%%%%%%%%%%%%%%%%%%%%%
%\begin{figure*}
%\begin{center}
%\includegraphics[width = 14cm]{figures/im151}
%\end{center}
%\caption{Velocity power of transverse waves in the chromosphere as a function of frequency determined
%from ROSA data for (a) magnetically active and (b) quiescent Sun regions. The coloured lines show
%photospheric velocity power spectra for the horizontal granular motions and bright points from various
%sources. These have been scaled up by factors of 15 - 70 for visualisation purposes. Reproduced from 
%\citet{2014ApJ...784...29M}. }
%\end{figure*}
%%%%%%%%%%%%%%%%%%%%%%%%%%%%%%%%%%%%%%%%%%%%%%%%%%%%%%%%%%%%%

Knowledge of the energy budget in the high-frequency domain is therefore crucial for atmospheric heating. EST observations will allow the simultaneous study of several photospheric and chromospheric spectral lines. This will provide the spectropolarimetric precision required to study the photospheric motions and their interaction with the magnetic field, and the process of energy transfer via MHD wave excitation and propagation.

See table on page \pageref{OP2.4.1table}.

\subsubsection{Time-dependent behaviour of chromospheric jets}\label{OP2.4.2}

\vspace{-2ex}
The chromosphere is constantly perturbed by short lived (1 -10 minutes), jet-like extrusions that reach heights of 2000 - 10000 km above the photosphere. These thin jets are formed in the vicinity of photospheric magnetic field concentrations. They can be a consequence of shock waves channeled along the magnetic field lines. It is not clear how different periodicities are created, and what is the role of the high-frequency waves likely excited in the photosphere. For this reason, this observing campaign will measure the dominant periods of waves depending on the strength and topology of the fields, concentrating on the longitudinally propagating waves. It will then be determined whether the shock behaviour is always present in such jets. Local energy dissipation and transfer to the thermal energy should be possible to trace via localized small-scale brightening in chromospheric lines.

See table on page \pageref{OP2.4.2table}.

\subsubsection{Network and plage oscillations}\label{OP2.4.3}

\vspace{-2ex}
There seems to be still no agreement on the explanation of the presence of ubiquitous long-period waves in the chromosphere above network bright points and plages. Different observations give evidences both for inclined and vertical propagation (both upward and downward), at least for the network elements, while theoretical models seem to point toward the inclined propagation as a dominant mechanism. Inclined propagation receives more observational support especially for the strong field plage regions where it is expected that the slow acoustic waves propagate field-aligned in the low-$\beta$ regime. This observing campaign addresses the propagation in weaker structures, accompanied by observations of more network and inter-network elements. The requirements are similar to the program ``Excitation mechanisms of sunspot waves''.

See table on page \pageref{OP2.4.3table}.

\paragraph{Alfv\'en wave detection depending on topology}

The aims and the instrumental setup of this campaign are similar to ``Alfv\'en waves in sunspots or pores'', except that the target areas are quiet areas of the Sun: network, internetwork and plage.

\paragraph{Transverse waves observed at the limb}

Theoretical models for acoustic halos and shadows indicate that both may have a similar origin, tracing mode transformation process, and fast mode refraction and reflection at the inclined field of the magnetic canopy. It remains to clarify the relation between the acoustic halos and glories, present on the same locations. Some theoretical models for halos suggest that the power enhancement in horizontal velocity component should be significantly stronger than in the vertical one. This campaign will study halos and shadows at locations off the disc centre to shed more light on their nature. Models of transformation to Alfv\'en waves suggest that they will be reinforced at the periphery of active regions. Conforming this by observations is an aim of this campaign.

\paragraph{Acoustic shadows and glories across the disc center}

The observed distributions of photospheric and chromospheric power of long- and short-period waves seems to be distinct over network and plage/facular regions. In the network, long-period 5 min waves seem to be transmitted in the close proximity of magnetic elements, while short-period 3 min waves are ``shadowed'' due to the interaction of magneto-acoustic waves with the more horizontal fields of the canopies. In the stronger-field plage regions, short-period (3 min, $\nu = 5-7$ mHz) halos dominate both in the photosphere and in the chromosphere, and the propagation of long-period waves is enhanced for inclined fields, but vertical propagation has also been reported. The reason for this different behaviour may be the variation of the height of the magnetic canopy $\beta = 1$ layer, being lower for stronger plage fields, thus modifying the spectrum of waves reaching the heights sampled by photospheric and chromospheric observations. 

This campaign will perform a high-resolution study, comparing regions with different magnetic fluxes, both at the disc centre and closer to the limb to get information about horizontal velocities. Simultaneous measurements of the magnetic field vector are also required.
 %\newpage \input{SG2_tables}
%-----------------------------------------------------------
\section{Chromospheric dynamics, magnetism, and heating} 
\label{sec_chromosphere}
{Authors: Jorrit Leenaarts,  Mats Carlsson, Peter G{\"o}m{\"o}ry, Christoph Kuckein, Ada Ortiz Carbonell} %\footnote{version: ({2018-Jun-07})}
%\section{Chromospheric dynamics, magnetism, and heating}
%{Jorrit Leenaarts,  Mats Carlsson, Peter G{\"o}m{\"o}ry, Christoph Kuckein, Ada Ortiz Carbonell}

%%%%%%%%%%%%%%%%%%%%%%%%%%%%%%%%%%%%%%%%%%%%%%%
%\subsection{Introduction}
%%%%%%%%%%%%%%%%%%%%%%%%%%%%%%%%%%%%%%%%%%%%%%%

The chromosphere is the interface between the photosphere and the corona. In the chromosphere the dynamics change from gas-pressure driving to magnetic-force driving, radiation transport changes from optically thick to optically thin, the gas state changes from neutral to ionised and from local thermodynamic equilibrium (LTE) to non-equilibrium conditions, and the MHD approximation is not sufficient to fully describe its physics. The magnetic field is the key quantity in the physics of the chromosphere. Understanding its physics thus requires determination of the magnetic field at all locations in the chromosphere. Major open questions are: How large is the non-thermal energy input into the chromosphere and how and where is it dissipated? How does the chromosphere regulate the mass and energy flow into the corona? How does magnetic flux rise through the chromosphere? It is also an astrophysical laboratory where MHD and plasma physics can be studied at scales inaccessible on Earth, and serves as a  test case for understanding chromospheres of other stars.

The chromosphere is relatively difficult to study compared to the photosphere: Evolution timescales are shorter, changes are observed down to a timescale of one second. Chromospheric lines are often broad and deep, and thus have a relatively small photon flux. The magnetic field in the chromosphere is volume filling, and the resulting flux densities and thus polarisation signals are weaker than in the photosphere. The fundamental spatial scales are expected to be as small or even smaller than the resolution of even a 4-meter telescope. Because the chromosphere is highly stratified, one cannot sample the entire chromosphere with a single spectral line. Simultaneous and co-spatial observations in multiple spectral lines are required to sample all heights.

Observations of the chromosphere will thus always require a trade-off. High signal-to-noise is desired in order to measure weak polarisation signals, but simultaneously the highest resolution and short integration times are needed to resolve the relevant spatial and temporal scales. Signal-to-noise is therefore crucial for observing the chromosphere. The overall requirements on EST  for studying the chromosphere are thus:
\begin{itemize}
\item High photon efficiency of the telescope-instrument-camera system. Experience with the Swedish 1-m Solar Telescope indicates that an image derotation system is not essential. Such a system should not be added if it would add additional mirrors or lenses.
\item Polarimetric accuracy of at least $10^{-4}$.
\item A spatial PSF that minimizes straylight. Straylight decreases the contrast at small spatial frequencies, and thus requires higher signal-to-noise to reach diffraction-limited resolution. 
\item A light distribution system that allows simultaneous observations in at least three and preferably four or five spectral lines.
\item The required spectral resolution of each instrument, in particular for FPs, should be considered carefully. A too high spectral resolution lowers the SNR but might not lead to an increased science output \citep[see for an example in \CaII\ 854.2 the tests by][]{2012A&A...543A..34D}.
\end{itemize}

%%%%%%%%%%%%%%%%%%%%%%%%%%%%%%%%%%%%%%%%%%%%%%%
\subsection{Magnetic structure at supergranular scales}
%%%%%%%%%%%%%%%%%%%%%%%%%%%%%%%%%%%%%%%%%%%%%%%

The magnetic structure in the photosphere at supergranular scales is relatively well understood
\citep[e.g.,][]{2014ApJ...797...49G}.
The magnetic field in the chromosphere is much less well constrained owing to the weaker polarization signal compared to photospheric lines. This is especially true outside active regions. 

\subsubsection{Magnetic field structure in the quiet Sun chromosphere}\label{OP3.1.1}

\vspace{-2ex}
See table on page \pageref{OP3.1.1table}.
%\label{OP3.1.1table}

This OP aims to answer the following questions:  How do network field and internetwork field look and connect? How does the magnetic carpet look in the chromosphere? At what heights is the magnetic canopy located? How does the height of the $\beta=1$ surface vary? To do so a full supergranule should be scanned with a slit spectrograph, covering many lines from the photosphere to the chromosphere, simultaneously and co-spatially to allow for multi-line inversions. Such data can be inverted using multi-line non-LTE inversion codes 
\citep{2016ApJ...830L..30D}
 to get full 3D magnetic and atmospheric structure. It can also be used for field extrapolations and construction of magnetic connectivity maps. The aim is to measure weak fields, so spatial resolution is of secondary concern, signal to noise and spectropolarimetric accuracy are more important. The \NaI\ lines are needed to get sensitivity in the upper photosphere and low chromosphere, in the sensitivity gap between \FeI\ 630.1/630.2 and \CaII\ 854.2.

%%%%%%%%%%%%%%%%%%%%%%%%%%%%%%%%%%%%%%%%%%%%%%%

\subsubsection{Fibrilar structure of the chromosphere}\label{OP3.1.2}

\vspace{-2ex}
See table on page \pageref{OP3.1.2table}.

One of the most intriguing features of the solar chromosphere is its fibrilar topology in almost all chromospheric structures. This behaviour, beautifully manifested in high resolution H$\alpha$ images, clearly originates from the presence of smallest-scale magnetic elements. Measurements at the diffraction limit of the Swedish 1-m Solar Telescope in the \CaII~K line show that these fibrilar structures cover a large fraction of the solar chromosphere: they extend from active regions far into the quiet Sun, where they become more diffuse and at some point too faint for detection 
\citep{2009A&A...502..647P}.
There is an indication that such quiet Sun fibrils can effectively suppress the dominant 3-minute oscillatory power in the chromosphere. Fibrils carry transverse waves that can propagate with Alfv{\'e}nic speeds 
\citep{2012NatCo...3E1315M}.
A thorough investigation of these fibrilar structures requires spectropolarimetric measurements at the highest attainable spatial resolution since the average width of fibrils is comparable to or smaller than the diffraction limit of a 1-m class telescope. This program focusses on their spatial structure at the highest resolution and dynamics on timescales of $\sim$30 s or longer.

%%%%%%%%%%%%%%%%%%%%%%%%%%%%%%%%%%%%%%%%%%%%%%%
\subsection{Spicules and jets}
%%%%%%%%%%%%%%%%%%%%%%%%%%%%%%%%%%%%%%%%%%%%%%%

The chromosphere produces a plethora of jet-like phenomena. The most ubiquitous are Type I and type II spicules
\citep{2007PASJ...59S.655D}. 
Type I spicules are slow-mode magneto-acoustic shocks, and are rather well modelled and understood
\citep{2007ApJ...655..624D}.
Type II spicules are faster, and not so well understood. 
\citet{2017Sci...356.1269M}
proposed a model for Type II spicule acceleration by magnetic tension forces. A crucial aspect in this model is the effect of ion-neutral interactions, which allows the magnetic field to diffuse from the photosphere into the chromosphere and amplifies the magnetic tension. Ion-neutral interactions also drive heating in the spicules. It is highly interesting to study Type II spicules to test the model predictions. Spicules are narrow ($<$200 km) and not very long (up to $\sim10$ arcsec) and their acceleration sites are compact ($<1$ arcsec). They evolve fast ($<2$ min) and flows reach high speeds ($~100$~\kms). They are thus best studied with 2D imaging spectropolarimetry. Imaging spectropolarimeters have the additional advantage that the S/N cadence trade-off can be made after the observation is made. Fabry-Perot instruments will likely be too slow to capture the evolution owing to the need for spectral scanning, while slit spectrographs are fast but require slow spatial scanning if imaging information is required.

\subsubsection{Type II spicule acceleration on disk} \label{SG3-spic-launch}\label{OP3.2.1}

\vspace{-2ex}
See table on page \pageref{OP3.2.1table}.

The aim is to observe the edge of a network or plage region that harbors Type II spicules, with sit and stare using 2D spectroscopy. Full-Stokes polarimetry in \FeI~630.1/630.2, \NaI~589 and Ca II 854.2 are required to get the magnetic field in the photosphere and low chromosphere. Spectroscopy in H$\alpha$ is used to follow the spicule to larger heights, spectroscopy in Ca II K should ideally be used to obtain images of the acceleration site at the highest spatial resolution.
%%%%%%%%%%%%%%%%%%%%%%%%%%%%%%%%%%%%%%%%%%%%%%%

\subsubsection{Type II spicule evolution on disk} \label{SG3-spic-evol}\label{OP3.2.2}

\vspace{-2ex}
See table on page \pageref{OP3.2.2table}.

Similar to \ref{SG3-spic-launch}, but now a focus on the further evolution after launching. Simulations indicate that the spicule axis and the magnetic field vector need not be parallel. \HeI~1083.0 observations are needed to measure magnetic field in the upper chromosphere. 

%%%%%%%%%%%%%%%%%%%%%%%%%%%%%%%%%%%%%%%%%%%%%%%

\subsubsection{Type II spicule evolution off-limb} \label{SG3-spic-offlimb}\label{OP3.2.3}

\vspace{-2ex}
See table on page \pageref{OP3.2.3table}.

This program is similar to \ref{SG3-spic-evol}, but here the aim is to observe at the limb. Off-limb observations are required to have less superposition of structures along the line of sight. Context imaging using a Fabry-Perot or broad-band imager is desirable. Coordination with space missions is highly desirable to trace spicule evolution at UV wavelengths to higher layers and temperatures. Example scientific questions that can be addressed with this OP are: how does spicule fine structure look and evolve? What kind of wave motions do spicules exhibit?  What is the energy flux and power spectrum of the waves? What is the magnetic field structure in spicules?

%%%%%%%%%%%%%%%%%%%%%%%%%%%%%%%%%%%%%%%%%%%%%%%

\subsubsection{Magnetic field of spicules}\label{OP3.2.4}

\vspace{-2ex}
See table on page \pageref{OP3.2.4table}.

Spicules are intrinsically magnetic structures. Measuring the magnetic field has been difficult owing to the small width, fast evolution and relatively weak fields in spicules. This OP focusses on measuring the magnetic field vector in many spicules using spectropolarimetric observations in the \HeI~587.6 nm \citep[e.g.][]{2005A&A...436..325L} and 1083 nm lines \citep[e.g.][]{2010ApJ...708.1579C}. Both lines form in the triplet system of \HeI, and observing both lines simultaneously constrains the magnetic field and thermodynamic state of spicules better than single-line observations
\citep[e.g.][]{2009ApJ...703..114C}.
  The large spectral separation of the lines requires atmospheric refraction compensation in the spectrograph to get strictly co-spatial and co-temporal spectra.
\subsection{Structure of small-scale chromospheric jets}

\subsubsection{Small-scale chromospheric jets}\label{OP3.3.1}

\vspace{-2ex}
See table on page \pageref{OP3.3.1table}.

Some numerical simulations indicate an occurrence of bidirectional or counterstreaming plasma flows within chromospheric jets, which have not yet been observationally confirmed. Fast plasma flows within jets may trigger increased turbulence and instabilities of the Kelvin-Helmholtz type at interface of the jet and the ambient plasma. However, the internal structure, dynamics and possible instabilities within the jets are beyond the resolution limit of current 1-m class telescopes. This OP aims to study the internal structure and dynamics of jets using imaging only.

%%%%%%%%%%%%%%%%%%%%%%%%%%%%%%%%%%%%%%%%%%%%%%%
\subsection{Wave propagation, mode conversion and wave damping}
%%%%%%%%%%%%%%%%%%%%%%%%%%%%%%%%%%%%%%%%%%%%%%%

The chromosphere is pervaded by compressive (magnetic-acoustic) waves, as well as transverse and torsonial Alfv{\'e}nic waves  
\citep[e.g.,][]{2015SSRv..190..103J}. 
There are indications for sausage-modes. A large fraction of these waves are excited in the photosphere by granular buffeting, and through motions of photospheric magnetic elements.  Mode conversion occurs at the $\beta=1$ surface, wave damping occurs, as well as wave reflection at the $\beta=1$ surface and the transition region. Acoustic waves have dominant periods of around 3 minutes and speeds of order 10-30 \kms, while Alfv{\'e}nic waves have been observed with periods down to 30~s, but waves with shorter periods are also present. The Alfv{\'e}n speed in the chromosphere can be much higher (up to a few hundred \kms) than the sound speed. Observations of Alfv{\'e}nic waves thus require much higher cadence than magneto-acoustic waves to resolve periods and track propagation.

%%%%%%%%%%%%%%%%%%%%%%%%%%%%%%%%%%%%%%%%%%%%%%%

\subsubsection{Alfv{\'e}n waves}\label{OP3.4.1}

\vspace{-2ex}
See table on page \pageref{OP3.4.1table}.

Transverse and torsional waves that propagate at Alfv{\'e}nic speeds are observed in spicules and fibrils. This OP aims to derive phase speeds, periods, amplitudes, damping lengths and  damping timescales. A large FOV is required to track propagation. High-cadence imaging in CaII K, H$\alpha$ is required to catch short period oscillations and oscillations with high phase speeds, but this comes at the price of sampling the lines at only a few wavelength positions. Spectropolarimetry at lower cadence in \FeI~630.2, \CaII~854.2 is required to constrain the magnetic field.

\subsubsection{Acoustic wave interaction with chromospheric field structure}\label{OP3.4.2}

\vspace{-2ex}
See table on page \pageref{OP3.4.2table}.

The aim of this OP  is to track how acoustic waves propagate upward and reflect, deflect, and/or mode-converse when interacting with the chromospheric magnetic field canopy, and determine the wave energy flux carried by the waves. Waves propagate both vertically and horizontally, so a decent FOV is required. Knowledge of the photospheric and chromospheric magnetic field is required, so spectropolarimetry in \FeI~630.2 and \CaII~854.2 nm is needed. Wave amplitudes are expected to be of order 5\,--\,20~\kms.

%%%%%%%%%%%%%%%%%%%%%%%%%%%%%%%%%%%%%%%%%%%%%%%
\subsection{Flux emergence and reconnection events }
%%%%%%%%%%%%%%%%%%%%%%%%%%%%%%%%%%%%%%%%%%%%%%%

\subsubsection{Highly variable phenomena in the chromosphere}\label{OP3.5.1}

\vspace{-2ex}
See table on page \pageref{OP3.5.1table}.

SST observations have shown numerous examples of small-scale, highly variable phenomena with associated plane-of-the-sky velocities well above 100 \kms
\citep{2012ApJ...747..129L, 2015ApJ...814..123S}.
These events appear to be related to re-configurations of the magnetic field, possibly as a result of reconnection. Rapid intensity variations can also be observed at sites of magnetic flux emergence. To resolve these phenomena, both high spatial resolution and high temporal resolution are needed. Fabry-Perots have large FOV but require spectral scanning and thus allow only a few wavelengths to fulfil the cadence requirement.

%%%%%%%%%%%%%%%%%%%%%%%%%%%%%%%%%%%%%%%%%%%%%%%

\subsubsection{Highly variable phenomena in the chromosphere}\label{OP3.5.2}

\vspace{-2ex}
See table on page \pageref{OP3.5.2table}.

As observing program 5, but now with 2D spectroscopy to get full spectral information. Horizontally propagating disturbances cannot be tracked over a large distance.

%%%%%%%%%%%%%%%%%%%%%%%%%%%%%%%%%%%%%%%%%%%%%%%

\subsubsection{Reconnection at different heights}\label{OP3.5.3}

\vspace{-2ex}
See table on page \pageref{OP3.5.3table}.

Reconnection of magnetic field lines is thought to be a way of releasing large amounts of magnetic energy into the solar atmosphere. That released magnetic energy can be transformed into heating of the chromosphere and/or kinetic energy of the accelerated plasma that may be expelled from the reconnection site. Ellerman bombs have been shown to be the signal of reconnection at upper-photospheric levels 
\citep[e.g.][]{2015ApJ...812...11V,2017A&A...601A.122D}
UV bursts
\citep[previously called IRIS bombs][]{2014Sci...346C.315P}
appear to be the signature of reconnection in the upper chromosphere and transition region
\citep{2017ApJ...839...22H}. 

\citet{2015ApJ...812...11V} also describe flaring active region fibrils that appear very bright in UV lines, but also visible in \HI~656.3, and likely also in \CaII\ 393.4. They are likely  signs of reconnection too, but possibly in a different geometry than Ellerman bombs.  Many unanswered questions remain: What are the fundamental differences / similarities between Ellerman Bombs,UV bursts, microflares, and flaring active-region fibrils? Do Ellerman bombs trigger reconnection higher up? How does reconnection happen as a function of height? Because reconnection is an inherently magnetic process, this OP requires high polarimetric sensitivity to constrain the magnetic field.

Such observations require a high cadence and a large spectral range because of the high speeds of reconnection jets and  the very wide line profiles observed in such reconnection events. Co-observing with space based UV instruments is essential to track the evolution of reconnection events at higher temperatures.

%%%%%%%%%%%%%%%%%%%%%%%%%%%%%%%%%%%%%%%%%%%%%%%
\subsection{Observational determination of electric currents}
%%%%%%%%%%%%%%%%%%%%%%%%%%%%%%%%%%%%%%%%%%%%%%%

Dissipation of electric currents are one of the candidates proposed to explain the heating of the upper solar atmosphere. Measuring not only the magnitude but also the direction of the electric currents is important to quantify the contribution of Ohmic dissipation to the overall heating. However, this is a very difficult measurement to make, and only a few attempts have been done 
\citep{2003Natur.425..692S,2005ApJ...633L..57S}
The electric current is determined by taking the curl of the magnetic field. The magnetic field must be determined by inversion of spectropolarimetric measurements. The curl operation amplifies any noise or errors made in the determination of the magnetic field. 

The measurements are best done by simultaneously obsvering a number of photospheric and chromospheric lines. Because the atmosphere and magnetic fields evolve, there is a trade-off between spatial resolution, time resolution, S/N and FOV. Various different observations with different trade-offs must be made.

\subsubsection{Electric currents and heating in active regions}\label{OP3.6.1}

\vspace{-2ex}
See table on page \pageref{OP3.6.1table}.

The magnetic field in the chromosphere of active regions is relatively strong (typically $\sim$100-500 G) and a lower S/N is sufficient, This allows for the use of Fabry-P{\'e}rot instruments with an acceptable time resolution.

%%%%%%%%%%%%%%%%%%%%%%%%%%%%%%%%%%%%%%%%%%%%%%%

\subsubsection{Electric currents and heating outside active regions} \label{internetwork_currents} \label{SG3-currents2}\label{OP3.6.2}

\vspace{-2ex}
See table on page \pageref{OP3.6.2table}.

Spectro-polarimetric scans of a quiet region (network and internetwork). Field strengths are lower than in active regions, so a higher S/N is required. In order to perform inversions (quasi-) simultaneous spectra are required. This cannot be done with narrowband filtergraphs, so a slit spectrograph is required.

%%%%%%%%%%%%%%%%%%%%%%%%%%%%%%%%%%%%%%%%%%%%%%%

\subsubsection{Electric currents and heating outside active regions at small spatial scales}\label{OP3.6.3}

\vspace{-2ex}
See table on page \pageref{OP3.6.3table}.

Similar to OP~\ref{SG3-currents2}. This program focusses on  small scale structure and fast time evolution using integral field units. 

%%%%%%%%%%%%%%%%%%%%%%%%%%%%%%%%%%%%%%%%%%%%%%%
\subsection{Temperature structure of the solar chromosphere}
%%%%%%%%%%%%%%%%%%%%%%%%%%%%%%%%%%%%%%%%%%%%%%%

The temperature of the solar chromosphere is time-varying and highly inhomogeneous. Using off-limb observations and the center-to-limb variation in the solar 4.7-$\mu$m rovibrational bands of carbon monoxide.
\citet{1996ApJ...460.1042A}
found evidence for a surprisingly cool atmospheric component at chromospheric heights, as tracked by the presence of CO molecules. Observations and inversions of optical and IR lines indicate temperatures down to 4000 K, but these methods are insensitive to lower temperatures because of the non-LTE scattering formation of the diagnostic spectral lines. Theoretical studies indicate that a non-magnetic solar chromosphere should cool down to temperatures below 1500  K 
\citep{2011A&A...530A.124L}.
Ion-neutral interactions can dissipate sufficient energy in cool bubbles to avoid cooling to such low temperatures 
\citep{2012ApJ...753..161M,2017Sci...356.1269M}
%(Martinez-Sykora et al., 2012, 2017). 
The location and temperature of such cool clouds is thus still unclear. Many of the previous observing programs can be used to infer temperatures using inversions
\citep[e.g.][]{2012A&A...543A..34D}.

There are more options to study temperatures and the next OPs use different techniques to study cool phases of the chromosphere.

%%%%%%%%%%%%%%%%%%%%%%%%%%%%%%%%%%%%%%%%%%%%%%%

\subsubsection{Measurement of CO clouds}\label{OP3.7.1}

Long-slit spectroscopy in the 4700 nm CO bands has been a proven tool to study cool chromospheric gas. The large diameter of EST allows such observations at previously unattainable resolution. The polarisation in the CO lines offers a unique window on the magnetic field in cool gas. As glas does not transmit 4700 nm, this OP would require an instrument in the Nasmyth focus, and therefore is not required by the Science Advisory Group.

%%%%%%%%%%%%%%%%%%%%%%%%%%%%%%%%%%%%%%%%%%%%%%%

\subsubsection{Temperature bifurcation diagnostics by scattering polarization in  \CaII~K}\label{OP3.7.2}

\vspace{-2ex}
See table on page \pageref{OP3.7.2table}.

Scattering polarization in  CaII K spectra taken at the solar limb has been used by
\citet{2006A&A...449L..41H}
to find further evidence of a temperature bifurcation in the chromosphere, independently from techniques based on CO spectroscopy or inversions. Observed polarization signals are of order 0.5\%-1\%. Slit spectroscopy can catch such signals in 30~s, allowing much better time resolution than the 30 min of the data in \citet{2006A&A...449L..41H}.

%%%%%%%%%%%%%%%%%%%%%%%%%%%%%%%%%%%%%%%%%%%%%%%
\subsection{Magnetic field measurements using Ca II\ H\&K}
%%%%%%%%%%%%%%%%%%%%%%%%%%%%%%%%%%%%%%%%%%%%%%%

\citet{1990ApJ...361L..81M} measured circular polarization in the \CaII\ H\&K lines for quiet sun, plage, a sunspot umbra and penumbra and a flare. A Stokes $V$ signal of order 5--15\% was measured in all but the quiet Sun observations. Polarization measurements in these lines could thus in principle be an interesting tool to infer magnetic fields in the formation height range between \CaII\ 854.2 nm and \HeI\ 1083 nm.
Spectropolarimetric data can be obtained with slit spectrographs, Fabry-Perot instruments and/or integral field units. All previous OPs that include the line could thus in principle be extended with polarimetry. However, as the line has a very low photon flux, it takes long integration times to get a high SNR. As an example program we include an observing program with a slit spectrograph aimed at measuring the magnetic field structure in plage. 

\subsubsection{Magnetic field determination in plage including \CaII\ H\&K}\label{OP3.8.1}

\vspace{-2ex}
See table on page \pageref{OP3.8.1table}.

The aim of this OP is to determine the 3D spatial variation of the magnetic field in plage using multiple lines, including \CaII~H\&K. 

\subsection{Summary of requested instrument capabilities}

Almost all observing programs require (near-) simultaneous multi-line observations. Popular lines are photospheric Fe I lines (557.6, 630.1/630.2) to get magnetic fields in the photosphere, Ca II 854.2 to get magnetic field in the middle chromosphere and He I 1083 for upper chromospheric magnetic fields. Imaging spectroscopy (but not spectropolarimetry) in the Ca II H and/or K line is often requested too, as is imaging in  H$\alpha$ 656.3. Other lines that are requested are  \HeI~587.6, \NaI~589.0/589.6, \SiI~1082.7, and the CO band at 4700 nm.

The FOV requested for imaging spectropolarimetry is typically $30\arcsec \times 30 \arcsec$, except where entire supergranules or active regions are observed, in which case a larger FOV of typically $60\arcsec \times 60\arcsec$ is requested. OPs focussing on events evolving on timescales of seconds typically request a small $10\arcsec \times10 \arcsec$ FOV. 

Ideally the light distribution system is thus split into at least four domains: $<500$~nm, 500-700~nm, 700-900~nm, and $>900$~nm channels, allowing four lines, from the photosphere to upper chromosphere to be observed simultaneously using FPs or IFUs. 

Popular instrument requests and capability per channel (we do not include the infrequent requests for $> 1100$ nm) can be divided as 

\begin{table}[H]
\centering
%\hline
\begin{tabular}{lllllll}
\hline
\hline
& IMSP & & FP& & SP & \\
& requested & polarimetry & requested & polarimetry & requested & polarimetry \\
$<500$~nm &yes & no&yes& no & yes & yes\\
 500-700~nm &yes& yes&yes& yes & yes & yes\\ 
 700-900~nm &yes& yes&yes& yes & yes & yes\\
 $>900$~nm &yes& yes&yes& yes & yes & yes\\
 \hline
 
\end{tabular}
\end{table}

%Some considerations: 
%\begin{itemize}
%\item fundamental limit on solar photon flux requires trade-off between integration time, spatial resolution, spectral resolution, and signal to noise.
%\item large-FOV (60 arcsec) Fabry-Perot interferometers are expensive and big, and might lead to low transmission like VTF@DKIST. Smaller FOV (30 arcsec?) might lead to cheaper instruments, and higher transmission.
%\item imaging spectropolarimetry is possible, but will likely have a small FOV ($<$10x10arcsec)
%\item slit scanning spectropolarimetry is possible, but requires atmospheric refraction compensation (apparently possible) if multiple lines are to be detected simultaneously.
%\item the spatial power spectrum of the Sun drops off towards smaller spatial scales. This implies that the higher the desired spatial resolution, the higher S/N is required to actually observe the weak variations at high spatial frequencies above the noise. This is a general statement, and might not be true in individual instances of sharp structures.
%\item Multi-line inversions require near-simultaneous observations to catch the atmosphere in a "frozen" state. This obviously depends on the evolution speed of the phenomenon under consideration. SST experience shows that 5 s -10 s is generally acceptable at 1-m class resolution.
%\item Full-Stokes spectropolarimetry  leads to a 3 (not 4 because I is the quadratic sum of Q, U and V) times smaller cadence than intensity-only observations at the same S/N
%\end{itemize}

 %\newpage \input{SG3_tables}
%-----------------------------------------------------------
\section{Large scale magnetic structures: sunspots, prominences and filaments} 
\label{sec_large_scale}
{Authors: Jan Jur\v{c}\'ak, Marian Mart\'{\i}nez Gonz\'alez, Luc Rouppe van der Voort} %\footnote{(version: {2018-Nov-30})}

%\section{Large scale magnetic structures: sunspots, prominences and filaments}
%Jan Jur\v{c}\'ak, Marian Mart\'{\i}nez Gonz\'alez, Luc Rouppe van der Voort\\
%\today\\

%%%%%%%%%%%%%%%%%%%%%%%%%%%%%%%%%%%%%%%%%%%%%%%%%%%%%%%%%%%%%%%%%%%%%%%%%%%%
%THE FOLLOWING PART SHOULD BE MOVED TO THE INTRODUCTION OF THE TOP-LEVEL SCIENCE GOALS\\
%%%
%Many structures and phenomena observed on the Sun are a result of the interaction between moving plasma and large-scale magnetic fields. The magnetic field gives rise to a variety of physical effects:
%\begin{itemize}
%\itemsep 0pt
% \item[--] it exerts a force, which may accelerate plasma or create structures
% \item[--] it stores energy, which may later be released as an eruption or solar flare
% \item[--] it acts as a thermal blanket, giving rise to sunspots or stabilizing a prominence
% \item[--] it channels fast particles and plasma
% \item[--] it drives instabilities and supports waves.
%\end{itemize}
%Each of these effects is crucial for solar activity related phenomena like sunspots, flares, and prominences. Novel high-resolution and high-polarimetric-sensitivity observations are necessary to understand the interaction between plasma motions and magnetic fields both in the solar photosphere and chromosphere.\\
%%%%%%%%%%%%%%%%%%%%%%%%%%%%%%%%%%%%%%%%%%%%%%%%%%%%%%%%%%%%%%%%%%%%%%%%%%%%%

%HERE IS THE BEGINNING OF THE SECTION\\

Sunspots and filaments (also known as prominences when observed off-limb) are large-scale magnetic structures that cover a significant fraction of the solar disk. 
EST is designed to excell at resolving the smallest spatial scales which inevitably implies that the field of view is limited as compared to the extent of the largest sunspots and filaments. 
However, fundamental physical processes that define these structures occur at spatial scales that are currently beyond the reach of present-day instrumentation.
The capabilities of EST will be crucial for the full characterisation of the smallest building blocks of these structures so that fundamental questions about the nature of sunspots and filaments can finally be addressed. 

Sunspots are large-scale concentrations of strong magnetic fields and the primary manifestation of solar activity. The umbra is the dark central area of a sunspot and harbors the strongest and most vertical magnetic field. It is surrounded by the filamentary penumbra where the magnetic field is more inclined and in part nearly horizontal. The strong magnetic field in sunspots inhibits the normal convective overturning motion that results in the granulation pattern found all over the solar surface outside active regions. However, advances in observations and numerical simulations have arrived at the view that fine-scale structures in sunspots, such as umbral dots, light bridges and penumbral filaments, result from a form of convective motions that is heavily affected by the strong magnetic field. 

For example, spectropolarimetric observations of sunspot penumbrae \citep{2012A&A...548A...5V, 2013A&A...557A..25T}, observations of orphan penumbrae \citep{2012A&A...539A.131K, 2014A&A...564A..91J, 2014ApJ...787...57Z}, and advanced MHD simulations of sunspots \citep[e.g.,][]{2011ApJ...729....5R} provide a coherent picture of sunspot penumbrae resulting from convection in an inclined magnetic field.  The penumbral filaments contain horizontal fields and carry a plasma flow, historically known as the Evershed flow. These regions with horizontal magnetic field are interlaced with more vertical and typically stronger field (so-called ``spines"). The penumbral topology of rapidly varying magnetic field inclination is often referred to as the ``uncombed" magnetic field \citep{1993A&A...275..283S}. 

Filaments consist essentially of chromospheric material found at great altitude as magnetically levitating clouds embedded in the hot corona \citep{1989ASSL..150...77L, 1995Sci...269..111T, 2002Natur.415..403T}. When observed off-limb, filaments are traditionally referred to as prominences. Broadly speaking, filaments can be segregated in active region (AR) and quiescent (QS) structures. Both lie above the polarity inversion line, an observationally defined line that separates the two dominant opposite polarity magnetic fields.

QS filaments are very long structures suspended at more than 10 Mm above the surface that often live for weeks or even months. AR filaments are formed in active regions, often in recurrent flaring areas, and are shorter in length and life time as compared to QS filaments. They are difficult to be seen as prominences at the limb because they lie lower in the atmosphere and are hidden behind other features along the line of sight. 
Though having different parameters, their nature is generally believed to be the same. In particular, the magnetic field plays a fundamental role in the formation, support, and eruption of filaments.

The diagnostic of magnetic fields in such structures is mainly based on spectro-polarimetry in the He I multiplet at 1083.0 nm and in the He\,{\sc I} D$_3$ line at 587.6 nm. The reason is that these lines are basically absent in the photospheric spectrum and its mere detection provides a fingerprint on the presence of dense chromospheric structures. 

The radiation produced in these lines (either in absorption or in emission) is linearly polarised due to scattering processes. When atoms are illuminated by the anisotropic radiation field from the underlying photosphere, it induces atomic-level polarisation (population imbalances and coherences between magnetic sublevels). This linear polarisation is modified via partial decoherence in the presence of a magnetic field (Hanle effect), and circular polarisation is typically generated by Zeeman splitting \citep[e.g.,][]{1989ASSL..150...77L, 2002Natur.415..403T}. Hence, to constrain the magnetic field vector in filaments we need to observe the full Stokes vector. These signals are typically weak, and we need a signal to noise ratio much better than 2000 to properly infer the magnetic configuration. 

\subsection{Stability of the umbra}

Various studies show that umbrae are stable if the vertical component of the magnetic field ($B_\textrm{ver}$) is stronger than a certain invariant value \citep[1867 G as estimated from Hinode spectropolarimetric data,][]{2018A&A...611L...4J}. If $B_\textrm{ver}$ is lower, these originally dark umbral areas are prone to be transformed into penumbrae, light bridges, or granulation depending on the magnetic field inclination and location within the umbra \citep{2015A&A...580L...1J, 2017A&A...597A..60J}. 
The theoretical study by \citet{1992ApJ...388..211W}, and the advanced MHD simulations by \citet{2011ApJ...729....5R}, suggest a mechanism to turn predominantly vertical field lines into horizontal ones: from the convectively unstable sub-photosphere, hot buoyant plasma pushes upwards along the umbral field lines; when cooling, the mass load exerted by the plasma will bend and incline field lines where the vertical field has reduced strength; only in areas with strong vertical magnetic fields is the magnetic tension as a restoring force strong enough to prevent the field lines from bending over and becoming horizontal. This scenario which explains the fundamental property of sunspot structure has yet to be confirmed by observations. This requires observations with high temporal and spatial resolution to measure the interplay between the convective motions and the magnetic field at the boundaries between umbra and penumbra, granulation, and light bridges both in the solar photosphere and chromosphere.

\subsubsection{Stability of the umbra - interplay between the convection and magnetic forces}
\label{bver_crit}\label{OP4.1.1}

\vspace{-2ex}
See table on page \pageref{OP4.1.1table}.

%%%%%%%%%%%%%%%%%%%%%%%%%%%%%%%%%%%%%%%%%%%%%%%

\subsection{Umbral dots}

Umbral dots are very fine-scale structures in sunspot umbrae and pores that have sub-structure that is barely resolved with present-day instrumentation. Umbral dots were successfully reproduced in advanced MHD simulations \citep{2006ApJ...641L..73S}, where they correspond to the top of magneto-convective cells in the strong and vertical magnetic field of the umbra. Current spectropolarimetric observations of highest spatial resolution are in agreement with the modeled structures \citep[see, e.g.,][]{2010ApJ...713.1282O}, but the observations still do not fully spatially resolve umbral dots. Spectropolarimetric observations of better spatial resolution achieved simultaneously in a number of spectral lines probing different heights in the atmosphere would further clarify the nature of umbral dots and their impact on the sunspot chromosphere. 

\subsubsection{Multi-wavelength analysis of umbral dots}\label{OP4.2.1}

\vspace{-2ex}
See table on page \pageref{OP4.2.1table}.

\subsection{Structure of cool sunspot umbrae}

The presence of molecular lines near the 630 nm Fe~I lines and the IR Fe~I lines at 1.56~$\mu$m hinder the reliable characterization of the magnetic and the thermal structure of cool sunspot umbrae. The Ti~I multiplet at 2.2~$\mu$m, sensitive to properties of cool plasma, is free from molecular blends and therefore represents an ideal tool for the diagnostics of the coolest regions on the solar surface. Additionally, this multiplet is extremely Zeeman sensitive allowing for a reliable determination of the magnetic structure of cool umbrae. Alternatively, cool sunspot umbrae can be studied using observations in the visible and 1.6~$\mu$m lines, but this requires to account for the molecular lines in the inversion codes that are used to interpret the Stokes profiles.

Such observations would allow us to investigate the possible presence of granular-scale patterns in sunspot umbra that were not yet detected, but their signatures are present in the MHD simulations of sunspot umbrae \citep{2006ApJ...641L..73S, 2011PhDT........83V}.

\subsubsection{Probing the structure of cool sunspot umbrae}\label{OP4.3.1}

\vspace{-2ex}
See table on page \pageref{OP4.3.1table}.

\subsection{Umbral flashes as a probe of fine structure in the umbra chromosphere}

Umbral flashes are sudden brightenings in the Ca II H \& K and IR triplet lines observed in the chromosphere above the sunspot umbra. They were first reported by \citet{1969SoPh....7..351B} and are commonly thought to be formed by magnetoacoustic waves that originate in the photosphere and steepen into shocks as they propagate to chromospheric heights \citep{2010ApJ...722..888B}. The increased brightness provided by the umbral flashes has revealed fine structure of both cool and hot material intermixed on very small spatial scales \citep{2009ApJ...696.1683S, 2013A&A...556A.115D}. These fine structures could be created by the shocks or become apparent when they are illuminated by the umbral flashes. The horizontal orientation and general stability of these small-scale fibrilar structures have questioned the general belief that the magnetic field topology in the umbra chromosphere is uniformly vertical \citep{2015A&A...574A.131H}. 
High-resolution imaging in Ca II K or H combined with full-Stokes polarimetry in other spectral lines will allow us to gain insight into the inhomogeneous nature of the chromosphere above the sunspot umbra. The lack of intensity in the umbra and the strong photospheric field make this project particularly challenging and well suited for the polarimetric capabilities of EST.

\subsubsection{Umbral flashes}\label{OP4.4.1}

\vspace{-2ex}
See table on page \pageref{OP4.4.1table}.

\subsection{Penumbral and umbral micro-jets}

Various chromospheric jet-like phenomena have been discovered in the sunspot umbra and penumbra. There are impulsive penumbral and umbral microjets \citep{2007Sci...318.1594K, 2013A&A...552L...1B}, and slower, longer-lived, and more ubiquitous sunspot dynamic fibrils \citep[in the umbra also referred to as umbral spikes,][]{2013ApJ...776...56R, 2014ApJ...787...58Y}. 

The latter are a shorter version of the well-known dynamic fibrils found in plage regions and are well understood \citep{2007ApJ...655..624D}. It is still unclear how the impulsive micro-jets are related or affected by the dynamic fibrils and large-scale sunspot waves (umbral flashes and running penumbral waves). It is likely that they are driven by magnetic reconnection. Studies of IRIS data suggest that some chromospheric microjets may be driven from above by reconnection in the transition region rather than from below \citep{2017ApJ...835L..19S}. With the unprecedented spatial resolution of EST, we will be able to better characterise the dynamic properties of sunspot jets. A study with the Swedish 1-meter Solar Telescope \citep{2017A&A...602A..80D} indicated that not all microjets were spatially resolved (the SST diffraction limit at Ca II 8542 is 130 km) so the high spatial resolution of EST is needed to establish the possible existence of a ubiquitous population of narrow microjets that can presently not be detected. Combined with high-sensitivity spectro-polarimetry, we will be able to address the formation mechanism of microjets. The strong field environment of sunspots, which gives strong signals in spectro-polarimetric diagnostics, makes sunspot microjets an ideal target to study the details of the fundamental process of magnetic reconnection.

\subsubsection{Penumbral and umbral micro-jets}\label{OP4.5.1}

\vspace{-2ex}
See table on page \pageref{OP4.5.1table}.

\subsection{Light bridges}
%this is partly based on text from Sect 5 ....

Light bridges are long bright lanes that cross the dark umbrae of sunspots. Light bridges display a large variety of morphology and fine structure: some harbor cells that resemble the granulation cells of the normal quiet Sun, some are intrusions of penumbral filaments, while others appear as the alignment of bright grains and seem to be associated with the isolated umbral dots found in the umbra \citep[see, e.g.,][]{2008ApJ...672..684R,2010ApJ...718L..78R,2014A&A...568A..60L,2016A&A...596A...7S}.

It is now firmly established that light bridges are regions that harbor convective flows of relatively weakly magnetized plasma in the strongly magnetized surroundings of the umbra. It will be of great interest to utilize the capabilities of EST to characterize in detail the conditions in the different types of light bridges to explain their morphology and fine structure. With these light bridge observations, together with the detailed observations of umbral dots and penumbral filaments as described in other observing programs, EST will provide a major advancement in the understanding of magneto-convection in the strong field regime. 

It has been noted that light bridges often show enhanced chromospheric activity, with (H$\alpha$) surges and chomospheric jets reported in a number of analyses \citep[see, e.g.,][]{1973SoPh...32..139R, 2009ApJ...696L..66S,2014A&A...567A..96L}. There have been strong indications that magnetic reconnection is driving these surges \citep{2016A&A...590A..57R, 2018A&A...609A..14R}. \citet{2018ApJ...854...92T} differentiate between two types of jets above light bridges: a short type that is virtually continuously present and are caused by upward leakage of magnetoacoustic waves from the photosphere, similar as for dynamic fibrils \citep{2007ApJ...655..624D}, and more occasionally occurring long and fast surges that are caused by intermittent reconnection. 
Sometimes, the periodically re-occurring jets above light bridges are referred to as light walls \citep{2017ApJ...843L..15Y,2017ApJ...848L...9H}.
The comprehensive study of the chromospheric and transition-region properties of light bridges by \citet{2018A&A...609A..73R} confirms that they are complex multi-temperature structures associated with enhanced energy deposition, but demonstrates the need for further multi-height investigation of the magnetic and dynamic structure through the photosphere and chromosphere, in order to identify heating mechanisms and the height at which energy is deposited.

\subsubsection{Structure and dynamics of light-bridges}\label{OP4.6.1}

\vspace{-2ex}
See table on page \pageref{OP4.6.1table}.

%%%%%%%%%%%%%%%%%%%%%%%%%%%%%%%%%%%%%%%%%%%%%%%

\subsection{Evolution of an individual penumbral filament}

There is general consensis about the structure of individual penumbral filaments, both from an observational and a theoretical point of view. Penumbral filaments are convective cells heavily influenced by the strong and inclined magnetic field of the penumbra \citep{2013A&A...557A..25T, 2011ApJ...729....5R}. However, there still remain questions about the details of filament formation and decay processes, their influence on the larger-scale magnetic field topology, and the dynamical evolution of the plasma flow field and the magnetic field within the filaments. This is closely related to the OP~\ref{bver_crit} that focus on a different aspect of the interplay between the convection and the magnetic field.

\subsubsection{Evolution of an individual penumbral filament}\label{OP4.7.1}

\vspace{-2ex}
See table on page \pageref{OP4.7.1table}.

\subsection{Formation and decay of sunspot penumbrae}

While individual penumbral filaments have a typical life time of less than an hour, the penumbra as a whole is generally a rather stable structure that remains present around a sunspot for days or even weeks. The formation and decay of the sunspot penumbra, on the other hand, is a process that only takes a few hours. For example, \citet{2010A&A...512L...1S} analysed the formation of a penumbra around a protospot that took only 4.5 hours. They concluded that this protospot developed a penumbra while accumulating magnetic flux from the active region flux emergence site. \citet{2012A&A...537A..19R} found signatures of a canopy at the photospheric level around the protospot before the penumbra appearance at the solar surface and this is in agreement with analyses of \citet{2012ApJ...747L..18S} and \citet{2013ApJ...771L...3R}, who found indications of a penumbral halo at the chromospheric level prior to the formation of a penumbra in the photosphere. \citet{2014PASJ...66S...3J} found an increase in magnetic field inclination at the protospot boundary prior to the onset of penumbra formation. \citet{2013ApJ...771L...3R}, \citet{2014ApJ...784...10R}, and \citet{2016ApJ...825...75M} put these findings into the context of magnetic flux at chromospheric height bending down to the photosphere leading to penumbra formation at the solar surface. They also found signatures of (counter Evershed) inflowing material similar to those reported by \citet{2011IAUS..273..134S}, and they discuss on the siphon nature of those in the context of falling flux tubes. 

From the modeling point of view, several studies help to put into context the observational findings: \citet{1970SoPh...13...85S}, \citet{1994A&A...290..295J}, and \citet{2000MNRAS.314..793H} found that with increasing magnetic flux, the inclination of the field becomes increasingly more horizontal. Numerical simulations by \citet{2016ApJ...831L...4M} show that magnetic canopies naturally form during the emergence of a twisted flux tube. According to \citet{2012ApJ...750...62R}, these canopies have sufficiently inclined magnetic field to form extended penumbrae. \citet{1992ApJ...388..211W} proposed a penumbral model fed by flux tubes fallen onto the photosphere owing to the upwelling of a mass flow in the inner footpoint (within the umbra) of field lines closing up (submerging) in the surroundings of a sunspot. 

A decrease in the size of the penumbra marks the beginning of the end for mature spots. The penumbra eventually disappears and only the umbra remains visible in white-light images. It has been suggested that this process contributes to the removal of flux from active regions \citep[see][and references therein]{2002AN....323..342M}. In that case, it would be an essential ingredient of the solar activity cycle. Penumbral decay may also be a source of localized chromospheric and coronal heating \citep{2004ApJ...601L.195W, 2005ApJ...623.1195D}. 

Despite its far-reaching implications, the disappearance of the penumbra is not well understood. \citet{2008ApJ...676..698B}, \citet{2014ApJ...796...77W}, and \citet{2018A&A...614A...2V} analysed the disappearance of penumbra and explained it by the rise of photospheric magnetic field lines into the chromosphere, i.e., an inverse mechanism to the predicted formation mechanism of the penumbra. Similar changes of the magnetic field orientation were also described in case of flare-induced rapid penumbral decay \citep{2004ApJ...601L.195W, 2005ApJ...623.1195D}.

For the moment, this idea remains speculative. To identify the exact mechanism responsible for the disappearance of the penumbra, EST should perform spectropolarimetric observations of mature spots throughout their decay phase. This requires high and stable data quality on long time scales. Furthermore, it is important to measure magnetic field orientation both in photosphere and in chromosphere with high spatial resolution. 

\subsubsection{Capturing the formation and decay of penumbrae}\label{OP4.8.1}

\vspace{-2ex}
See table on page \pageref{OP4.8.1table}.

%%%%%%%%%%%%%%%%%%%%%%%%%%%%%%%%%%%%%%%%%%%%%%%
\subsection{Relation between moat flows and sunspot decay}
\label{moat}
%%%%%%%%%%%%%%%%%%%%%%%%%%%%%%%%%%%%%%%%%%%%%%%

Mature sunspots are usually surrounded by a so-called moat region that harbors a conspicuous flow pattern with radially outward moving magnetic features \citep[MMFs,][]{1972SoPh...25...98S, 2005ApJ...635..659H}.
The sizes of sunspot moats are typically close to supergranular cell sizes and the moat flow velocities reach several hundred meters per second. The origin of sunspot moats is not well understood. Observations \citep[e.g.,][]{2008ApJ...679..900V} show that the appearance of moat flows is closely related to the presence of a penumbra. However, this view has been challenged by \citet{2013A&A...551A.105L} as well as by \citet{2015ApJ...814..125R} who find that the presence of the moat flow does not depend on the existence of an Evershed flow and/or penumbra around the umbra.

It has also been suggested that the MMFs are extensions of penumbral filaments which turn back below the visible solar surface at the sunspot magnetopauses and reappear as MMFs in the moats. Observations by \citet{2008A&A...481L..21S} indeed support this so-called sea-serpent model \citep{2002AN....323..303S}. As argued by \citet{2002AN....323..342M}, MMFs, while probably part of the decay process, have an origin not related to the decay process itself. However, there are also studies relating the observed decay rates of the sunspot magnetic flux to the flux carried by MMFs \citep{2005ApJ...635..659H, 2008ApJ...686.1447K}.

Detailed observations of the relationship between sunspots and their surrounding quiet regions are required to solve the puzzling appearance of sunspot moats. Particularly, high spatial and temporal resolution observations of the dynamics of sunspot penumbrae and MMFs, including full magnetic field vector and flow velocity measurements, as will be provided by the EST, shall lead to substantial progress in the understanding of large-scale dynamic processes in and around sunspots and how these processes influence sunspot decay. The observational setup is comparable to the study of penumbra formation and decay, but need to capture a larger FOV.

\subsubsection{Observations of moat flow properties and its impact on sunspot decay}\label{OP4.9.1}

\vspace{-2ex}
See table on page \pageref{OP4.9.1table}.

%%%%%%%%%%%%%%%%%%%%%%%%%%%%%%%%%%%%%%%%%%%%%%%%%%%%%%%%%%%%%%%

\subsection{Fine structure of prominences and filaments}

A very important issue, with respect to expected capabilities of the EST, is the fine structure of the magnetic field in filaments (and prominences but we'll use the word filament since both are the same phenomenon observed with different geometries). Direct imaging shows that filaments appear to be highly dynamic at different spatio-temporal scales, tend to form very fine filamentary structures, and show counterintuitive behaviors arising from the competition between the vertical gravitational stratification and magnetic forces operating in any other direction. 
However, such fine structure has never been reported in magnetic field measurements. In fact, the magnetic fields inferred with the best spatial resolution achievable at present (0.5”) show a very smooth field \citep{2003ApJ...598L..67C, 2014A&A...566A..46O, 2015ApJ...802....3M}. Understanding if the fields follow the intensity structures or not is of vital importance to understand how these structures are sustained against gravity and insulated from the million degree corona. In particular, various theoretical models predict magnetic dips in which the filament plasma is condensed, but such dips have not been observationally identified so far. Confirming or denying this will support the works that claim that magnetic fields in filaments are horizontal to the surface and unrelated to what we observe in intensity \citep[e.g.,][]{2014A&A...566A..46O, 2016ApJ...818...31L} or aligned with the plasma structures \citep[e.g.,][]{1998Natur.396..440Z, 2015ApJ...802....3M}. Spectro-polarimetric data with much higher spatial resolution and better polarimetric accuracy than achieved today are needed for this. \\

\subsubsection{Determining the evolution of the magnetic and dynamic fine structure of prominences and filaments}\label{OP4.10.1}

\vspace{-2ex}
See table on page \pageref{OP4.10.1table}.

\subsubsection{Determining the magnetic and dynamic fine structure of prominences and filaments}\label{OP4.10.2}

\vspace{-2ex}
See table on page \pageref{OP4.10.2table}.

\subsection{Are quiescent and active region filaments the same phenomenon?}

Active region filaments as well as quiescent ones are seen as dark absorption features in the core of many chromospheric spectral lines. Their properties seem to differ mainly in the life time and the heights they can reach. However, they are assumed to have the same origin, the only difference being the region where they appear, that somehow has an impact in their lifetimes and heights of formation. 

At present, the few existing measurements of the full magnetic field vector of QS filaments show that a controversy exists concerning the magnetic properties of both structures. Spectropolarimetric observations of quiescent filaments have revealed magnetic fields strengths of the order of a tens of Gauss \citep{2006ApJ...642..554M, 2014A&A...566A..46O, 2015ApJ...802....3M}, while AR filaments have been found with magnetic fields from 100G up to 700G \citep{2007ASPC..368..467S, 2011A&A...526A..42S, 2012ApJ...749..138X, 2012A&A...539A.131K}. However, the strongest fields have been put in doubt by \cite{2016ApJ...822...50D}. In light of these results it is clear that, from the observational point of view, the precise topology of magnetic fields in filaments is still a matter of debate. The unprecedented spatial resolution and polarimetric accuracy expected of the EST will help overcoming the observational challenges that restrain the study of these structures in detail. In particular, the main aim with the EST observations will be to unveil if both QS and AR filaments harbor weak fields or not. In other words, EST observations will help unveil if QS and AR chromospheric structures share the same origin and are the same magnetic phenomenon seen in different perspectives. 

It has been shown from empirical data of AR filaments that these show complex Stokes profiles, revealing that gradients of the physical properties (including the magnetic field) happen along the line of sight \citep{2019A&A...625A.129D}. However, the analysis of such signals using more realistic models including gradients or many atmospheric components has not been successful because of the low signal to noise ratio. Improving the polarimetric accuracy by collecting more photons will definitely help a proper inference of the gradients of the magnetic and thermodynamical properties in AR filaments.

\subsubsection{Comparison of the magnetic filed properties in QS prominences and AR filaments}\label{OP4.11.1}

\vspace{-2ex}
See table on page \pageref{OP4.11.1table}.

\subsection{Magnetic field and dynamics of tornado prominences}

The name of solar tornadoes was first introduced by \cite{1932ApJ....76....9P} to label a specific kind of solar prominences that appeared like ”vertical spirals or tightly twisted ropes”. High spatial resolution and high time cadence and continuity of the coronal data provided by the Atmospheric Imaging Assembly (AIA) onboard the Solar Dynamics Observatory (SDO) has reawaken the interest on the study of solar tornadoes. Solar tornadoes are nowadays associated to vertical funnel-shaped dark structures in the coronal \ion{Fe}{xii} line at 17.1 nm. These structures are found to be hosted in the legs of some quiescent prominences \citep[(e.g.,][]{2012ApJ...756L..41S, 2013ApJ...774..123W}, and are directly linked to the evolution and final fate of the prominence.

The magnetic and dynamic properties of solar tornadoes are still a matter of debate and a hot topic in the literature. Concerning the magnetic configuration, \citet{2015ApJ...802....3M, 2016ApJ...825..119M} argue that they harbour vertical, helical fields that connect the main body of the prominence with the underlying surface. These findings are in agreement with previous claims of vertical fields in filament barbs \citep[(or prominence legs; Zirker et al. 1998)][]{1998Natur.396..440Z}. On the contrary, \citet{2015IAUS..305..275S} and \citet{2016ApJ...826..164L} infer magnetic fields that are almost parallel to the surface, in agreement with the magnetic field measured in a photospheric barb endpoint \citep{2006A&A...456..725L}. A major advancement in observational capabilities such as EST will provide is required to settle the debate. 

Concerning the dynamics of tornado prominences, a debate exists on the real or apparent rotation of the structure. \citet{2012ApJ...756L..41S} reported on the oscillatory pattern in the plane-of-the-sky motions of solar tornadoes as seen in SDO/AIA 17.1 nm, claiming that these structures are rotating with periods of about 50 min. Subsequent works \citep{2012ApJ...761L..25O, 2013ApJ...774..123W, 2014ApJ...785L...2S, 2015A&A...582A..27L} confirm the rotation scenario measuring opposite sign Doppler shifts at both sides of prominence legs. \citet{2014SoPh..289..603P} propose that the rotation of solar tornadoes as seen in SDO/AIA 17.1 nm is only apparent and could be explained by oscillations. In particular, \citet{2016A&A...593A..64L} propose that this apparent rotational motion can be caused by large amplitude oscillations along the magnetic field. The work of \citet{2015ApJ...810...89M} explores both the rotation and the wave scenario and find both compatible with SDO/AIA 17.1 nm data of solar tornadoes. They also propose that, most likely, both rotation and waves are at work, identifying the oscillation as a magnetohydrodynamic kink mode. Consistent with this view, \citet{2016ApJ...825..119M} argue that (1) if rotation exists, it is intermittent, lasting no more than one hour, and (2) the observed velocity pattern is also consistent with an oscillatory velocity pattern.

The true nature of tornadoes and their relationship to prominences requires an understanding of their magnetic structure and dynamical properties. In particular for the dynamics, the EST spatial resolution and photon collector capabilities will be the key to understand if tornadoes are really rotating structures or if the rotation is just a visual illusion due to waves. 

\subsubsection{Magnetism and dynamics of tornado prominences}\label{OP4.12.1}

\vspace{-2ex}
See table on page \pageref{OP4.12.1table}.

%%%%%%%%%%%%%%%%%%%%%%%%%%%%%%%%%%%%%%%%%%%%%%%

%\bibliographystyle{aa}
%\bibliography{SG4.bib}
%\end{document}

 %\newpage \input{SG4_tables}
%-----------------------------------------------------------
\section{Coronal Science} 
\label{sec_corona}
{Authors: Sarah Matthews, Robertus Erdelyi, Mihalis Mathioudakis}%\footnote{(version: {2018-Jun-04})}
%\section{Coronal Science}

EST is not optimised to carry out coronal science as one of its primary objectives, but it is well suited to providing complementary observations of the underlying photosphere and chromosphere that will significantly advance our understanding of coronal physics in a number of areas. In order to achieve this, co-ordination with space-based facilities will be required, and on the timescale of EST first-light the new space facilities that are envisaged to be operating are Solar Orbiter and Solar C EUVST. Co-ordination with Solar Orbiter will require detailed planning due to the complexity of its orbit. EST will be a critical resource for Solar C EUVST which, as a single instrument spacecraft, will provide high spatial, spectral and temporal resolution spectroscopic measurements of the transition region and corona, but have no on-board magnetic field capability.  

\subsection{Sunspot light-bridges/light walls}

Sunspot light-bridges are one topic where EST combined with space observations can provide new insights into the magnetic and dynamic structure through the photosphere and chromosphere, and up into the overlying transitions region, allowing us to identify heating mechanisms and the height at which energy is deposited. Further details on a proposed observing programme to determine the variation of the magnetic field and dynamics of light-bridges can be found in Section 4 of this document.

\subsection{Light Walls}
One particular aspect of activity in light-bridges are surges, which have been found to form light walls that may penetrate deeper into the low corona, and also present a possible application for the SMS techniques, assuming asymmetric waveguides. Whilst their true nature is uncertain, the main features of light walls are the observed groups of surges, brightenings, and other magnetic structures in the chromosphere rooted in sunspot light bridges. They have been demonstrated to guide MHD waves driven by nearby disturbances \citep{Yan2015} and provide an excellent testbed for the theory of asymmetric magnetic slabs (\citep{Alcock2017}; \citep{Zsamberger2018}) allowing the 
actual geometry of light walls to be determined. Coupled with observations from space by Solar Orbiter and Solar C EUVST it may also be possible to
determine where the plasma surrounding light walls has a turning point from high-beta to low-beta.

\subsection{Origins of the solar wind}

Hinode EIS has demonstrated that active regions upflows are common, but combined studies with in situ observations have showed clearly that not all upflows will become outflows measurable in the solar wind, and debates continue about where in the atmosphere the upflows originate. Solar Orbiter SPICE/SWA and Solar C EUVST will provide multi-vantage points measurements of coronal outflows, from the chromosphere to the corona in the case of Solar C EUVST, but need corresponding chromospheric magnetic field measurements in order to determine the processes
that produce them, e.g. low-altitude reconnection, presence of small-scale open field at active region boundaries.

\subsubsection{Determining which coronal upflows become outflows}\label{OP5.3.1}

\vspace{-2ex}
See table on page \pageref{OP5.3.1table}.
 
This observing plan aims to explore the role of small-scale photospheric and chromopsheric dynamics in driving the solar wind at the edges of active regions and coronal holes where persistent upflows are observed in coronal lines. It will measure the magnetic field strength and direction from the photosphere to the upper chromosphere using spectropolarimetry, and the plasma dynamics using 2D spectroscopy to determine intensity, line widths, and line of sight velocities. Measurements should be coordinated with space platforms to provide the link to coronal spectroscopy, and ideally with in-situ measurements of plasma composition.

\subsection{Probing pre-flare triggers}

Spectroscopy of the transition region and corona indicates significant dynamics can be present prior to the onset of
flares and eruptions (e.g. \citep{Woods2017} and references therein), which are likely signatures of flux emergence and/or cancellation, tether cutting reconnection, and flux rope formation. Imaging spectroscopy indicates that these dynamics typically occur on small-scales and can subsequently lead to large-scale destabilization and explosive energy release. While these higher atmospheric signatures will be well-probed with Solar Orbiter EUI, SPICE and STIX, or with Solar C EUVST, the small scale chromospheric dynamics and magnetic field evolution that
are critical to answering questions about how the flare/eruption is triggered, and when and where in the atmosphere flux ropes are formed, requires high resolution observations of the chromospheric magnetic field and associated dynamics.

\subsubsection{Measuring the photospheric and chromospheric dynamics before flares}\label{OP5.4.1}

\vspace{-2ex}
See table on page \pageref{OP5.4.1table}.

This observing plan aims to explore the role of small-scale photospheric and chromopsheric dynamics in the period before flare/CME onset and its relationship to TR and coronal dynamics. It will measure the magnetic field strength and direction from the photosphere to the upper chromosphere using spectropolarimetry, and the plasma dynamics using 2D spectroscopy to determine intensity, line widths, and line of sight velocities. Measurements should be coordinated with space platforms to provide the link to coronal spectroscopy.

\subsection{Macrospicules/spicules and Transition Regions Quakes (TRQs)} 

Spicules and macrospicules are most likely generated in the lower solar
atmosphere. Their impact on the solar corona is fundamentally important, as e.g. a mere 1\% of
spicule material may be sufficient to supply the mass needed for solar wind. Therefore, the
diagnostics from their footpoints in the photosphere/low chromosphere up to the corona is an
important question. Again, Joint EST and space observations may be able to identify and track MHD
waves in these structures, enabling us to construct the magnetic skeleton of these jet structures. Details of observing programmes related to spicules and macro-spicules can be found in Section 2.

TRQS (\citep{Scullion2011}) are energetic waves that have been detected in the transition region, and evolve in a similar manner to waves on 2D elastic waveguides. The question of what drives these TRQs remains open, with one possible explanation being that they are driven by (macro) spicules. Here, combining EST with space observations of the TR and corona will provide new insights into the effects of spicules on the TR and low corona, allowing us to identify the low coronal signatures of TRQs. To achieve this, a large sample of TRQs will be constructed, using  IRIS/Solar Oribter/Solar C EUVST in imaging mode, observed jointly with EST. The discovery of a link between these waves and RBEs identified in CRISP H$\alpha$ data is potentially very far reaching. Work by \citep{Henriques2016} has already found convincing evidence of links between RBEs and coronal transient events. However, they did not study whether their coronal features corresponded to coherent waves (i.e., TRQs manifesting as propagating wave fronts in IRIS 140 nm data). This study on TRQs is vital for expanding our knowledge of the coupling between the lower solar atmosphere and the TR/low corona, and in particular searching for alternative drivers of TRQs (e.g. shocks) using chromospheric and NIR line profiles.

\subsubsection{Rising spicule signatures}\label{OP5.5.1}

\vspace{-2ex}
See table on page \pageref{OP5.5.1table}.

\subsection{Ellerman bombs}

Ellerman bombs (EBs; \citep{Ellerman1917}) have been widely studied within ARs over
the past century (\citep{Nelson2015}) due to the hypothesis that they are driven by
photospheric magnetic reconnection. Research highlighting similar events in the quiet
Sun (QSEBs; \citep{2016A&A...592A.100R}) suggests that magnetic reconnection can happen almost continuously throughout the photosphere. Using H$\alpha$ datasets (multi-positions and cadences $~$5 s) sampled at varying latitudes at the solar limb
(where identification of QSEBs is less ambiguous) could lead to fundamental improvements in our
understanding of QSEBs through analysis of their signatures in both EST and IRIS. EST will
investigate the spatial coverage of QSEBs with respect to latitude to understand whether these
events (and potentially small-scale magnetic reconnection in general) occur over the whole
solar disk, including the polar regions, or whether they are confined to active latitudes (like
sunspots). Further, obtaining a large sample of QSEBs will allow us to discover any links to
jet-like events (e.g., spicules, macro-spicules) that penetrate the upper solar atmosphere using
EST/Solar Orbiter/Solar C EUVST data. Section 2 outlines sample observing programmes for Ellerman bombs in more detail.

\subsubsection{Ellerman bombs in the lower solar atmosphere}\label{OP5.6.1}

\vspace{-2ex}
See table on page \pageref{OP5.6.1table}.

 %\newpage \input{SG5_tables}
%-----------------------------------------------------------
\section{Solar Flares and Eruptive Events} %
\label{sec_events}
{Authors: Lyndsay Fletcher, Francesca Zuccarello, Christoph Kuckein, K\'evin Dalmasse, Paulo Sim\~oes}
%\footnote{version: ({2018-Nov-26})}

%\section{Solar Flares and Eruptive Events}

%\subsection{Introduction}% - all (but LF to start draft)

In a solar flare the reconfiguration of the current-carrying coronal magnetic field, in response to the development of a coronal magnetic instability, results in the liberation of stored magnetic energy. This is converted to thermal energy, kinetic energy of accelerated particles, and ultimately radiated energy, and mass motion of the flare is accompanied by a coronal mass ejection (CME). To understand the 3-D magnetic configuration and electrical current distribution before, during and after a solar eruptive event requires measurement of the photospheric and chromospheric vector magnetic field. As most of the radiated energy in a flare originates in the chromosphere, a concerted study of this region is necessary also to understand flare energetics and to diagnose flare energy transport.

The energy input from a solar flare has a profound impact on the thermal structure of the chromosphere, and may also influence the photosphere. Different mechanisms of energy input (particle beams, MHD waves, thermal conduction) and dissipation will lead to different temperature, density and velocity structures. This gives, in principle, some power to discriminate between models \citep{2016ApJ...827..101K}. Spectroscopic observations in multiple lines, and continua, through the solar chromosphere, and at the highest possible spatial and temporal resolution are needed to track the evolution of the flaring chromosphere as it responds to intense energy input. It is already well-known from lightcurves in radio, X-rays and H$\alpha$ that while timescales of 10s of seconds are characteristic of major bursts of energy release in a flare \citep{2005ApJ...625L.143G}, they also exhibit variability on shorter timescales \citep{2007A&A...461..303R}. Spatial structures on sub-arcsecond scales have also been detected in flare chromospheric ribbons and photospheric sources \citep{2007PASJ...59S.807I,2014ApJ...788L..18S}. The high spatial and temporal resolution offered by EST is therefore required to probe the elementary energy release, transport and dissipation processes in a flare. 

Particularly in larger events, an eruption of CME plasma away from the Sun, driven by unbalanced Lorentz and gravitational forces, accompanies the flare. Analysis of coronal magnetic configurations prone to eruption are currently based on photospheric magnetograms, giving non-force-free boundary conditions for force-free field extrapolations. Measurements of the chromospheric field, which is closer to force-free, will significantly advance our ability to identify the conditions for eruptions, when used in state-of-the-art extrapolations and in data-driven simulations. Also of interest are changes that take place in the photospheric and chromospheric  magnetic field in response to the rapid field reconfiguration happening above. 

Electric currents are fundamental to flares and CMEs. At large scales, the magnetic field component associated with the current appears to store the free magnetic energy for a flare \citep{2017SoPh..292..159K}. At very small scales, they provide localised regions for dissipating and releasing the free energy through magnetic reconnection or Joule dissipation \citep{2014ApJ...788L..18S}. By providing vector magnetic field measurements at different heights from the photosphere to the chromosphere, EST will for the first time enable the detailed analysis of the 3D properties of electric currents in the lower solar atmosphere, as well as the study of their evolution during solar flares and CMEs. 

Observations and simulations strongly suggest that one particular current-carrying structure is often at the heart of a flare and CME - a highly sheared or twisted flux rope. In many cases this supports material in the form of a prominence, which can be analysed spectroscopically to reveal aspects of the distribution of mass and magnetic field. This prominence material can constitute the inner part, or bright core, of a coronal mass ejection (CME), that can have important consequences for Space Weather conditions. Therefore the ability to map out the prominence field prior to the eruption will improve boundary conditions for the evolving field distribution of a CME in space. Similar phenomena are now seen to happen in simulations and observations of much smaller jet-like events as well.

These different aspects of flares and eruptive events demonstrates a need, which will be met by EST, for chromospheric radiation and magnetic diagnostics, at high spatial and temporal resolution, but also with the ability to observe or construct large (active-region scale) fields-of-view on a longer timescale pre-eruption.

\subsection{Thermal structure of flare chromosphere and photosphere: line observations} %- LF/CK?
The nominal association between the formation temperature of a particular ion, and height in the atmosphere, as well as the difference in optical depth in line core and wings, provides a means to determine the atmospheric properties as a function of height \citep[e.g.][]{2017ApJ...846....9K}, and potentially also time. In a converging chromospheric magnetic field, the physical size of sources emitted by different atoms and ions, and at different wavelengths is also expected to vary with height \citep{2012ApJ...750L...7X}. To address the evolution of the flaring chromosphere requires that we observe in spectral lines formed at different chromospheric heights, with as high a cadence as possible. In the upper chromosphere flows can be strong so a broad spectral window is necessary, while as large as possible a FOV optimises the chance of catching a flare. Only Stokes I is required. These high-cadence programmes should be sequenced with programmes to map the vector field, to identify associations with e.g. electric current density, Lorentz forces, and other 3-D magnetic properties.
 
\subsubsection{Thermal structure of flare: line observations}\label{OP6.1.1}

\vspace{-2ex}
See table on page \pageref{OP6.1.1table}.

\subsection{Thermal structure of flare chromosphere and photosphere: continuum observations}
`White light' emission is often detectable in highly energetic flare events, but is also a feature of some much smaller flares \citep{2008ApJ...688L.119J}
The significance of this broad-band optical emission is that it embodies a very large fraction of the radiated energy \citep{2014ApJ...793...70M}. The most likely source for flare-enhanced WL emission is recombination emission in the hydrogen Balmer and Paschen continuum (at UV and EUV wavelengths enhancement of the Lyman continuum of H and He is clearly seen). This indicates a central role for ionisation and recombination in the thermodynamics of the flaring chromosphere; the change of state from a neutral to an ionised plasma species is an extremely efficient `thermostat' - until ionisation is complete. The ionisation of both H and He will have significant roles to play in flare thermodynamics, and can be examined by following the continuum emission at different parts of the optical spectrum. Broad-band continuum observations, even at low spectral resolution, will be capable of diagnosing this basic physical process.  
One way to distinguish between chromospheric and photospheric origins of flare white light emission is to look at differences in the continuum intensity around the H recombination edges. Though the Balmer jump  has been detected before \citep{1982SoPh...80..113H} it did not resolve this question. The Paschen jump (near 815nm) has been observed only once in flare \citep{1984SoPh...92..217N} but should provide a robust observational test for the white-light emission mechanism. The relatively simple measurements proposed here should suffice to finally settle the white-light flare mystery and allow us to use future continuum data for diagnostics of the flaring atmosphere.

However, observations in the optical are always going to be challenging, particularly in small flares, because of the small contrast compared to the photosphere (though this is alleviated if the flare ribbon extends into a sunspot). According to modeling \citep{2017A&A...605A.125S} extension into the infrared beyond 2$\mu$m - provides higher contrast than in optical flares, and is synchronised in time with flare energy input to within 0.2s. The primary signature of flare energy - the HXRs that are generated by flare-accelerated electrons, cannot in the foreseeable future be imaged with the spatial and temporal resolution achievable with the EST so observations in the infrared will provide a unique window into flare temporal and spatial evolution.

These high-cadence programmes should be sequenced with programmes to map the vector field, to identify associations with e.g. electric current density, Lorentz forces, and other 3-D magnetic properties.

\subsubsection{Thermal structure of flare: continuum observations} \label{OP6.2.1}

\vspace{-2ex}
See table on page \pageref{OP6.2.1table}.

\subsubsection{Thermal structure of flare: detection of the H Paschen Jump}\label{OP6.2.2}

\vspace{-2ex}
See table on page \pageref{OP6.2.2table}.

\subsection{Velocity structure of the flaring atmosphere} %- LF
If the energy that is input to the lower solar atmosphere cannot be radiated or conducted rapidly away then the plasma will expand. Strongly confined by the magnetic field, the flows will be predominantly along the field. The emission and absorption by upwards and downwards moving material leads to complex line profiles which must be interpreted with care; for example absorption in upflowing material can lead to an apparently red-shifted profile \citep{2015ApJ...813..125K,2017ApJ...850...36C}. Features in H$\alpha$ formed in the upper chromosphere indicate flows reaching  tens of $\rm{km~s^{-1}}$; 
%\sout{in Ca 8542 formed lower down the inferred flows are smaller}.
The possibility also exists of using multi-line observations, together with overall understanding of spectral line formation gleaned from models, to map out the bulk flow structure of the flare-excited chromosphere. 

The presence of plasma turbulence in the flare chromosphere is also to be expected, in association with intense heating, shocks and rapid distortion of the field as it responds to the coronal reconfiguration. Spectral lines emitted in turbulent regions are expected to be broadened either by  magnetohydrodynamic ``macroturbulence'', or because of out-of-equilibrium distributions of emitting ions, having non-Maxwellian velocity distributions, offering the possibility of examining the development of non-equilibrium velocity distributions evolving on rapid timescales. These high-cadence programmes should be sequenced with programmes to map the vector field, to identify associations with e.g. electric current density, Lorentz forces, and other 3-D magnetic properties.

\subsubsection{Velocity structure of the flaring atmosphere}\label{OP6.3.1}

\vspace{-2ex}
See table on page \pageref{OP6.3.1table}.
%- LF/CK?

\vspace*{10pt}

%\begin{itemize}
%\item Spectral line profiles and plasma flows (i.e. evaporation and condensation)
%\item Line broadening
%\end{itemize}

\subsection{Diagnostics for non-thermal particles} % LF
Chromospheric hard X-ray radiation is non-thermal bremsstrahlung, diagnosing flare-accelerated electrons. The dominant model of energy transport places the source of these electrons in the corona, with the electrons arriving in a beam directed along the magnetic field. However HXR spectra interpreted as having two components - a directly-viewed component and an albedo component - are not consistent with an anisotropic distribution. While Coulomb collisions or wave-particle interactions may effectively isotropise the electron distribution, a possible - and long-sought- diagnostic of anisotropic particle distributions is the collisional linear impact polarization generated when appropriate angular momentum sub-states are preferentially excited according to the angular distribution of the exciter particles \citep{1990JQSRT..44..193H}. Claims of observations of linear polarization in lines such as H$\alpha$ remain contentious \citep[e.g.][]{2005A&A...434.1183B}, as non-simultaneous measurements of Stokes parameters can easily lead to spurious apparent polarization signatures (as, in certain lines, can opacity effects \citep{2015ApJ...814..100J}). The search for a flare-associated linear polarization signature, as well as measurement of its direction with respect to the solar radial direction (which can also distinguish between electron-generated and proton-generated emission) must therefore be a priority. 
%Possible topics:
%\begin{itemize}
%\item Impact polarization
%\item Diagnostics based on collisional excitation by non-thermals
%\end{itemize}
\subsubsection{Diagnostics for non-thermal particles}\label{OP6.4.1}

\vspace{-2ex}
See table on page \pageref{OP6.4.1table}.
 %- LF/CK?

\subsection{Oscillatory phenomena in flares} %FZ

Observations \citep{2009SSRv..149..119N,2016SoPh..291.3143V} provide indications of quasi-periodic emission during flares, observed in all the electromagnetic spectrum, from radio waves, through H$_{\alpha}$, WL, EUV, SXR and HXR wavelengths.  Among these phenomena, the so called quasi-periodic pulsations (QPPs), which owe their name to well-observed modulations in amplitude and period, can be characterised by periods ranging from fraction of seconds to minutes. However, it is also important to stress that statistical studies have shown that the distribution of the observed periodicities might be affected by the time resolution of the instruments used for their detection \citep{2018SSRv..214...45M}. 

It is not clear whether QPPs with different periods and modulations are different manifestations of the same physical process or whether the different periods indicate different physical mechanisms. Other questions concern whether QPPs detected in different phases of flares are due to different mechanisms or whether QPPs detected in thermal and non-thermal emission are different.

According to some authors, QPPs characterized by periods ranging from few seconds to several minutes can be attributed to MHD waves; QPPs with periods of fraction of seconds to fast sausage modes, while for longer periods other mechanisms, like slow magnetoacoustic and fast kink mode have been invoked \citep{2005A&A...440L..59F,2011ApJ...740...90V,2012ApJ...755..113S}%

Another mechanism that has been taken into account is based on the so-called oscillatory reconnection, occurring in distinct bursts that results in periodic emission when particles are accelerated and the plasma is heated. In this context, there are some studies \citep[e.g.][]{2009A&A...494..329M,2018SSRv..214...45M} based on the assumption that oscillatory reconnection can be caused by magnetic flux emergence in a pre-existing field, indicating that there should be a relationship between the initial submerged magnetic field strength and the observed QPPs periods: longer periods should be related to stronger magnetic fields.

\subsubsection{QPPs in different phases of flares}\label{OP6.5.1}

\vspace{-2ex}
See table on page \pageref{OP6.5.1table}.

\subsubsection{QPPs and oscillatory reconnection}\label{OP6.5.2}

\vspace{-2ex}
See table on page \pageref{OP6.5.2table}.

\subsection{Sunquakes}
Sunquakes are phenomena characterised by the propagation of acoustic waves that refract deep in the convection zone and appear in photospheric images as ripples, centered around sites hosting M- or X-class flares, and often associated with CMEs \citep{2011SSRv..158..451D}. They have energy between 10$^{27}$ and 10$^{29}$ erg, originate within an area of the order of 10 Mm$^{2}$ and show traveling speeds of the order of tens of km s$^{-1}$. Observations sometimes indicate a co-spatiality of the ripple center of the sunquake with hard X-ray, white light, gamma-ray emission, as well as with sites showing abrupt and permanent changes in the photospheric line-of-sight magnetic field.

The proposed mechanisms causing the sunquakes can be divided into two categories: those based on impulsive heating and those related to the action of the Lorentz force. In the first category the acoustic wave is generated by impulsive heating that might be related to: i) thick-target heating of the chromosphere by energetic electrons \citep{2006SoPh..238....1K}; ii) heating of the photosphere due to backwarming or deeply penetrating protons \citep{2007ApJ...664..573Z}; iii) wave heating of the photosphere and chromosphere \citep{2013ApJ...765...81R}. In the second category, the driver is the Lorentz force acting on the photospheric plasma \citep{2012SoPh..277...59F}. This arises from rapid changes in the photospheric magnetic field observed in many major flares,  caused by restructuring in the coronal field. It is often observed during flares associated with sunquakes. 

Moreover, another issue concerning sunquakes is related to the fact that the propagation of the acoustic wave in the solar interior seems to be triggered by a process of energy release occurring in the corona, and in order to drive an acoustic disturbance in the solar interior, the energy must propagate through nine pressure scale heights. Therefore, it would also be important to investigate how this energy transfer can take place. 

\subsubsection{Sunquakes initiated by impulsive heating during flares}\label{OP6.6.1}

\vspace{-2ex}
See table on page \pageref{OP6.6.1table}.

\subsubsection{Sunquakes initiated by changes in the Lorentz force during flares}\label{OP6.6.2}

\vspace{-2ex}
See table on page \pageref{OP6.6.2table}.

%\section{The magnetic structure and evolution of flaring active regions}

\subsection{Large-scale structure and evolution of the magnetic field} %FZ
It is now established that the energy released during flares (in the form of thermal, kinetic and radiative energy) is previously stored in stressed, non-potential magnetic field configurations until magnetic reconnection takes place. Before the flare occurrence, there is a slow phase of energy storage, which can be due to emergence of new magnetic flux from the subphotospheric layers or to shearing motions that can contribute to a continuous increase of the magnetic energy through build-up of electric currents, on a timescale much longer than the Alfv\'en timescale in the solar atmosphere. This eventually leads to force-free field configurations, with the coronal magnetic field (at least in ARs) appearing to evolve through a succession of quasi-force-free states

Only in a few cases do the WL images show changes in the sunspot structure after the flare occurrence, generally limited to major (WL) flares. However, it is difficult to single out whether these changes depend on the magnetic field global topology, on the magnetic energy released during the event, on the layer where the magnetic reconnection takes place, or it is mainly related to the spatial resolution of the instruments used.

During the last decade, there has been increasing evidence of rapid, irreversible enhancements of both the longitudinal and horizontal components of the photospheric magnetic field, as well as of magnetic shear at some sections of the magnetic neutral line. In this respect, it would be important to verify whether the reconnection that causes the coronal field reconfiguration leads to similar abrupt, non-reversing changes also in the chromospheric field, in concomitance with the passage of the flare ribbons.

\subsubsection{Changes in the magnetic field configuration during/after flares}\label{OP6.7.1}

\vspace{-2ex}
See table on page \pageref{OP6.7.1table}.

Spectropolarimetric data taken by SDO/HMI used to study the properties of the large-scale vector field, and in particular the evolution of the magnetic helicity flux, indicate that differences in the accumulation of the magnetic helicity flux in the corona can been attributed to the overall magnetic configuration and to the location of flux ropes in active regions. However, differing scenarios exist. \cite{2014ApJ...794..118R} find that energy injection provided by the shearing motions, pointed out by high values of the shear and the dip angles along the main polarity inversion line, is more significant than the energy injected by magnetic flux emergence. However \cite{2012ApJ...761..105L} report that the energy flux associated with magnetic flux emergence contributes to 60\% of the total energy, with only 40\% arising from shearing.  New observations are important to determine the relative importance of these. With appropriate lines chosen throughout the photosphere and chromosphere, it will be possible to assess the helicity flux and accumulation at different heights, to understand how magnetic helicity is transported through the atmosphere.

\subsubsection{Determine magnetic helicity accumulation, dip and shear angle in flaring active regions}\label{OP6.7.2}

\vspace{-2ex}
See table on page \pageref{OP6.7.2table}.

\subsection{Filaments in flaring active regions} %CK

Filaments are large scale objects embedded in the corona and chromosphere of the Sun. They are located on top of polarity inversion lines (PILs) which separate the positive and negative polarities of the photospheric magnetic field. Especially active region filaments are often associated with very narrow PIL channels which often produce flares. As a consequence, the filament can get ejected into space. However, it is still not understood why some filaments are ejected while others remain stable after the flare. The magnetic environment in such a scenario is extremely difficult to interpret. \cite{2014A&A...561A..98S} found up to five different magnetic components associated with the flaring filament in the chromosphere. The authors used ground-based spectropolarimetric observations in the He I 10830 spectral range, which is often used to infer the magnetic field in the chromosphere. The derived LOS velocities of the flaring filament were supersonic in both directions (toward and against the solar surface). 

\subsubsection{Filaments in flaring active regions}\label{OP6.8.1}

\vspace{-2ex}
See table on page \pageref{OP6.8.1table}.

\subsection{Coronal Mass Ejections}
During a coronal mass ejection an amount of magnetised plasma of the order of 10$^{14}$ – 10$^{16}$ g is expelled from the Sun at velocities ranging between 10$^{2}$ – 10$^{3}$ km s$^{-1}$. The energy involved in these phenomena is of the order of 10$^{28}$ – 10$^{32}$ erg.

There are still some difficulties in determining the relationship between flares and CMEs, due to the use of different instruments needed to observe phenomena on the disk and in the outer corona. 
This circumstance can have different implications, like for instance the fact that there can be a loss of information during the time when the plasma is travelling within the region covered by the occulting disk of the coronagraph. Also, it is not always straightforward to spatially associate a CME with a flare, i.e., to determine the exact location on the solar disk where the eruptive phenomenon started. Another issue arises from the fact that the statistical analyses show that in some cases the flare occurs before the CME, while in others the opposite is true, posing some problems on the initiation models.    

\subsubsection{CME's sources and temporal relation with flares}\label{OP6.9.1}

\vspace{-2ex}
See table on page \pageref{OP6.9.1table}.

\subsection{Structure and evolution of the magnetic field at small scales} %SD (CK)
As a complement to the large-scale observation of the magnetic field,  high temporal resolution observations of small regions are also required, for at least two purposes. Firstly, simultaneous sampling of magnetically sensitive photospheric and chromospheric spectral lines at high cadence would enable us to study how the magnetic field changes, albeit only in limited areas, during the most dynamic phase of the flare, when the timescale of evolution is seconds. Such observations will also allow a search for signatures of flare-associated reconnection not only in the corona, but also in the lower atmosphere (including component reconnection). Secondly, there are small-scale analogues to the flare process in the form of jets, which have similar magnetic topologies but a much smaller scale, at typically 40" - 60", and can involve the ejection of a mini-filament \citep{2016SSRv..201....1R,2015ApJ...806...11M,2016ApJ...821..100S,2017Natur.544..452W}. This suggests that these smaller and more readily-observed events share some of the magnetic and MHD properties of larger flare and CME events, and may also have similar chromospheric radiation and non-thermal particle populations. Selected OPs from above can be run with smaller FOVs at higher cadence, and the possibility of capturing an entire event simultaneously.
 %\newpage \input{SG6_tables}
%-----------------------------------------------------------
\section{Coupling in partially ionized solar plasma}
\label{sec_partial}
{Authors: Elena Khomenko, Robertus Erdelyi, Mihalis Mathioudakis, Sarrah Mattews, Lyndsay Fletcher}%\footnote{(version: {2018-Dec-08})}
%%%%%%%%%%%%%%%%%%%%%%%%%%%%%%%%%%%%%%%%%%%%%%%%%%%%%%%%%%
%\section{Coupling in partially ionized solar plasma}
%%%%%%%%%%%%%%%%%%%%%%%%%%%%%%%%%%%%%%%%%%%%%%%%%%%%%%%%%%

%split as a separate section. Ask for more scientific cases (flares, prominences, instabilities, fibrils, spicules). Waves to RE.

The solar photosphere and chromosphere are only partially ionized. The ionization fraction is below 10$^{-3}$ in the photosphere, increasing to about 0.5 in the chromosphere (with 1 being completely ionized plasma).  The importance of this fact has not been considered to its full extent in the past in the solar physics community. In the first decades of our century, when powerful computing techniques and codes have become accessible, we have started to be in a position to simulate complex partial ionization effects and understand their profound consequences. The influence of partial ionization of the solar plasma on its dynamics has been considered in analytical and numerical models by e.g. \citet{2007ApJ...666..541A, 2010A&A...512A..28S, 2012A&A...544A.143Z, 2012ApJ...760..109L, 2014PhPl...21i2901K, 2012ApJ...753..161M, 2016ApJ...831L...1M, 2016ApJ...819L..11S}. 

The great majority of the studies of photospheric, chromospheric and coronal plasma dynamics uses MHD as the main tool for quite successfully understanding the complex structure and dynamical processes of these solar atmospheric layers. Nevertheless,  the MHD approach overlooks a number of dissipative and dispersive non-ideal mechanisms associated with ion-neutral interactions in weakly ionized and weakly collisional solar plasmas, especially important under chromospheric conditions. It has been repeatedly demonstrated that processes related to the non-ideal plasma behaviour due to neutrals may be the key to solve the problem of chromospheric heating, dynamics and fine structure \citep[e.g.,][]{2012ApJ...753..161M, 2012ApJ...747...87K, 2016ApJ...831L...1M, 2016ApJ...819L..11S}. 

A suitable alternative to the MHD approach is a multi-fluid approach where all the plasma species are considered as separate fluids interacting by collisions. A multi-fluid treatment is essential for the weakly collisionally coupled chromosphere because the relevant processes for energy transport and conversion happen at spatial and temporal scales similar to ion-neutral collisional scales \citep{2018SSRv..214...58B}. For example, the use of the multi-fluid formalism allows to write an equation for the drift velocity between ions and neutrals \citep{Bittencourt}. The existence of these drifts between species is a direct consequence of the partial ionization, and reflects that the coupling between the fluids is not strong enough to ensure a behaviour as a single fluid. 

The processes related to partial ionization happen at short spatial and temporal scales.  Nevertheless, these scales are not as small as typical plasma scales of the fully ionized plasmas, because they are related to ion-neutral, and not ion-electron collisional scales.  The values of these scales depend on the details of the process and the values of the physical parameters. Estimates, and simulations, made for the case of the solar atmosphere in e.g. \citet{2012ApJ...747...87K, 2012ApJ...753..161M, 2014PhPl...21i2901K} suggest that temporal resolution of a fraction of a second is necessary to start resolving these effects. 

Observations of partial ionization effects lag significantly behind the theoretical developments.  Thanks to the powerful combination of instruments, allowing simultaneous observations in several spectral lines, and to the high photon efficiency, the EST will be the only telescope in the world able to push for direct observations of partially ionized plasma processes, allowing our community for the fist time to observationally test the limits of classical magneto-hydrodynamics.

\subsection{Dynamics of partially ionized prominence plasma}

One way of detecting partial ionization effects is the direct measurements of the dynamics of the different components of the plasma. Simultaneous measurements using spectral lines of elements with different degrees of ionization, or different atomic mass,  may provide indications of different dynamics of plasma components. The major obstacle for interpreting such measurement is the opacity of the solar plasma. When measured over the solar disc, spectral lines of different elements can form over different height ranges, and therefore provide information of dynamics of different volumes of the plasma. Measurements off-limb may help palliating this problem. The material of solar prominences can frequently be considered optically thin. In addition, the physical conditions in prominences are expected to give rise to significant partial ionization effects, with a considerable amount of neutral atoms. Therefore, different dynamical events in prominence plasmas can be used to test the existence of deviations from classical magneto-hydrodynamics.

Processes related to destabilisation of prominences are not well understood. Studying wave dissipation, instabilities and reconnection in weakly ionised prominence plasmas will allow confirming or rejecting the role of neutrals in destabilisation of prominences.

\subsubsection{Waves in prominences observed in neutral and ionized spectral lines}\label{OP7.1.1.1}\label{OP7.1.1.2}

\vspace{-2ex}
See tables OP 7.1.1.1 and OP 7.1.1.2 on page \pageref{OP7.1.1.1table}.

Velocities typically measured in solar prominences reveal the existence of important mass flows and waves with different periodicities, as well as instabilities. Instabilities may also lead to mass motions that can have a different impact on neutrals and ions \citep{2012ApJ...749..163S, 2014A&A...565A..45K}.  Several publications report simultaneous measurements of prominence dynamics in ionized and neutral spectral lines \citep{2016ApJ...823..132K, 2017A&A...601A.103A}. While \citet{2016ApJ...823..132K} show a detection of small differences in the ionized \CaII\ 854.2 nm and in the neutral \HeI\ 1083.0 nm velocities in the observed prominence, \citet{2017A&A...601A.103A} concluded that similar differences exist also between the velocities of atoms of the same species emitted by different spectral lines (\HI\ 397 nm, \HI\ 434 nm, \CaII\ 397 nm, and \CaII\ 854). 

With this campaign the study of ion-neutral effects in prominences will be expanded to the comparison of velocities measured in spectral lines of several neutral and ionized spectral lines strictly simultaneously. These measurements will be accompanied by measurements of the magnetic field vector in order to find possible relations between drift velocities and the magnetic field strength and orientation. \\

\subsubsection{Prominence-corona transition region (PCTR) dynamics}\label{OP7.1.2}

\vspace{-2ex}
See table on page \pageref{OP7.1.2table}.

The prominence-corona interface is a thin layer separating prominence material from the corona, where the transition from coronal to prominence values in all thermodynamic parameters happen. This region is subject to perturbations, instabilities and turbulence flows. Numerical simulations of Rayleigh-Taylor instability (RTI) at PCTR from \citet{ 2014A&A...565A..45K} suggest that the drift velocity between ions and neutrals acquire maximum values at this layer. Another conclusion from similar studies \citep{2012ApJ...749..163S} is that, thanks to the presence of the destabilising effect by neutral atoms,  Rayleigh-Taylor instability at PCTR can develop at the smallest scales possible despite the stabilising effect of the magnetic field. The purpose of this campaign is to push the capabilities of EST to observe the dynamics of PCTR at the highest spatial and temporal resolutions possible, using simultaneous measurements in neutral and ionized spectral lines. 

\subsubsection{Draining of neutral material from prominences and filaments}\label{OP7.1.3}

\vspace{-2ex}
See table on page \pageref{OP7.1.3table}.

Theoretical work by \citet{2002ApJ...577..464G} suggest that cross-field diffusion of neutral material in filaments/ prominences is an important mechanism of their mass loss. As a consequence of incomplete collisional coupling of prominence plasma, neutral helium and hydrogen drains out of prominences.  Since the atomic mass of helium and hydrogen is different, the elements have different diffusion speeds, and it takes different times for them to cross the prominence structure. \citet{2007ApJ...671..978G} found observational confirmation of this effect, by measuring the relative He/H abundance in filaments across the disc. They found a relative helium deficit in the upper parts of the prominence compared to a relative helium surplus in the lower regions. This was attributed to be a consequence of the large loss timescale for neutral helium compared to neutral hydrogen. 

In this campaign we will observe filaments and prominences at different positions of the solar disc with the purpose to study the relative  abundance of He, H and \CaII.  The emphasis will be on high spatial resolution and large fields of view covering all of the structure. The time resolution is less important since the drainage is a slow process. Magnetic field will be measured simultaneously to find the relation between the relative abundances of He, H and \CaII\ and the magnetic field structure. 

\subsection{Influence of partial ionization on spicules}

Spicules observed off-limb, as well as their on-disc counterparts are good candidates to show non-MHD effects. Spicules are made of partially ionized chromospheric material. These thin and elongated structures show dynamics on rather short time scales. Theoretical works and numerical simulations suggest that partial ionization effects play an important role in the formation of spicules \citep{1998A&A...338..729D, 2017Sci...356.1269M}.

\subsubsection{Dynamics of spicules observed in neutral and ionized spectral lines}\label{OP7.2.1}

\vspace{-2ex}
See table on page \pageref{OP7.2.1table}.

This observing program will measure velocities of ionized and neutral species associated with spicules at the highest cadence possible in order to verify how similar they are. Simultaneously, the magnetic field vector in spicules needs to be measured.

\subsubsection{Alignment between the magnetic field in fibrils and disc counterparts of spicules}\label{OP7.2.2}

\vspace{-2ex}
See table on page \pageref{OP7.2.2table}.

One is often making the assumption that linear features seen in observations in spectral lines formed in the chromosphere are aligned with the 
magnetic field vector. This is expected in the case of ideal MHD where the magnetic field is frozen in and particles can only move along the magnetic field. When ion-neutral effects are taken into account, this is no longer necessarily true. Numerical simulations show that the magnetic field is often not well aligned with chromospheric features \citep{2016ApJ...831L...1M}. There are also observations that indicate occasional misalignment between the magnetic field and chromospheric features \citep{2011A&A...527L...8D}. An observing program to address the alignment between the magnetic field and linear chromospheric features needs imaging in a chromospheric line and full vector magnetic field measurements in the chromosphere. High signal to noise is crucial for the accuracy of the magnetic field determination while temporal cadence and field-of-view is less critical.

\subsection{Detection of partial ionization effects in the photosphere}

\subsubsection{Neutral and ionized material Evershed flow in sunspots}\label{OP7.3.1}

\vspace{-2ex}
See table on page \pageref{OP7.3.1table}.

This campaign aims at detecting ion-neutral effects in the photosphere. Strong magnetic field of sunspots, through the increase of the ion gyro frequency, might lead to stronger decoupling effects. These effects, if detectable, should be present on very small spatial and temporal scales. \citet{2015A&A...584A..66K} have  measured the amplitudes of the Evershed flow using pairs of carefully selected \FeI\ and \FeII\ spectral lines. They compared azimuthally averaged amplitudes of the Evershed flow extracted from neutral and ion lines and found measurable differences in the radial component of the flow. All five pairs of lines show the same tendency, with a few hundred m/s larger amplitude of the flow measured from \FeI\ lines compared to \FeII\ lines. This tendency is preserved at all photospheric heights and radial distances in the penumbra. The origin of this effect is not entirely clear. Spatially and temporally resolved measurements of the Evershed flow, together with measurements of the magnetic field are necessary to conclude about the presence of ion-neutral effects in the photosphere of sunspots. 

\subsection{Multi-fluid physics of chromospheric waves, shocks and swirls}

Waves in the solar atmosphere are usually classified as slow and fast magneto-acoustic waves and Alfv\'en waves, assuming the complexity of the waveguide allows this simplification. The propagation of magneto-acoustic waves, driven by the combination of total pressure and magnetic tension, is guided by the magnetic field, though other physical quantities may also influence wave propagation. For example, in the upper solar atmosphere the magnetic field may be considered as locally uniform, however, density may vary across the waveguide. Alfv\'en waves are driven by the magnetic tension and carry energy along the magnetic field. The low ionization fraction of the solar photosphere and chromosphere, and the strong change of the ionization fraction with height strongly affects the propagation of magneto-acoustic waves and Alfv\'en waves. The motion of the neutrals in the partially ionized atmosphere will not follow the motion of the magnetic flux, questioning the use of atmospheric structures to constrain magnetic field extrapolation methods \citep{2016ApJ...831L...1M}. In a partially ionized plasma (even without structuring or stratification) the wave characteristics are different from those of their counterparts in a plasma that is described by the MHD approximation. Such characteristics may be the wave modes themselves, their frequencies or their spatial behavior (e.g. amplitudes, wave eigenfunctions). Theory has already predicted these changes at various degrees  \citep{2013ApJ...767..171S, 2018A&A...610A..56M}. The high temporal cadence and spatial resolution of EST measurements will be used to pick up these deviations in the wave characteristics caused by the presence of partial ionization. In general, it is necessary to accurately measure the wave frequencies in individual waveguides, as well as their eigenfunctions (i.e. spatial wavelength), and amplitudes. Once measured, these quantities can be used a diagnostic even for the degree of partial ionization using methods of solar magneto-seismology \citep{2009SSRv..149..199R}.
 
\subsubsection{Observation of magnetic swirls under 2-fluid condition}\label{OP7.4.1}

\vspace{-2ex}
See table on page \pageref{OP7.4.1table}.

We propose to further investigate photospheric intensity swirls under the two-fluid condition. Statistical results in \cite{2018ApJ...tmp...ToL} have shown that most ($> 70\%$) swirls detected have lifetime less than 6 sec. Extremely high cadence (0.1 sec) will be needed.  Spatial resolution plays vital influence in the number and parameters of swirls detected. The swirls are usually located in intergranular lanes. The 2D field of view of IFU is essential in the observation of these features.

\subsection{Flares and energetic events}

\subsubsection{Observations of reconnection and plasmoids in partially ionized plasma}\label{OP7.5.1}

\vspace{-2ex}
See table on page \pageref{OP7.5.1table}.

It is widely accepted that magnetic reconnection plays a critical role in flares and CMEs, in particular for abruptly and efficiently releasing magnetic energy (and converting it into kinetic and thermal energy), but also as a possible driver of these events. Most of our (incomplete) understanding of magnetic reconnection is based on observations, theory and simulations of reconnection in the fully ionized corona, but it is becoming clear that reconnection in the partially ionized chromosphere is also important, and simulations indicate that partial ionization effects lead to the development of fast reconnection as the result of the onset of the tearing mode instability without the need for anomalous resistivity  \citep[e.g.,][]{2012ApJ...760..109L, 2015PASJ...67...96S, 2015ApJ...799...79N}. \citet{2015PASJ...67...96S} in particular predict that the reconnection will proceed through different stages based on the level of ionization present, and that in very strong chromospheric field conditions kinetic scales can be reached. While simulations seem to show good agreement with the properties of small-scale chromoshperic jets, little work has so far been done to investigate the role and evolution of chromospheric reconnection in solar flares. Observations by \citet{2015ApJ...813...86I} and \citet{2017ApJ...851L...6R} indicate that both imaging and spectroscopy can be used to infer the presence of plasmoids produced by the onset of the tearing mode instability, but accompanying magnetic field measurements are critical to help confirm the magnetic nature of the islands.

\subsubsection{Wave damping by ion-neutral friction as a possible cause of flare chromosphere heating, wight-light flares, and sunquakes}\label{OP7.5.2}

\vspace{-2ex}
See table on page \pageref{OP7.5.2table}.

According to the thick-target model, the heating of the low atmospheric layers and the consequent chromospheric evaporation during flares is caused by particles that are accelerated at the reconnection site and then travel towards the lower atmosphere. However, there are still some open questions concerning this mechanism, like for instance the large number of particles that should be involved, when compared to coronal densities. 

Taking into account that Alfv\'en waves produced during reconnection can deliver concentrated Poynting flux to the chromosphere, the role of MHD waves during flares has been investigated \citep{2013ApJ...765...81R, 2016ApJ...818L..20R}. In particular, simulations of Alfv\'enic waves propagating from the corona to the chromosphere have shown that these waves can be damped due to ion-neutral friction and that the energy lost by these waves can heat the chromosphere and contribute to the evaporation of the chromospheric plasma. The damping of the Alfv\'en waves in the chromosphere strongly depends on the wave frequency and for periods of 1 s or fractions of s, an amount of energy in the range 37 \% - 100 \% that enters the chromosphere can be damped by ion-neutral friction. Moreover, if coronal waves are trapped in closed coronal structures, the total transmission results to be higher, due to multiple incidences.

These simulations also indicate that the propagation of Alfv\'en waves during flares could be important to explain WL flares (they in fact provide a direct mechanism for heating the temperature minimum region, where the bulk of white light flare emission is formed) and sunquakes (waves with periods of several seconds can enter the chromosphere and pass undamped into the solar interior). 

Because the atmospheric response is nearly identical, the heating due to Alfv\'en waves can appear extremely similar to that caused by electron beams. Therefore up to now it has not been possible to distinguish observationally these two mechanisms. 

 %\newpage \input{SG10_tables}
%-----------------------------------------------------------
\section{Scattering physics and Hanle-Zeeman diagnostics}
\label{sec_scattering}
{Authors: Luca Belluzzi, Alex Feller,  Rafael Manso Sainz, Javier Trujillo Bueno, Franziska Zeuner}
%\section{Scattering physics and Hanle-Zeeman diagnostics}
%\label{sec_scattering}
% -----------------------------------------------------------------------------
% Introduction
% -----------------------------------------------------------------------------
% Inputs from:
% Alex 2017-11-07 (mail)
% Alex 2018-02-06
% Luca 2018-03-22
% Luca 2018-05
%
% 2018-11-06 Alex: editorial changes
% 2018-11-14 Luca: editorial changes
% 2018-11-21 Alex: editorial changes (entire document)
%				   removed all colored text (comments, change suggestions)
%2019-12-9 Javier: some extra improvements
% -----------------------------------------------------------------------------

The great diagnostic potential of scattering polarization and the Hanle effect
is today widely recognized.
During the last decades, a series of novel diagnostic methods based on these
physical mechanisms have been developed and successfully applied, especially
for the investigation of the magnetism of the solar atmosphere in domains that
are not accessible through the standard Zeeman-effect techniques
\citep{trujillo+2006,trujillo+17,casini+08,trujillo2009,trujillo10,casini+17}.

The birth and rapid growth of this new research field has certainly been
possible thanks to the technological advances, which allowed the
development of new polarimeters with increasingly high sensitivities
(presently up to $10^{-5}$ in the degree of polarization).
On the other hand, it is clear that in order to fully exploit the
potentialities of these new diagnostic methods, a deep and solid understanding
of the physics of scattering polarization is necessary.
Unfortunately, the theoretical modeling of scattering polarization has turned
out to be a very complicated task. A solid description of matter-radiation
interaction within the framework of Quantum Electrodynamics is actually
necessary, a series of atomic physics aspects need to be taken into account,
and efficient numerical techniques for the solution of the radiative transfer
(RT) problem in non-local thermodynamic equilibrium (NLTE) conditions, in the
presence of polarization phenomena (NLTE problem of the 2$^{\rm nd}$ kind),
need to be applied.

The clearest manifestation of scattering polarization is certainly the
so-called Second Solar Spectrum, namely the linearly polarized spectrum of
the solar radiation coming from quiet regions close to the limb
\citep{stenflo+al1997,gandorfer2000,gandorfer2002,gandorfer2005}.
This spectrum is rich of signals and spectral details of double scientific
interest: on the one hand, they encode a wealth of information about the
physical conditions present in the solar atmosphere. On the other hand, the
physical origin of several of these signals remains unclear, and they thus
represent precious observational signatures (often not reproducible in
laboratory plasmas) of physical phenomena that still need to be completely
understood.
The Second Solar Spectrum is in fact a precious window also for improving our
understanding of the physics of scattering polarization, and since many
years it represents a key test-bench for the theories of the generation and
transfer of polarized radiation.
The observation of many details of the Second Solar Spectrum requires rather
high polarimetric sensitivities, which today can generally be achieved only
by completely sacrificing the spatial and temporal resolution.
With a 4~m aperture telescope like EST, it will finally be possible to observe
the spectral details of the Second Solar Spectrum with an unprecedented spatial
and temporal resolution, thus further increasing the potentialities of
this spectrum as a diagnostic tool, and possibly revealing new aspects of the
physics of scattering.

In this chapter, we present a series of observing programs (OPs) focused on a selection of spectral lines that produce scattering polarization signals of particular interest, either because of their diagnostic potential for the investigation of the properties of the solar atmosphere, or because their physical origin is still unclear. The OPs are specifically designed so as to exploit the specific advantages of the EST, in particular the combination of high polarimetric sensitivity and high spatio-temporal resolution.
The OPs are arranged on the basis of their main scientific goal as follows:
\begin{enumerate}
	\item{{\it Investigation of the small-scale magnetism of the quiet solar
		photosphere via the Hanle effect in atomic and molecular lines.}
		\begin{itemize}
			\item{OP~\ref{sec_sri}: Spatial fluctuations of scattering
				polarization in \ion{Sr}{i} 4607\,\AA}
			\item{OP~\ref{sec_molecules}: Spatial fluctuations of scattering
				polarization in C$_2$ molecular lines around 5140\,\AA}
			\item{OP~\ref{sec_tii}: Spatial fluctuations of scattering
				polarization in the Ti~{\sc i} multiplet around 4530\,\AA}
			\item{OP~\ref{sec_simul}: Simultaneous observations in \ion{Sr}{i}
				4607\,{\AA} and C$_2$ around 5140\,\AA}
		\end{itemize}
		}
	\item{{\it Investigation of the magnetism of the chromosphere via the
		combined action of Hanle, Zeeman, and magneto-optical effects in
		strong resonance lines.}
		\begin{itemize}
			\item{OP~\ref{sec_cai}: Ca~{\sc i} 4227\,{\AA}} resonance line
			\item{OP~\ref{sec_caiiHK}: Ca~{\sc ii} K and H (3934, 3968\,{\AA})} resonance lines
			\item{OP~\ref{sec_caiiIR}: Ca~{\sc ii} IR triplet (8498, 8542,
				8662\,{\AA})}
		\end{itemize}}
	\item{{\it OPs of interest for deepening our understanding of the physics
		of scattering polarization}
		\begin{itemize}
			\item{OP~\ref{sec_nai}: Na~{\sc i} D$_2$ and D$_1$ (5890,
				5896\,{\AA})} lines
		\end{itemize}}
\end{enumerate}
Some of the above-mentioned scientific goals have already been addressed in
previous sections of this document.
The reason for gathering in a separate section all the OPs focused on
scattering polarization signals is that they involve common dedicated
observational and instrumental requirements.
These include:
\begin{itemize}
	\item{increased polarimetric sensitivity (generally higher than what is
		needed for Zeeman measurements);}
	\item{interest to perform observations closer to the limb (or even off-
		limb), where the use of adaptive optics (AO) is generally more difficult due to the
		reduced continuum intensity contrast;}
	\item{observations in particular spectral lines and molecular bands,
		which are normally not used for Zeeman diagnostics;}
	\item{interest for the blue and near-UV region of the solar spectrum, where
		the scattering amplitudes are typically larger;}
	\item{simultaneous observation of lines with different sensitivities
		to the Hanle effect (differential Hanle effect techniques).}
\end{itemize}

% 2018-11-06 Alex: editorial changes; added concluding sentence

\subsection{Investigation of the small-scale magnetism of the quiet solar
photosphere}
\label{sec_hidden_mag}
% -----------------------------------------------------------------------------
% Introduction
% -----------------------------------------------------------------------------
% Inputs from:
% Alex 2018-01-26
% Luca 2018-11-14
% -----------------------------------------------------------------------------
The quiet solar photosphere is permeated by small-scale ($<$ 1 arcsec)
magnetic fields, which interact strongly with the turbulent convection and
typically evolve on a short time scale of tens of seconds, like the solar
granulation.
The individual quiet Sun magnetic structures are small in extent, compared to
active regions, but ubiquitous on the Sun, and are therefore thought to play,
in their entirety, a major role in solar activity and in the energetic coupling
of the various height layers of the solar atmosphere.
Some fundamental questions are related to these fields, like the existence and
influence of a small-scale dynamo driven by the turbulent motions, or the
question about the total amount of magnetic flux on the Sun.
Multiple approaches can be taken to answer these questions making use of the
fact that the Zeeman and Hanle effects probe magnetic fields in a complementary
way.
The Hanle effect, in particular, being sensitive to tangled magnetic fields
showing opposite polarities below the resolution element (to which the Zeeman
effect is practically blind), is a key tool for investigating
this aspect of the solar magnetism \citep{stenflo1982,faurobert+al1995,trujillo+2004}.

One of the main difficulties of Hanle effect diagnostics is that the amplitude
of a given scattering polarization signal depends on several factors, the most
important one being the symmetry properties of the pumping radiation field.
Such symmetry properties, on the other hand, strongly depend on the thermal
and dynamic structure of the solar atmosphere, and they are very sensitive to the presence of horizontal inhomogeneities in the solar atmospheric plasma. As this kind of information is not 
known a priori, it is necessary to develop suitable techniques to disentangle the impact of the
magnetic field from that of other ``symmetry-breaking'' causes.

There are basically two strategies to overcome this difficulty.
The first one is to calculate theoretically the amplitude of a given scattering
polarization signal by solving the radiative transfer problem,
both taking into account and neglecting the magnetic
field, using three-dimensional (3D) numerical models of the solar atmosphere \citep{trujillo+2004}.
The computational cost of 3D radiative transfer investigations is very significant,
but they allow us to clearly identify and
disentangle the impact of the various physical mechanisms affecting the observed
scattering polarization, and in particular to obtain the zero-field
reference signal.
This method is necessarily model-dependent, but thanks to the availability
of increasingly detailed and realistic simulations of the solar atmosphere,
it is today a reliable approach.
One of the most suitable and exploited spectral lines for this kind of
diagnostics is the \ion{Sr}{i} 4607\,{\AA} line. An OP focused on this line is described in Sect.~\ref{sec_sri}.

A second approach is the so-called {\it differential Hanle effect technique}
\citep{stenflo+1998,trujillo2003}. This approach consists in considering two (or
more) spectral lines having similar formation properties but different
sensitivities to the Hanle effect, and to consider amplitude ratios of the
corresponding scattering polarization signals.
Indeed, if the lines form under the same conditions, the dependence on
several critical parameters cancels out in the ratios, and these latter can
be used to directly infer information on the magnetic field, without the need
of a theoretical zero-field reference signal.
Spectral lines that are particularly suitable for the application of this
technique are those produced by molecules, such as C$_2$ and MgH \citep{trujillo2003}.
These are the main advantages:
\begin{itemize}
	\item{molecular lines are generally weak in the intensity spectrum and
		their radiative transfer modeling is relatively simple (which
		simplifies the application of the differential Hanle effect technique);}
	\item{molecular lines pertaining to the same branch have very similar
		formation properties and fall within a rather small frequency interval
		(they can be easily observed simultaneously);}
	\item{molecular lines pertaining to the same branch may have significantly
		different Land\'e factors, and therefore very different Hanle effect
		sensitivities;}
	\item{interestingly, several molecular lines produce clear scattering
		polarization signals in the Second Solar Spectrum.}
\end{itemize}
In the past decades, differential Hanle effect diagnostics has been carried
out exploiting various sets of molecular lines. Particular attention has
been paid to lines of C$_2$ \citep{trujillo2003,trujillo+2004,berdyugina+2004,kleint+2010,kleint+2011}, MgH \citep{asensio+2005}, and CN
\citep{shapiro+2011}.
An OP focused on the C$_2$ lines at 5140\,{\AA} is described in
Sect.~\ref{sec_molecules}

The differential Hanle effect technique can of course be applied also to
atomic lines. The main disadvantage with respect to molecular lines is that
atomic lines are generally stronger and their radiative transfer modeling is
more complicated. A particularly nice example of a set of atomic lines suitable
for differential Hanle effect diagnostics is a titanium multiplet at around
4530\,{\AA} \citep{manso+2004}.
An OP dedicated to this multiplet is described in Sect.~\ref{sec_tii}.

The possibility of performing differential Hanle effect diagnostics has been generally limited by the necessity of considering lines falling within the single and relatively narrow spectral windows of today's solar spectrographs. The new multi-wavelength capabilities of the EST instrumentation will significantly extent the wavelength range for differential diagnostics and thus increase the potentialities of this technique.

Another important aspect that has to be taken into account when applying
Hanle effect diagnostics is the impact of depolarizing collisions with neutral
perturbers (hydrogen and helium atoms).
Such collisions sensibly affect the amplitude of the scattering polarization signals, and must be carefully taken into account.
This is especially true in the lower layers of the solar atmosphere, where
the plasma density is higher.
Detailed theoretical calculations of the collisional rates for the \ion{Sr}{i}
4607\,{\AA} line are today available
\citep[e.g.][]{faurobert+al1995, manso+2014}.
The availabilty of precise estimates of these rates is crucial because relatively small variations of these quantities can have a clear impact on the amplitude of the synthetic scattering polarization profiles, and therefore on the inferred magnetic field \citep[see][]{delPino+18}. Unfortunately, the \ion{Sr}{i} line is a kind of exception, and in most cases only very rough estimates of the depolarizing rates are available.
Differential Hanle diagnostics can be suitably exploited to reduce the uncertainties related to collisional depolarization, especially concerning the number density of perturbers. This quantity, which is not known a priori, cancels out when combining scattering polarization signals observed in different spectral lines with similar formation properties. Such a multi-line observing technique applied with the EST will thus allow to reduce the uncertainties related to the interpretation of Hanle depolarization.

The Hanle effect has been successfully applied in the past for investigating
the small-scale magnetism of the solar photosphere \citep[e.g.,][and references
therein]{trujillo+2006}.
However, all such works were based on observing data with very low spatial and
temporal resolution.
On the other hand, sizable variations of the amplitude of the scattering
polarization signals of photospheric lines are expected at sub-granular spatial scales
\citep[see][]{trujillo+al2007,delPino+18}.
Detecting such fluctuations would allow us to deepen our understanding of the
various mechanisms affecting scattering polarization and, at the same time, to study the small-scale magnetic fields of the intergranular
plasma.
This is a typical goal for a 4m-class telescope like EST, and it is the basic
objective of the OPs described in the following.

% 2018-11-06 Alex: OP table revised; editorial changes in text

\subsubsection{Spatial fluctuations of scattering polarization in
\ion{Sr}{i} 4607\,\AA}
\label{sec_sri}\label{OP8.1.1}

\vspace{-2ex}
See table on page \pageref{OP8.1.1table}.

% -----------------------------------------------------------------------------
% Input from:
% Alex, 2018-01-26
% Franziska, 2025-12-12: added recent publications and achievements
% -----------------------------------------------------------------------------

The goal of this observing program is to study spatial fluctuations of the
scattering polarization signal of the \ion{Sr}{i} 4607.4\,{\AA} line, on
subgranular spatial scales.

In observations with low spatio-temporal resolution, the \ion{Sr}{i} line shows
a relatively large scattering polarization amplitude
\citep[above 1{\%} at $\mu=0.1$, e.g.][]{stenflo+al1997,gandorfer2002,Zeuner+2022}.
Further, theoretical predictions \citep{trujillo+al2007, delPino+18,DelPino+2021} anticipate
spatial fluctuations of the polarization signal in the order of several 0.1\%,
depending on limb distance and resolution of the instrument.
The latter has a large effect on the signal amplitudes and so far observations
are just at the limit of the critical combination of high spatial resolution
and increased polarimetric sensitivity \citep[e.g.,][]{Bianda+18, Zeuner+18, Zeuner+2020}.
%(see Table below).
The EST with its large telescope aperture will overcome those limitations and
is therefore ideally suitable for this type of measurements \citep[for a demonstration of the scientific potential of large aperture telescopes for this line, see][]{Zeuner+2025}.
 A key concept of this observing program is to refrain from strictly diffraction limited
observations but to choose an optimum slightly larger spatial and temporal
sampling, in order to reach the required signal-to-noise ratio in the best
possible way, in the presence of solar evolution (cf.
Appendix~\ref{sec_optres}).

% 2018-11-07 Alex: OP table revised

\subsubsection{Spatial fluctuations of scattering polarization in C$_2$
molecular lines at 5140\,{\AA}}
\label{sec_molecules}\label{OP8.1.2}

\vspace{-2ex}
See table on page \pageref{OP8.1.2table}.

Molecular lines of particular interest for differential Hanle effect
diagnostics are the C$_2$ lines of the Swan system (d $^3\Pi_u$ - a $^3\Pi_g$),
with their R and P branches \citep{trujillo2003,trujillo+2006}. Of particular
interest are those around 5140\,{\AA}, proposed by \citet{berdyugina+2004}.
In particular, the three lines of the R branch are perfectly resolved and free
of blends with other lines. They produce weak $(10^{-3})$ but clear
signals in the Second Solar Spectrum \citep[see][]{gandorfer2002}, and are
characterized by the following critical fields for the onset of the Hanle
effect (hereafter Hanle critical field): ${\rm B_H}{\approx}4$\,G for the
${\rm R_1}(J=14)$ and ${\rm R_3}(J=12)$ lines and ${\rm B_H}{\approx}40$\,G
for the ${\rm R_2}(J=13)$ line. Their drawback with respect to the C$_2$
lines with $J>20$ used by other authors, namely that a line with
${\rm B_H}{\approx}40$\,G cannot really be considered as an ideal reference
line given the significant magnetization of the quiet Sun photosphere, is
compensated for by the fact that the ${\rm R_2}(J=13)$ and ${\rm R_1}(J=14)$
lines are not blended. These lines have been extensively used for
synoptic programs aimed at monitoring possible long-term variations of weak
photospheric magnetic fields with the solar cycle \citep{kleint+2010,kleint+2011}. For a review, see \citep{2019ASPC..526..283R}.

Interestingly, the average magnetic field intensity inferred by means of
C$_2$ molecular lines is sensibly smaller than that revealed by the Hanle
effect in the Sr~{\sc i} 4607\,{\AA} line.
A possible solution of this apparent disagreement was proposed by
\citet{trujillo2003}, who pointed out from numerical simulations that the
scattering polarization signals produced by weak molecular lines should mainly
come from the upflowing granular cell centers, while both granular and
intergranular regions should contribute to the signal of the Sr~{\sc i} line
\citep[see also][]{trujillo+2004}. This further stresses the interest of
observing scattering polarization in molecular and atomic lines to detect
magnetic fields below the spatial resolution limit of the EST, where Zeeman
diagnostics is known to suffer from potential signal cancellation. Apart from
this particular aspect, the general detection of fluctuations in the scattering
polarization signals produced by the above-mentioned C$_2$ lines at 5140\,{\AA}
is the main goal of this OP.

%% 2018-11-08 Alex: OP table revised; editorial changes in text

\subsubsection{Spatial fluctuations of scattering polarization in the \ion{Ti}{i} multiplet around 4530\,\AA}
\label{sec_tii}\label{OP8.1.3}

\vspace{-2ex}
See table on page \pageref{OP8.1.3table}.

% -----------------------------------------------------------------------------
% Input from:
% Rafa, 2018-01-29
% -----------------------------------------------------------------------------

Some of the most remarkable features of the Second Solar Spectrum
are the conspicuous spectral patterns produced by relatively low abundance atomic species (rare earths, transition metals --- other than Fe) and molecules.
This is important because it offers the possibility of studying tangled magnetic fields through the Hanle effect, at the lowest layers of the photosphere (e.g., \ion{Ce}{ii} forms barely above the continuum), just where magnetoconvection simulations can be best constrained.

\ion{Ti}{i} is an important case in point; there are about 20 lines in the
visible showing relatively large scattering polarization well above the
continuum level. Several of them correspond to the singlet system of
\ion{Ti}{i}. They show some of the largest signals and are also relatively
simple to model, which simplifies their analysis and the possibility of
inversion. On the other hand, there are many multiplets from the quintet
system. These are interesting mainly because lines of the same multiplet can
be modeled similarly and consistently, which offers the possibility of
differential analysis among different lines.
Further, some of these multiplets lie on a relatively narrow spectral region,
which allows observing if not all, at least many of them simultaneously.
Probably, the most interesting case in this regard is the multiplet
$a\, {^5 \! F} - y \, {^5 \! F^{\rm o}}$ around 453\,nm, which has a line
(453.6\,nm) which is completely unaffected by magnetic fields (the Land\'e
factor of both levels is zero), and serves as a reference for Hanle effect
depolarization, while 8 lines of the same multiplet lie within less than 1\,nm,
which allows observing all of them simultaneously \citep[see][]{manso+2004}.

\subsubsection{Simultaneous observations in \ion{Sr}{i} and C$_2$}
\label{sec_simul}

The OPs described above are designed for observations in an individual line (\ion{Sr}{i}) or in groups of lines located within a given spectrograph FOV (C$_2$, \ion{Ti}{i}). The advantage of individual observations is the higher throughput of the instrument. For some science cases involving differential diagnostics we also want to perform observations in the C$_2$ molecular band and in \ion{Sr}{i} simultaneously. In this case the requirements in terms of spectral resolution, FOV and sensitivity remain the same as in the case of individual observations. We accept a degradation of the spatio-temporal resolution though, as an inevitable consequence of the reduced instrument throughput. According to the EST photon budget the throughput will be decreased by about a factor 4 compared to individual observations, which results in a degradation of the optimum spatial sampling and cadence by about a factor $4^{1/3} \approx 1.6$ respectively (cf. Appendix~\ref{sec_optres}). The coarser sampling will be achieved by a post-facto numerical binning of the data and thus does not impose any additional requirements on the instrumentation.

% 2018-11-15 Alex: editorial changes in text

\subsection{Investigation of the magnetism of the solar chromosphere}
% -----------------------------------------------------------------------------
% Introduction
% -----------------------------------------------------------------------------
% Input from:
% Javier, 2018-02-06
% -----------------------------------------------------------------------------
The information on the magnetic field of the solar chromosphere is encoded in
the polarization of the spectral line radiation that originates in a highly
inhomogeneous and dynamic atmospheric region, where hydrodynamic and
magnetic forces compete for dominance. Relatively few spectral lines whose core
originates in the chromosphere can be observed with ground-based telescopes.
Here we focus on the strongest lines of Ca~{\sc i} and Ca~{\sc ii} and, in
particular, on the diagnostic potential of their polarization signals produced
by the joint action of scattering processes and the Hanle and Zeeman effects.

As mentioned above, the rarefied plasma of the solar chromosphere is very
dynamic and highly inhomogeneous, and the magnetic field is thought to play a
key role on the formation, evolution, and destruction of the observed plasma
structures. We are still far from understanding the physics of the solar
chromosphere, and precisely for this reason we need EST. This telescope will
provide novel observations of the intensity and polarization of solar spectral
lines, with unprecedented spatial, temporal, and spectral resolution.
By confronting the Stokes profiles observed in chromospheric lines with those
calculated in increasingly realistic numerical models of the solar chromosphere
we may hope to achieve new breakthroughs in solar physics. By the time EST will
be operative, such three-dimensional (3D) numerical models of the solar
chromosphere will be much more realistic than nowadays, given that they will
include a number of possibly key physical ingredients such as non-equilibrium
ionization and ambipolar diffusion (interactions between ions and neutrals in
the presence of magnetic fields).
Likewise, ongoing synergistic efforts suggest that by the time EST will be
available, we will have completed the development of a computer program capable
of solving with massively parallel computers the (NLTE) problem of the
generation and transfer of polarized radiation in such 3D models of the solar
atmosphere, taking into account frequency correlation effects between
the incoming and outgoing photons in the scattering events, as well as the
joint action of the Hanle and Zeeman effects.
Obviously, in order to validate or discard such numerical models we need to
confront measured and calculated observable quantities sensitive to the
thermal, dynamic, and magnetic time-dependent structure of the solar
chromosphere (i.e., the Stokes profiles of chromospheric spectral lines). 
Comparing only the intensity of the spectral line radiation is not
sufficient, since the Stokes $I$ profile of spectral lines is practically
insensitive to the strength and structure of the magnetic field.

% -----------------------------------------------------------------------------

%% 2018-11-15 Alex: OP table revised; editorial changes in text

\subsubsection{The Ca~{\sc i} 4227\,\AA\ resonance line}
\label{sec_cai}\label{OP8.2.1}

\vspace{-2ex}
See table on page \pageref{OP8.2.1table}.
% -----------------------------------------------------------------------------
% Input from:
% Javier, 2018-02-06
% Franziska, 2025-12-12: updated most recent publications
% -----------------------------------------------------------------------------

Outside sunspots, this spectral line shows very sizable linear polarization
signals produced by scattering processes, of the order of a few percent when
observing close to the limb \citep[e.g.,][]{gandorfer2002} and of the order of
0.1\% when observing moderately magnetized regions at the solar disk center
\citep[e.g.,][]{Bianda+11}. Therefore, it is a very suitable line for observing
with EST its scattering polarization with unprecedented spatial and temporal
resolution.

A correct modeling of the linear polarization produced by scattering processes
in this line requires taking into account the effects of correlations between
the frequencies of the incoming and outgoing photons in the scattering events 
\citep[for advances in modeling, see][]{Janett+2021,Jaume+2021,Belluzzi+2024}.
For lines of sights (LOS) pointing to quiet regions located away from the solar
disk center, such partial frequency redistribution (PRD) effects produce in the
Ca~{\sc i} 4227\,\AA\ line a (typically) triple-peak fractional linear
polarization pattern with a central peak and sizable lobes in the blue and red
wings of the line.

In semi-empirical models of the solar atmosphere the height where the
line-center optical depth is unity varies between about 900~km (disk center
LOS) and 1200~km (close to the limb LOS), while the radiation of the extended
wings of this line stems from increasingly deeper layers when going from the
line center to the far wings (e.g., at ${\pm}0.5$\,\AA\ from line center the
height where the ensuing optical depth is unity is about 200~km for a close
to the limb line of sight).

Only one of the stable isotopes of calcium has non-zero nuclear spin, but its
abundance is only 0.135\%. Therefore, we can assume that hyperfine structure
is a negligible physical ingredient for modeling the scattering polarization
of calcium lines. The Ca~{\sc i} 4227\,\AA\ resonance line results from a
transition with a lower level with angular momentum $J_l=0$ and and upper
level with $J_u=1$, and the Hanle critical field in this line is $B_H=25$~G.
Fortunately, a two-level model atom is a suitable approximation for modeling
its polarization. This is good news, especially because the radiative transfer
numerical modeling of lines for which PRD phenomena are important is complex
and time-consuming. Is also good news that depolarization by elastic
collisions with neutral hydrogen atoms is practically negligible.

Theoretical developments have shown that the polarization of this line,
whose core originates in the lower chromosphere, is due to the following
physical mechanisms \citep{Alsina+18,Capozzi+2020}:

\begin{itemize}

\item the familiar Zeeman effect, which can produce measurable circular
	polarization for longitudinal field components of at least 10 G.

\item the linear polarization caused by scattering processes in the line-core,
	along with its modification by the Hanle effect (which operates in the line
	core). Therefore, this line-core polarization is sensitive to the magnetism
	of the lower solar chromosphere.

\item the linear polarization caused by scattering processes and PRD effects
	in the line wings, along with its magnetic sensitivity through the
	$\rho_V$ magneto-optical (MO) terms of the Stokes-vector transfer equation.
	Such MO terms produce sizable $U/I$ wing signals as well
	as a sensitivity of both the $Q/I$ and $U/I$ wings to the
	presence of magnetic fields in the region of formation of the line wings
	(the photosphere). Magnetic fields with strengths as low as 10~G are
	sufficient to produce a measurable impact on the wings of both $Q/I$ and
	$U/I$.

\end{itemize}

These mechanisms have been demonstrated to be inferable with the inversion code 
of \citet{Janett+2025} and applicable to observations (Janett et al. 2026, in prep.).

In summary, the Ca~{\sc i} 4227\,\AA\ resonance line is of high scientific
interest for EST because

\begin{itemize}

	\item scattering processes produce very significant linear polarization
	signals, which are sensitive to the magnetization of the lower solar
	chromosphere (via the Hanle effect in the line core) and of the underlying
	photosphere (via the joint action of PRD and MO effects). Magnetic fields
	as weak as 10 G are sufficient to have a measurable impact on such linear
	polarization signals.

%\item EST will be able to map such linear polarization signals with a spatial
%	resolution better than 1~arcsec and a temporal resolution of the order of
%	10~seconds. This has never been achieved before.

\item the confrontation of the Stokes profiles observed by EST with those
	calculated by solving the problem of the generation and transfer of
	polarized radiation in 3D numerical models (achieved in this line by 
	Benedusi et al. 2025, submitted) of the solar atmosphere, will
	allow us to probe the thermal, dynamic, and magnetic structure of the lower
	solar chromosphere and its coupling with the underlying layers.

\item with the telescopes that are presently available spectropolarimetric
	observations of this line have always required to seriously sacrifice the
	spatio-temporal resolution.

\end{itemize}

%% 2018-11-16 Alex: OP table revised; editorial changes in text

\subsubsection{The Ca~{\sc ii} H \& K resonance lines}
\label{sec_caiiHK}\label{OP8.2.2}

\vspace{-2ex}
See table on page \pageref{OP8.2.2table}.
% -----------------------------------------------------------------------------
% Input from:
% Javier, 2018-02-06
% -----------------------------------------------------------------------------

The Ca~{\sc ii} H \& K resonance lines at 3934\,\AA\ and 3969\,\AA\ are the
strongest chromospheric lines that can be observed from ground-based
facilities.
In fact, in quiet regions of the solar atmosphere most of the calcium atoms
are in the form of Ca~{\sc ii}. In models of the quiet solar atmosphere the
height where the line-center optical depth is unity is only about
200--300\,km below the corrugated surface that delineates the
chromosphere-corona transition region.
Therefore, especially for close to the limb LOS the line-center radiation of
these resonance lines stems mainly from the upper chromosphere.

The PRD effects are essential for understanding the Stokes profiles of these
resonance lines. A rigorous modeling of their spectral line radiation requires
using at least a 5-level model atom, with the H \& K resonance lines and the
IR triplet of Ca~{\sc ii}.
An additional important physical ingredient is quantum mechanical interference
between the sublevels pertaining to the upper level of the K line, which has
angular momentum $J_u=3/2$, and those of the upper level of the H line whose
$J_u=1/2$. In spite of the fact that the H and K lines are separated by
35\,\AA, in the (optically thick) plasma of the solar atmosphere such $J$-state
interference effects produce observable signatures in the scattering fractional
linear polarization pattern, especially in their wings \citep{Stenflo80}.
In semi-empirical models of the quiet solar atmosphere the $Q/I$ pattern shows
positive peaks in the blue wing of the K line and in the red wing of the H line
(of the order of 1\%, but with the K blue peak larger than the H red peak) and
a negative peak between the K and H lines, in qualitative agreement with the
observations of \citet{Stenflo80}. Such wing signals are caused by the joint
action of PRD and $J$-state interference, exactly as it occurs with the
theoretical $Q/I$ pattern of the Mg {\sc ii} $h$ \& $k$ lines
\citep[see][]{belluzzi+2012}.

In these resonance lines, the Hanle effect operates only in the core of the K
line (critical Hanle field $B_H{\approx}12$\,G). On the other hand, given that
the joint action of PRD and $J$-state interference produce sizable signals in
the wings of the $Q/I$ profiles, the MO terms of the
Stokes-vector transfer equation already introduced in section \ref{sec_cai}
can produce significant signals in the wings of the $U/I$ pattern and an
interesting magnetic sensitivity in the wings of both $Q/I$ and $U/I$ 
\citep{2025A&A...704A.173J}, exactly as it happens with the wings of
the Mg~{\sc ii} k line \citep{Alsina+16}, with the wings of the Mg~{\sc ii}
h \& k lines \citep{delPino+16}, with the wings of the Ca~{\sc i} 4227\,\AA\
line \citep{Alsina+18}, and with the wings of the hydrogen Ly-$\alpha$ line
\citep{2019ApJ...880...85A}.

The photons of the far wings of the H \& K lines stem from the solar
photosphere, and via the MO effects photospheric fields as low as 20\,G can
therefore have a significant impact in the wings of $Q/I$ and $U/I$.
The same happens with the scattering polarization wings of the Mg~{\sc ii}
h \& k lines \citep{Alsina+16,delPino+16}.
Finally, it should be noted that the circular polarization is dominated by
the familiar Zeeman effect, which can produce measurable signals for
longitudinal field components of at least 10\,G.

In summary, the Ca~{\sc ii} resonance lines are of high scientific interest
for EST because

\begin{itemize}

\item Their line-core radiation stems from the upper solar chromosphere,
	only about 200--300\,km below the corrugated surface
	that delineates the chromosphere-corona transition region.

\item The line-core of the K line is sensitive to the Hanle effect. Therefore,
	it reacts to the presence of magnetic fields in the upper solar
	chromosphere. The critical Hanle field of the K line is 12\,G, so we may
	expect good sensitivity to field strengths between 2 and 50\,G,
	approximately.

\item The wings of the fractional linear polarization profiles of the H and K
	lines are sensitive to the presence of magnetic fields as low as 20\,G.
	This magnetic sensitivity in the wings of the $Q/I$ and $U/I$ profiles,
	which extends all through the solar atmosphere, is caused by the MO terms
	$\rho_VQ$ and $\rho_VU$ of the transfer equations for Stokes $Q$ and
	$U$.

\item the familiar Zeeman effect can produce measurable circular polarization
	if the longitudinal field component is sufficiently large (e.g., for a
	longitudinal field of 10\,G the Stokes-$V$ amplitude of the Ca~{\sc ii} H
	line is about 0.2\%).

\item with the telescopes that are presently available spectropolarimetric
	observations of these lines have always required to dramatically sacrifice
	the spatio-temporal resolution.

\end{itemize}

\subsubsection{The Ca~{\sc ii} IR triplet}
\label{sec_caiiIR}\label{OP8.2.3}

\vspace{-2ex}
See table on page \pageref{OP8.2.3table}.

The strongest line of the Ca~{\sc ii} IR triplet is the 8542\,\AA\ line,
followed by the 8662\,\AA\ and 8498\,\AA\ lines. Together, they encode
information over a significant range of heights in the middle solar
chromosphere.
\cite{MansoTrujillo03} demonstrated quantitatively that the physical origin of
the enigmatic scattering polarization observed by \cite{Stenflo+00} in the
solar disk radiation of the Ca~{\sc ii} lines at 8662\,\AA\ and 8542\,\AA\ is
``zero-field dichroism" (i.e., differential absorption of polarization
components caused by the presence of a significant amount of atomic
polarization in their metastable lower levels).
The scattering polarization of the 8498\,\AA\ line, the weakest of the triplet,
has contributions from the atomic polarization in its upper and lower levels.

The magnetic sensitivity of the scattering polarization in the Ca~{\sc ii} IR
triplet has been theoretically investigated by \cite{MansoTrujillo10}.
They showed that the linear polarization profiles produced by scattering in the
Ca~{\sc ii} IR triplet have thermal and magnetic sensitivities potentially of
great diagnostic value. The scattering polarization in the 8498\,\AA\ line
shows a strong sensitivity to inclined magnetic fields with strengths between
0.001 and 10\,G, while the scattering polarization in the 8542\,\AA\ and
8662\,\AA\ lines is mainly sensitive to magnetic fields with strengths between
0.001 and 0.1\,G. The reason for this peculiar behavior is that the scattering
polarization of the 8662\,\AA\ and 8542\,\AA\ lines, unlike the Ca~{\sc ii}
8498\,\AA\ line, is controlled mainly by the lower level Hanle effect.
Therefore, in regions with magnetic strengths sensibly larger than 1\,G, their
Stokes $Q$ and $U$ profiles are sensitive only to the orientation of the
magnetic field vector. \cite{MansoTrujillo10} also found that the sign of the
emergent Stokes $Q/I$ and $U/I$ profiles of the 8662\,\AA\ and 8542\,\AA\ lines
is rather insensitive to the chromospheric thermal structure, while the sign of
the linear polarization profiles of the 8498\,\AA\ line turns out to be very
sensitive to the thermal structure of the lower chromosphere. They concluded
that spectropolarimetric observations providing information on the relative
scattering polarization amplitudes of the Ca~{\sc ii} IR triplet could be very
useful to improve our empirical understanding of the thermal and magnetic
structure of the quiet chromosphere. The required radiative transfer
modeling for reliably inferring information from confrontations with future
high-spatial resolution spectropolarimetric observations is however more
complex. On the one hand, the line-core of the Ca~{\sc ii} IR triplet
originates in a shock-dominated region of the solar chromosphere and the
ensuing macroscopic velocity gradients may have a significant impact on the
anisotropy of the spectral line radiation and, therefore, on the shape and
amplitude of the observed scattering line polarization (\citealt{Carlin+12}).
On the other hand, the breaking of the axial symmetry of the pumping radiation
field caused by the three-dimensional thermal, dynamical and magnetic structure
of the solar chromosphere has to be taken into account
\citep{StepanTrujillo16}.

In general, the emergent $Q/I$ and $U/I$ profiles are produced by the joint
action of atomic level polarization, and the Hanle and Zeeman effects.
Atomic polarization and the Hanle effect dominate the emergent linear
polarization profiles for inclined magnetic fields with strengths weaker than
$B_0$, where the $B_0$ value depends on the scattering geometry. In the forward
scattering geometry of a disk center observation ($\mu=1$), the linear
polarization of the Ca~{\sc ii} IR triplet is dominated by the Hanle effect if
the magnetic field is weaker than about 10\,G. In fact, while the contribution
of the Zeeman effect to the linear polarization is negligible for
$0{<}B{<}10$\,G the Hanle effect creates weak but measurable fractional linear
polarization signals already for horizontal magnetic fields as low as 0.1\,G
\citep{MansoTrujillo10,StepanTrujillo16}.
For magnetic strengths $10{\lesssim}B{\lesssim}100$\,G the contribution of the
transverse Zeeman effect to the linear polarization observed at the solar disk
center should not be neglected. Detection of $Q/I$ and $U/I$ disk center
signals caused either by the Hanle effect alone (if $B{<}10$\,G) or by the
joint action of the Hanle and transverse Zeeman effects (if
$10{\lesssim}B{\lesssim}100$\,G) requires very high polarimetric sensitivity
together with a spatial and temporal resolution sufficient to resolve the
magnetic field azimuth. Finally, it is important to note that the circular
polarization of the Ca~{\sc ii} IR triplet is dominated by the Zeeman effect,
and that the weak field approximation that relates the Stokes $V$ profile with
the wavelength derivative of the Stokes $I$ profile can provide a reasonable
estimation of the longitudinal component of the magnetic field vector
\citep[e.g.,][]{Centeno18}.

%%\subsubsection{The O~{\sc i} triplet at 7770~{\AA}}
%%\label{sec_oi}

%% 2018-11-13 Alex: OP table revised; editorial changes in text

\subsection{Deepening our understanding of the physics of scattering
polarization}
% -----------------------------------------------------------------------------
% Introduction
% -----------------------------------------------------------------------------

\subsubsection{The physics and diagnostic potential of the \ion{Na}{i} D$_1$
and D$_2$ lines}
\label{sec_nai}\label{OP8.3.1}

\vspace{-2ex}
See table on page \pageref{OP8.3.1table}.

% -----------------------------------------------------------------------------
% Input from:
% Luca, 2018-02-06
% -----------------------------------------------------------------------------
The theoretical interpretation of the rich variety of signals of the Second
Solar Spectrum has played a key role in the development of new theoretical
approaches for the description of scattering polarization.
In particular, the interpretation of the signals produced by the \ion{Na}{i}
D$_1$ and D$_2$ lines at 5895.9\,{\AA} and 5890.0\,{\AA} respectively,
has been challenging scientists for more than twenty years.
Indeed, these signals show the signatures of several different physical
mechanisms, and not by chance they have always represented a key test bench
for the theory.

The \ion{Na}{i} D$_1$ and D$_2$ lines originate from the transitions between
the ground level of sodium, $^2 {\rm S}_{1/2}$, and the upper levels
$^2 {\rm P}^{\rm \, o}_{1/2}$ and $^2 {\rm P}^{\rm \, o}_{3/2}$, respectively.
Sodium has a single stable isotope, $^{23}$Na, which has hyperfine structure
(HFS; nuclear spin $I=3/2$).
The \ion{Na}{i} D-lines are among the strongest spectral lines of the visible
solar spectrum, and encode information on the physical properties of the solar
atmosphere ranging from the low chromosphere to the photosphere.
In semi-empirical models of the solar atmosphere, the line-core of these lines
forms at about 900~km for an observation at $\mu=1$, and at about 1200~km for
an observation at $\mu=0.1$.

In the Second Solar Spectrum, the D$_2$ line shows a peculiar triplet-peak
$Q/I$ profile. At $\mu=0.1$, the central peak reaches an amplitude of about
0.35\%, thus representing one of the largest signals of the Second Solar
Spectrum \citep[see][]{gandorfer2000}.
The triplet-peak structure of this signal is ultimately due to frequency
correlation effects between the incoming and outgoing photons in the scattering
processes: PRD effects thus need to be taken into account for modeling this
signal, as well as the depolarizing effect of HFS.
The central peak is sensitive to the Hanle effect, the critical field being
of the order of 5~G.

Besides the triplet-peak structure profile of the D$_2$ line, the scattering
polarization signal produced by the sodium doublet shows an overall pattern
across the two lines, with a sign reversal between D$_1$ and D$_2$, and an
anti-symmetric structure across D$_1$, with a negative dip in the blue wing,
and a positive bump in the red wing \citep{gandorfer2000}.
This overall pattern is the result of quantum interference between the upper
levels of D$_1$ and D$_2$ ($J$-state interference), and it is fully analogous
to the one observed across the H and K lines of \mbox{\ion{Ca}{ii}}.
Notably, these effects take place in the far wings of the spectral lines, where
the emissivity is extremely low. This explains why the signatures of $J$-state
interference cannot be observed in laboratory experiments, but only on the
Sun, where the low emissivity is compensated by the optical thickness of the
solar plasma.

\citet{stenflo+al1997} first observed a conspicuous scattering polarization
signal also in the core of D$_1$ ($Q/I \approx 0.15\%$ at 5~arcsec from the
limb).
This signal was totally unexpected, as this line is produced by a $1/2-1/2$
transition, and it was considered therefore as intrinsically unpolarizable.
A first possible explanation was proposed by \citet{Landi1998}, who could
reproduce the observed signal by taking PRD effects and HFS into account, and
by assuming that a substantial amount of atomic polarization was present in
the ground level of sodium.
However, as pointed out by the same author, the required amount of lower level
polarization is incompatible with the presence in the lower solar chromosphere
of inclined magnetic fields stronger than 0.01~G, in apparent contradiction
with the results obtained from other type of observations
\citep[e.g.][]{bianda+1998,stenflo+1998}.
This circumstance opened a sort of ``sodium paradox'', which is still
largely debated.
\citet{belluzzi_trujillo2013} and \citet{belluzzi+2015} proposed a
new mechanism that may explain the presence of such a signal, without the need
of atomic polarization in the lower level.
According to these works, the observed signals are ultimately due to small
variations of the anisotropy of the chromospheric radiation, over spectral
intervals as small as the separation among the HFS components of this line.
This picture seems to be in agreement with the results of new observations
carried out with the Zurich Imaging Polarimeter \citep[ZIMPOL,
see][]{ramelli+2010},  which have unequivocally confirmed the presence of
such signals, and shown a rich diversity of profiles.

Up to now, the polarimetric accuracy and spectral resolution necessary to
detect the polarimetric signals described above could only be reached by
completely sacrificing the temporal and spatial resolution of the observation.
The great challenge today is to observe these signals at high spatial and
temporal resolution, so as to understand whether and how they change depending
on the plasma structure that is observed.
This is a scientific target for which the use of a 4~m aperture telescope,
such as EST, is required.
Observing the sodium doublet by combining high polarimetric accuracy and increased spectral, spatial, and temporal resolution will be of great scientific interest, both for getting more insights about the physics of scattering polarization, and for investigating and possibly exploiting the diagnostic potential of the signals produced by these lines.
 %\newpage\input{SG7_tables}
%-----------------------------------------------------------

%-----------------------------------------------------------
\section{Solar science exploration between 350 and 400 nm}
\label{sec_nasmyth}
{Authors: S. Mathews, J. Leenaarts, L. Bellot Rubio, A. Feller, T. Riethmueller, M. Mathioudakis}%\footnote{(version: {2018-Jun-08})}
%\section{Nasmyth focus science}
%\begin{center}
%J. Leenaarts, L. Bellot Rubio, A. Feller, T. Riethmueller, M. Mathioudakis, S. Matthews
%\end{center}

\subsection{White-light emission from flares - Continuum diagnostics in
vicinity of Balmer jump}

White-light continuum emission in the visible and NUV carries a significant fraction of the solar flare
radiative energy losses. The white-light continuum is usually attributed to hydrogen recombination
radiation to n=2 and n=3 (Balmer and Paschen) and to a lesser extent negative hydrogen (H-) bound-free
and free-free emission with the local plasma conditions determining the processes that
dominate. Radiative hydrodynamic (RHD) simulations of stellar and solar flares have shown that the
320nm - 400nm wavelength range contains important diagnostics that are very sensitive to the flare
heating processes. If we consider flare energy transport by non-thermal electrons, the colour
temperature of the NUV continuum and Balmer jump ratio (364.6nm), are very sensitive to the non-thermal
electron flux. High resolution spectroscopy combined with RHD simulations can be used to
determine the photospheric/chromospheric flare electron density and allow us to disentangle the
atomic processes that contribute to form the observed shape of the flare white-light continua.
The ratio of the relative strength of the higher Balmer lines (i.e. Balmer decrement) can also be used
as a diagnostic for the optical depth, temperature and density of the flaring atmosphere.
Despite a wealth of stellar flare observations in the visible and NUV, solar flare observations in
vicinity of the Balmer jump are currently scarce \citep{2015ApJ...798..107K, 2016ApJ...820...95K}. This is
due to the lack of suitable solar instrumentation. A high-resolution spectrograph in the NUV will
allow us to access important line and continuum diagnostics and hence determine the atmospheric
conditions, dynamics and heating processes that lead to the formation of the white-light flare.

\subsection{Coronal forbidden lines in the visible}

Coronal forbidden lines offer many important diagnostic capabilities, including the potential to
measure magnetic fields for those in the NIR \citep[e.g., ]{2014LRSP...11....2P, 1998ApJ...500.1009J}. 
As highlighted in \citet{2018ApJ...852...52D} they also offer diagnostic capability in the form of plasma densities, flows
and line widths, with the latter in particular providing potential constraints for coronal heating
models, as well as the prospect of detecting the presence of non-Maxwellian electron distributions
\citep{2014A&A...570A.124D}, in combination with measurements in the UV or EUV.
The photospheric brightness is such that coronal lines in both visible and NIR wavelengths are most
readily detected above the limb with coronagraphs. However, there are many lines in the visible
range \citep[see Table 2 in ][]{2018ApJ...852...52D} including the Fe XIII line at 3388.1 A, for which
there has been a positive on disk detection during a flare on the M dwarfs CN Leo and LHS 2076,
with the line displaying high variability \citep{2003A&A...403..247F}, confirming the detection by
\citet{2001Natur.412..508S}. Indeed, Fuhrmeister et al. find variability of the Fe XIII outside of the
flare on timescales that they suggest may be consistent with microflaring.
Table 2 of Del Zanna \& DeLuca (2017) provides estimates of observed and predicted radiance for an
AR at the limb, that indicate the Ca XIII 4087 A, Ni XIII 5116 A, Ca XV 5696 A and Fe X 6374 A lines
have radiances of similar order of magnitude. The Fe XIV 5304 A line is an order of magnitude
brighter. A proper assessment of predicted on-disk intensities is needed, but a high-resolution
spectrograph could potentially allow the measurement of the variability of these coronal lines
(including measurement of velocities and line widths) during flares at the same time as
chromospheric and white-light measurements, allowing complete tracking of the energy throughout
the atmosphere.

\subsection{Ca II H\&K spectroscopy and spectropolarimetry} %\ion{Ca}{ii}\,H\&K spectroscopy and spectropolarimetry}

High resolution imaging spectroscopy taking with CHROMIS at the SST shows the great
potential of the \ion{Ca}{ii}\,H\&K lines 
\citep{2018A&A...612A..28L}\footnote{\verb| http://iopscience.iop.org/article/10.3847/2041-8213/aa99dd/pdf |}, providing high spatial resolution
owing to their short wavelength and great temperature sensitivity.
Spectropolarimetric measurements in the same lines\footnote{\verb| http://adsabs.harvard.edu/abs/1990ApJ...361L..81M |}
show Stokes V signal in sunspots, plage and
flares of the order of ~5\%-10\%, while quiet Sun signals appear weaker than ~0.2\%.
The lines have a higher opacity than Ca II 854.2 and 
Halpha\footnote{\verb| https://arxiv.org/pdf/1306.0671.pdf,https://arxiv.org/pdf/1712.01045.pdf |}, and thus provide
diagnostic information at larger heights in the chromosphere than any other line available from the
ground (maybe except for the He I 587.6 and 1083.0 lines).
The drawback of the lines are the low photon fluxes. At a total efficiency of 2\% (=EST coude room)
and R=5e4 one needs to integrate ~60 s to reach a S/N of 1000 in a critically sampled, diffraction-limited
pixel in the line core in the quiet sun.
Putting an instrument in the Nasmyth focus would increase the total efficiency, perhaps as high as
10\%-15\%, which would lead to a factor 5-7 shorter integration time. Sampling at half the diffraction
limit decreases integration time by another factor 4.
High resolution imaging requires an AO system.
A Fabry-Perot or 2D imaging spectropolarimeter or even a slit spectropolarimeter would provide a
unique window on the solar chromosphere. DKIST first light instruments will not have a narrow-band
imaging instrument in the \ion{Ca}{ii}\,H\&K lines, only a slit spectropolarimeter.

\subsection{Multi-line spectropolarimetry}

Our knowledge of the lower solar atmosphere is mainly obtained from spectropolarimetric
observations, which are often carried out in the red or infrared spectral range and almost always
cover only a single or a few spectral lines. In the short-wavelength range, below 4300\,\AA{}, the line
density but also the photon noise are considerably higher than in the red. This wavelength range
was measured during the third science flight of the balloon-borne solar telescope Sunrise. It is expected
to be a rewarding science target for the EST as well. As a first step, the spectropolarimetric
diagnostic potential of this spectral range has been studied based on simulations \citep{2019A&A...622A..36R}. For an
ensemble of state-of-the-art magneto-hydrodynamical atmospheres, exemplary spectral regions
around, 4080\,\AA{} (328 lines), and, for comparison, around 6302\,\AA{} (111 lines) have been
synthesised assuming a spectral resolving power of 150\,000 as a reasonable trade-off between
spectral FOV and spectral line resolution. The synthetic Stokes profiles are degraded with typical
photon noise and then inverted. In the 4080\,\AA{} region the simulations show longitudinal
Zeeman signal amplitudes up to 20\% and transversal Zeeman amplitudes of order 1-3\%. A
polarimetric sensitivity of order 0.1\% is thus recommended for Stokes Q, U. For Stokes V the
required noise level is comparatively lower, but the instrument design shall be driven by the
highest sensitivity requirement. The atmospheric parameters of the inversion are compared
with the original noise-free MHD quantities. We find that from many-line inversions significantly
more information can be obtained than from a traditionally used inversion of only a few lines. We
further find that information about the upper photosphere can be significantly more reliably
obtained at short wavelengths. In terms of sensitivity, in the mid and lower photosphere, the
many-line approach at 4080\,\AA{} provides equally good results than at 6302\,\AA{} for
the magnetic field strength and the line-of-sight (LOS) velocity, while the temperature
determination is more precise by a factor of three. We conclude from our first theoretical
modeling results that many-line spectropolarimetry at short wavelengths offers high
potential in solar physics.

%-----------------------------------------------------------
\newpage
\section{Tables for Observing Progammes}
In the observing programme tables, we distinguish between the following instruments:
\begin{enumerate}\item[(1)] \label{def_bbi}
BBI: Broad Band Imagers, which take images with exposure times shorter than 1 ms. Images are taken at a fixed wavelength and can be centerred at spectral continua or in spectral lines with pass bands as narrow as about 0.1\,nm.
\item[(2)] \label{def_nbi}
NBI: Narrow-Band Imagers, which scan a spectral range in wavelength. We assume that there will be three NBIs for three different wavelength regimes: blue visible, red visible, IR. For cost reasons, we assume a FOV of about 40 arcsec in diameter or 30 arcsec square side length.
\item[(3)] \label{def_sp}
SP: classical long-slit spectrograph, which scans the solar image. Slit length of 60 arcsec is assumed. Since IFUs are superior to SPs as image reconstruction techniques can be applied to their spectral images, only one SP from blue visible to IR is assumed.
\item[(4)] \label{def_ifu}
IFU: Integral Field Units, which record spectrum and image simultaneuosly for a relatively small FOV of about 10 arcsec. We assume that IFUs are available for the spectral range from 392nm to 1600 nm. At this point of development it is unknown whether micro-lense systems (which are restricted to short wavelength ranges) or reflective image slicers (which can cover large wavelength regimes) are to be preferred.
\end{enumerate}

All instruments, except for BBIs, are assumed to be operated in polarimetric mode, unless stated otherwise.

In this section we list the tables of a total 97 observing programmes that were addressed in the previous section. The electronic pdf of this document contains hyperlinks that relate the description of the science cases with the tables of the observing programme.

Spectral resolution is defined as $R = \delta \lambda / \lambda $.

The SNR (Signal-to-Noise) values in the OP tables correspond to the continuum of the intensity signal.

In the table of the Observing Programmes, a "$+$" sign denotes goals, i.e. observations that are not required but would help to achieve the objectives of the particular Observing Programme.

In the table of the Observing Programmes, a "$^*$" sign denotes the requirement on the spatial pixel sampling.

\newpage

\begin{table}[H]
\OPtabletitleCustom{1.1.1}{Formation and evolution of intense flux tubes in the solar atmosphere}{OP1.1.1table}{OP1.1.1}
\centering
\vspace{4mm}
\vspace{2mm}
\begin{flushleft}
Duration of the observations: 1 hour, to cover the lifetime of newly formed magnetic flux tubes. The program should be repeated several times to build up a significant statistical sample. Light distribution: All instruments work simultaneously and receive 100\% of the light at the indicated wavelengths.
\end{flushleft}
\vspace{2mm}
\begin{adjustbox}{max width=\textwidth}
\begin{tabular}{|l|c|c|c|c|c|c|c|c|c|c|c|}
\hline
\textbf{Goal:} & \multicolumn{11}{p{17cm}|}{TIS/FBIs: Detection of rapid flows and intensification of the field in photosphere and chromosphere over large FOVs, to build up statistically significant sample. Response of chromosphere to convective collapse and possible heating events. Context information. IFS: Detailed study of physical processes in and around flux tubes undergoing convective collapse} \\
\hline
\makecell[t]{\textbf{Instrument}} & \makecell[t]{\textbf{Optical}\\ \textbf{Arm}} & \makecell[t]{\textbf{Atm}\\ \textbf{Layer}} & \makecell[t]{\textbf{CWL}\\ \textbf{{[nm]}}} & \makecell[t]{\textbf{Spectral}\\ \textbf{Range}\\ \textbf{{[nm]}}} & \makecell[t]{\textbf{Mode}} & \makecell[t]{\textbf{FOV}\\ \textbf{{[\arcsec]}}} & \makecell[t]{\textbf{Sampling}\\ \textbf{{[\arcsec /px]}}} & \makecell[t]{\textbf{R}} & \makecell[t]{\textbf{SNR}} & \makecell[t]{\textbf{\# WL}\\ \textbf{Points}} & \makecell[t]{\textbf{Cadence}\\ \textbf{[s]}} \\
\hline
TIS/FBI & B & Photosphere & 430 & 1.00 & img & 60 & $^{*}$0.01 & N/A & 500 & N/A & 15 \\
 &  & Chromosphere & $^{+}$396.8 & 0.40 & spec & 60 & $^{*}$0.01 & 50000 & 500 & 15 & 15 \\
TIS/FBI & V & Chromosphere & 517.3 & 0.20 & pol & 60 & $^{*}$0.01 & 100000 & 500 & 12 & 15 \\
 &  & Chromosphere & $^{+}$656.3 & 0.30 & spec & 60 & $^{*}$0.02 & 100000 & 500 & 20 & 15 \\
IFS-M & B & Chromosphere & $^{+}$393.4 & 0.75 & pol & 7x7 & 0.05 & 75000 & 3000 & N/A & 15 \\
IFS-M & V & Photosphere & 630.2 & 0.75 & pol & 7x7 & 0.05 & 100000 & 3000 & N/A & 15 \\
IFS-M & R & Chromosphere & 854.2 & 1.00 & pol & 7x7 & 0.05 & 80000 & 3000 & N/A & 15 \\
EMBER & IR & Chromosphere & 1083 & 1.50 & pol & 7x7 & 0.05 & 100000 & 3000 & N/A & 15 \\
 &  & Photosphere & 1565 & 1.50 & pol & 7x7 & 0.05 & 100000 & 3000 & N/A & 15 \\
\hline
\textbf{Notes:} & \multicolumn{11}{p{17cm}|}{IFS: Allows tiles to be reconstructed and minimizes differential refraction effects} \\
\hline
\end{tabular}%
\end{adjustbox}%
\end{table}

\begin{table}[H]
\OPtabletitleCustom{1.2.1}{Internal structure and evolution of magnetic elements}{OP1.2.1table}{OP1.2.1}
\centering
\vspace{4mm}
\vspace{2mm}
\begin{flushleft}
Duration of the observations: 1 hour, to cover the lifetime of magnetic flux tubes. The program should be repeated several times to build up a statistically significant sample. Light distribution: All instruments work simultaneously and receive 100\% of the light at the indicated wavelengths.
\end{flushleft}
\vspace{2mm}
\begin{adjustbox}{max width=\textwidth}
\begin{tabular}{|l|c|c|c|c|c|c|c|c|c|c|c|}
\hline
\textbf{Goal:} & \multicolumn{11}{p{17cm}|}{TIS/FBIs: Response of chromosphere to flows and waves. Context information (horizontal motions) IFS: Detect spatial variations of field and flows across magnetic elements as a function of height in the atmosphere. Study temporal evolution of fields, flows, and waves} \\
\hline
\makecell[t]{\textbf{Instrument}} & \makecell[t]{\textbf{Optical}\\ \textbf{Arm}} & \makecell[t]{\textbf{Atm}\\ \textbf{Layer}} & \makecell[t]{\textbf{CWL}\\ \textbf{{[nm]}}} & \makecell[t]{\textbf{Spectral}\\ \textbf{Range}\\ \textbf{{[nm]}}} & \makecell[t]{\textbf{Mode}} & \makecell[t]{\textbf{FOV}\\ \textbf{{[\arcsec]}}} & \makecell[t]{\textbf{Sampling}\\ \textbf{{[\arcsec /px]}}} & \makecell[t]{\textbf{R}} & \makecell[t]{\textbf{SNR}} & \makecell[t]{\textbf{\# WL}\\ \textbf{Points}} & \makecell[t]{\textbf{Cadence}\\ \textbf{{[s]}}} \\
\hline
TIS/FBI & B & Photosphere & 430 & 1.00 & img & 30 & $^{*}$0.01 & N/A & 500 & N/A & 10 \\
 &  & Chromosphere & 396.8 & 0.40 & spec & 30 & $^{*}$0.01 & 50000 & 500 & 10 & 10 \\
TIS/FBI & V & Photosphere & 525 & 0.20 & pol & 30 & $^{*}$0.01 & 100000 & 500 & 10 & 10 \\
IFS-M & B & Chromosphere & $^{+}$393.4 & 0.75 & pol & 7x7 & 0.03 & 75000 & 2000 & N/A & 15 \\
IFS-M & V & Chromosphere & 517.3 & 0.75 & pol & 7x7 & 0.03 & 100000 & 2000 & N/A & 15 \\
IFS-M & R & Chromosphere & 854.2 & 1.00 & pol & 7x7 & 0.03 & 80000 & 2000 & N/A & 15 \\
EMBER & IR & Chromosphere & $^{+}$1083 & 1.50 & pol & 7x7 & $^{*}$0.03 & 100000 & 2000 & N/A & 15 \\
 &  & Photosphere & 1565 & 1.50 & pol & 7x7 & $^{*}$0.04 & 100000 & 2000 & N/A & 15 \\
\hline
\textbf{Notes:} & \multicolumn{11}{p{17cm}|}{N/A} \\
\hline
\end{tabular}%
\end{adjustbox}%
\end{table}

\begin{table}[H]
\OPtabletitleCustom{1.3.1}{Magnetic bright points}{OP1.3.1table}{OP1.3.1}
\centering
\vspace{4mm}
\vspace{2mm}
\begin{flushleft}
Duration of the observations: Single co-temporal shots of several broad-band filtergrams to study morphologic appearance of MBPs from the photosphere to the chromosphere together with IFU for detailed magnetic field distribution characterization. The program should be repeated several times to build up a significant statistical sample. Light distribution: IFS spectroplarimeters and TIS/FBIs work simultaneously with major (80\%) fraction of light going to IFS.
\end{flushleft}
\vspace{2mm}
\begin{adjustbox}{max width=\textwidth}
\begin{tabular}{|l|c|c|c|c|c|c|c|c|c|c|c|}
\hline
\textbf{Goal:} & \multicolumn{11}{p{17cm}|}{TIS/FBIs: Morphologic appearance of MBPs: size distribution and shape of MBPs at various atmospheric heights. IFS: Detailed study of the magnetic field strength distribution of MBPs at several atmospheric heights} \\
\hline
\makecell[t]{\textbf{Instrument}} & \makecell[t]{\textbf{Optical}\\ \textbf{Arm}} & \makecell[t]{\textbf{Atm}\\ \textbf{Layer}} & \makecell[t]{\textbf{CWL}\\ \textbf{{[nm]}}} & \makecell[t]{\textbf{Spectral}\\ \textbf{Range}\\ \textbf{{[nm]}}} & \makecell[t]{\textbf{Mode}} & \makecell[t]{\textbf{FOV}\\ \textbf{{[\arcsec]}}} & \makecell[t]{\textbf{Sampling}\\ \textbf{{[\arcsec /px]}}} & \makecell[t]{\textbf{R}} & \makecell[t]{\textbf{SNR}} & \makecell[t]{\textbf{\# WL}\\ \textbf{Points}} & \makecell[t]{\textbf{Cadence}\\ \textbf{{[s]}}} \\
\hline
TIS/FBI & B & Photosphere & 430 & 1.00 & img & 60 & $^{*}$0.01 & N/A & 500 & N/A & 10 \\
 &  & Chromosphere & 396.8 & 0.40 & spec & 60 & $^{*}$0.01 & 50000 & 500 & 10 & 10 \\
TIS/FBI & V & Chromosphere & 517.3 & 0.20 & pol & 60 & $^{*}$0.01 & 100000 & 1000 & 15 & 30 \\
IFS-M & V & Photosphere & 630.2 & 0.75 & pol & 21x21 & 0.05 & 100000 & 1000 & N/A & 108 \\
IFS-M & R & Chromosphere & 854.2 & 0.75 & pol & 21x21 & 0.05 & 80000 & 1000 & N/A & 108 \\
EMBER & IR & Chromosphere & $^{+}$1083 & 1.50 & pol & 21x21 & 0.05 & 100000 & 1000 & N/A & 108 \\
 &  & Photosphere & 1565 & 1.50 & pol & 21x21 & 0.05 & 100000 & 1000 & N/A & 108 \\
\hline
\textbf{Notes:} & \multicolumn{11}{p{17cm}|}{IFS: Allows tiles to be reconstructed and minimizes differential refraction effects} \\
\hline
\end{tabular}%
\end{adjustbox}%
\end{table}

\begin{table}[H]
\OPtabletitleCustom{1.3.2}{Dynamic parameters and evolution of MBPs}{OP1.3.2table}{OP1.3.2}
\centering
\vspace{4mm}
\vspace{2mm}
\begin{flushleft}
Duration of the observations: 15 minutes minimum up to 1 hour, to cover the lifetime of newly formed MBPs. The program should be repeated several times to build up a significant statistical sample. Light distribution: TIS/FBIs imagers work simultaneously with 100\% of the light going to each instrument at the indicated wavelength.
\end{flushleft}
\vspace{2mm}
\begin{adjustbox}{max width=\textwidth}
\begin{tabular}{|l|c|c|c|c|c|c|c|c|c|c|c|}
\hline
\textbf{Goal:} & \multicolumn{11}{p{17cm}|}{TIS/FBIs: Detection of physical parameters in and around MBP during its evolution. Intensity and shape of MBPs during their evolution in various heights} \\
\hline
\makecell[t]{\textbf{Instrument}} & \makecell[t]{\textbf{Optical}\\ \textbf{Arm}} & \makecell[t]{\textbf{Atm}\\ \textbf{Layer}} & \makecell[t]{\textbf{CWL}\\ \textbf{{[nm]}}} & \makecell[t]{\textbf{Spectral}\\ \textbf{Range}\\ \textbf{{[nm]}}} & \makecell[t]{\textbf{Mode}} & \makecell[t]{\textbf{FOV}\\ \textbf{{[\arcsec]}}} & \makecell[t]{\textbf{Sampling}\\ \textbf{{[\arcsec /px]}}} & \makecell[t]{\textbf{R}} & \makecell[t]{\textbf{SNR}} & \makecell[t]{\textbf{\# WL}\\ \textbf{Points}} & \makecell[t]{\textbf{Cadence}\\ \textbf{{[s]}}} \\
\hline
TIS/FBI & B & Photosphere & 430 & 1.00 & img & 60 & $^{*}$0.01 & N/A & 500 & N/A & 10 \\
 &  & Chromosphere & 396.8 & 0.40 & spec & 60 & $^{*}$0.01 & 50000 & 500 & 10 & 10 \\
TIS/FBI & V & Chromosphere & 517.3 & 0.20 & pol & 60 & $^{*}$0.01 & 100000 & 500 & 15 & 10 \\
 &  & Photosphere & 617.3 & 0.15 & pol & 60 & $^{*}$0.02 & 100000 & 500 & 10 & 10 \\
IFS-M & V & Photosphere & 525 & 0.75 & pol & 10.5 x 7 & 0.05 & 100000 & 700 & N/A & 10 \\
\hline
\textbf{Notes:} & \multicolumn{11}{p{17cm}|}{N/A} \\
\hline
\end{tabular}%
\end{adjustbox}%
\end{table}

\begin{table}[H]
\OPtabletitleCustom{1.4.1}{Emergence and evolution of magnetic fields in granular convection}{OP1.4.1table}{OP1.4.1}
\centering
\vspace{4mm}
\vspace{2mm}
\begin{flushleft}
Duration of the observations: 3-4 hours, to study temporal variation of appearance rates and build up a statistically significant sample. Light distribution: All instruments work simultaneously and receive 100\% of the light at the indicated wavelengths.
\end{flushleft}
\vspace{2mm}
\begin{adjustbox}{max width=\textwidth}
\begin{tabular}{|l|c|c|c|c|c|c|c|c|c|c|c|}
\hline
\textbf{Goal:} & \multicolumn{11}{p{17cm}|}{TIS/FBIs: Evaluate frequency of appearance of small-scale magnetic loops and unipolar flux features in quiet Sun. Estimate magnetic fluxes. Assess influence on chromospheric layers. Resolve substructure of emerging flux concentrations. Detect small-scale chromospheric heating events. Context information (horizontal motions).} \\
\hline
\makecell[t]{\textbf{Instrument}} & \makecell[t]{\textbf{Optical}\\ \textbf{Arm}} & \makecell[t]{\textbf{Atm}\\ \textbf{Layer}} & \makecell[t]{\textbf{CWL}\\ \textbf{{[nm]}}} & \makecell[t]{\textbf{Spectral}\\ \textbf{Range}\\ \textbf{{[nm]}}} & \makecell[t]{\textbf{Mode}} & \makecell[t]{\textbf{FOV}\\ \textbf{{[\arcsec]}}} & \makecell[t]{\textbf{Sampling}\\ \textbf{{[\arcsec /px]}}} & \makecell[t]{\textbf{R}} & \makecell[t]{\textbf{SNR}} & \makecell[t]{\textbf{\# WL}\\ \textbf{Points}} & \makecell[t]{\textbf{Cadence}\\ \textbf{{[s]}}} \\
\hline
TIS/FBI & B & Photosphere & 430 & 1.00 & img & 60 & $^{*}$0.01 & N/A & 500 & N/A & 10 \\
 &  & Chromosphere & 396.8 & 0.40 & spec & 60 & $^{*}$0.01 & 50000 & 500 & 10 & 10 \\
TIS/FBI & V & Chromosphere & 517.3 & 0.15 & pol & 60 & 0.10 & 100000 & 2000 & 15 & 15 \\
 &  & Photosphere & 630.2 & 0.15 & pol & 60 & 0.10 & 100000 & 2000 & 15 & 15 \\
 &  & Chromosphere & 656.3 & 0.15 & spec & 60 & 0.10 & 100000 & 1000 & 20 & 15 \\
TIS/FBI & R & Chromosphere & 854.2 & 0.40 & pol & 60 & 0.10 & 80000 & 2000 & 23 & 15 \\
EMBER & IR & Photosphere & 1565 & 1.50 & pol & 24.5 x 3.5 & 0.10 & 100000 & 1500 & N/A & 15 \\
\hline
\textbf{Notes:} & \multicolumn{11}{p{17cm}|}{TIS/FBIs: ame FOV as narrow-band filtergraphs} \\
\hline
\end{tabular}%
\end{adjustbox}%
\end{table}

\begin{table}[H]
\OPtabletitleCustom{1.4.2}{Properties of magnetic fields emerging in the quiet Sun}{OP1.4.2table}{OP1.4.2}
\centering
\vspace{4mm}
\vspace{2mm}
\begin{flushleft}
Duration of the observations: 1 hour, to cover full emergence process and subsequent evolution and build up a statistically significant sample. Light distribution: All instruments work simultaneously and receive 100\% of the light at the indicated wavelengths.
\end{flushleft}
\vspace{2mm}
\begin{adjustbox}{max width=\textwidth}
\begin{tabular}{|l|c|c|c|c|c|c|c|c|c|c|c|}
\hline
\textbf{Goal:} & \multicolumn{11}{p{17cm}|}{TIS/FBIs: Resolve internal structure of emerging flux concentrations. Detect small-scale chromospheric heating events. Context information (horizontal motions, large-scale magnetic topology at emergence site). IFS: Determine magnetic field topology and dynamics of emerging loops as a function of height. Search for opposite polarities in unipolar flux emergence processes. Detailed study of the interaction of emerging fields with pre-existing flux in photosphere and chromosphere. Resolve polarity inversion lines.} \\
\hline
\makecell[t]{\textbf{Instrument}} & \makecell[t]{\textbf{Optical}\\ \textbf{Arm}} & \makecell[t]{\textbf{Atm}\\ \textbf{Layer}} & \makecell[t]{\textbf{CWL}\\ \textbf{{[nm]}}} & \makecell[t]{\textbf{Spectral}\\ \textbf{Range}\\ \textbf{{[nm]}}} & \makecell[t]{\textbf{Mode}} & \makecell[t]{\textbf{FOV}\\ \textbf{{[\arcsec]}}} & \makecell[t]{\textbf{Sampling}\\ \textbf{{[\arcsec /px]}}} & \makecell[t]{\textbf{R}} & \makecell[t]{\textbf{SNR}} & \makecell[t]{\textbf{\# WL}\\ \textbf{Points}} & \makecell[t]{\textbf{Cadence}\\ \textbf{{[s]}}} \\
\hline
TIS/FBI & B & Photosphere & 430 & 1.00 & img & 60 & $^{*}$0.01 & N/A & 500 & N/A & 10 \\
 &  & Chromosphere & 396.8 & 0.40 & spec & 60 & $^{*}$0.01 & 50000 & 500 & 10 & 10 \\
TIS/FBI & V & Chromosphere & 517.3 & 0.20 & pol & 60 & 0.03 & 100000 & 2000 & 15 & 40 \\
IFS-M & B & Chromosphere & $^{+}$393.4 & 0.75 & pol & 7x7 & 0.03 & 75000 & 2000 & N/A & 40 \\
IFS-M & V & Photosphere & 630.2 & 0.75 & pol & 7x7 & 0.03 & 100000 & 2000 & N/A & 40 \\
IFS-M & R & Chromosphere & 854.2 & 1.00 & pol & 7x7 & 0.03 & 80000 & 2000 & N/A & 40 \\
EMBER & IR & Chromosphere & $^{+}$1083 & 1.50 & pol & 7x7 & $^{*}$0.03 & 100000 & 2000 & N/A & 40 \\
 &  & Photosphere & 1565 & 1.50 & pol & 7x7 & $^{*}$0.04 & 100000 & 2000 & N/A & 40 \\
\hline
\textbf{Notes:} & \multicolumn{11}{p{17cm}|}{TIS/FBIs: FOV should cover at least one full supergranule. Ca ii K instead of H to allow IFS to observe Ca ii H} \\
\hline
\end{tabular}%
\end{adjustbox}%
\end{table}

\begin{table}[H]
\OPtabletitleCustom{1.5.1}{Magnetic field topology, dynamics and energy release at flux cancellation sites}{OP1.5.1table}{OP1.5.1}
\centering
\vspace{4mm}
\vspace{2mm}
\begin{flushleft}
Duration of the observations: 1 hour, to study evolution of cancelling features as they approach each other and postcancellation effects. The program should be repeated several times to build up a statistically significant sample.
\end{flushleft}
\vspace{2mm}
\begin{adjustbox}{max width=\textwidth}
\begin{tabular}{|l|c|c|c|c|c|c|c|c|c|c|c|}
\hline
\textbf{Goal:} & \multicolumn{11}{p{17cm}|}{IFS: Determine magnetic field topology of cancelling features as a function of height in the atmosphere. Detect magnetic field configuration at polarity inversion line. Study dynamical effects and possible energy release in upper atmospheric layers.} \\
\hline
\makecell[t]{\textbf{Instrument}} & \makecell[t]{\textbf{Optical}\\ \textbf{Arm}} & \makecell[t]{\textbf{Atm}\\ \textbf{Layer}} & \makecell[t]{\textbf{CWL}\\ \textbf{{[nm]}}} & \makecell[t]{\textbf{Spectral}\\ \textbf{Range}\\ \textbf{{[nm]}}} & \makecell[t]{\textbf{Mode}} & \makecell[t]{\textbf{FOV}\\ \textbf{{[\arcsec]}}} & \makecell[t]{\textbf{Sampling}\\ \textbf{{[\arcsec /px]}}} & \makecell[t]{\textbf{R}} & \makecell[t]{\textbf{SNR}} & \makecell[t]{\textbf{\# WL}\\ \textbf{Points}} & \makecell[t]{\textbf{Cadence}\\ \textbf{{[s]}}} \\
\hline
TIS/FBI & B & Chromosphere & 396.8 & 0.40 & spec & 60 & 0.02 & 50000 & 1000 & 21 & 20 \\
TIS/FBI & V & Chromosphere & 517.3 & 0.20 & pol & 60 & 0.03 & 100000 & 2000 & 7 & 40 \\
 &  & Chromosphere & 656.3 & 0.30 & spec & 60 & $^{*}$0.02 & 100000 & 500 & 20 & 40 \\
IFS-M & B & Photosphere & 492.3 & 0.75 & pol & 7x7 & 0.03 & 75000 & 2000 & N/A & 40 \\
IFS-M & V & Photosphere & 630.2 & 0.75 & pol & 7x7 & 0.03 & 100000 & 2000 & N/A & 40 \\
IFS-M & R & Chromosphere & 854.2 & 1.00 & pol & 7x7 & 0.03 & 80000 & 2000 & N/A & 40 \\
EMBER & IR & Chromosphere & $^{+}$1083 & 1.50 & pol & 7x7 & $^{*}$0.03 & 100000 & 2000 & N/A & 40 \\
 &  & Photosphere & 1565 & 1.50 & pol & 7x7 & $^{*}$0.04 & 100000 & 2000 & N/A & 40 \\
\hline
\textbf{Notes:} & \multicolumn{11}{p{17cm}|}{IFS: Four tiles of 3.8"x3.8" with pixel size of 0.03" needed to cover FOV Goal priority: Cadence, SNR, FOV} \\
\hline
\end{tabular}%
\end{adjustbox}%
\end{table}

\begin{table}[H]
\OPtabletitleCustom{1.6.1}{Physical properties of internetwork magnetic fields}{OP1.6.1table}{OP1.6.1}
\centering
\vspace{4mm}
\vspace{2mm}
\begin{flushleft}
Duration of the observations: The time needed to take a single map. However, it would be useful to keep observing for 1-2 hours to have a few maps and study short-term variations within the same supergranular cell.
\end{flushleft}
\vspace{2mm}
\begin{adjustbox}{max width=\textwidth}
\begin{tabular}{|l|c|c|c|c|c|c|c|c|c|c|c|}
\hline
\textbf{Goal:} & \multicolumn{11}{p{17cm}|}{IFS: Uncover linear polarisation signals. Determine distribution of field strengths, field inclinations, and flows in internetwork regions at highest spatial resolution. Determine height variation of field properties and flows. Resolve polarity inversion lines. Compare Zeeman and Hanle views of internetwork magnetism. Determine total flux content of supergranular cells.} \\
\hline
\makecell[t]{\textbf{Instrument}} & \makecell[t]{\textbf{Optical}\\ \textbf{Arm}} & \makecell[t]{\textbf{Atm}\\ \textbf{Layer}} & \makecell[t]{\textbf{CWL}\\ \textbf{{[nm]}}} & \makecell[t]{\textbf{Spectral}\\ \textbf{Range}\\ \textbf{{[nm]}}} & \makecell[t]{\textbf{Mode}} & \makecell[t]{\textbf{FOV}\\ \textbf{{[\arcsec]}}} & \makecell[t]{\textbf{Sampling}\\ \textbf{{[\arcsec /px]}}} & \makecell[t]{\textbf{R}} & \makecell[t]{\textbf{SNR}} & \makecell[t]{\textbf{\# WL}\\ \textbf{Points}} & \makecell[t]{\textbf{Cadence}\\ \textbf{{[s]}}} \\
\hline
TIS/FBI & V & Photosphere & 525 & 0.20 & pol & 60 & 0.05 & 100000 & 3000 & 12 & 50 \\
IFS-M & B & Photosphere & 460.7 & 0.75 & pol & 40x40 & 0.05 & 75000 & 10000 & N/A & 3600 \\
IFS-M & V & Photosphere & 630.2 & 0.75 & pol & 40x40 & 0.05 & 100000 & 10000 & N/A & 3600 \\
IFS-M & R & Chromosphere & 854.2 & 1.00 & pol & 40x40 & 0.05 & 80000 & 10000 & N/A & 3600 \\
EMBER & IR & Photosphere & 1565 & 1.50 & pol & 40x40 & 0.05 & 100000 & 10000 & N/A & 3600 \\
\hline
\textbf{Notes:} & \multicolumn{11}{p{17cm}|}{IFS: FOV should cover at least one entire supergranule. Program can also be done with long-slit spectrograph, but IFU allows to reconstruct tiles and minimizes differential refraction effects. 44 tiles of 600"x 600" with pixel size of 0.05" needed to cover full FOV.} \\
\hline
\end{tabular}%
\end{adjustbox}%
\end{table}

\begin{table}[H]
\OPtabletitleCustom{1.6.2}{Short-term evolution of internetwork fields}{OP1.6.2table}{OP1.6.2}
\centering
\vspace{4mm}
\vspace{2mm}
\begin{flushleft}
Duration of the observations: 2 hours, to determine flux appearance and disappearance rates and their variations with time. Light distribution: All instruments work simultaneously and receive 100\% of the light at the indicated wavelengths.
\end{flushleft}
\vspace{2mm}
\begin{adjustbox}{max width=\textwidth}
\begin{tabular}{|l|c|c|c|c|c|c|c|c|c|c|c|}
\hline
\textbf{Goal:} & \multicolumn{11}{p{17cm}|}{TIS/FBIs: Study appearance, evolution, and disappearance of IN fields. Determine connectivity between photosphere and chromosphere.  Detection of small-scale heating events at different heights in the atmosphere. Provide context information (horizontal motions, large scale structure of supergranule)} \\
\hline
\makecell[t]{\textbf{Instrument}} & \makecell[t]{\textbf{Optical}\\ \textbf{Arm}} & \makecell[t]{\textbf{Atm}\\ \textbf{Layer}} & \makecell[t]{\textbf{CWL}\\ \textbf{{[nm]}}} & \makecell[t]{\textbf{Spectral}\\ \textbf{Range}\\ \textbf{{[nm]}}} & \makecell[t]{\textbf{Mode}} & \makecell[t]{\textbf{FOV}\\ \textbf{{[\arcsec]}}} & \makecell[t]{\textbf{Sampling}\\ \textbf{{[\arcsec /px]}}} & \makecell[t]{\textbf{R}} & \makecell[t]{\textbf{SNR}} & \makecell[t]{\textbf{\# WL}\\ \textbf{Points}} & \makecell[t]{\textbf{Cadence}\\ \textbf{{[s]}}} \\
\hline
TIS/FBI & B & Chromosphere & 396.8 & 0.40 & pol & 60 & 0.05 & 50000 & 1000 & 25 & 15 \\
TIS/FBI & V & Chromosphere & 517.3 & 0.20 & pol & 60 & 0.05 & 100000 & 2000 & 10 & 30 \\
 &  & Photosphere & 617.3 & 0.15 & pol & 60 & 0.05 & 100000 & 2000 & 15 & 30 \\
TIS/FBI & R & Chromosphere & 854.2 & 0.40 & pol & 60 & 0.06 & 80000 & 2000 & 18 & 30 \\
EMBER & IR & Photosphere & 1565 & 1.50 & pol & 10.5 x 7 & 0.07 & 100000 & 2000 & N/A & 30 \\
\hline
\textbf{Notes:} & \multicolumn{11}{p{17cm}|}{TIS/FBIs: FOV should cover one full supergranule. FOV should cover at least a supergranule. Ca ii K to allow filtergraphs to observe Ca ii H. Really necessary when filtergraphs have wide-band cameras?} \\
\hline
\end{tabular}%
\end{adjustbox}%
\end{table}

\begin{table}[H]
\OPtabletitleCustom{1.7.1}{Structure of the polar faculae}{OP1.7.1table}{OP1.7.1}
\centering
\vspace{4mm}
\vspace{2mm}
\begin{flushleft}
Duration of the observations: 1 hour, to cover lifetime of individual structures and build up a statistically significant sample. Light distribution: All instruments work simultaneously and receive 100\% of the light at the indicated wavelengths.
\end{flushleft}
\vspace{2mm}
\begin{adjustbox}{max width=\textwidth}
\begin{tabular}{|l|c|c|c|c|c|c|c|c|c|c|c|}
\hline
\textbf{Goal:} & \multicolumn{11}{p{17cm}|}{TIS/FBIs:Resolve substructure of flux concentrations. Detect small-scale chromospheric events (jets, surges, waves). Context information IFS: Determine height variation of field properties and flows in isolated strong flux concentrations near the poles.} \\
\hline
\makecell[t]{\textbf{Instrument}} & \makecell[t]{\textbf{Optical}\\ \textbf{Arm}} & \makecell[t]{\textbf{Atm}\\ \textbf{Layer}} & \makecell[t]{\textbf{CWL}\\ \textbf{{[nm]}}} & \makecell[t]{\textbf{Spectral}\\ \textbf{Range}\\ \textbf{{[nm]}}} & \makecell[t]{\textbf{Mode}} & \makecell[t]{\textbf{FOV}\\ \textbf{{[\arcsec]}}} & \makecell[t]{\textbf{Sampling}\\ \textbf{{[\arcsec /px]}}} & \makecell[t]{\textbf{R}} & \makecell[t]{\textbf{SNR}} & \makecell[t]{\textbf{\# WL}\\ \textbf{Points}} & \makecell[t]{\textbf{Cadence}\\ \textbf{{[s]}}} \\
\hline
TIS/FBI & B & Chromosphere & 396.8 & 0.40 & spec & 60 & $^{*}$0.01 & 50000 & 700 & 9 & 40 \\
 &  & Photosphere & 430 & 1.00 & img  & 60 & $^{*}$0.01 & N/A & 100 & N/A & 40 \\
TIS/FBI & V & Chromosphere & 517.3 & 0.20 & pol & 60 & 0.03 & 100000 & 1000 & 16 & 40 \\
IFS-M & B & Chromosphere & $^{+}$393.4 & 0.75 & spec & 7x7 & 0.03 & 75000 & 1500 & N/A & 40 \\
IFS-M & V & Photosphere & 630.2 & 0.75 & pol & 7x7 & 0.03 & 100000 & 1500 & N/A & 40 \\
IFS-M & R & Chromosphere & 854.2 & 1.00 & pol & 7x7 & 0.03 & 80000 & 1500 & N/A & 40 \\
EMBER & IR & Photosphere & 1565 & 1.50 & pol & 7x7 & $^{*}$0.04 & 100000 & 1500 & N/A & 40 \\
\hline
\textbf{Notes:} & \multicolumn{11}{p{17cm}|}{IFS: Highest spatial resolution needed to partially compensate for projection effects. Four tiles of 3.8"x3.8" with pixel size of 0.03" needed to cover FOV.} \\
\hline
\end{tabular}%
\end{adjustbox}%
\end{table}

\begin{table}[H]
\OPtabletitleCustom{1.7.2}{Properties, distribution, and evolution of polar magnetic fields }{OP1.7.2table}{OP1.7.2}
\centering
\vspace{4mm}
\vspace{2mm}
\begin{flushleft}
Duration of the observations: 2 hour, to cover lifetime of individual structures and build up a statistically significant sample. If FOV is small, different pointings may be necessary to observe a wide range of latitudes. Light distribution: All instruments work simultaneously and receive 100\% of the light at the indicated wavelengths.
\end{flushleft}
\vspace{2mm}
\begin{adjustbox}{max width=\textwidth}
\begin{tabular}{|l|c|c|c|c|c|c|c|c|c|c|c|}
\hline
\textbf{Goal:} & \multicolumn{11}{p{17cm}|}{TIS/FBIs: Determine field strength, field inclination, and flux distribution as a function of latitude near the poles. Study temporal evolution of flux concentrations and interactions between them. Detection of fast transient events produced by release of magnetic energy in photosphere/chromosphere. Context information.} \\
\hline
\makecell[t]{\textbf{Instrument}} & \makecell[t]{\textbf{Optical}\\ \textbf{Arm}} & \makecell[t]{\textbf{Atm}\\ \textbf{Layer}} & \makecell[t]{\textbf{CWL}\\ \textbf{{[nm]}}} & \makecell[t]{\textbf{Spectral}\\ \textbf{Range}\\ \textbf{{[nm]}}} & \makecell[t]{\textbf{Mode}} & \makecell[t]{\textbf{FOV}\\ \textbf{{[\arcsec]}}} & \makecell[t]{\textbf{Sampling}\\ \textbf{{[\arcsec /px]}}} & \makecell[t]{\textbf{R}} & \makecell[t]{\textbf{SNR}} & \makecell[t]{\textbf{\# WL}\\ \textbf{Points}} & \makecell[t]{\textbf{Cadence}\\ \textbf{{[s]}}} \\
\hline
TIS/FBI & B & Chromosphere & $^{+}$396.8 & 0.40 & spec & 60 & $^{*}$0.01 & 50000 & 500 & 14 & 30 \\
 &  & Photosphere & 430 & 1.00 & img  & 60 & $^{*}$0.01 & N/A & 1000 & N/A & 30 \\
TIS/FBI & V & Photosphere & 617.3 & 0.15 & pol & 60 & 0.03 & 100000 & 1000 & 10 & 30 \\
 &  & Chromosphere & $^{+}$656.3 & 0.30 & spec & 60 & 0.03 & 100000 & 500 & 20 & 30 \\
IFS-M & V & Chromosphere & 517.3 & 0.75 & pol & 10.5 x 7 & 0.03 & 100000 & 1000 & N/A & 30 \\
IFS-M & R & Chromosphere & 854.2 & 1.00 & pol & 10.5 x 7 & 0.03 & 80000 & 1000 & N/A & 30 \\
EMBER & IR & Photosphere & 1565 & 1.50 & pol & 10.5 x 7 & $^{*}$0.04 & 100000 & 1000 & N/A & 30 \\
\hline
\textbf{Notes:} & \multicolumn{11}{p{17cm}|}{IFS-M\&EMBER: FOV should cover at least one full supergranule. Highest spatial resolution needed to partially compensate for projection effects. FOV matching that of filtergraphs.} \\
\hline
\end{tabular}%
\end{adjustbox}%
\end{table}

\begin{table}[H]
\OPtabletitleCustom{1.8.1}{Network evolution and dynamics }{OP1.8.1table}{OP1.8.1}
\centering
\vspace{4mm}
\begin{adjustbox}{max width=\textwidth}
\begin{tabular}{|l|c|c|c|c|c|c|c|c|c|c|c|}
\hline
\textbf{Goal:} & \multicolumn{11}{p{17cm}|}{TIS/FBIs: Tracking of internetwork flux elements in supergranules. Detection of small-scale magnetic elements. IFS: Detailed study of physical processes during the merging of internetwork elements with network elements} \\
\hline
\makecell[t]{\textbf{Instrument}} & \makecell[t]{\textbf{Optical}\\ \textbf{Arm}} & \makecell[t]{\textbf{Atm}\\ \textbf{Layer}} & \makecell[t]{\textbf{CWL}\\ \textbf{{[nm]}}} & \makecell[t]{\textbf{Spectral}\\ \textbf{Range}\\ \textbf{{[nm]}}} & \makecell[t]{\textbf{Mode}} & \makecell[t]{\textbf{FOV}\\ \textbf{{[\arcsec]}}} & \makecell[t]{\textbf{Sampling}\\ \textbf{{[\arcsec /px]}}} & \makecell[t]{\textbf{R}} & \makecell[t]{\textbf{SNR}} & \makecell[t]{\textbf{\# WL}\\ \textbf{Points}} & \makecell[t]{\textbf{Cadence}\\ \textbf{{[s]}}} \\
\hline
TIS/FBI & B & Chromosphere & 396.8 & 0.40 & spec & 60 & $^{*}$0.01 & 50000 & 500 & 9 & 10 \\
 &  & Photosphere & 430 & 1.00 & img  & 60 & $^{*}$0.01 & N/A & 100 & N/A & 10 \\
TIS/FBI & V & Photosphere & 617.3 & 0.15 & pol & 60 & $^{*}$0.02 & 100000 & 1000 & 10 & 30 \\
TIS/FBI & R & Photosphere & 709 & 0.15 & spec & 60 & $^{*}$0.02 & 80000 & 500 & 20 & 30 \\
 &  & Chromosphere & 854.2 & 0.40 & pol & 60 & $^{*}$0.02 & 80000 & 1000 & 12 & 30 \\
IFS-M & B & Chromosphere & $^{+}$393.4 & 0.75 & spec & 10.5x14 & 0.03 & 75000 & 1000 & N/A & 30 \\
IFS-M & V & Chromosphere & 517.3 & 0.75 & pol & 10.5x14 & 0.03 & 100000 & 1000 & N/A & 30 \\
EMBER & IR & Photosphere & 1565 & 1.50 & pol & 10.5x14 & $^{*}$0.04 & 100000 & 1000 & N/A & 30 \\
\hline
\textbf{Notes:} & \multicolumn{11}{p{17cm}|}{TIS/FBIs: The lines Fe I 630.25 and Ca II 854.2 nm should be observed both with TIS/FBIs and IFS, but not simultaneously. IFS: allows to reconstruct tiles and minimizes differential refraction effects.} \\
\hline
\end{tabular}%
\end{adjustbox}%
\end{table}

\begin{table}[H]
\OPtabletitleCustom{2.1.1}{Sausage and kink oscillations in MBPs}{OP2.1.1table}{OP2.1.1}
\centering
\vspace{4mm}
\begin{adjustbox}{max width=\textwidth}
\begin{tabular}{|l|c|c|c|c|c|c|c|c|c|c|c|}
\hline
\textbf{Goal:} & \multicolumn{11}{p{17cm}|}{TIS/FBIs: High-resolution imaging of MBPs at very high time cadence. High-resolution spectroscopy of MBPs. IFS: High-resolution spectropolarimetry of MBPs.} \\
\hline
\makecell[t]{\textbf{Instrument}} & \makecell[t]{\textbf{Optical}\\ \textbf{Arm}} & \makecell[t]{\textbf{Atm}\\ \textbf{Layer}} & \makecell[t]{\textbf{CWL}\\ \textbf{{[nm]}}} & \makecell[t]{\textbf{Spectral}\\ \textbf{Range}\\ \textbf{{[nm]}}} & \makecell[t]{\textbf{Mode}} & \makecell[t]{\textbf{FOV}\\ \textbf{{[\arcsec]}}} & \makecell[t]{\textbf{Sampling}\\ \textbf{{[\arcsec /px]}}} & \makecell[t]{\textbf{R}} & \makecell[t]{\textbf{SNR}} & \makecell[t]{\textbf{\# WL}\\ \textbf{Points}} & \makecell[t]{\textbf{Cadence}\\ \textbf{{[s]}}} \\
\hline
TIS/FBI & B & Photosphere & 430 & 1.00 & img & 40 & 0.02 & N/A & 500 & N/A & 1 \\
IFS-M & B & Chromosphere & 393.3 & 0.30 & pol & 8x8 & 0.02 & 75000 & 2000 & N/A & 10 \\
IFS-M & V & Photosphere & 630.2 & 0.30 & pol & 8x8 & 0.03 & 100000 & 2000 & N/A & 10 \\
TIS/FBI & V & Upper photosphere & 589.6 & 0.40 & spec & 40 & 0.03 & 100000 & 500 & 15 & 4 \\
TIS/FBI & R & Chromosphere & 854.2 & 0.30 & spec & 40 & 0.05 & 80000 & 500 & 21 & 5 \\
EMBER & IR & Photosphere & 1565 & 1.00 & pol & 8x8 & 0.10 & 100000 & 2000 & N/A & 10 \\
\hline
\textbf{Notes:} & \multicolumn{11}{p{17cm}|}{N/A} \\
\hline
\end{tabular}%
\end{adjustbox}%
\end{table}

\begin{table}[H]
\OPtabletitleCustom{2.1.2}{Magneto-acoustic waves in spicules on disk}{OP2.1.2table}{OP2.1.2}
\centering
\vspace{4mm}
\begin{adjustbox}{max width=\textwidth}
\begin{tabular}{|l|c|c|c|c|c|c|c|c|c|c|c|}
\hline
\textbf{Goal:} & \multicolumn{11}{p{17cm}|}{TIS/FBIs: High-resolution G-band imaging at very high cadence to determine wave signatures in RBEs/RREs. High-resolution spectroscopy to determine wave signatures in RBEs/RREs. IFS: High- resolution spectropolarimetry to determine wave signatures in RBEs/RREs.} \\
\hline
\makecell[t]{\textbf{Instrument}} & \makecell[t]{\textbf{Optical}\\ \textbf{Arm}} & \makecell[t]{\textbf{Atm}\\ \textbf{Layer}} & \makecell[t]{\textbf{CWL}\\ \textbf{{[nm]}}} & \makecell[t]{\textbf{Spectral}\\ \textbf{Range}\\ \textbf{{[nm]}}} & \makecell[t]{\textbf{Mode}} & \makecell[t]{\textbf{FOV}\\ \textbf{{[\arcsec]}}} & \makecell[t]{\textbf{Sampling}\\ \textbf{{[\arcsec /px]}}} & \makecell[t]{\textbf{R}} & \makecell[t]{\textbf{SNR}} & \makecell[t]{\textbf{\# WL}\\ \textbf{Points}} & \makecell[t]{\textbf{Cadence}\\ \textbf{{[s]}}} \\
\hline
TIS/FBI & B & Chromosphere & 393.3 & 0.30 & spec & 40 & 0.02 & 50000 & 500 & 15 & 4 \\
IFS-M & V & Photosphere & 630.2 & 0.40 & pol & 8x8 & 0.03 & 100000 & 2000 & N/A & 10 \\
TIS/FBI & V & Chromosphere & 656.3 & 0.40 & spec & 40 & 0.03 & 100000 & 500 & 21 & 5 \\
TIS/FBI & R & Chromosphere & 854.2 & 0.30 & spec & 40 & 0.05 & 80000 & 500 & 21 & 5 \\
EMBER & IR & Chromosphere & 1083 & 0.40 & pol & 8x8 & 0.10 & 100000 & 2000 & N/A & 10 \\
\hline
\textbf{Notes:} & \multicolumn{11}{p{17cm}|}{N/A} \\
\hline
\end{tabular}%
\end{adjustbox}%
\end{table}

\begin{table}[H]
\OPtabletitleCustom{2.1.3}{Magneto-acoustic waves in spicules at limb}{OP2.1.3table}{OP2.1.3}
\centering
\vspace{4mm}
\begin{adjustbox}{max width=\textwidth}
\begin{tabular}{|l|c|c|c|c|c|c|c|c|c|c|c|}
\hline
\textbf{Goal:} & \multicolumn{11}{p{17cm}|}{TIS/FBIs: High-resolution high cadence imaging to determine waves in spicules at the limb. High- resolution spectro-polarimetry to determine waves in spicules at the limb. IFS: High-resolution spectro-polarimetry to determine waves in spicules at the limb.} \\
\hline
\makecell[t]{\textbf{Instrument}} & \makecell[t]{\textbf{Optical}\\ \textbf{Arm}} & \makecell[t]{\textbf{Atm}\\ \textbf{Layer}} & \makecell[t]{\textbf{CWL}\\ \textbf{{[nm]}}} & \makecell[t]{\textbf{Spectral}\\ \textbf{Range}\\ \textbf{{[nm]}}} & \makecell[t]{\textbf{Mode}} & \makecell[t]{\textbf{FOV}\\ \textbf{{[\arcsec]}}} & \makecell[t]{\textbf{Sampling}\\ \textbf{{[\arcsec /px]}}} & \makecell[t]{\textbf{R}} & \makecell[t]{\textbf{SNR}} & \makecell[t]{\textbf{\# WL}\\ \textbf{Points}} & \makecell[t]{\textbf{Cadence}\\ \textbf{{[s]}}} \\
\hline
TIS/FBI & B & Chromosphere & 393.3 & 0.30 & spec & 40 & 0.02 & 50000 & 500 & 15 & 4 \\
TIS/FBI & V & Chromosphere & 656.3 & 0.40 & spec & 40 & 0.03 & 100000 & 500 & 21 & 5 \\
TIS/FBI & R & Chromosphere & 854.2 & 0.30 & spec & 40 & 0.05 & 80000 & 500 & 21 & 5 \\
EMBER & IR & Chromosphere & 1083 & 0.40 & pol & 8x8 & 0.10 & 100000 & 2000 & N/A & 10 \\
\hline
\textbf{Notes:} & \multicolumn{11}{p{17cm}|}{TIS/FBIs: Wavelength range: Doppler velocities in order of  100 km/s.} \\
\hline
\end{tabular}%
\end{adjustbox}%
\end{table}

\begin{table}[H]
\OPtabletitleCustom{2.1.4}{Torsional Alfv\'en wave (TAW) propagation in spicules}{OP2.1.4table}{OP2.1.4}
\centering
\vspace{4mm}
\begin{adjustbox}{max width=\textwidth}
\begin{tabular}{|l|c|c|c|c|c|c|c|c|c|c|c|}
\hline
\textbf{Goal:} & \multicolumn{11}{p{17cm}|}{IFS: High-resolution, high cadence Fe I 630.2 and Ca II spectro-polarimetry to determine TAWs in spicules.} \\
\hline
\makecell[t]{\textbf{Instrument}} & \makecell[t]{\textbf{Optical}\\ \textbf{Arm}} & \makecell[t]{\textbf{Atm}\\ \textbf{Layer}} & \makecell[t]{\textbf{CWL}\\ \textbf{{[nm]}}} & \makecell[t]{\textbf{Spectral}\\ \textbf{Range}\\ \textbf{{[nm]}}} & \makecell[t]{\textbf{Mode}} & \makecell[t]{\textbf{FOV}\\ \textbf{{[\arcsec]}}} & \makecell[t]{\textbf{Sampling}\\ \textbf{{[\arcsec /px]}}} & \makecell[t]{\textbf{R}} & \makecell[t]{\textbf{SNR}} & \makecell[t]{\textbf{\# WL}\\ \textbf{Points}} & \makecell[t]{\textbf{Cadence}\\ \textbf{{[s]}}} \\
\hline
TIS/FBI & B & Photosphere & 430 & 10.00 & img & 40 & 0.02 & N/A & 200 & N/A & 1 \\
IFS-M & B & Chromosphere & 393.3 & 0.30 & pol & 8x8 & 0.02 & 75000 & 2000 & N/A & 8 \\
IFS-M & V & Photosphere & 630.2 & 0.50 & pol & 8x8 & 0.03 & 100000 & 2000 & N/A & 8 \\
TIS/FBI & V & Chromosphere & 656.3 & 0.40 & spec & 40 & 0.03 & 100000 & 500 & 21 & 5 \\
IFS-M & R & Chromosphere & 854.2 & 0.30 & pol & 8x8 & 0.05 & 80000 & 2000 & N/A & 8 \\
EMBER & IR & Chromosphere & 1083 & 0.40 & pol & 8x8 & 0.10 & 100000 & 2000 & N/A & 10 \\
\hline
\textbf{Notes:} & \multicolumn{11}{p{17cm}|}{N/A} \\
\hline
\end{tabular}%
\end{adjustbox}%
\end{table}

\begin{table}[H]
\OPtabletitleCustom{2.2.1}{Magnetic torsion and torsional oscillations of pores or sunspots}{OP2.2.1table}{OP2.2.1}
\centering
\vspace{4mm}
\begin{adjustbox}{max width=\textwidth}
\begin{tabular}{|l|c|c|c|c|c|c|c|c|c|c|c|}
\hline
\textbf{Goal:} & \multicolumn{11}{p{17cm}|}{IFS: High-resolution, high cadence Fe I 630.2, Ca II and He I spectro-polarimetry to determine TAWs in pores (or sunspots).} \\
\hline
\makecell[t]{\textbf{Instrument}} & \makecell[t]{\textbf{Optical}\\ \textbf{Arm}} & \makecell[t]{\textbf{Atm}\\ \textbf{Layer}} & \makecell[t]{\textbf{CWL}\\ \textbf{{[nm]}}} & \makecell[t]{\textbf{Spectral}\\ \textbf{Range}\\ \textbf{{[nm]}}} & \makecell[t]{\textbf{Mode}} & \makecell[t]{\textbf{FOV}\\ \textbf{{[\arcsec]}}} & \makecell[t]{\textbf{Sampling}\\ \textbf{{[\arcsec /px]}}} & \makecell[t]{\textbf{R}} & \makecell[t]{\textbf{SNR}} & \makecell[t]{\textbf{\# WL}\\ \textbf{Points}} & \makecell[t]{\textbf{Cadence}\\ \textbf{{[s]}}} \\
\hline
TIS/FBI & B & Photosphere & 430 & 10.00 & img & 40 & 0.02 & N/A & 200 & N/A & 1 \\
IFS-M & B & Chromosphere & 393.3 & 0.50 & pol & 15x15 & 0.02 & 75000 & 2000 & N/A & 30 \\
IFS-M & V & Photosphere & 630.2 & 0.50 & pol & 15x15 & 0.03 & 100000 & 2000 & N/A & 30 \\
TIS/FBI & R & Chromosphere & 854.2 & 0.50 & pol & 40 & 0.05 & 80000 & 500 & 21 & 16 \\
EMBER & IR & Photosphere & 1565 & 0.50 & pol & 15x15 & 0.10 & 100000 & 2000 & N/A & 30 \\
\hline
\textbf{Notes:} & \multicolumn{11}{p{17cm}|}{N/A} \\
\hline
\end{tabular}%
\end{adjustbox}%
\end{table}

\begin{table}[H]
\OPtabletitleCustom{2.2.2}{Observation of magnetic vortexes and tornadoes}{OP2.2.2table}{OP2.2.2}
\centering
\vspace{4mm}
\begin{adjustbox}{max width=\textwidth}
\begin{tabular}{|l|c|c|c|c|c|c|c|c|c|c|c|}
\hline
\textbf{Goal:} & \multicolumn{11}{p{17cm}|}{TIS/FBIs: Statistical results in Liu et al. (2018) have shown that most ($>$ 70\%) swirls detected have lifetime less than 6 sec. High cadence is needed. Spatial resolution plays vital influence in the number and parameters of swirls detected. Very high cadence observations with few line positions are needed to study the coupling and propagation of very short-lived intensity and magnetic swirls. Extremely high cadence observations in Fe I line centre. More lines are desirable but high cadence is essential.} \\
\hline
\makecell[t]{\textbf{Instrument}} & \makecell[t]{\textbf{Optical}\\ \textbf{Arm}} & \makecell[t]{\textbf{Atm}\\ \textbf{Layer}} & \makecell[t]{\textbf{CWL}\\ \textbf{{[nm]}}} & \makecell[t]{\textbf{Spectral}\\ \textbf{Range}\\ \textbf{{[nm]}}} & \makecell[t]{\textbf{Mode}} & \makecell[t]{\textbf{FOV}\\ \textbf{{[\arcsec]}}} & \makecell[t]{\textbf{Sampling}\\ \textbf{{[\arcsec /px]}}} & \makecell[t]{\textbf{R}} & \makecell[t]{\textbf{SNR}} & \makecell[t]{\textbf{\# WL}\\ \textbf{Points}} & \makecell[t]{\textbf{Cadence}\\ \textbf{{[s]}}} \\
\hline
TIS/FBI & B & Photosphere & 430 & 10.00 & img & 40 & 0.02 & N/A & 100 & N/A & 1 \\
IFS-M & B & Chromosphere & 393.3 & 0.30 & pol & 8x8 & 0.02 & 75000 & 2000 & N/A & 10 \\
IFS-M & V & Photosphere & 630.2 & 0.30 & pol & 8x8 & 0.03 & 100000 & 2000 & N/A & 10 \\
TIS/FBI & V & Chromosphere & 656.3 & 0.40 & spec & 40 & 0.03 & 100000 & 500 & 21 & 5 \\
IFS-M & R & Chromosphere & 854.2 & 0.30 & pol & 8x8 & 0.05 & 80000 & 2000 & N/A & 10 \\
EMBER & IR & Photosphere & 1565 & 0.50 & pol & 8x8 & 0.10 & 100000 & 2000 & N/A & 10 \\
\hline
\textbf{Notes:} & \multicolumn{11}{p{17cm}|}{N/A} \\
\hline
\end{tabular}%
\end{adjustbox}%
\end{table}

\begin{table}[H]
\OPtabletitleCustom{2.2.3}{Formation of magnetic swirls in intergranular lanes}{OP2.2.3table}{OP2.2.3}
\centering
\vspace{4mm}
\begin{adjustbox}{max width=\textwidth}
\begin{tabular}{|l|c|c|c|c|c|c|c|c|c|c|c|}
\hline
\textbf{Goal:} & \multicolumn{11}{p{17cm}|}{TIS/FBIs and IFS IFU simultaneous imaging in G-Band. High cadence and resolution observations of the intergranular lanes.} \\
\hline
\makecell[t]{\textbf{Instrument}} & \makecell[t]{\textbf{Optical}\\ \textbf{Arm}} & \makecell[t]{\textbf{Atm}\\ \textbf{Layer}} & \makecell[t]{\textbf{CWL}\\ \textbf{{[nm]}}} & \makecell[t]{\textbf{Spectral}\\ \textbf{Range}\\ \textbf{{[nm]}}} & \makecell[t]{\textbf{Mode}} & \makecell[t]{\textbf{FOV}\\ \textbf{{[\arcsec]}}} & \makecell[t]{\textbf{Sampling}\\ \textbf{{[\arcsec /px]}}} & \makecell[t]{\textbf{R}} & \makecell[t]{\textbf{SNR}} & \makecell[t]{\textbf{\# WL}\\ \textbf{Points}} & \makecell[t]{\textbf{Cadence}\\ \textbf{{[s]}}} \\
\hline
TIS/FBI & B & Photosphere & 430 & 10.00 & img & 40 & 0.02 & N/A & 100 & N/A & 1 \\
IFS-M & B & Chromosphere & 393.3 & 0.30 & pol & 8x8 & 0.02 & 75000 & 2000 & N/A & 10 \\
IFS-M & V & Photosphere & 630.2 & 0.40 & pol & 8x8 & 0.03 & 100000 & 2000 & N/A & 10 \\
TIS/FBI & V & Chromosphere & 656.3 & 0.40 & spec & 40 & 0.03 & 100000 & 500 & 21 & 5 \\
TIS/FBI & R & Chromosphere & 854.2 & 0.30 & spec & 40 & 0.05 & 80000 & 500 & 15 & 4 \\
EMBER & IR & Chromosphere & 1083 & 0.50 & pol & 8x8 & 0.10 & 100000 & 2000 & N/A & 10 \\
\hline
\textbf{Notes:} & \multicolumn{11}{p{17cm}|}{N/A} \\
\hline
\end{tabular}%
\end{adjustbox}%
\end{table}

\begin{table}[H]
\OPtabletitleCustom{2.2.4}{Vortex flows in the lower solar atmosphere}{OP2.2.4table}{OP2.2.4}
\centering
\vspace{4mm}
\begin{adjustbox}{max width=\textwidth}
\begin{tabular}{|l|c|c|c|c|c|c|c|c|c|c|c|}
\hline
\textbf{Goal:} & \multicolumn{11}{p{17cm}|}{TIS/FBIs: Detection of vortex flows in the photosphere and tracking their evolution into the chromosphere. Detection of vertical flows within vortex motions and swirls. IFS: Detailed study of physical processes in and around vortex flows and their interaction with magnetic fields.} \\
\hline
\makecell[t]{\textbf{Instrument}} & \makecell[t]{\textbf{Optical}\\ \textbf{Arm}} & \makecell[t]{\textbf{Atm}\\ \textbf{Layer}} & \makecell[t]{\textbf{CWL}\\ \textbf{{[nm]}}} & \makecell[t]{\textbf{Spectral}\\ \textbf{Range}\\ \textbf{{[nm]}}} & \makecell[t]{\textbf{Mode}} & \makecell[t]{\textbf{FOV}\\ \textbf{{[\arcsec]}}} & \makecell[t]{\textbf{Sampling}\\ \textbf{{[\arcsec /px]}}} & \makecell[t]{\textbf{R}} & \makecell[t]{\textbf{SNR}} & \makecell[t]{\textbf{\# WL}\\ \textbf{Points}} & \makecell[t]{\textbf{Cadence}\\ \textbf{{[s]}}} \\
\hline
TIS/FBI & B & Photosphere & 430 & 10.00 & img & 40 & 0.02 & N/A & 100 & N/A & 1 \\
IFS-M & B & Chromosphere & 396.9 & 0.40 & pol & 8x8 & 0.02 & 75000 & 2000 & N/A & 10 \\
IFS-M & V & Photosphere & 630.2 & 0.50 & pol & 8x8 & 0.03 & 100000 & 2000 & N/A & 10 \\
TIS/FBI & V & Chromosphere & 517.3 & 0.40 & spec & 40 & 0.03 & 100000 & 500 & 15 & 4 \\
TIS/FBI & R & Photosphere & 709 & 0.40 & spec & 40 & 0.05 & 80000 & 500 & 15 & 4 \\
IFS-M & R & Chromosphere & 854.2 & 0.40 & pol & 8x8 & 0.05 & 80000 & 2000 & N/A & 10 \\
EMBER & IR & Chromosphere & 1083 & 0.50 & spec & 8x8 & 0.10 & 100000 & 2000 & N/A & 10 \\
\hline
\textbf{Notes:} & \multicolumn{11}{p{17cm}|}{IFS: allows to reconstruct tiles and minimizes differential refraction effects.} \\
\hline
\end{tabular}%
\end{adjustbox}%
\end{table}

\begin{table}[H]
\OPtabletitleCustom{2.3.1.1}{Excitation mechanisms of sunspot waves}{OP2.3.1.1table}{OP2.3.1.1}
\centering
\vspace{4mm}
\begin{adjustbox}{max width=\textwidth}
\begin{tabular}{|l|c|c|c|c|c|c|c|c|c|c|c|}
\hline
\textbf{Goal:} & \multicolumn{11}{p{17cm}|}{TIS/FBIs: Measure velocity and intensity oscillations in sunspots and nearby quiet Sun with high cadence, moderate spatial resolution, but large field of view. Long stable time series are required. Require simultaneously photospheric and chromospheric lines, spectroscopic measurements, not sensitive to magnetic field (non-polarimetric).} \\
\hline
\makecell[t]{\textbf{Instrument}} & \makecell[t]{\textbf{Optical}\\ \textbf{Arm}} & \makecell[t]{\textbf{Atm}\\ \textbf{Layer}} & \makecell[t]{\textbf{CWL}\\ \textbf{{[nm]}}} & \makecell[t]{\textbf{Spectral}\\ \textbf{Range}\\ \textbf{{[nm]}}} & \makecell[t]{\textbf{Mode}} & \makecell[t]{\textbf{FOV}\\ \textbf{{[\arcsec]}}} & \makecell[t]{\textbf{Sampling}\\ \textbf{{[\arcsec /px]}}} & \makecell[t]{\textbf{R}} & \makecell[t]{\textbf{SNR}} & \makecell[t]{\textbf{\# WL}\\ \textbf{Points}} & \makecell[t]{\textbf{Cadence}\\ \textbf{{[s]}}} \\
\hline
TIS/FBI & B & Photosphere & 430 & N/A & img & 40 & 0.05 & N/A & 500 & N/A & 10 \\
IFS-M & B & Upper photosphere & 455.4 & 0.50 & spec & 40x40 & 0.05 & 75000 & 500 & N/A & 10 \\
IFS-M & V & Photosphere & 543.4 & 0.50 & spec & 40x40 & 0.05 & 100000 & 500 & N/A & 10 \\
EMBER & IR & Chromosphere & 1083 & 0.50 & spec & 40x40 & 0.05 & 100000 & 500 & N/A & 10 \\
IFS-M & R & Photosphere & 709 & 0.50 & spec & 40x40 & 0.05 & 80000 & 500 & N/A & 10 \\
TIS/FBI & R & Chromosphere & 854.2 & 1.00 & spec & 40 & 0.05 & 80000 & 500 & 30 & 10 \\
TIS/FBI & V & Chromosphere & 656.3 & 2.00 & spec & 40 & 0.05 & 100000 & 500 & 30 & 10 \\
\hline
\textbf{Notes:} & \multicolumn{11}{p{17cm}|}{TIS/FBIs: Large field of view and simultaneous photospheric and chromospheric lines.} \\
\hline
\end{tabular}%
\end{adjustbox}%
\end{table}

\begin{table}[H]
\OPtabletitleCustom{2.3.1.2}{Excitation mechanisms of sunspot waves}{OP2.3.1.2table}{OP2.3.1.2}
\centering
\vspace{4mm}
\begin{adjustbox}{max width=\textwidth}
\begin{tabular}{|l|c|c|c|c|c|c|c|c|c|c|c|}
\hline
\textbf{Goal:} & \multicolumn{11}{p{17cm}|}{IFS: Measure magnetic field vector in the photosphere and chromosphere to complement the velocity data} \\
\hline
\makecell[t]{\textbf{Instrument}} & \makecell[t]{\textbf{Optical}\\ \textbf{Arm}} & \makecell[t]{\textbf{Atm}\\ \textbf{Layer}} & \makecell[t]{\textbf{CWL}\\ \textbf{{[nm]}}} & \makecell[t]{\textbf{Spectral}\\ \textbf{Range}\\ \textbf{{[nm]}}} & \makecell[t]{\textbf{Mode}} & \makecell[t]{\textbf{FOV}\\ \textbf{{[\arcsec]}}} & \makecell[t]{\textbf{Sampling}\\ \textbf{{[\arcsec /px]}}} & \makecell[t]{\textbf{R}} & \makecell[t]{\textbf{SNR}} & \makecell[t]{\textbf{\# WL}\\ \textbf{Points}} & \makecell[t]{\textbf{Cadence}\\ \textbf{{[s]}}} \\
\hline
EMBER & IR & Photosphere & 1564.8 & 1.00 & pol & 40x40 & 0.05 & 100000 & 2000 & N/A & 45 \\
IFS-M & R & Chromosphere & 854.2 & 1.00 & pol & 40x40 & 0.05 & 80000 & 2000 & N/A & 45 \\
\hline
\textbf{Notes:} & \multicolumn{11}{p{17cm}|}{IFS: Enough polarimetric precision to derive the magnetic field vector.} \\
\hline
\end{tabular}%
\end{adjustbox}%
\end{table}

\begin{table}[H]
\OPtabletitleCustom{2.3.2.1}{Alfv\'en waves in sunspots}{OP2.3.2.1table}{OP2.3.2.1}
\centering
\vspace{4mm}
\begin{adjustbox}{max width=\textwidth}
\begin{tabular}{|l|c|c|c|c|c|c|c|c|c|c|c|}
\hline
\textbf{Goal:} & \multicolumn{11}{p{17cm}|}{IFS: Study oscillations of velocity and magnetic field in sunspots at several positions across the disc to determine if they are caused by Alfv\'en waves. Measurements in photospheric Fe I 1564 nm lines provide magnetic field in the deep layers, while He I 1083, and Ca II 854 nm provide magnetic field in the chromosphere. High polarimetric precision is crucial.} \\
\hline
\makecell[t]{\textbf{Instrument}} & \makecell[t]{\textbf{Optical}\\ \textbf{Arm}} & \makecell[t]{\textbf{Atm}\\ \textbf{Layer}} & \makecell[t]{\textbf{CWL}\\ \textbf{{[nm]}}} & \makecell[t]{\textbf{Spectral}\\ \textbf{Range}\\ \textbf{{[nm]}}} & \makecell[t]{\textbf{Mode}} & \makecell[t]{\textbf{FOV}\\ \textbf{{[\arcsec]}}} & \makecell[t]{\textbf{Sampling}\\ \textbf{{[\arcsec /px]}}} & \makecell[t]{\textbf{R}} & \makecell[t]{\textbf{SNR}} & \makecell[t]{\textbf{\# WL}\\ \textbf{Points}} & \makecell[t]{\textbf{Cadence}\\ \textbf{{[s]}}} \\
\hline
EMBER & IR & Photosphere & 1564.8 & 1.00 & pol & 30x30 & 0.05 & 100000 & 2000 & N/A & 60 \\
IFS-M & R & Chromosphere & 854.2 & 1.00 & pol & 30x30 & 0.05 & 80000 & 2000 & N/A & 60 \\
\hline
\textbf{Notes:} & \multicolumn{11}{p{17cm}|}{N/A} \\
\hline
\end{tabular}%
\end{adjustbox}%
\end{table}

\begin{table}[H]
\OPtabletitleCustom{2.3.2.2}{Alfv\'en waves in sunspots}{OP2.3.2.2table}{OP2.3.2.2}
\centering
\vspace{4mm}
\begin{adjustbox}{max width=\textwidth}
\begin{tabular}{|l|c|c|c|c|c|c|c|c|c|c|c|}
\hline
\textbf{Goal:} & \multicolumn{11}{p{17cm}|}{TIS/FBIs: Bi-dimensional spectra to determine velocity oscillations at different heights from photosphere to chromosphere; phase speed of waves to determine their direction of propagation and heights of reflection. Spectroscopic (non-polarimetric mode).
EMBER: Study oscillations of velocity and magnetic field in sunspots at several positions across the disc to determine if they are caused by Alfv\'en waves. Measurements in photospheric Fe I 1564 nm lines provide magnetic field in the deep layers, while He I 1083 provides magnetic field in the chromosphere.} \\
\hline
\makecell[t]{\textbf{Instrument}} & \makecell[t]{\textbf{Optical}\\ \textbf{Arm}} & \makecell[t]{\textbf{Atm}\\ \textbf{Layer}} & \makecell[t]{\textbf{CWL}\\ \textbf{{[nm]}}} & \makecell[t]{\textbf{Spectral}\\ \textbf{Range}\\ \textbf{{[nm]}}} & \makecell[t]{\textbf{Mode}} & \makecell[t]{\textbf{FOV}\\ \textbf{{[\arcsec]}}} & \makecell[t]{\textbf{Sampling}\\ \textbf{{[\arcsec /px]}}} & \makecell[t]{\textbf{R}} & \makecell[t]{\textbf{SNR}} & \makecell[t]{\textbf{\# WL}\\ \textbf{Points}} & \makecell[t]{\textbf{Cadence}\\ \textbf{{[s]}}} \\
\hline
EMBER & IR & Chromosphere & 1082.7 & 1.00 & spec & 30x30 & 0.05 & 100000 & 500 & N/A & 30 \\
TIS/FBI & R & Chromosphere & 854.2 & 1.00 & spec & 30 & 0.05 & 80000 & 500 & 30 & 30 \\
TIS/FBI & V & Chromosphere & 656.3 & 2.00 & spec & 30 & 0.05 & 100000 & 500 & 30 & 30 \\
TIS/FBI & B & Upper photosphere & 455.4 & 0.50 & spec & 30 & 0.05 & 40000 & 500 & 30 & 30 \\
\hline
\textbf{Notes:} & \multicolumn{11}{p{17cm}|}{} \\
\hline
\end{tabular}%
\end{adjustbox}%
\end{table}

\begin{table}[H]
\OPtabletitleCustom{2.3.3}{Magnetic field oscillations in sunspots}{OP2.3.3table}{OP2.3.3}
\centering
\vspace{4mm}
\begin{adjustbox}{max width=\textwidth}
\begin{tabular}{|l|c|c|c|c|c|c|c|c|c|c|c|}
\hline
\textbf{Goal:} & \multicolumn{11}{p{17cm}|}{IFS: Measure precisely oscillations of magnetic field, together with velocity oscillations, across sunspot umbra and penumbra. Time series of 2D spectropolarimetric data in several lines with different temperature and magnetic field sensitivity simultaneously.} \\
\hline
\makecell[t]{\textbf{Instrument}} & \makecell[t]{\textbf{Optical}\\ \textbf{Arm}} & \makecell[t]{\textbf{Atm}\\ \textbf{Layer}} & \makecell[t]{\textbf{CWL}\\ \textbf{{[nm]}}} & \makecell[t]{\textbf{Spectral}\\ \textbf{Range}\\ \textbf{{[nm]}}} & \makecell[t]{\textbf{Mode}} & \makecell[t]{\textbf{FOV}\\ \textbf{{[\arcsec]}}} & \makecell[t]{\textbf{Sampling}\\ \textbf{{[\arcsec /px]}}} & \makecell[t]{\textbf{R}} & \makecell[t]{\textbf{SNR}} & \makecell[t]{\textbf{\# WL}\\ \textbf{Points}} & \makecell[t]{\textbf{Cadence}\\ \textbf{{[s]}}} \\
\hline
EMBER & IR & Photosphere & 1564.8 & 1.00 & pol & 5x30 & 0.08 & 100000 & 8000 & N/A & 30 \\
TIS/FBI & V & Photosphere & 630.2 & 0.50 & pol & 5x30 & 0.08 & 100000 & 8000 & 5 & 30 \\
IFS-M & R & Chromosphere & 854.2 & 1.00 & pol & 5x30 & 0.08 & 80000 & 8000 & N/A & 30 \\
\hline
\textbf{Notes:} & \multicolumn{11}{p{17cm}|}{IFS: High polarimetric precision and low noise is crucial.} \\
\hline
\end{tabular}%
\end{adjustbox}%
\end{table}

\begin{table}[H]
\OPtabletitleCustom{2.3.4.1}{Sunspot penumbral waves in the photosphere and above}{OP2.3.4.1table}{OP2.3.4.1}
\centering
\vspace{4mm}
\begin{adjustbox}{max width=\textwidth}
\begin{tabular}{|l|c|c|c|c|c|c|c|c|c|c|c|}
\hline
\textbf{Goal:} & \multicolumn{11}{p{17cm}|}{TIS/FBIs: Measure 2D velocity field over all sunspot simultaneously. Spectroscopic mode. IFS: Measure 2D magnetic field vector in the photosphere and chromosphere to complement the velocity data.} \\
\hline
\makecell[t]{\textbf{Instrument}} & \makecell[t]{\textbf{Optical}\\ \textbf{Arm}} & \makecell[t]{\textbf{Atm}\\ \textbf{Layer}} & \makecell[t]{\textbf{CWL}\\ \textbf{{[nm]}}} & \makecell[t]{\textbf{Spectral}\\ \textbf{Range}\\ \textbf{{[nm]}}} & \makecell[t]{\textbf{Mode}} & \makecell[t]{\textbf{FOV}\\ \textbf{{[\arcsec]}}} & \makecell[t]{\textbf{Sampling}\\ \textbf{{[\arcsec /px]}}} & \makecell[t]{\textbf{R}} & \makecell[t]{\textbf{SNR}} & \makecell[t]{\textbf{\# WL}\\ \textbf{Points}} & \makecell[t]{\textbf{Cadence}\\ \textbf{{[s]}}} \\
\hline
TIS/FBI & B & Upper photosphere & 455.4 & 0.50 & spec & 40 & 0.05 & 40000 & 500 & 10 & 4 \\
IFS-M & V & Photosphere & 543.4 & 0.50 & spec & 40x40 & 0.05 & 100000 & 500 & N/A & 60 \\
IFS-M & R & Photosphere & 709 & 0.50 & spec & 40x40 & 0.05 & 80000 & 500 & N/A & 60 \\
TIS/FBI & R & Chromosphere & 854.2 & 1.00 & spec & 40 & 0.05 & 80000 & 500 & 10 & 4 \\
TIS/FBI & V & Chromosphere & 656.3 & 2.00 & spec & 40 & 0.05 & 100000 & 500 & 10 & 4 \\
\hline
\textbf{Notes:} & \multicolumn{11}{p{17cm}|}{IFS: High polarimetric precision and low noise is crucial.} \\
\hline
\end{tabular}%
\end{adjustbox}%
\end{table}

\begin{table}[H]
\OPtabletitleCustom{2.3.4.2}{Sunspot penumbral waves in the photosphere and above}{OP2.3.4.2table}{OP2.3.4.2}
\centering
\vspace{4mm}
\begin{adjustbox}{max width=\textwidth}
\begin{tabular}{|l|c|c|c|c|c|c|c|c|c|c|c|}
\hline
\textbf{Goal:} & \multicolumn{11}{p{17cm}|}{TIS/FBIs: Measure 2D velocity field over all sunspot simultaneously. Spectroscopic mode. IFS: Measure 2D magnetic field vector in the photosphere and chromosphere to complement the velocity data.} \\
\hline
\makecell[t]{\textbf{Instrument}} & \makecell[t]{\textbf{Optical}\\ \textbf{Arm}} & \makecell[t]{\textbf{Atm}\\ \textbf{Layer}} & \makecell[t]{\textbf{CWL}\\ \textbf{{[nm]}}} & \makecell[t]{\textbf{Spectral}\\ \textbf{Range}\\ \textbf{{[nm]}}} & \makecell[t]{\textbf{Mode}} & \makecell[t]{\textbf{FOV}\\ \textbf{{[\arcsec]}}} & \makecell[t]{\textbf{Sampling}\\ \textbf{{[\arcsec /px]}}} & \makecell[t]{\textbf{R}} & \makecell[t]{\textbf{SNR}} & \makecell[t]{\textbf{\# WL}\\ \textbf{Points}} & \makecell[t]{\textbf{Cadence}\\ \textbf{{[s]}}} \\
\hline
EMBER & IR & Photosphere & 1564.8 & 1.00 & pol & 40x40 & 0.05 & 100000 & 2000 & N/A & 60 \\
IFS-M & R & Chromosphere & 854.2 & 1.00 & pol & 40x40 & 0.05 & 80000 & 2000 & N/A & 60 \\
\hline
\textbf{Notes:} & \multicolumn{11}{p{17cm}|}{IFS: High polarimetric precision and low noise is crucial.} \\
\hline
\end{tabular}%
\end{adjustbox}%
\end{table}

\begin{table}[H]
\OPtabletitleCustom{2.3.5}{Sausage and kink oscillations in pores}{OP2.3.5table}{OP2.3.5}
\centering
\vspace{4mm}
\begin{adjustbox}{max width=\textwidth}
\begin{tabular}{|l|c|c|c|c|c|c|c|c|c|c|c|}
\hline
\textbf{Goal:} & \multicolumn{11}{p{17cm}|}{TIS/FBIs and IFS: High-resolution G-band images and LoS magnetic field to determine sausage and kink waves in pores.} \\
\hline
\makecell[t]{\textbf{Instrument}} & \makecell[t]{\textbf{Optical}\\ \textbf{Arm}} & \makecell[t]{\textbf{Atm}\\ \textbf{Layer}} & \makecell[t]{\textbf{CWL}\\ \textbf{{[nm]}}} & \makecell[t]{\textbf{Spectral}\\ \textbf{Range}\\ \textbf{{[nm]}}} & \makecell[t]{\textbf{Mode}} & \makecell[t]{\textbf{FOV}\\ \textbf{{[\arcsec]}}} & \makecell[t]{\textbf{Sampling}\\ \textbf{{[\arcsec /px]}}} & \makecell[t]{\textbf{R}} & \makecell[t]{\textbf{SNR}} & \makecell[t]{\textbf{\# WL}\\ \textbf{Points}} & \makecell[t]{\textbf{Cadence}\\ \textbf{{[s]}}} \\
\hline
TIS/FBI & B & Photosphere & 430 & N/A & img & 10 & $^{*}$0.04 & N/A & 800 & N/A & 1 \\
IFS-M & B & Chromosphere & $^{+}$393.3 & 0.40 & pol & 8 & 0.04 & 75000 & 2000 & N/A & 10 \\
IFS-M & V & Upper photosphere & $^{+}$589 & 0.40 & pol & 8 & 0.04 & 100000 & 2000 & N/A & 10 \\
TIS/FBI & V & Photosphere & $^{+}$630.2 & 0.50 & spec & 10 & 0.03 & 100000 & 8000 & 2 & 30 \\
TIS/FBI & R & Upper photosphere & $^{+}$769.9 & 0.40 & spec & 10 & 0.03 & 80000 & 8000 & 2 & 30 \\
IFS-M & R & Chromosphere & $^{+}$854.2 & 1.00 & pol & 8 & 0.04 & 80000 & 2000 & N/A & 10 \\
EMBER & IR & Photosphere & 1565 & 1.00 & pol & 10 & $^{*}$0.04 & 100000 & 800 & N/A & 10 \\
\hline
\textbf{Notes:} & \multicolumn{11}{p{17cm}|}{Extremely high-cadence is desired.} \\
\hline
\end{tabular}%
\end{adjustbox}%
\end{table}

\begin{table}[H]
\OPtabletitleCustom{2.3.6}{(Torsional) Alfv\'en waves in pores}{OP2.3.6table}{OP2.3.6}
\centering
\vspace{4mm}
\begin{adjustbox}{max width=\textwidth}
\begin{tabular}{|l|c|c|c|c|c|c|c|c|c|c|c|}
\hline
\textbf{Goal:} & \multicolumn{11}{p{17cm}|}{TIS/FBIs: Spectra to determine velocity oscillations at different heights from photosphere to chromosphere; phase speed of waves to determine their direction of propagation and heights of reflection. IFS: Study oscillations of velocity and magnetic field in sunspots at several positions across the disc to determine if they are caused by TAWs. Measurements in photospheric Fe I 1564 nm lines provide magnetic field in the deep layers, while He I 1083, and Ca II 854 nm provide magnetic field in the chromosphere.} \\
\hline
\makecell[t]{\textbf{Instrument}} & \makecell[t]{\textbf{Optical}\\ \textbf{Arm}} & \makecell[t]{\textbf{Atm}\\ \textbf{Layer}} & \makecell[t]{\textbf{CWL}\\ \textbf{{[nm]}}} & \makecell[t]{\textbf{Spectral}\\ \textbf{Range}\\ \textbf{{[nm]}}} & \makecell[t]{\textbf{Mode}} & \makecell[t]{\textbf{FOV}\\ \textbf{{[\arcsec]}}} & \makecell[t]{\textbf{Sampling}\\ \textbf{{[\arcsec /px]}}} & \makecell[t]{\textbf{R}} & \makecell[t]{\textbf{SNR}} & \makecell[t]{\textbf{\# WL}\\ \textbf{Points}} & \makecell[t]{\textbf{Cadence}\\ \textbf{{[s]}}} \\
\hline
TIS/FBI & B & Photosphere & $^{+}$430 & N/A & img & 40 & $^{*}$0.03 & N/A & 800 & N/A & 3 \\
 &  & Upper photosphere & 455.4 & 0.50 & spec & 40 & 0.03 & 40000 & 300 & 3 & 20 \\
IFS-M & V & Photosphere & $^{+}$630.2 & 0.50 & pol & 8 & 0.03 & 100000 & 8000 & N/A & 20 \\
TIS/FBI & V & Chromosphere & 656.3 & 2.00 & spec & 40 & 0.03 & 100000 & 300 & 3 & 20 \\
IFS-M & R & Chromosphere & 854.2 & 1.00 & pol & 8 & 0.03 & 80000 & 2000 & N/A & 3 \\
TIS/FBI & R & Upper photosphere & 769.9 & 0.40 & spec & 40 & 0.03 & 80000 & 300 & 3 & 20 \\
EMBER & IR & Chromosphere & 1083 & 0.50 & pol & 10 & $^{*}$0.03 & 100000 & 2000 & N/A & 3 \\
 &  & Photosphere & 1565 & 1.00 & pol & 10 & $^{*}$0.04 & 100000 & 2000 & N/A & 3 \\
\hline
\textbf{Notes:} & \multicolumn{11}{p{17cm}|}{TIS/FBIs: Bi-dimensional field of view. IFS: High polarimetric precision is crucial.} \\
\hline
\end{tabular}%
\end{adjustbox}%
\end{table}

\begin{table}[H]
\OPtabletitleCustom{2.4.1}{High-frequency wave propagation and dissipation from the photosphere to the chromosphere}{OP2.4.1table}{OP2.4.1}
\centering
\vspace{4mm}
\begin{adjustbox}{max width=\textwidth}
\begin{tabular}{|l|c|c|c|c|c|c|c|c|c|c|c|}
\hline
\textbf{Goal:} & \multicolumn{11}{p{17cm}|}{TIS/FBIs: Measure 2D velocity field over quiet and plage regions with simultaneous G-Band imaging. IFS: Measure 2D magnetic field vector in the photosphere and chromosphere to complement the velocity data.} \\
\hline
\makecell[t]{\textbf{Instrument}} & \makecell[t]{\textbf{Optical}\\ \textbf{Arm}} & \makecell[t]{\textbf{Atm}\\ \textbf{Layer}} & \makecell[t]{\textbf{CWL}\\ \textbf{{[nm]}}} & \makecell[t]{\textbf{Spectral}\\ \textbf{Range}\\ \textbf{{[nm]}}} & \makecell[t]{\textbf{Mode}} & \makecell[t]{\textbf{FOV}\\ \textbf{{[\arcsec]}}} & \makecell[t]{\textbf{Sampling}\\ \textbf{{[\arcsec /px]}}} & \makecell[t]{\textbf{R}} & \makecell[t]{\textbf{SNR}} & \makecell[t]{\textbf{\# WL}\\ \textbf{Points}} & \makecell[t]{\textbf{Cadence}\\ \textbf{{[s]}}} \\
\hline
TIS/FBI & B & Photosphere & 430 & 1.00 & img & 30 & $^{*}$0.03 & N/A & 300 & N/A & 3 \\
IFS-M & B & Upper photosphere & 455.4 & 0.50 & pol & 30 & 0.03 & 75000 & 2000 & N/A & 45 \\
IFS-M & V & Photosphere & 630.2 & 0.30 & pol & 30 & 0.03 & 100000 & 2000 & N/A & 45 \\
TIS/FBI & V & Chromosphere & 517.3 & 0.40 & spec & 30 & 0.03 & 100000 & 300 & 4 & 3 \\
IFS-M & R & Photosphere & 709 & 0.20 & pol & 30 & 0.03 & 80000 & 2000 & N/A & 45 \\
TIS/FBI & R & Chromosphere & 854.2 & 0.50 & spec & 30 & 0.03 & 80000 & 300 & 6 & 3 \\
EMBER & IR & Chromosphere & 1083 & 0.50 & spec & 30 & $^{*}$0.03 & 100000 & 2000 & N/A & 45 \\
\hline
\textbf{Notes:} & \multicolumn{11}{p{17cm}|}{TIS/FBIs: High resolution and high cadence in 2D field of view in spectroscopic mode. IFS: High polarimetric precision and low noise.} \\
\hline
\end{tabular}%
\end{adjustbox}%
\end{table}

\begin{table}[H]
\OPtabletitleCustom{2.4.2}{Time-dependent behaviour of chromospheric jets}{OP2.4.2table}{OP2.4.2}
\centering
\vspace{4mm}
\begin{adjustbox}{max width=\textwidth}
\begin{tabular}{|l|c|c|c|c|c|c|c|c|c|c|c|}
\hline
\textbf{Goal:} & \multicolumn{11}{p{17cm}|}{TIS/FBIs: Measure 2D velocity and intensity oscillations over network magnetic field concentrations. IFS: Measure 2D magnetic field vector in the photosphere and chromosphere to complement the velocity data.} \\
\hline
\makecell[t]{\textbf{Instrument}} & \makecell[t]{\textbf{Optical}\\ \textbf{Arm}} & \makecell[t]{\textbf{Atm}\\ \textbf{Layer}} & \makecell[t]{\textbf{CWL}\\ \textbf{{[nm]}}} & \makecell[t]{\textbf{Spectral}\\ \textbf{Range}\\ \textbf{{[nm]}}} & \makecell[t]{\textbf{Mode}} & \makecell[t]{\textbf{FOV}\\ \textbf{{[\arcsec]}}} & \makecell[t]{\textbf{Sampling}\\ \textbf{{[\arcsec /px]}}} & \makecell[t]{\textbf{R}} & \makecell[t]{\textbf{SNR}} & \makecell[t]{\textbf{\# WL}\\ \textbf{Points}} & \makecell[t]{\textbf{Cadence}\\ \textbf{{[s]}}} \\
\hline
TIS/FBI & B & Photosphere & 430 & 1.00 & img & 20 & $^{*}$0.03 & N/A & 500 & N/A & 20 \\
IFS-M & B & Chromosphere & 393.3 & 0.50 & pol & 8 & 0.03 & 75000 & 2000 & N/A & 4 \\
TIS/FBI & V & Upper photosphere & 589 & 0.40 & spec & 20 & 0.03 & 100000 & 500 & 20 & 4 \\
IFS-M & V & Photosphere & 630.2 & 0.30 & pol & 8 & 0.03 & 100000 & 2000 & N/A & 4 \\
IFS-M & R & Chromosphere & 854.2 & 0.50 & pol & 8 & 0.03 & 80000 & 2000 & N/A & 4 \\
EMBER & IR & Chromosphere & 1083 & 0.50 & pol & 10 & $^{*}$0.03 & 100000 & 2000 & N/A & 4 \\
\hline
\textbf{Notes:} & \multicolumn{11}{p{17cm}|}{TIS/FBIs: High resolution and high cadence in 2D field of view in spectroscopic mode. IFS: High polarimetric precision and low noise.} \\
\hline
\end{tabular}%
\end{adjustbox}%
\end{table}

\begin{table}[H]
\OPtabletitleCustom{2.4.3}{Network and plage oscillations}{OP2.4.3table}{OP2.4.3}
\centering
\vspace{4mm}
\begin{adjustbox}{max width=\textwidth}
\begin{tabular}{|l|c|c|c|c|c|c|c|c|c|c|c|}
\hline
\textbf{Goal:} & \multicolumn{11}{p{17cm}|}{TIS/FBIs: Measure long-period velocity oscillations in the network, high spatial resolution, medium- large field of view. Long stable time series are required. Require simultaneously photospheric and chromospheric lines best suited for velocity measurements. IFS: Measure magnetic field vector in the photosphere and chromosphere to complement the velocity data.} \\
\hline
\makecell[t]{\textbf{Instrument}} & \makecell[t]{\textbf{Optical}\\ \textbf{Arm}} & \makecell[t]{\textbf{Atm}\\ \textbf{Layer}} & \makecell[t]{\textbf{CWL}\\ \textbf{{[nm]}}} & \makecell[t]{\textbf{Spectral}\\ \textbf{Range}\\ \textbf{{[nm]}}} & \makecell[t]{\textbf{Mode}} & \makecell[t]{\textbf{FOV}\\ \textbf{{[\arcsec]}}} & \makecell[t]{\textbf{Sampling}\\ \textbf{{[\arcsec /px]}}} & \makecell[t]{\textbf{R}} & \makecell[t]{\textbf{SNR}} & \makecell[t]{\textbf{\# WL}\\ \textbf{Points}} & \makecell[t]{\textbf{Cadence}\\ \textbf{{[s]}}} \\
\hline
TIS/FBI & B & Photosphere & 430 & 1.00 & img & 40 & $^{*}$0.03 & N/A & 500 & N/A & 4 \\
IFS-M & B & Upper photosphere & 455.4 & 0.50 & pol & 40 & 0.03 & 75000 & 2000 & N/A & 50 \\
TIS/FBI & V & Upper photosphere & 589 & 0.50 & spec & 40 & 0.03 & 100000 & 500 & 20 & 30 \\
IFS-M & V & Photosphere & 630.2 & 0.30 & pol & 40 & 0.03 & 100000 & 2000 & N/A & 50 \\
IFS-M & R & Photosphere & 709 & 0.20 & pol & 40 & 0.03 & 80000 & 2000 & N/A & 50 \\
TIS/FBI & R & Chromosphere & 854.2 & 0.50 & spec & 40 & 0.03 & 80000 & 500 & 20 & 30 \\
EMBER & IR & Photosphere & 1564 & 0.50 & spec & 40 & $^{*}$0.04 & 100000 & 500 & N/A & 50 \\
\hline
\textbf{Notes:} & \multicolumn{11}{p{17cm}|}{TIS/FBIs: Large field of view and simultaneous photospheric and chromospheric lines. IFS: High polarimetric precision to derive the magnetic field vector in network.} \\
\hline
\end{tabular}%
\end{adjustbox}%
\end{table}

\begin{table}[H]
\OPtabletitleCustom{3.1.1}{Magnetic field structure in the quiet Sun chromosphere}{OP3.1.1table}{OP3.1.1}
\centering
\vspace{4mm}
\begin{adjustbox}{max width=\textwidth}
\begin{tabular}{|l|c|c|c|c|c|c|c|c|c|c|c|}
\hline
\textbf{Goal:} & \multicolumn{11}{p{17cm}|}{} \\
\hline
\makecell[t]{\textbf{Instrument}} & \makecell[t]{\textbf{Optical}\\ \textbf{Arm}} & \makecell[t]{\textbf{Atm}\\ \textbf{Layer}} & \makecell[t]{\textbf{CWL}\\ \textbf{{[nm]}}} & \makecell[t]{\textbf{Spectral}\\ \textbf{Range}\\ \textbf{{[nm]}}} & \makecell[t]{\textbf{Mode}} & \makecell[t]{\textbf{FOV}\\ \textbf{{[\arcsec]}}} & \makecell[t]{\textbf{Sampling}\\ \textbf{{[\arcsec /px]}}} & \makecell[t]{\textbf{R}} & \makecell[t]{\textbf{SNR}} & \makecell[t]{\textbf{\# WL}\\ \textbf{Points}} & \makecell[t]{\textbf{Cadence}\\ \textbf{{[s]}}} \\
\hline
TIS/FBI & V & Photosphere & 589.0 & 0.3 & pol & 40 & $^{*}$0.01 & 50000 & 2000 & 6 & 30 \\
IFS-M & V & Photosphere & 630.2 & 0.1 & pol & 60x60 & $^{*}$0.02 & 100000 & 10000 & N/A & 3600 \\
IFS-M & R & Chromosphere & 854.2 & 0.6 & pol & 60x60 & $^{*}$0.02 & 100000 & 10000 & N/A & 3600 \\
EMBER & IR & Chromosphere & 1083.0 & 0.6 & pol & 60x60 & $^{*}$0.03 & 100000 & 10000 & N/A & 3600 \\ 
& & Photosphere & 1082.7 & 0.6 & pol & 60x60 & $^{*}$0.03 & 100000 & 10000 & N/A & 3600 \\ 
\hline
\textbf{Notes:} & \multicolumn{11}{p{17cm}|}{N/A} \\
\hline
\end{tabular}%
\end{adjustbox}%
\end{table}

\begin{table}[H]
\OPtabletitleCustom{3.1.2}{Fibrilar structure of the chromosphere}{OP3.1.2table}{OP3.1.2}
\centering
\vspace{4mm}
\begin{adjustbox}{max width=\textwidth}
\begin{tabular}{|l|c|c|c|c|c|c|c|c|c|c|c|}
\hline
\textbf{Goal:} & \multicolumn{11}{p{17cm}|}{TIS/FBIs: Observe fibrillar structure of the chromosphere at the highest resolution.} \\
\hline
\makecell[t]{\textbf{Instrument}} & \makecell[t]{\textbf{Optical}\\ \textbf{Arm}} & \makecell[t]{\textbf{Atm}\\ \textbf{Layer}} & \makecell[t]{\textbf{CWL}\\ \textbf{{[nm]}}} & \makecell[t]{\textbf{Spectral}\\ \textbf{Range}\\ \textbf{{[nm]}}} & \makecell[t]{\textbf{Mode}} & \makecell[t]{\textbf{FOV}\\ \textbf{{[\arcsec]}}} & \makecell[t]{\textbf{Sampling}\\ \textbf{{[\arcsec /px]}}} & \makecell[t]{\textbf{R}} & \makecell[t]{\textbf{SNR}} & \makecell[t]{\textbf{\# WL}\\ \textbf{Points}} & \makecell[t]{\textbf{Cadence}\\ \textbf{{[s]}}} \\
\hline
TIS/FBI & B & Chromosphere & 393.4 & 0.07 & spec & 30 & $^{*}$0.01 & 50000 & 250 & 10 & 15 \\
TIS/FBI & V & Photosphere & 617.3 & 0.10 & pol & 30 & $^{*}$0.02 & 100000 & 1000 & 7 & 15 \\
 &  & Chromosphere & 656.3 & 0.20 & spec & 30 & $^{*}$0.02 & 100000 & 500 & 7 & 15 \\
EMBER & IR & Chromosphere & $^{+}$1083 & 0.03 & pol & 10x10 & $^{*}$0.03 & 100000 & 3000 & N/A & 3 \\
\hline
\textbf{Notes:} & \multicolumn{11}{p{17cm}|}{N/A} \\
\hline
\end{tabular}%
\end{adjustbox}%
\end{table}

\begin{table}[H]
\OPtabletitleCustom{3.2.1}{Type II spicule acceleration on disk}{OP3.2.1table}{OP3.2.1}
\centering
\vspace{4mm}
\begin{adjustbox}{max width=\textwidth}
\begin{tabular}{|l|c|c|c|c|c|c|c|c|c|c|c|}
\hline
\textbf{Goal:} & \multicolumn{11}{p{17cm}|}{IFS: Catch spicule acceleration on-disk at highest spatial and temporal resolution with polarimetry.} \\
\hline
\makecell[t]{\textbf{Instrument}} & \makecell[t]{\textbf{Optical}\\ \textbf{Arm}} & \makecell[t]{\textbf{Atm}\\ \textbf{Layer}} & \makecell[t]{\textbf{CWL}\\ \textbf{{[nm]}}} & \makecell[t]{\textbf{Spectral}\\ \textbf{Range}\\ \textbf{{[nm]}}} & \makecell[t]{\textbf{Mode}} & \makecell[t]{\textbf{FOV}\\ \textbf{{[\arcsec]}}} & \makecell[t]{\textbf{Sampling}\\ \textbf{{[\arcsec /px]}}} & \makecell[t]{\textbf{R}} & \makecell[t]{\textbf{SNR}} & \makecell[t]{\textbf{\# WL}\\ \textbf{Points}} & \makecell[t]{\textbf{Cadence}\\ \textbf{{[s]}}} \\
\hline
IFS-M & V & Photosphere & 617.3 & 0.10 & pol & 10x10 & $^{*}$0.02 & 100000 & 3000 & N/A & 20 \\
IFS-M & R & Chromosphere & 854.2 & 0.28 & pol & 10x10 & 0.03 & 80000 & 3000 & N/A & 20 \\
IFS-M & B & Chromosphere & 393.4 & 0.13 & spec & 10x10 & $^{*}$0.01 & 75000 & 500 & N/A & 10 \\
IFS-M & V & Upper photosphere & $^{+}$589 & 0.19 & pol & 10x10 & $^{*}$0.01 & 100000 & 1000 & N/A & 20 \\
\hline
\textbf{Notes:} & \multicolumn{11}{p{17cm}|}{N/A} \\
\hline
\end{tabular}%
\end{adjustbox}%
\end{table}

\begin{table}[H]
\OPtabletitleCustom{3.2.2}{Type II spicule evolution on disk}{OP3.2.2table}{OP3.2.2}
\centering
\vspace{4mm}
\begin{adjustbox}{max width=\textwidth}
\begin{tabular}{|l|c|c|c|c|c|c|c|c|c|c|c|}
\hline
\textbf{Goal:} & \multicolumn{11}{p{17cm}|}{IFS: Catch spicule evolution on disk at the highest spatial and temporal resolution with polarimetry.} \\
\hline
\makecell[t]{\textbf{Instrument}} & \makecell[t]{\textbf{Optical}\\ \textbf{Arm}} & \makecell[t]{\textbf{Atm}\\ \textbf{Layer}} & \makecell[t]{\textbf{CWL}\\ \textbf{{[nm]}}} & \makecell[t]{\textbf{Spectral}\\ \textbf{Range}\\ \textbf{{[nm]}}} & \makecell[t]{\textbf{Mode}} & \makecell[t]{\textbf{FOV}\\ \textbf{{[\arcsec]}}} & \makecell[t]{\textbf{Sampling}\\ \textbf{{[\arcsec /px]}}} & \makecell[t]{\textbf{R}} & \makecell[t]{\textbf{SNR}} & \makecell[t]{\textbf{\# WL}\\ \textbf{Points}} & \makecell[t]{\textbf{Cadence}\\ \textbf{{[s]}}} \\
\hline
IFS-M & V & Photosphere & $^{+}$617.3 & 0.10 & pol & 10x10 & $^{*}$0.02 & 100000 & 3000 & N/A & 20 \\
IFS-M & R & Chromosphere & 854.2 & 0.43 & pol & 10x10 & 0.03 & 80000 & 3000 & N/A & 20 \\
IFS-M & B & Chromosphere & $^{+}$393.4 & 0.20 & spec & 10x10 & $^{*}$0.01 & 75000 & 500 & N/A & 10 \\
IFS-M & V & Chromosphere & 656.3 & 0.33 & spec & 10x10 & $^{*}$0.02 & 100000 & 500 & N/A & 1 \\
EMBER & IR & Chromosphere & 1083 & 0.03 & pol & 10x10 & $^{*}$0.03 & 100000 & 3000 & N/A & 10 \\
\hline
\textbf{Notes:} & \multicolumn{11}{p{17cm}|}{N/A} \\
\hline
\end{tabular}%
\end{adjustbox}%
\end{table}

\begin{table}[H]
\OPtabletitleCustom{3.2.3}{Type II spicule evolution off-limb}{OP3.2.3table}{OP3.2.3}
\centering
\vspace{4mm}
\begin{adjustbox}{max width=\textwidth}
\begin{tabular}{|l|c|c|c|c|c|c|c|c|c|c|c|}
\hline
\textbf{Goal:} & \multicolumn{11}{p{17cm}|}{IFS: Catch spicule evolution o  limb at highest spatial and temporal resolution with polarimetry.} \\
\hline
\makecell[t]{\textbf{Instrument}} & \makecell[t]{\textbf{Optical}\\ \textbf{Arm}} & \makecell[t]{\textbf{Atm}\\ \textbf{Layer}} & \makecell[t]{\textbf{CWL}\\ \textbf{{[nm]}}} & \makecell[t]{\textbf{Spectral}\\ \textbf{Range}\\ \textbf{{[nm]}}} & \makecell[t]{\textbf{Mode}} & \makecell[t]{\textbf{FOV}\\ \textbf{{[\arcsec]}}} & \makecell[t]{\textbf{Sampling}\\ \textbf{{[\arcsec /px]}}} & \makecell[t]{\textbf{R}} & \makecell[t]{\textbf{SNR}} & \makecell[t]{\textbf{\# WL}\\ \textbf{Points}} & \makecell[t]{\textbf{Cadence}\\ \textbf{{[s]}}} \\
\hline
IFS-M & V & Photosphere & $^{+}$617.3 & 0.10 & pol & 10x10 & $^{*}$0.02 & 100000 & 3000 & N/A & 20 \\
IFS-M & R & Chromosphere & 854.2 & 0.43 & pol & 10x10 & 0.03 & 80000 & 3000 & N/A & 20 \\
IFS-M & B & Chromosphere & 393.4 & 0.20 & spec & 10x10 & $^{*}$0.01 & 75000 & 500 & N/A & 10 \\
IFS-M & V & Chromosphere & 656.3 & 0.33 & spec & 10x10 & $^{*}$0.02 & 100000 & 500 & N/A & 1 \\
EMBER & IR & Chromosphere & 1083 & 0.05 & pol & 10x10 & $^{*}$0.03 & 100000 & 3000 & N/A & 10 \\
\hline
\textbf{Notes:} & \multicolumn{11}{p{17cm}|}{N/A} \\
\hline
\end{tabular}%
\end{adjustbox}%
\end{table}

\begin{table}[H]
\OPtabletitleCustom{3.2.4}{Magnetic field of spicules}{OP3.2.4table}{OP3.2.4}
\centering
\vspace{4mm}
\begin{adjustbox}{max width=\textwidth}
\begin{tabular}{|l|c|c|c|c|c|c|c|c|c|c|c|}
\hline
\textbf{Goal:} & \multicolumn{11}{p{17cm}|}{IFS: Measure magnetic fields in spicules on disk using He I lines.} \\
\hline
\makecell[t]{\textbf{Instrument}} & \makecell[t]{\textbf{Optical}\\ \textbf{Arm}} & \makecell[t]{\textbf{Atm}\\ \textbf{Layer}} & \makecell[t]{\textbf{CWL}\\ \textbf{{[nm]}}} & \makecell[t]{\textbf{Spectral}\\ \textbf{Range}\\ \textbf{{[nm]}}} & \makecell[t]{\textbf{Mode}} & \makecell[t]{\textbf{FOV}\\ \textbf{{[\arcsec]}}} & \makecell[t]{\textbf{Sampling}\\ \textbf{{[\arcsec /px]}}} & \makecell[t]{\textbf{R}} & \makecell[t]{\textbf{SNR}} & \makecell[t]{\textbf{\# WL}\\ \textbf{Points}} & \makecell[t]{\textbf{Cadence}\\ \textbf{{[s]}}} \\
\hline
EMBER & IR & Chromosphere & 1083 & 0.05 & pol & 10x10 & $^{*}$0.03 & 100000 & 3000 & N/A & 10 \\
IFS-M & V & Chromosphere & 587.6 & 0.03 & pol & 10x10 & 0.02 & 100000 & 3000 & N/A & 10 \\
\hline
\textbf{Notes:} & \multicolumn{11}{p{17cm}|}{IFS: requires atmospheric refraction compensation.} \\
\hline
\end{tabular}%
\end{adjustbox}%
\end{table}

\begin{table}[H]
\OPtabletitleCustom{3.3.1}{Small-scale chromospheric jets}{OP3.3.1table}{OP3.3.1}
\centering
\vspace{4mm}
\begin{adjustbox}{max width=\textwidth}
\begin{tabular}{|l|c|c|c|c|c|c|c|c|c|c|c|}
\hline
\textbf{Goal:} & \multicolumn{11}{p{17cm}|}{TIS/FBIs: Measure the fine structure of small-scale chromospheric jets.} \\
\hline
\makecell[t]{\textbf{Instrument}} & \makecell[t]{\textbf{Optical}\\ \textbf{Arm}} & \makecell[t]{\textbf{Atm}\\ \textbf{Layer}} & \makecell[t]{\textbf{CWL}\\ \textbf{{[nm]}}} & \makecell[t]{\textbf{Spectral}\\ \textbf{Range}\\ \textbf{{[nm]}}} & \makecell[t]{\textbf{Mode}} & \makecell[t]{\textbf{FOV}\\ \textbf{{[\arcsec]}}} & \makecell[t]{\textbf{Sampling}\\ \textbf{{[\arcsec /px]}}} & \makecell[t]{\textbf{R}} & \makecell[t]{\textbf{SNR}} & \makecell[t]{\textbf{\# WL}\\ \textbf{Points}} & \makecell[t]{\textbf{Cadence}\\ \textbf{{[s]}}} \\
\hline
TIS/FBI & B & Chromosphere & 393.4 & 0.13 & spec & 30 & $^{*}$0.01 & 50000 & 250 & 20 & 10 \\
TIS/FBI & R & Chromosphere & 854.2 & 0.29 & spec & 30 & 0.03 & 80000 & 1000 & 20 & 20 \\
TIS/FBI & V & Chromosphere & 656.3 & 0.22 & spec & 30 & $^{*}$0.02 & 100000 & 1000 & 20 & 20 \\
\hline
\textbf{Notes:} & \multicolumn{11}{p{17cm}|}{TIS/FBIs: requires atmospheric refraction compensation.} \\
\hline
\end{tabular}%
\end{adjustbox}%
\end{table}

\begin{table}[H]
\OPtabletitleCustom{3.4.1}{Alfv\'en waves}{OP3.4.1table}{OP3.4.1}
\centering
\vspace{4mm}
\begin{adjustbox}{max width=\textwidth}
\begin{tabular}{|l|c|c|c|c|c|c|c|c|c|c|c|}
\hline
\textbf{Goal:} & \multicolumn{11}{p{17cm}|}{TIS/FBIs: Study the propagation and damping of Alfv\'en waves in the chromosphere.} \\
\hline
\makecell[t]{\textbf{Instrument}} & \makecell[t]{\textbf{Optical}\\ \textbf{Arm}} & \makecell[t]{\textbf{Atm}\\ \textbf{Layer}} & \makecell[t]{\textbf{CWL}\\ \textbf{{[nm]}}} & \makecell[t]{\textbf{Spectral}\\ \textbf{Range}\\ \textbf{{[nm]}}} & \makecell[t]{\textbf{Mode}} & \makecell[t]{\textbf{FOV}\\ \textbf{{[\arcsec]}}} & \makecell[t]{\textbf{Sampling}\\ \textbf{{[\arcsec /px]}}} & \makecell[t]{\textbf{R}} & \makecell[t]{\textbf{SNR}} & \makecell[t]{\textbf{\# WL}\\ \textbf{Points}} & \makecell[t]{\textbf{Cadence}\\ \textbf{{[s]}}} \\
\hline
TIS/FBI & B & Chromosphere & 393.4 & 0.01 & spec & 30 & $^{*}$0.01 & 50000 & 250 & 5 & 5 \\
TIS/FBI & R & Chromosphere & 854.2 & 0.21 & pol & 30 & 0.03 & 80000 & 1000 & 10 & 20 \\
TIS/FBI & V & Chromosphere & 656.3 & 0.16 & spec & 30 & $^{*}$0.02 & 100000 & 1000 & 5 & 13 \\
 &  & Photosphere & 617.3 & 0.10 & pol & 30 & $^{*}$0.02 & 100000 & 1000 & 7 & 13 \\
EMBER & IR & Chromosphere & $^{+}$1083 & 0.05 & pol & 10 & $^{*}$0.03 & 100000 & 3000 & N/A & 10 \\
\hline
\textbf{Notes:} & \multicolumn{11}{p{17cm}|}{N/A} \\
\hline
\end{tabular}%
\end{adjustbox}%
\end{table}

\begin{table}[H]
\OPtabletitleCustom{3.4.2}{Acoustic wave interaction with chromospheric field structure}{OP3.4.2table}{OP3.4.2}
\centering
\vspace{4mm}
\begin{adjustbox}{max width=\textwidth}
\begin{tabular}{|l|c|c|c|c|c|c|c|c|c|c|c|}
\hline
\textbf{Goal:} & \multicolumn{11}{p{17cm}|}{TIS/FBIs: Study the interaction of acoustic waves with the chromospheric magnetic field.} \\
\hline
\makecell[t]{\textbf{Instrument}} & \makecell[t]{\textbf{Optical}\\ \textbf{Arm}} & \makecell[t]{\textbf{Atm}\\ \textbf{Layer}} & \makecell[t]{\textbf{CWL}\\ \textbf{{[nm]}}} & \makecell[t]{\textbf{Spectral}\\ \textbf{Range}\\ \textbf{{[nm]}}} & \makecell[t]{\textbf{Mode}} & \makecell[t]{\textbf{FOV}\\ \textbf{{[\arcsec]}}} & \makecell[t]{\textbf{Sampling}\\ \textbf{{[\arcsec /px]}}} & \makecell[t]{\textbf{R}} & \makecell[t]{\textbf{SNR}} & \makecell[t]{\textbf{\# WL}\\ \textbf{Points}} & \makecell[t]{\textbf{Cadence}\\ \textbf{{[s]}}} \\
\hline
TIS/FBI & B & Chromosphere & 393.4 & 0.01 & spec & 30 & $^{*}$0.01 & 50000 & 250 & 15 & 15 \\
TIS/FBI & V & Photosphere & 617.3 & 0.10 & pol & 30 & $^{*}$0.02 & 100000 & 1000 & 7 & 20 \\
 &  & Chromosphere & 656.3 & 0.16 & spec & 30 & $^{*}$0.02 & 100000 & 1000 & 15 & 20 \\
TIS/FBI & R & Chromosphere & 854.2 & 0.21 & pol & 30 & 0.03 & 80000 & 1000 & 10 & 20 \\
\hline
\textbf{Notes:} & \multicolumn{11}{p{17cm}|}{N/A} \\
\hline
\end{tabular}%
\end{adjustbox}%
\end{table}

\begin{table}[H]
\OPtabletitleCustom{3.5.1}{Highly variable phenomena in the chromosphere}{OP3.5.1table}{OP3.5.1}
\centering
\vspace{4mm}
\begin{adjustbox}{max width=\textwidth}
\begin{tabular}{|l|c|c|c|c|c|c|c|c|c|c|c|}
\hline
\textbf{Goal:} & \multicolumn{11}{p{17cm}|}{TIS/FBIs: Study highly variable phenomena in the chromosphere.} \\
\hline
\makecell[t]{\textbf{Instrument}} & \makecell[t]{\textbf{Optical}\\ \textbf{Arm}} & \makecell[t]{\textbf{Atm}\\ \textbf{Layer}} & \makecell[t]{\textbf{CWL}\\ \textbf{{[nm]}}} & \makecell[t]{\textbf{Spectral}\\ \textbf{Range}\\ \textbf{{[nm]}}} & \makecell[t]{\textbf{Mode}} & \makecell[t]{\textbf{FOV}\\ \textbf{{[\arcsec]}}} & \makecell[t]{\textbf{Sampling}\\ \textbf{{[\arcsec /px]}}} & \makecell[t]{\textbf{R}} & \makecell[t]{\textbf{SNR}} & \makecell[t]{\textbf{\# WL}\\ \textbf{Points}} & \makecell[t]{\textbf{Cadence}\\ \textbf{{[s]}}} \\
\hline
TIS/FBI & V & Photosphere & 617.3 & 0.10 & pol & 30 & $^{*}$0.02 & 100000 & 3000 & 7 & 20 \\
TIS/FBI & B & Chromosphere & 393.4 & 0.20 & img & 30 & $^{*}$0.01 & 50000 & 500 & N/A & 1 \\
TIS/FBI & R & Chromosphere & 854.2 & 0.43 & spec & 30 & 0.03 & 80000 & 1000 & 3 & 3 \\
TIS/FBI & V & Chromosphere & 656.3 & 0.33 & img & 30 & $^{*}$0.02 & 50000 & 500 & N/A & 1 \\
\hline
\textbf{Notes:} & \multicolumn{11}{p{17cm}|}{N/A} \\
\hline
\end{tabular}%
\end{adjustbox}%
\end{table}

\begin{table}[H]
\OPtabletitleCustom{3.5.2}{Highly variable phenomena in the chromosphere}{OP3.5.2table}{OP3.5.2}
\centering
\vspace{4mm}
\begin{adjustbox}{max width=\textwidth}
\begin{tabular}{|l|c|c|c|c|c|c|c|c|c|c|c|}
\hline
\textbf{Goal:} & \multicolumn{11}{p{17cm}|}{IFS: Study highly variable phenomena in the chromosphere.} \\
\hline
\makecell[t]{\textbf{Instrument}} & \makecell[t]{\textbf{Optical}\\ \textbf{Arm}} & \makecell[t]{\textbf{Atm}\\ \textbf{Layer}} & \makecell[t]{\textbf{CWL}\\ \textbf{{[nm]}}} & \makecell[t]{\textbf{Spectral}\\ \textbf{Range}\\ \textbf{{[nm]}}} & \makecell[t]{\textbf{Mode}} & \makecell[t]{\textbf{FOV}\\ \textbf{{[\arcsec]}}} & \makecell[t]{\textbf{Sampling}\\ \textbf{{[\arcsec /px]}}} & \makecell[t]{\textbf{R}} & \makecell[t]{\textbf{SNR}} & \makecell[t]{\textbf{\# WL}\\ \textbf{Points}} & \makecell[t]{\textbf{Cadence}\\ \textbf{{[s]}}} \\
\hline
IFS-M & V & Photosphere & $^{+}$617.3 & 0.10 & pol & 5 & $^{*}$0.02 & 100000 & 3000 & N/A & 20 \\
IFS-M & B & Chromosphere & 393.4 & 0.20 & spec & 5 & $^{*}$0.01 & 75000 & 500 & N/A & 1 \\
IFS-M & R & Chromosphere & 854.2 & 0.43 & pol & 5 & 0.03 & 80000 & 1000 & N/A & 2 \\
IFS-M & V & Chromosphere & 656.3 & 0.33 & spec & 5 & $^{*}$0.02 & 100000 & 500 & N/A & 1 \\
EMBER & IR & Chromosphere & $^{+}$1083 & 0.05 & pol & 5 & $^{*}$0.03 & 100000 & 3000 & N/A & 10 \\
\hline
\textbf{Notes:} & \multicolumn{11}{p{17cm}|}{N/A} \\
\hline
\end{tabular}%
\end{adjustbox}%
\end{table}

\begin{table}[H]
\OPtabletitleCustom{3.5.3}{Reconnection at different heights}{OP3.5.3table}{OP3.5.3}
\centering
\vspace{4mm}
\begin{adjustbox}{max width=\textwidth}
\begin{tabular}{|l|c|c|c|c|c|c|c|c|c|c|c|}
\hline
\textbf{Goal:} & \multicolumn{11}{p{17cm}|}{IFS: Study reconnection at different heights.} \\
\hline
\makecell[t]{\textbf{Instrument}} & \makecell[t]{\textbf{Optical}\\ \textbf{Arm}} & \makecell[t]{\textbf{Atm}\\ \textbf{Layer}} & \makecell[t]{\textbf{CWL}\\ \textbf{{[nm]}}} & \makecell[t]{\textbf{Spectral}\\ \textbf{Range}\\ \textbf{{[nm]}}} & \makecell[t]{\textbf{Mode}} & \makecell[t]{\textbf{FOV}\\ \textbf{{[\arcsec]}}} & \makecell[t]{\textbf{Sampling}\\ \textbf{{[\arcsec /px]}}} & \makecell[t]{\textbf{R}} & \makecell[t]{\textbf{SNR}} & \makecell[t]{\textbf{\# WL}\\ \textbf{Points}} & \makecell[t]{\textbf{Cadence}\\ \textbf{{[s]}}} \\
\hline
IFS-M & V & Photosphere & 617.3 & 0.10 & pol & 5 & $^{*}$0.02 & 100000 & 3000 & N/A & 20 \\
IFS-M & B & Chromosphere & 393.4 & 0.20 & spec & 5 & $^{*}$0.01 & 75000 & 250 & N/A & 5 \\
IFS-M & R & Chromosphere & 854.2 & 0.43 & pol & 5 & 0.03 & 80000 & 3000 & N/A & 5 \\
IFS-M & V & Chromosphere & 656.3 & 0.33 & spec & 5 & $^{*}$0.02 & 100000 & 500 & N/A & 5 \\
EMBER & IR & Chromosphere & 1083 & 0.05 & pol & 10 & $^{*}$0.03 & 100000 & 3000 & N/A & 10 \\
\hline
\textbf{Notes:} & \multicolumn{11}{p{17cm}|}{N/A} \\
\hline
\end{tabular}%
\end{adjustbox}%
\end{table}

\begin{table}[H]
\OPtabletitleCustom{3.6.1}{Electric currents and heating in active regions}{OP3.6.1table}{OP3.6.1}
\centering
\vspace{4mm}
\begin{adjustbox}{max width=\textwidth}
\begin{tabular}{|l|c|c|c|c|c|c|c|c|c|c|c|}
\hline
\textbf{Goal:} & \multicolumn{11}{p{17cm}|}{TIS/FBIs: Determine electric currents in active regions.} \\
\hline
\makecell[t]{\textbf{Instrument}} & \makecell[t]{\textbf{Optical}\\ \textbf{Arm}} & \makecell[t]{\textbf{Atm}\\ \textbf{Layer}} & \makecell[t]{\textbf{CWL}\\ \textbf{{[nm]}}} & \makecell[t]{\textbf{Spectral}\\ \textbf{Range}\\ \textbf{{[nm]}}} & \makecell[t]{\textbf{Mode}} & \makecell[t]{\textbf{FOV}\\ \textbf{{[\arcsec]}}} & \makecell[t]{\textbf{Sampling}\\ \textbf{{[\arcsec /px]}}} & \makecell[t]{\textbf{R}} & \makecell[t]{\textbf{SNR}} & \makecell[t]{\textbf{\# WL}\\ \textbf{Points}} & \makecell[t]{\textbf{Cadence}\\ \textbf{{[s]}}} \\
\hline
TIS/FBI & V & Photosphere & 617.3 & 0.10 & pol & 30 & $^{*}$0.02 & 100000 & 3000 & 7 & 20 \\
 &  & Upper photosphere & 589 & 0.15 & pol & 30 & $^{*}$0.01 & 100000 & 1000 & 10 & 20 \\
TIS/FBI & R & Chromosphere & 854.2 & 0.30 & pol & 30 & 0.03 & 80000 & 1000 & 10 & 20 \\
 &  & Chromosphere & 849.8 & 0.30 & pol & 30 & 0.03 & 80000 & 1000 & 10 & 20 \\
EMBER & IR & Chromosphere & $^{+}$1083 & 0.05 & pol & 10 & $^{*}$0.03 & 100000 & 3000 & N/A & 20 \\
\hline
\textbf{Notes:} & \multicolumn{11}{p{17cm}|}{N/A} \\
\hline
\end{tabular}%
\end{adjustbox}%
\end{table}

\begin{table}[H]
\OPtabletitleCustom{3.6.2}{Electric currents and heating outside active regions}{OP3.6.2table}{OP3.6.2}
\centering
\vspace{4mm}
\begin{adjustbox}{max width=\textwidth}
\begin{tabular}{|l|c|c|c|c|c|c|c|c|c|c|c|}
\hline
\textbf{Goal:} & \multicolumn{11}{p{17cm}|}{IFS: Determine electric currents outside active regions} \\
\hline
\makecell[t]{\textbf{Instrument}} & \makecell[t]{\textbf{Optical}\\ \textbf{Arm}} & \makecell[t]{\textbf{Atm}\\ \textbf{Layer}} & \makecell[t]{\textbf{CWL}\\ \textbf{{[nm]}}} & \makecell[t]{\textbf{Spectral}\\ \textbf{Range}\\ \textbf{{[nm]}}} & \makecell[t]{\textbf{Mode}} & \makecell[t]{\textbf{FOV}\\ \textbf{{[\arcsec]}}} & \makecell[t]{\textbf{Sampling}\\ \textbf{{[\arcsec /px]}}} & \makecell[t]{\textbf{R}} & \makecell[t]{\textbf{SNR}} & \makecell[t]{\textbf{\# WL}\\ \textbf{Points}} & \makecell[t]{\textbf{Cadence}\\ \textbf{{[s]}}} \\
\hline
TIS/FBI & V & Photosphere & 617.3 & 0.10 & pol & 60 & $^{*}$0.02 & 100000 & 3000 & 7 & 80 \\
 &  & Upper photosphere & 589 & 0.15 & pol & 60 & $^{*}$0.01 & 100000 & 1000 & 10 & 80 \\
TIS/FBI & R & Chromosphere & 854.2 & 0.30 & pol & 60 & 0.03 & 80000 & 1000 & 10 & 80 \\
 &  & Chromosphere & 849.8 & 0.30 & pol & 60 & 0.03 & 80000 & 1000 & 10 & 80 \\
EMBER & IR & Chromosphere & $^{+}$1083 & 0.05 & pol & 20 & $^{*}$0.03 & 100000 & 3000 & N/A & 80 \\
\hline
\textbf{Notes:} & \multicolumn{11}{p{17cm}|}{N/A} \\
\hline
\end{tabular}%
\end{adjustbox}%
\end{table}

\begin{table}[H]
\OPtabletitleCustom{3.6.3}{Electric currents and heating outside active regions at small spatial scales}{OP3.6.3table}{OP3.6.3}
\centering
\vspace{4mm}
\begin{adjustbox}{max width=\textwidth}
\begin{tabular}{|l|c|c|c|c|c|c|c|c|c|c|c|}
\hline
\textbf{Goal:} & \multicolumn{11}{p{17cm}|}{IFS: Determine time-varying electric currents outside active regions at small spatial scales.} \\
\hline
\makecell[t]{\textbf{Instrument}} & \makecell[t]{\textbf{Optical}\\ \textbf{Arm}} & \makecell[t]{\textbf{Atm}\\ \textbf{Layer}} & \makecell[t]{\textbf{CWL}\\ \textbf{{[nm]}}} & \makecell[t]{\textbf{Spectral}\\ \textbf{Range}\\ \textbf{{[nm]}}} & \makecell[t]{\textbf{Mode}} & \makecell[t]{\textbf{FOV}\\ \textbf{{[\arcsec]}}} & \makecell[t]{\textbf{Sampling}\\ \textbf{{[\arcsec /px]}}} & \makecell[t]{\textbf{R}} & \makecell[t]{\textbf{SNR}} & \makecell[t]{\textbf{\# WL}\\ \textbf{Points}} & \makecell[t]{\textbf{Cadence}\\ \textbf{{[s]}}} \\
\hline
IFS-M & V & Photosphere & 617.3 & 0.10 & pol & 5 & $^{*}$0.02 & 100000 & 3000 & N/A & 10 \\
IFS-M & R & Chromosphere & 854.2 & 0.43 & pol & 5 & 0.03 & 80000 & 3000 & N/A & 10 \\
EMBER & IR & Chromosphere & 1083 & 0.05 & pol & 5 & $^{*}$0.03 & 100000 & 3000 & N/A & 10 \\
IFS-M & V & Upper photosphere & 589 & 0.15 & pol & 5 & $^{*}$0.01 & 100000 & 3000 & N/A & 10 \\
IFS-M & R & Chromosphere & 849.8 & 0.43 & pol & 5 & 0.03 & 80000 & 3000 & N/A & 10 \\
\hline
\textbf{Notes:} & \multicolumn{11}{p{17cm}|}{N/A} \\
\hline
\end{tabular}%
\end{adjustbox}%
\end{table}

\begin{table}[H]
\OPtabletitleCustom{3.7.2}{Temperature bifurcation diagnostics by scattering polarization in Ca II K}{OP3.7.2table}{OP3.7.2}
\centering
\vspace{4mm}
\begin{adjustbox}{max width=\textwidth}
\begin{tabular}{|l|c|c|c|c|c|c|c|c|c|c|c|}
\hline
\textbf{Goal:} & \multicolumn{11}{p{17cm}|}{IFS: Study properties of cool clouds in the chromosphere.} \\
\hline
\makecell[t]{\textbf{Instrument}} & \makecell[t]{\textbf{Optical}\\ \textbf{Arm}} & \makecell[t]{\textbf{Atm}\\ \textbf{Layer}} & \makecell[t]{\textbf{CWL}\\ \textbf{{[nm]}}} & \makecell[t]{\textbf{Spectral}\\ \textbf{Range}\\ \textbf{{[nm]}}} & \makecell[t]{\textbf{Mode}} & \makecell[t]{\textbf{FOV}\\ \textbf{{[\arcsec]}}} & \makecell[t]{\textbf{Sampling}\\ \textbf{{[\arcsec /px]}}} & \makecell[t]{\textbf{R}} & \makecell[t]{\textbf{SNR}} & \makecell[t]{\textbf{\# WL}\\ \textbf{Points}} & \makecell[t]{\textbf{Cadence}\\ \textbf{{[s]}}} \\
\hline
IFS-M & B & Chromosphere & 393.4 & 1.00 & pol & 5x60 & 0.15 & 75000 & 1000 & N/A & as needed for SNR \\
\hline
\textbf{Notes:} & \multicolumn{11}{p{17cm}|}{N/A} \\
\hline
\end{tabular}%
\end{adjustbox}%
\end{table}

\begin{table}[H]
\OPtabletitleCustom{3.8.1}{Magnetic field determination in plage including Ca II H\&K}{OP3.8.1table}{OP3.8.1}
\centering
\vspace{4mm}
\begin{adjustbox}{max width=\textwidth}
\begin{tabular}{|l|c|c|c|c|c|c|c|c|c|c|c|}
\hline
\textbf{Goal:} & \multicolumn{11}{p{17cm}|}{IFS: Determine magnetic fields in plage, with a focus on the upper chromosphere.} \\
\hline
\makecell[t]{\textbf{Instrument}} & \makecell[t]{\textbf{Optical}\\ \textbf{Arm}} & \makecell[t]{\textbf{Atm}\\ \textbf{Layer}} & \makecell[t]{\textbf{CWL}\\ \textbf{{[nm]}}} & \makecell[t]{\textbf{Spectral}\\ \textbf{Range}\\ \textbf{{[nm]}}} & \makecell[t]{\textbf{Mode}} & \makecell[t]{\textbf{FOV}\\ \textbf{{[\arcsec]}}} & \makecell[t]{\textbf{Sampling}\\ \textbf{{[\arcsec /px]}}} & \makecell[t]{\textbf{R}} & \makecell[t]{\textbf{SNR}} & \makecell[t]{\textbf{\# WL}\\ \textbf{Points}} & \makecell[t]{\textbf{Cadence}\\ \textbf{{[s]}}} \\
\hline
IFS-M & B & Chromosphere & $^{+}$393.4 & 0.20 & pol & 60x20 & $^{*}$0.01 & 75000 & 1000 & N/A & as needed for SNR \\
 &  & Chromosphere & 396.8 & 0.20 & pol & 60x20 & $^{*}$0.01 & 75000 & 1000 & N/A & as needed for SNR \\
IFS-M & V & Photosphere & $^{+}$617.3 & 0.10 & pol & 60x20 & $^{*}$0.02 & 100000 & 3000 & N/A & as needed for SNR \\
 &  & Upper photosphere & 589 & 0.15 & pol & 60x20 & $^{*}$0.01 & 100000 & 3000 & N/A & as needed for SNR \\
IFS-M & R & Chromosphere & 854.2 & 0.43 & pol & 60x20 & 0.03 & 80000 & 3000 & N/A & as needed for SNR \\
EMBER & IR & Chromosphere & 1083 & 0.05 & pol & 60x20 & $^{*}$0.03 & 100000 & 3000 & N/A & as needed for SNR \\
\hline
\textbf{Notes:} & \multicolumn{11}{p{17cm}|}{IFS: SNR is set for Ca II 393.4 and Ca II 396.8, other lines might have different requirements.} \\
\hline
\end{tabular}%
\end{adjustbox}%
\end{table}

\begin{table}[H]
\OPtabletitleCustom{4.1.1}{Stability of the umbra - interplay between the convection and magnetic forces}{OP4.1.1table}{OP4.1.1}
\centering
\vspace{4mm}
\begin{adjustbox}{max width=\textwidth}
\begin{tabular}{|l|c|c|c|c|c|c|c|c|c|c|c|}
\hline
\textbf{Goal:} & \multicolumn{11}{p{17cm}|}{TIS/FBIs: Context information about magnetic field configuration in a larger FOV. Context information, for LCT and feature tracking. IFS: Spectropolarimetric observations of the umbral boundary to study the detailed processes in the magnetic field evolution.} \\
\hline
\makecell[t]{\textbf{Instrument}} & \makecell[t]{\textbf{Optical}\\ \textbf{Arm}} & \makecell[t]{\textbf{Atm}\\ \textbf{Layer}} & \makecell[t]{\textbf{CWL}\\ \textbf{{[nm]}}} & \makecell[t]{\textbf{Spectral}\\ \textbf{Range}\\ \textbf{{[nm]}}} & \makecell[t]{\textbf{Mode}} & \makecell[t]{\textbf{FOV}\\ \textbf{{[\arcsec]}}} & \makecell[t]{\textbf{Sampling}\\ \textbf{{[\arcsec /px]}}} & \makecell[t]{\textbf{R}} & \makecell[t]{\textbf{SNR}} & \makecell[t]{\textbf{\# WL}\\ \textbf{Points}} & \makecell[t]{\textbf{Cadence}\\ \textbf{{[s]}}} \\
\hline
TIS/FBI & B & Chromosphere & 396 & 0.10 & img & 40 & 0.02 & N/A & N/A & N/A & 10 \\
 &  & Photosphere & 430 & 0.80 & img & 40 & 0.02 & N/A & N/A & N/A & 10 \\
IFS-M & V & Photosphere & 630.2 & 0.25 & pol & 8 & 0.04 & 100000 & 2000 & N/A & 10 \\
IFS-M & R & Chromosphere & 854.2 & 0.60 & pol & 8 & 0.04 & 80000 & 2000 & N/A & 10 \\
EMBER & IR & All & 1083.2 & 1.60 & pol & 8 & 0.04 & 100000 & 2000 & N/A & 10 \\
\hline
\textbf{Notes:} & \multicolumn{11}{p{17cm}|}{N/A} \\
\hline
\end{tabular}%
\end{adjustbox}%
\end{table}

\begin{table}[H]
\OPtabletitleCustom{4.2.1}{Multi-wavelength analysis of umbral dots}{OP4.2.1table}{OP4.2.1}
\centering
\vspace{4mm}
\begin{adjustbox}{max width=\textwidth}
\begin{tabular}{|l|c|c|c|c|c|c|c|c|c|c|c|}
\hline
\textbf{Goal:} & \multicolumn{11}{p{17cm}|}{TIS/FBIs: Context information about magnetic field configuration from other spectral lines. Context information. IFS: Spectropolarimetric observations of the umbral dots for detailed study of magnetic field configuration and temporal evolution.} \\
\hline
\makecell[t]{\textbf{Instrument}} & \makecell[t]{\textbf{Optical}\\ \textbf{Arm}} & \makecell[t]{\textbf{Atm}\\ \textbf{Layer}} & \makecell[t]{\textbf{CWL}\\ \textbf{{[nm]}}} & \makecell[t]{\textbf{Spectral}\\ \textbf{Range}\\ \textbf{{[nm]}}} & \makecell[t]{\textbf{Mode}} & \makecell[t]{\textbf{FOV}\\ \textbf{{[\arcsec]}}} & \makecell[t]{\textbf{Sampling}\\ \textbf{{[\arcsec /px]}}} & \makecell[t]{\textbf{R}} & \makecell[t]{\textbf{SNR}} & \makecell[t]{\textbf{\# WL}\\ \textbf{Points}} & \makecell[t]{\textbf{Cadence}\\ \textbf{{[s]}}} \\
\hline
TIS/FBI & B & Chromosphere & 396 & 0.10 & img & 40 & 0.02 & N/A & N/A & N/A & 10 \\
 &  & Photosphere & 430 & 0.80 & img & 40 & 0.02 & N/A & N/A & N/A & 10 \\
IFS-M & V & Photosphere & 630.2 & 0.25 & pol & 10x10 & 0.04 & 100000 & 2000 & N/A & 10 \\
IFS-M & R & Chromosphere & 854.2 & 0.60 & pol & 10x10 & 0.04 & 80000 & 2000 & N/A & 10 \\
EMBER & IR & Photosphere & 1565.5 & 3.00 & pol & 10x10 & 0.05 & 100000 & 2000 & N/A & 10 \\
\hline
\textbf{Notes:} & \multicolumn{11}{p{17cm}|}{N/A} \\
\hline
\end{tabular}%
\end{adjustbox}%
\end{table}

\begin{table}[H]
\OPtabletitleCustom{4.3.1}{Probing the structure of cool sunspot umbrae}{OP4.3.1table}{OP4.3.1}
\centering
\vspace{4mm}
\begin{adjustbox}{max width=\textwidth}
\begin{tabular}{|l|c|c|c|c|c|c|c|c|c|c|c|}
\hline
\textbf{Goal:} & \multicolumn{11}{p{17cm}|}{IFS: High-precision spectropolarimetry of darkest umbral regions.} \\
\hline
\makecell[t]{\textbf{Instrument}} & \makecell[t]{\textbf{Optical}\\ \textbf{Arm}} & \makecell[t]{\textbf{Atm}\\ \textbf{Layer}} & \makecell[t]{\textbf{CWL}\\ \textbf{{[nm]}}} & \makecell[t]{\textbf{Spectral}\\ \textbf{Range}\\ \textbf{{[nm]}}} & \makecell[t]{\textbf{Mode}} & \makecell[t]{\textbf{FOV}\\ \textbf{{[\arcsec]}}} & \makecell[t]{\textbf{Sampling}\\ \textbf{{[\arcsec /px]}}} & \makecell[t]{\textbf{R}} & \makecell[t]{\textbf{SNR}} & \makecell[t]{\textbf{\# WL}\\ \textbf{Points}} & \makecell[t]{\textbf{Cadence}\\ \textbf{{[s]}}} \\
\hline
TIS/FBI & R & Photosphere & 706 & 0.50 & img & 40 & 0.05 & N/A & N/A & N/A & 10 \\
IFS-M & V & Photosphere & 630.2 & 0.25 & pol & 20x20 & 0.05 & 100000 & 4000 & N/A & 10 \\
EMBER & IR & Photosphere & 1565.5 & 3.00 & pol & 20x20 & 0.05 & 100000 & 4000 & N/A & 10 \\
\hline
\textbf{Notes:} & \multicolumn{11}{p{17cm}|}{IFS: The observed FOV should cover the whole umbral region.} \\
\hline
\end{tabular}%
\end{adjustbox}%
\end{table}

\begin{table}[H]
\OPtabletitleCustom{4.4.1}{Umbral flashes}{OP4.4.1table}{OP4.4.1}
\centering
\vspace{4mm}
\begin{adjustbox}{max width=\textwidth}
\begin{tabular}{|l|c|c|c|c|c|c|c|c|c|c|c|}
\hline
\textbf{Goal:} & \multicolumn{11}{p{17cm}|}{TIS/FBIs: Additional spectropolarimetric observations using other spectral lines. Context information. IFS: Spectropolarimetric observations of the umbra to study the magnetic field configuration.} \\
\hline
\makecell[t]{\textbf{Instrument}} & \makecell[t]{\textbf{Optical}\\ \textbf{Arm}} & \makecell[t]{\textbf{Atm}\\ \textbf{Layer}} & \makecell[t]{\textbf{CWL}\\ \textbf{{[nm]}}} & \makecell[t]{\textbf{Spectral}\\ \textbf{Range}\\ \textbf{{[nm]}}} & \makecell[t]{\textbf{Mode}} & \makecell[t]{\textbf{FOV}\\ \textbf{{[\arcsec]}}} & \makecell[t]{\textbf{Sampling}\\ \textbf{{[\arcsec /px]}}} & \makecell[t]{\textbf{R}} & \makecell[t]{\textbf{SNR}} & \makecell[t]{\textbf{\# WL}\\ \textbf{Points}} & \makecell[t]{\textbf{Cadence}\\ \textbf{{[s]}}} \\
\hline
TIS/FBI & B & Chromosphere & 396 & 0.10 & img & 40 & 0.02 & N/A & N/A & N/A & 5 \\
TIS/FBI & V & Chromosphere & 656.3 & 0.10 & img & 40 & $^{*}$0.02 & N/A & N/A & N/A & 5 \\
IFS-M & R & Chromosphere & 854.2 & 0.60 & pol & 20x20 & 0.05 & 80000 & 2000 & N/A & 10 \\
EMBER & IR & All & 1083.2 & 1.60 & pol & 20x20 & 0.05 & 100000 & 2000 & N/A & 10 \\
\hline
\textbf{Notes:} & \multicolumn{11}{p{17cm}|}{N/A} \\
\hline
\end{tabular}%
\end{adjustbox}%
\end{table}

\begin{table}[H]
\OPtabletitleCustom{4.5.1}{Penumbral and umbral micro-jets}{OP4.5.1table}{OP4.5.1}
\centering
\vspace{4mm}
\begin{adjustbox}{max width=\textwidth}
\begin{tabular}{|l|c|c|c|c|c|c|c|c|c|c|c|}
\hline
\textbf{Goal:} & \multicolumn{11}{p{17cm}|}{TIS/FBIs: Observe the dynamical evolution of micro-jets. G-band or continuum is for context information, for LCT and feature tracking.Additional spectropolarimetric observations for context information.  IFS: Determine the magnetic topology in and around micro-jets. High-cadence spectropolarimetric observations of a small FOV.} \\
\hline
\makecell[t]{\textbf{Instrument}} & \makecell[t]{\textbf{Optical}\\ \textbf{Arm}} & \makecell[t]{\textbf{Atm}\\ \textbf{Layer}} & \makecell[t]{\textbf{CWL}\\ \textbf{{[nm]}}} & \makecell[t]{\textbf{Spectral}\\ \textbf{Range}\\ \textbf{{[nm]}}} & \makecell[t]{\textbf{Mode}} & \makecell[t]{\textbf{FOV}\\ \textbf{{[\arcsec]}}} & \makecell[t]{\textbf{Sampling}\\ \textbf{{[\arcsec /px]}}} & \makecell[t]{\textbf{R}} & \makecell[t]{\textbf{SNR}} & \makecell[t]{\textbf{\# WL}\\ \textbf{Points}} & \makecell[t]{\textbf{Cadence}\\ \textbf{{[s]}}} \\
\hline
TIS/FBI & B & Chromosphere & 396 & 0.10 & img & 40 & 0.02 & N/A & N/A & N/A & 0.1 \\
IFS-M & V & Photosphere & 630.2 & 0.25 & pol & 10x10 & 0.04 & 100000 & 1000 & N/A & 2 \\
IFS-M & R & Chromosphere & 854.2 & 0.60 & pol & 10x10 & 0.04 & 80000 & 1000 & N/A & 2 \\
EMBER & IR & All & 1083.2 & 1.60 & pol & 10x10 & 0.04 & 100000 & 1000 & N/A & 2 \\
\hline
\textbf{Notes:} & \multicolumn{11}{p{17cm}|}{TIS/FBIs: Extreme high cadence is needed to observe the onset and evolution of microjets. IFS: Core of this OP.} \\
\hline
\end{tabular}%
\end{adjustbox}%
\end{table}

\begin{table}[H]
\OPtabletitleCustom{4.6.1}{Structure and dynamics of light bridges}{OP4.6.1table}{OP4.6.1}
\centering
\vspace{4mm}
\begin{adjustbox}{max width=\textwidth}
\begin{tabular}{|l|c|c|c|c|c|c|c|c|c|c|c|}
\hline
\textbf{Goal:} & \multicolumn{11}{p{17cm}|}{TIS/FBIs: Observe the dynamical evolution of light bridges and jets in the chromosphere above light bridges. G-band or continuum is for context information and feature tracking. Additional spectropolarimetric observations for context information. IFS: Determine the magnetic topology in and around light bridges. High-cadence spectropolarimetric observations of a small FOV. This program aims to study the details of magneto-convection and reconnection above light bridges.} \\
\hline
\makecell[t]{\textbf{Instrument}} & \makecell[t]{\textbf{Optical}\\ \textbf{Arm}} & \makecell[t]{\textbf{Atm}\\ \textbf{Layer}} & \makecell[t]{\textbf{CWL}\\ \textbf{{[nm]}}} & \makecell[t]{\textbf{Spectral}\\ \textbf{Range}\\ \textbf{{[nm]}}} & \makecell[t]{\textbf{Mode}} & \makecell[t]{\textbf{FOV}\\ \textbf{{[\arcsec]}}} & \makecell[t]{\textbf{Sampling}\\ \textbf{{[\arcsec /px]}}} & \makecell[t]{\textbf{R}} & \makecell[t]{\textbf{SNR}} & \makecell[t]{\textbf{\# WL}\\ \textbf{Points}} & \makecell[t]{\textbf{Cadence}\\ \textbf{{[s]}}} \\
\hline
TIS/FBI & B & Chromosphere & 396 & 0.10 & img & 40 & 0.02 & N/A & N/A & N/A & 5 \\
TIS/FBI & V & Chromosphere & 656.3 & 0.10 & img & 40 & $^{*}$0.02 & N/A & N/A & N/A & 5 \\
IFS-M & V & Photosphere & 630.2 & 0.25 & pol & 20x20 & 0.04 & 100000 & 1000 & N/A & 10 \\
IFS-M & R & Chromosphere & 854.2 & 0.60 & pol & 20x20 & 0.04 & 80000 & 1000 & N/A & 10 \\
EMBER & IR & All & 1083.2 & 1.60 & pol & 20x20 & 0.04 & 100000 & 1000 & N/A & 10 \\
\hline
\textbf{Notes:} & \multicolumn{11}{p{17cm}|}{TIS/FBIs: Extreme high cadence is needed to observe the onset and evolution of jets above light bridges. Combination with UV/EUV imaging from space (e.g. Solar Orbiter, Solar-C) allows to study the impact of these jets on the Transition Region and Corona. IFS: Core of this OP.} \\
\hline
\end{tabular}%
\end{adjustbox}%
\end{table}

\begin{table}[H]
\OPtabletitleCustom{4.7.1}{Evolution of an individual penumbral filament}{OP4.7.1table}{OP4.7.1}
\centering
\vspace{4mm}
\begin{adjustbox}{max width=\textwidth}
\begin{tabular}{|l|c|c|c|c|c|c|c|c|c|c|c|}
\hline
\textbf{Goal:} & \multicolumn{11}{p{17cm}|}{TIS/FBIs: Context information, for LCT and feature tracking. Additional spectropolarimetric observations using different spectral lines. IFS: Spectropolarimetric observations of the umbral boundary to study the detailed processes in the magnetic field evolution.} \\
\hline
\makecell[t]{\textbf{Instrument}} & \makecell[t]{\textbf{Optical}\\ \textbf{Arm}} & \makecell[t]{\textbf{Atm}\\ \textbf{Layer}} & \makecell[t]{\textbf{CWL}\\ \textbf{{[nm]}}} & \makecell[t]{\textbf{Spectral}\\ \textbf{Range}\\ \textbf{{[nm]}}} & \makecell[t]{\textbf{Mode}} & \makecell[t]{\textbf{FOV}\\ \textbf{{[\arcsec]}}} & \makecell[t]{\textbf{Sampling}\\ \textbf{{[\arcsec /px]}}} & \makecell[t]{\textbf{R}} & \makecell[t]{\textbf{SNR}} & \makecell[t]{\textbf{\# WL}\\ \textbf{Points}} & \makecell[t]{\textbf{Cadence}\\ \textbf{{[s]}}} \\
\hline
TIS/FBI & B & Chromosphere & 396 & 0.10 & img & 40 & 0.02 & N/A & N/A & N/A & 5 \\
 &  & Photosphere & 430 & 0.80 & img & 40 & 0.02 & N/A & N/A & N/A & 5 \\
TIS/FBI & V & Photosphere & 630.2 & 0.25 & pol & 40 & $^{*}$0.02 & 100000 & 1000 & 10 & 10 \\
IFS-M & R & Chromosphere & 854.2 & 0.60 & pol & 10x10 & 0.04 & 80000 & 2000 & N/A & 10 \\
EMBER & IR & All & 1083.2 & 1.60 & pol & 10x10 & 0.04 & 100000 & 2000 & N/A & 10 \\
\hline
\textbf{Notes:} & \multicolumn{11}{p{17cm}|}{N/A} \\
\hline
\end{tabular}%
\end{adjustbox}%
\end{table}

\begin{table}[H]
\OPtabletitleCustom{4.8.1}{Capturing the formation and decay of penumbrae}{OP4.8.1table}{OP4.8.1}
\centering
\vspace{4mm}
\begin{adjustbox}{max width=\textwidth}
\begin{tabular}{|l|c|c|c|c|c|c|c|c|c|c|c|}
\hline
\textbf{Goal:} & \multicolumn{11}{p{17cm}|}{TIS/FBIs: Context information, for LCT and feature tracking. Additional spectropolarimetric observations using other spectral lines. IFS: Spectropolarimetric observations of the sunspot to study the magnetic field configuration.} \\
\hline
\makecell[t]{\textbf{Instrument}} & \makecell[t]{\textbf{Optical}\\ \textbf{Arm}} & \makecell[t]{\textbf{Atm}\\ \textbf{Layer}} & \makecell[t]{\textbf{CWL}\\ \textbf{{[nm]}}} & \makecell[t]{\textbf{Spectral}\\ \textbf{Range}\\ \textbf{{[nm]}}} & \makecell[t]{\textbf{Mode}} & \makecell[t]{\textbf{FOV}\\ \textbf{{[\arcsec]}}} & \makecell[t]{\textbf{Sampling}\\ \textbf{{[\arcsec /px]}}} & \makecell[t]{\textbf{R}} & \makecell[t]{\textbf{SNR}} & \makecell[t]{\textbf{\# WL}\\ \textbf{Points}} & \makecell[t]{\textbf{Cadence}\\ \textbf{{[s]}}} \\
\hline
TIS/FBI & B & Chromosphere & 396 & 0.10 & img & 60 & 0.02 & N/A & N/A & N/A & 60 \\
 &  & Photosphere & 430 & 0.80 & img & 60 & 0.02 & N/A & N/A & N/A & 60 \\
TIS/FBI & V & Photosphere & 630.2 & 0.25 & pol & 60 & 0.05 & 100000 & 1000 & 20 & 20 \\
IFS-M & R & Chromosphere & 854.2 & 0.60 & pol & 60x60 & 0.05 & 80000 & 1000 & N/A & 120 \\
EMBER & IR & All & 1083.2 & 1.60 & pol & 60x60 & 0.05 & 100000 & 1000 & N/A & 120 \\
\hline
\textbf{Notes:} & \multicolumn{11}{p{17cm}|}{IFS: The goal is to cover the full process of penumbral formation and decay lasting several hours and the main requirement will be stable conditions on these time scales. In principle, short periods of worse seeing conditions can be tolerated as long as high quality spectral profiles are acquired every 3 minutes.} \\
\hline
\end{tabular}%
\end{adjustbox}%
\end{table}

\begin{table}[H]
\OPtabletitleCustom{4.9.1}{Observations of moat flow properties and its impact on sunspot decay}{OP4.9.1table}{OP4.9.1}
\centering
\vspace{4mm}
\begin{adjustbox}{max width=\textwidth}
\begin{tabular}{|l|c|c|c|c|c|c|c|c|c|c|c|}
\hline
\textbf{Goal:} & \multicolumn{11}{p{17cm}|}{TIS/FBIs: Context information, for LCT and feature tracking. Additional spectropolarimetric observations using other spectral lines. IFS: Spectropolarimetric observations of a sunspot and its surroundings to study the magnetic field configuration.} \\
\hline
\makecell[t]{\textbf{Instrument}} & \makecell[t]{\textbf{Optical}\\ \textbf{Arm}} & \makecell[t]{\textbf{Atm}\\ \textbf{Layer}} & \makecell[t]{\textbf{CWL}\\ \textbf{{[nm]}}} & \makecell[t]{\textbf{Spectral}\\ \textbf{Range}\\ \textbf{{[nm]}}} & \makecell[t]{\textbf{Mode}} & \makecell[t]{\textbf{FOV}\\ \textbf{{[\arcsec]}}} & \makecell[t]{\textbf{Sampling}\\ \textbf{{[\arcsec /px]}}} & \makecell[t]{\textbf{R}} & \makecell[t]{\textbf{SNR}} & \makecell[t]{\textbf{\# WL}\\ \textbf{Points}} & \makecell[t]{\textbf{Cadence}\\ \textbf{{[s]}}} \\
\hline
TIS/FBI & B & Chromosphere & 396 & 0.10 & img & 120 & 0.02 & N/A & N/A & N/A & 60 \\
 &  & Photosphere & 430 & 0.80 & img & 120 & 0.02 & N/A & N/A & N/A & 60 \\
TIS/FBI & V & Photosphere & 630.2 & 0.25 & pol & 120 & 0.05 & 100000 & 1000 & 15 & 120 \\
IFS-M & R & Chromosphere & 854.2 & 0.60 & pol & 120x120 & 0.05 & 80000 & 1000 & N/A & 180 \\
EMBER & IR & All & 1083.2 & 1.60 & pol & 120x120 & 0.05 & 100000 & 1000 & N/A & 180 \\
\hline
\textbf{Notes:} & \multicolumn{11}{p{17cm}|}{N/A} \\
\hline
\end{tabular}%
\end{adjustbox}%
\end{table}
\begin{table}[H]
\OPtabletitleCustom{4.10.1}{Determining the evolution of the magnetic and dynamic fine structure of prominences and filaments}{OP4.10.1table}{OP4.10.1}
\centering
\vspace{4mm}
\begin{adjustbox}{max width=\textwidth}
\begin{tabular}{|l|c|c|c|c|c|c|c|c|c|c|c|}
\hline
\textbf{Goal:} & \multicolumn{11}{p{17cm}|}{IFS: Spectropolarimetric observations of prominences/filaments to study the evolution of the magnetic field configuration and the dynamics at the smallest spatial scales possible, with the higher cadence possible. Type of observations: time series of a 20"x20" area.} \\
\hline
\makecell[t]{\textbf{Instrument}} & \makecell[t]{\textbf{Optical}\\ \textbf{Arm}} & \makecell[t]{\textbf{Atm}\\ \textbf{Layer}} & \makecell[t]{\textbf{CWL}\\ \textbf{{[nm]}}} & \makecell[t]{\textbf{Spectral}\\ \textbf{Range}\\ \textbf{{[nm]}}} & \makecell[t]{\textbf{Mode}} & \makecell[t]{\textbf{FOV}\\ \textbf{{[\arcsec]}}} & \makecell[t]{\textbf{Sampling}\\ \textbf{{[\arcsec /px]}}} & \makecell[t]{\textbf{R}} & \makecell[t]{\textbf{SNR}} & \makecell[t]{\textbf{\# WL}\\ \textbf{Points}} & \makecell[t]{\textbf{Cadence}\\ \textbf{{[s]}}} \\
\hline
EMBER & IR & Chromosphere & 1083 & 0.11 & pol & 10x10 & 0.10 & 100000 & 40000 & N/A & 20 \\
IFS-M & V & Chromosphere & 587.6 & 0.06 & pol & 10x10 & $^{*}$0.10 & 100000 & 40000 & N/A & 80 \\
\hline
\textbf{Notes:} & \multicolumn{11}{p{17cm}|}{IFS: The use of the adaptive optics on an off-limb target is necessary in case of prominence observations. High stability is required to reduce spurious signals down to 10-5, in terms of the continuum intensity. Pointing error of the order of 1".} \\
\hline
\end{tabular}%
\end{adjustbox}%
\end{table}

\begin{table}[H]
\OPtabletitleCustom{4.10.2}{Determining the magnetic and dynamic fine structure of prominences and filaments}{OP4.10.2table}{OP4.10.2}
\centering
\vspace{4mm}
\begin{adjustbox}{max width=\textwidth}
\begin{tabular}{|l|c|c|c|c|c|c|c|c|c|c|c|}
\hline
\textbf{Goal:} & \multicolumn{11}{p{17cm}|}{IFS: Spectropolarimetric observations of prominences/filaments to study the magnetic field configuration at the smallest spatial scales possible. Type of observations: Scan of a large area} \\
\hline
\makecell[t]{\textbf{Instrument}} & \makecell[t]{\textbf{Optical}\\ \textbf{Arm}} & \makecell[t]{\textbf{Atm}\\ \textbf{Layer}} & \makecell[t]{\textbf{CWL}\\ \textbf{{[nm]}}} & \makecell[t]{\textbf{Spectral}\\ \textbf{Range}\\ \textbf{{[nm]}}} & \makecell[t]{\textbf{Mode}} & \makecell[t]{\textbf{FOV}\\ \textbf{{[\arcsec]}}} & \makecell[t]{\textbf{Sampling}\\ \textbf{{[\arcsec /px]}}} & \makecell[t]{\textbf{R}} & \makecell[t]{\textbf{SNR}} & \makecell[t]{\textbf{\# WL}\\ \textbf{Points}} & \makecell[t]{\textbf{Cadence}\\ \textbf{{[s]}}} \\
\hline
EMBER & IR & Chromosphere & $^{+}$1083 & 0.11 & pol & 40x20 & 0.10 & 100000 & 40000 & N/A & 15 \\
IFS-M & V & Chromosphere & $^{+}$587.6 & 0.06 & pol & 40x20 & $^{*}$0.10 & 100000 & 40000 & N/A & 60 \\
\hline
\textbf{Notes:} & \multicolumn{11}{p{17cm}|}{IFS: The use of the adaptive optics on an off-limb target is necessary in case of prominence observations. High stability is required to reduce spurious signals down to 10-5, in terms of the continuum intensity. Pointing error of the order of 1".} \\
\hline
\end{tabular}%
\end{adjustbox}%
\end{table}
\begin{table}[H]
\OPtabletitleCustom{4.11.1}{Comparison of the magnetic filed properties in QS prominences and AR filaments}{OP4.11.1table}{OP4.11.1}
\centering
\vspace{4mm}
\begin{adjustbox}{max width=\textwidth}
\begin{tabular}{|l|c|c|c|c|c|c|c|c|c|c|c|}
\hline
\textbf{Goal:} & \multicolumn{11}{p{17cm}|}{IFS: Spectropolarimetric observations to study the magnetic field differences between QS prominences and AR filaments. Type of observations: Scan of a large area.} \\
\hline
\makecell[t]{\textbf{Instrument}} & \makecell[t]{\textbf{Optical}\\ \textbf{Arm}} & \makecell[t]{\textbf{Atm}\\ \textbf{Layer}} & \makecell[t]{\textbf{CWL}\\ \textbf{{[nm]}}} & \makecell[t]{\textbf{Spectral}\\ \textbf{Range}\\ \textbf{{[nm]}}} & \makecell[t]{\textbf{Mode}} & \makecell[t]{\textbf{FOV}\\ \textbf{{[\arcsec]}}} & \makecell[t]{\textbf{Sampling}\\ \textbf{{[\arcsec /px]}}} & \makecell[t]{\textbf{R}} & \makecell[t]{\textbf{SNR}} & \makecell[t]{\textbf{\# WL}\\ \textbf{Points}} & \makecell[t]{\textbf{Cadence}\\ \textbf{{[s]}}} \\
\hline
EMBER & IR & Chromosphere & 1083 & 0.11 & pol & 10x10 & 0.20 & 100000 & 40000 & N/A & 20 \\
IFS-M & V & Chromosphere & 587.6 & 0.06 & pol & 10x10 & 0.20 & 100000 & 40000 & N/A & 80 \\
IFS-M & R & Chromosphere & 854.2 & 0.09 & pol & 10x10 & 0.20 & 80000 & 40000 & N/A & 150 \\
\hline
\textbf{Notes:} & \multicolumn{11}{p{17cm}|}{IFS: The use of the adaptive optics on an off-limb target is necessary in case of prominence observations. High stability is required to reduce spurious signals down to 10-5, in terms of the continuum intensity. Pointing error of the order of 1".} \\
\hline
\end{tabular}%
\end{adjustbox}%
\end{table}

\begin{table}[H]
\OPtabletitleCustom{4.12.1}{Magnetism and dynamics of tornado prominences}{OP4.12.1table}{OP4.12.1}
\centering
\vspace{4mm}
\begin{adjustbox}{max width=\textwidth}
\begin{tabular}{|l|c|c|c|c|c|c|c|c|c|c|c|}
\hline
\textbf{Goal:} & \multicolumn{11}{p{17cm}|}{TIS/FBIs: Spectroscopic observations to study the dynamics of tornado prominences. Type of observations: Time series of a 50"x50" area. These observations are simultaneous to the spectro-polarimetric ones detailed in the table. IFS: Spectro-polarimetric observations to study the magnetic field configuration of tornado prominences. Type of observations: time series of a 20"x20" area.} \\
\hline
\makecell[t]{\textbf{Instrument}} & \makecell[t]{\textbf{Optical}\\ \textbf{Arm}} & \makecell[t]{\textbf{Atm}\\ \textbf{Layer}} & \makecell[t]{\textbf{CWL}\\ \textbf{{[nm]}}} & \makecell[t]{\textbf{Spectral}\\ \textbf{Range}\\ \textbf{{[nm]}}} & \makecell[t]{\textbf{Mode}} & \makecell[t]{\textbf{FOV}\\ \textbf{{[\arcsec]}}} & \makecell[t]{\textbf{Sampling}\\ \textbf{{[\arcsec /px]}}} & \makecell[t]{\textbf{R}} & \makecell[t]{\textbf{SNR}} & \makecell[t]{\textbf{\# WL}\\ \textbf{Points}} & \makecell[t]{\textbf{Cadence}\\ \textbf{{[s]}}} \\
\hline
EMBER & IR & Chromosphere & $^{+}$1083 & 0.11 & spec & 40x40 & 0.10 & 100000 & 4000 & N/A & 10 \\
IFS-M & V & Chromosphere & $^{+}$587.6 & 0.06 & spec & 40x40 & 0.10 & 100000 & 4000 & N/A & 10 \\
 &  & Chromosphere & $^{+}$656.3 & 0.06 & spec & 40x40 & 0.10 & 100000 & 4000 & N/A & 10 \\
IFS-M & B & Chromosphere & $^{+}$486.1 & 0.06 & spec & 40x40 & 0.10 & 75000 & 4000 & N/A & 10 \\
\hline
\textbf{Notes:} & \multicolumn{11}{p{17cm}|}{TIS/FBIs and IFS: The use of the adaptive optics on an off-limb target is necessary in case of prominence observations. High stability is required to reduce spurious signals down to 10-5, in terms of the continuum intensity. Pointing error of the order of 1".} \\
\hline
\end{tabular}%
\end{adjustbox}%
\end{table}

\begin{table}[H]
\OPtabletitleCustom{5.3.1}{Determining which coronal upflows become outflows}{OP5.3.1table}{OP5.3.1}
\centering
\vspace{4mm}
\begin{adjustbox}{max width=\textwidth}
\begin{tabular}{|l|c|c|c|c|c|c|c|c|c|c|c|}
\hline
\textbf{Goal:} & \multicolumn{11}{p{17cm}|}{TIS/FBIs: Detect small-scale photospheric and chromospheric activity. Provide context information on the surrounding AR/CH. IFS: Determine magnetic field topology and dynamics of the photosphere and chromosphere at active region and coronal hole boundaries as a function of height. Combine with UV/EUV imaging and spectroscopic observations of TR and coronal response using Solar Orbiter EUI/SPICE and/or Solar C EUVST.} \\
\hline
\makecell[t]{\textbf{Instrument}} & \makecell[t]{\textbf{Optical}\\ \textbf{Arm}} & \makecell[t]{\textbf{Atm}\\ \textbf{Layer}} & \makecell[t]{\textbf{CWL}\\ \textbf{{[nm]}}} & \makecell[t]{\textbf{Spectral}\\ \textbf{Range}\\ \textbf{{[nm]}}} & \makecell[t]{\textbf{Mode}} & \makecell[t]{\textbf{FOV}\\ \textbf{{[\arcsec]}}} & \makecell[t]{\textbf{Sampling}\\ \textbf{{[\arcsec /px]}}} & \makecell[t]{\textbf{R}} & \makecell[t]{\textbf{SNR}} & \makecell[t]{\textbf{\# WL}\\ \textbf{Points}} & \makecell[t]{\textbf{Cadence}\\ \textbf{{[s]}}} \\
\hline
IFS-M & B & Chromosphere & 396.8 & 0.20 & pol & 10x10 & 0.02 & 75000 & 2000 & N/A & 10 \\
TIS/FBI & B & Photosphere & 430 & N/A & img & 40 & 0.02 & N/A & 500 & N/A & 1 \\
IFS-M & V & Photosphere & 630.2 & 0.20 & pol & 10x10 & 0.03 & 100000 & 2000 & N/A & 10 \\
TIS/FBI & V & Chromosphere & 656.3 & 0.30 & spec & 40 & 0.03 & 100000 & 500 & 21 & 5 \\
IFS-M & R & Chromosphere & 854.2 & 0.30 & pol & 10x10 & 0.05 & 80000 & 2000 & N/A & 10 \\
EMBER & IR & Chromosphere & 1083 & 0.60 & pol & 10x10 & 0.10 & 100000 & 2000 & N/A & 10 \\
\hline
\textbf{Notes:} & \multicolumn{11}{p{17cm}|}{TIS/FBIs: See below for co-ordination requirements. IFS: These observations need to be co-ordinated with space platforms to enable the connection to the TR and corona to be made. For Solar Orbiter this should be done at conjunction. This OP links to OP 3.3.1 on spicule acceleration on disk, so could potentially be merged, and is very similar to DKIST SUC-61 created by Louise Harra. Appropriate Solar Orbiter SOOPs, as listed in SUC-61 are L BOTH HIRES+LCAD CH Boundary Expansion; L SMALL HIRES HCAD SlowWindConnection; L SMALL HRES HCAD Fast Wind; L SMALL HRES HCAD SlowWindConenction; R SMALL HRES MCAD PolarObservations. EST can potentially provide greater multi- height coverage.} \\
\hline
\end{tabular}%
\end{adjustbox}%
\end{table}

\begin{table}[H]
\OPtabletitleCustom{5.4.1}{Measuring the photospheric and chromospheric dynamics before flares}{OP5.4.1table}{OP5.4.1}
\centering
\vspace{4mm}
\begin{adjustbox}{max width=\textwidth}
\begin{tabular}{|l|c|c|c|c|c|c|c|c|c|c|c|}
\hline
\textbf{Goal:} & \multicolumn{11}{p{17cm}|}{TIS/FBIs: Detect small-scale photospheric and chromospheric activity around the neutral line, including flux emegernce, cancellation. Provide context information on the surrounding AR.} \\
\hline
\makecell[t]{\textbf{Instrument}} & \makecell[t]{\textbf{Optical}\\ \textbf{Arm}} & \makecell[t]{\textbf{Atm}\\ \textbf{Layer}} & \makecell[t]{\textbf{CWL}\\ \textbf{{[nm]}}} & \makecell[t]{\textbf{Spectral}\\ \textbf{Range}\\ \textbf{{[nm]}}} & \makecell[t]{\textbf{Mode}} & \makecell[t]{\textbf{FOV}\\ \textbf{{[\arcsec]}}} & \makecell[t]{\textbf{Sampling}\\ \textbf{{[\arcsec /px]}}} & \makecell[t]{\textbf{R}} & \makecell[t]{\textbf{SNR}} & \makecell[t]{\textbf{\# WL}\\ \textbf{Points}} & \makecell[t]{\textbf{Cadence}\\ \textbf{{[s]}}} \\
\hline
IFS-M & B & Chromosphere & 396.8 & 0.20 & pol & 20x20 & 0.02 & 75000 & 2000 & N/A & 100 \\
TIS/FBI & B & Photosphere & 430 & N/A & img & 40 & 0.02 & N/A & 500 & N/A & 1 \\
IFS-M & V & Photosphere & 630.2 & 0.20 & pol & 20x20 & 0.03 & 100000 & 2000 & N/A & 100 \\
TIS/FBI & V & Chromosphere & 656.3 & 0.30 & spec & 40 & 0.03 & 100000 & 500 & 21 & 10 \\
IFS-M & R & Chromosphere & 854.2 & 0.30 & pol & 20x20 & 0.05 & 80000 & 2000 & N/A & 100 \\
EMBER & IR & Photosphere & 1565 & 0.60 & pol & 20x20 & 0.10 & 100000 & 2000 & N/A & 100 \\
\hline
\textbf{Notes:} & \multicolumn{11}{p{17cm}|}{TIS/FBIs: See below for coordination requirements. IFS: These observations would have to be a TOO but need to be coordinated with space platforms to enable the connection to the TR and corona to be made. For Solar Orbiter this could be done either at conjunction or quadrature, with quadrature providing a 3D view of the overlying corona, particularly if combined with Solar C EUVST. Relevant SOOPs: L FULL HRES HCAD Coronal Dynamics; R SMALL HRES HCAD AR Dynamics; R SMALL HRES HCAD RSburst; L FULL HRES HCAD Eruption Watch} \\
\hline
\end{tabular}%
\end{adjustbox}%
\end{table}

\begin{table}[H]
\OPtabletitleCustom{5.5.1}{Rising spicule signatures}{OP5.5.1table}{OP5.5.1}
\centering
\vspace{4mm}
\begin{adjustbox}{max width=\textwidth}
\begin{tabular}{|l|c|c|c|c|c|c|c|c|c|c|c|}
\hline
\textbf{Goal:} & \multicolumn{11}{p{17cm}|}{All: High-resolution spectroscopy and spectropolarimetry combined with G-band imaging to determine the associated rising spicule signatures.} \\
\hline
\makecell[t]{\textbf{Instrument}} & \makecell[t]{\textbf{Optical}\\ \textbf{Arm}} & \makecell[t]{\textbf{Atm}\\ \textbf{Layer}} & \makecell[t]{\textbf{CWL}\\ \textbf{{[nm]}}} & \makecell[t]{\textbf{Spectral}\\ \textbf{Range}\\ \textbf{{[nm]}}} & \makecell[t]{\textbf{Mode}} & \makecell[t]{\textbf{FOV}\\ \textbf{{[\arcsec]}}} & \makecell[t]{\textbf{Sampling}\\ \textbf{{[\arcsec /px]}}} & \makecell[t]{\textbf{R}} & \makecell[t]{\textbf{SNR}} & \makecell[t]{\textbf{\# WL}\\ \textbf{Points}} & \makecell[t]{\textbf{Cadence}\\ \textbf{{[s]}}} \\
\hline
TIS/FBI & B & Photosphere & 430 & N.A & img & 40 & 0.02 & N/A & 1000 & N/A & 1 \\
IFS-M & V & Photosphere & 630.2 & 0.20 & pol & 10x10 & 0.03 & 100000 & 1000 & N/A & 2 \\
TIS/FBI & V & Chromosphere & 656.3 & 0.30 & spec & 40 & 0.03 & 100000 & 500 & 11 & 2 \\
TIS/FBI & R & Chromosphere & 854.2 & 0.30 & pol & 40 & 0.05 & 80000 & 1000 & 11 & 2 \\
EMBER & IR & Chromosphere & 1083 & 0.60 & pol & 10x10 & 0.10 & 100000 & 2500 & N/A & 10 \\
\hline
\textbf{Notes:} & \multicolumn{11}{p{17cm}|}{The short lifetime of RBEs/RREs requires the highest cadence possible.} \\
\hline
\end{tabular}%
\end{adjustbox}%
\end{table}

\begin{table}[H]
\OPtabletitleCustom{5.6.1}{Ellerman bombs in the lower solar atmosphere }{OP5.6.1table}{OP5.6.1}
\centering
\vspace{4mm}
\begin{adjustbox}{max width=\textwidth}
\begin{tabular}{|l|c|c|c|c|c|c|c|c|c|c|c|}
\hline
\textbf{Goal:} & \multicolumn{11}{p{17cm}|}{All: High-resolution spectro-polarimetry to determine signatures of EBs in the lower solar atmosphere.} \\
\hline
\makecell[t]{\textbf{Instrument}} & \makecell[t]{\textbf{Optical}\\ \textbf{Arm}} & \makecell[t]{\textbf{Atm}\\ \textbf{Layer}} & \makecell[t]{\textbf{CWL}\\ \textbf{{[nm]}}} & \makecell[t]{\textbf{Spectral}\\ \textbf{Range}\\ \textbf{{[nm]}}} & \makecell[t]{\textbf{Mode}} & \makecell[t]{\textbf{FOV}\\ \textbf{{[\arcsec]}}} & \makecell[t]{\textbf{Sampling}\\ \textbf{{[\arcsec /px]}}} & \makecell[t]{\textbf{R}} & \makecell[t]{\textbf{SNR}} & \makecell[t]{\textbf{\# WL}\\ \textbf{Points}} & \makecell[t]{\textbf{Cadence}\\ \textbf{{[s]}}} \\
\hline
TIS/FBI & B & Chromosphere & 393.37 & 0.20 & spec & 40 & 0.02 & 50000 & 500 & 21 & 10 \\
TIS/FBI & V & Chromosphere & 656.3 & 0.30 & spec & 40 & 0.03 & 100000 & 500 & 21 & 10 \\
TIS/FBI & R & Chromosphere & 854.2 & 0.30 & pol & 40 & 0.05 & 80000 & 1000 & 21 & 10 \\
EMBER & IR & Chromosphere & 1083 & 0.60 & pol & 10x10 & 0.10 & 100000 & 2500 & N/A & 10 \\
\hline
\textbf{Notes:} & \multicolumn{11}{p{17cm}|}{High-cadence is essential. Spectral range (km/s)   100 for each line} \\
\hline
\end{tabular}%
\end{adjustbox}%
\end{table}

\begin{table}[H]
\OPtabletitleCustom{6.1.1}{Thermal structure of flare: line observations }{OP6.1.1table}{OP6.1.1}
\centering
\vspace{4mm}
\begin{adjustbox}{max width=\textwidth}
\begin{tabular}{|l|c|c|c|c|c|c|c|c|c|c|c|}
\hline
\textbf{Goal:} & \multicolumn{11}{p{17cm}|}{TIS/FBIs: Observe spectral lines formed at different heights in the chromosphere, from which temperature profile with column mass can be determined. High temporal and spatial resolution context observations of the flare ribbons.} \\
\hline
\makecell[t]{\textbf{Instrument}} & \makecell[t]{\textbf{Optical}\\ \textbf{Arm}} & \makecell[t]{\textbf{Atm}\\ \textbf{Layer}} & \makecell[t]{\textbf{CWL}\\ \textbf{{[nm]}}} & \makecell[t]{\textbf{Spectral}\\ \textbf{Range}\\ \textbf{{[nm]}}} & \makecell[t]{\textbf{Mode}} & \makecell[t]{\textbf{FOV}\\ \textbf{{[\arcsec]}}} & \makecell[t]{\textbf{Sampling}\\ \textbf{{[\arcsec /px]}}} & \makecell[t]{\textbf{R}} & \makecell[t]{\textbf{SNR}} & \makecell[t]{\textbf{\# WL}\\ \textbf{Points}} & \makecell[t]{\textbf{Cadence}\\ \textbf{{[s]}}} \\
\hline
TIS/FBI & B & Photosphere & 430 & N/A & img & 40 & 0.02 & N/A & 100 & N/A & 5 \\
 &  & Chromosphere & 396.9 & 0.30 & spec & 40 & 0.02 & 50000 & 500 & 15 & 5 \\
TIS/FBI & V & Photosphere & 543.5 & 0.30 & spec & 40 & 0.03 & 100000 & 500 & 15 & 8 \\
 &  & Chromosphere & 656.2 & 0.40 & spec & 40 & 0.03 & 100000 & 500 & 15 & 8 \\
TIS/FBI & R & Chromosphere & 854.2 & 0.30 & spec & 40 & 0.05 & 80000 & 500 & 15 & 4 \\
EMBER & IR & Chromosphere & 1083 & 0.50 & pol & 8x8 & 0.10 & 100000 & 2000 & N/A & 10 \\
\hline
\textbf{Notes:} & \multicolumn{11}{p{17cm}|}{TIS/FBIs: The focus here is only on I, hence low SNR requirement. To capture dynamical changes need to prioritise cadence. This observing program requires stable observations with AO locked, at different   values. Photospheric line chosen as it is formed quite high in photosphere, bearing in mind likely range in height of strong perturbations caused by flare heating. For FOV, the tradeo  is between having a good chance of catching a flare (FOV) and highest possible spatial and temporal resolutions. The programme requires stable observations with AO locked at different values.} \\
\hline
\end{tabular}%
\end{adjustbox}%
\end{table}

\begin{table}[H]
\OPtabletitleCustom{6.2.1}{Thermal structure of flare: continuum observations}{OP6.2.1table}{OP6.2.1}
\centering
\vspace{4mm}
\begin{adjustbox}{max width=\textwidth}
\begin{tabular}{|l|c|c|c|c|c|c|c|c|c|c|c|}
\hline
\textbf{Goal:} & \multicolumn{11}{p{17cm}|}{TIS/FBIs: High temporal and spatial resolution observations of a flare at several points in the continuum, to track the evolution of energy input via the hydrogen continuum emission (“white-light flare”).} \\
\hline
\makecell[t]{\textbf{Instrument}} & \makecell[t]{\textbf{Optical}\\ \textbf{Arm}} & \makecell[t]{\textbf{Atm}\\ \textbf{Layer}} & \makecell[t]{\textbf{CWL}\\ \textbf{{[nm]}}} & \makecell[t]{\textbf{Spectral}\\ \textbf{Range}\\ \textbf{{[nm]}}} & \makecell[t]{\textbf{Mode}} & \makecell[t]{\textbf{FOV}\\ \textbf{{[\arcsec]}}} & \makecell[t]{\textbf{Sampling}\\ \textbf{{[\arcsec /px]}}} & \makecell[t]{\textbf{R}} & \makecell[t]{\textbf{SNR}} & \makecell[t]{\textbf{\# WL}\\ \textbf{Points}} & \makecell[t]{\textbf{Cadence}\\ \textbf{{[s]}}} \\
\hline
TIS/FBI & B & Photosphere & 430 & N/A & img & 40 & 10.00 & N/A & 200 & N/A & 1 \\
TIS/FBI & V & Chromosphere & 656.2 & 0.03 & img & 40 & 10.00 & N/A & 200 & N/A & 1 \\
\hline
\textbf{Notes:} & \multicolumn{11}{p{17cm}|}{TIS/FBIs: For FOV, the trade-off  is between having a good chance of catching a flare (FOV) and highest possible spatial and temporal resolutions. The programme requires stable observations with AO locked at different values.} \\
\hline
\end{tabular}%
\end{adjustbox}%
\end{table}

\begin{table}[H]
\OPtabletitleCustom{6.2.2}{Thermal structure of flare: detection of the H Paschen Jump}{OP6.2.2table}{OP6.2.2}
\centering
\vspace{4mm}
\begin{adjustbox}{max width=\textwidth}
\begin{tabular}{|l|c|c|c|c|c|c|c|c|c|c|c|}
\hline
\textbf{Goal:} & \multicolumn{11}{p{17cm}|}{TIS/FBIs or IFS in spectroscopic mode: Identification of the H Paschen jump at 815.1nm in flares. TIS/FBIs in spectroscopic mode: Identification of the H Paschen jump at 815.1nm in flares.} \\
\hline
\makecell[t]{\textbf{Instrument}} & \makecell[t]{\textbf{Optical}\\ \textbf{Arm}} & \makecell[t]{\textbf{Atm}\\ \textbf{Layer}} & \makecell[t]{\textbf{CWL}\\ \textbf{{[nm]}}} & \makecell[t]{\textbf{Spectral}\\ \textbf{Range}\\ \textbf{{[nm]}}} & \makecell[t]{\textbf{Mode}} & \makecell[t]{\textbf{FOV}\\ \textbf{{[\arcsec]}}} & \makecell[t]{\textbf{Sampling}\\ \textbf{{[\arcsec /px]}}} & \makecell[t]{\textbf{R}} & \makecell[t]{\textbf{SNR}} & \makecell[t]{\textbf{\# WL}\\ \textbf{Points}} & \makecell[t]{\textbf{Cadence}\\ \textbf{{[s]}}} \\
\hline
IFS-M & R & Chromosphere & 815.1 & 20.00 & pol & 8x8 & 0.05 & 80000 & 2000 & N/A & 10 \\
\hline
\textbf{Notes:} & \multicolumn{11}{p{17cm}|}{As an alternative to a spectrometer, 2 or 3 narrowband-filter imager could be used, capturing the continuum levels between the Ca and H Paschen spectral lines in the range 800-900nm. The filters would have to be designed in a way to avoid stray light from other spectral ranges, especially from regions densely populated by spectral lines. Ideally, a high-resolution imaging spectrometer in the range 800-900nm, with high spatial/temporal resolutions, should provide enough information about the continuum emission along flare ribbons to track the evolution of this emission and allow the detection of the Paschen jump (or lack thereof!). This setup would also provide observations of the Ca and H Paschen lines in the range.} \\
\hline
\end{tabular}%
\end{adjustbox}%
\end{table}

\begin{table}[H]
\OPtabletitleCustom{6.3.1}{Velocity structure of the flaring atmosphere}{OP6.3.1table}{OP6.3.1}
\centering
\vspace{4mm}
\begin{adjustbox}{max width=\textwidth}
\begin{tabular}{|l|c|c|c|c|c|c|c|c|c|c|c|}
\hline
\textbf{Goal:} & \multicolumn{11}{p{17cm}|}{TIS/FBIs: Observations to track flows in the chromosphere at a variety of optical depths. High temporal and spatial resolution context observations of the flare ribbons.} \\
\hline
\makecell[t]{\textbf{Instrument}} & \makecell[t]{\textbf{Optical}\\ \textbf{Arm}} & \makecell[t]{\textbf{Atm}\\ \textbf{Layer}} & \makecell[t]{\textbf{CWL}\\ \textbf{{[nm]}}} & \makecell[t]{\textbf{Spectral}\\ \textbf{Range}\\ \textbf{{[nm]}}} & \makecell[t]{\textbf{Mode}} & \makecell[t]{\textbf{FOV}\\ \textbf{{[\arcsec]}}} & \makecell[t]{\textbf{Sampling}\\ \textbf{{[\arcsec /px]}}} & \makecell[t]{\textbf{R}} & \makecell[t]{\textbf{SNR}} & \makecell[t]{\textbf{\# WL}\\ \textbf{Points}} & \makecell[t]{\textbf{Cadence}\\ \textbf{{[s]}}} \\
\hline
TIS/FBI & B & Photosphere & 430 & N/A & img & 40 & 0.02 & N/A & 100 & N/A & 6 \\
 &  & Chromosphere & 396.9 & 0.30 & spec & 40 & 0.02 & 50000 & 500 & 21 & 6 \\
TIS/FBI & V & Photosphere & 557.6 & 0.30 & spec & 40 & 0.03 & 100000 & 500 & 21 & 10 \\
 &  & Chromosphere & 656.2 & 0.40 & spec & 40 & 0.03 & 100000 & 500 & 21 & 10 \\
TIS/FBI & R & Chromosphere & 854.2 & 0.40 & spec & 40 & 0.05 & 80000 & 500 & 21 & 5 \\
EMBER & IR & Chromosphere & 1083 & 0.50 & pol & 8x8 & 0.10 & 100000 & 2000 & N/A & 10 \\
\hline
\textbf{Notes:} & \multicolumn{11}{p{17cm}|}{The focus here is only on I, hence low SNR requirement. Based on observations e.g. with IRIS and in H , can expect large flows, with also an optically thin chromospheric component possible in the nominally ‘photospheric’ lines. So a wide spectral range is necessary. This observing program requires stable observations with AO locked, at different values. For FOV, the tradeo  is between having a good chance of catching a flare (FOV) and highest possible spatial and temporal resolutions. The programme requires stable observations with AO locked at different values.} \\
\hline
\end{tabular}%
\end{adjustbox}%
\end{table}

\begin{table}[H]
\OPtabletitleCustom{6.4.1}{Diagnostics for non-thermal particles}{OP6.4.1table}{OP6.4.1}
\centering
\vspace{4mm}
\begin{adjustbox}{max width=\textwidth}
\begin{tabular}{|l|c|c|c|c|c|c|c|c|c|c|c|}
\hline
\textbf{Goal:} & \multicolumn{11}{p{17cm}|}{TIS/FBIs: Linear polarization measurements in strong spectral lines, in a search for evidence of impact polarization generated by low-energy proton beams. Detect flare ribbons at chromospheric heights. Context image information.} \\
\hline
\makecell[t]{\textbf{Instrument}} & \makecell[t]{\textbf{Optical}\\ \textbf{Arm}} & \makecell[t]{\textbf{Atm}\\ \textbf{Layer}} & \makecell[t]{\textbf{CWL}\\ \textbf{{[nm]}}} & \makecell[t]{\textbf{Spectral}\\ \textbf{Range}\\ \textbf{{[nm]}}} & \makecell[t]{\textbf{Mode}} & \makecell[t]{\textbf{FOV}\\ \textbf{{[\arcsec]}}} & \makecell[t]{\textbf{Sampling}\\ \textbf{{[\arcsec /px]}}} & \makecell[t]{\textbf{R}} & \makecell[t]{\textbf{SNR}} & \makecell[t]{\textbf{\# WL}\\ \textbf{Points}} & \makecell[t]{\textbf{Cadence}\\ \textbf{{[s]}}} \\
\hline
TIS/FBI & B & Chromosphere & 396.9 & 0.30 & spec & 40 & 0.02 & 50000 & 500 & 21 & 5 \\
TIS/FBI & V & Upper Photosphere & 589.6 & 0.30 & spec & 40 & 0.03 & 100000 & 500 & 21 & 10 \\
 &  & Chromosphere & 656.2 & 0.40 & spec & 40 & 0.03 & 100000 & 500 & 21 & 10 \\
TIS/FBI & R & Chromosphere & 854.2 & 0.30 & spec & 40 & 0.05 & 80000 & 500 & 21 & 5 \\
EMBER & IR & Chromosphere & 1083 & 0.50 & pol & 8x8 & 0.10 & 100000 & 2000 & N/A & 10 \\
\hline
\textbf{Notes:} & \multicolumn{11}{p{17cm}|}{TIS/FBIs: The focus here is on I, Q, U in the core of the spectral lines only to try and isolate the upper part of the chromosphere where the particle beams would reach. FP is needed to capture different parts of a flare ribbon under different conditions. I, Q, U to be taken in as quick succession as possible. Note, it is expected that the spectral lines will brighten during a flare so short exposures should be possible. This observing program requires stable observations with AO locked.} \\
\hline
\end{tabular}%
\end{adjustbox}%
\end{table}

\begin{table}[H]
\OPtabletitleCustom{6.5.1}{QPPs in different phases of flares}{OP6.5.1table}{OP6.5.1}
\centering
\vspace{4mm}
\begin{adjustbox}{max width=\textwidth}
\begin{tabular}{|l|c|c|c|c|c|c|c|c|c|c|c|}
\hline
\textbf{Goal:} & \multicolumn{11}{p{17cm}|}{TIS/FBIs: Detect QPPs during flares and verify whether those detected in different phases of the flare have different characteristics. Determine the magnetic field configuration of the flaring site, estimate magnetic fluxes and search for any variation correlated with the flare occurrence. Detect flare signatures at photospheric and chromospheric heights. Context information.} \\
\hline
\makecell[t]{\textbf{Instrument}} & \makecell[t]{\textbf{Optical}\\ \textbf{Arm}} & \makecell[t]{\textbf{Atm}\\ \textbf{Layer}} & \makecell[t]{\textbf{CWL}\\ \textbf{{[nm]}}} & \makecell[t]{\textbf{Spectral}\\ \textbf{Range}\\ \textbf{{[nm]}}} & \makecell[t]{\textbf{Mode}} & \makecell[t]{\textbf{FOV}\\ \textbf{{[\arcsec]}}} & \makecell[t]{\textbf{Sampling}\\ \textbf{{[\arcsec /px]}}} & \makecell[t]{\textbf{R}} & \makecell[t]{\textbf{SNR}} & \makecell[t]{\textbf{\# WL}\\ \textbf{Points}} & \makecell[t]{\textbf{Cadence}\\ \textbf{{[s]}}} \\
\hline
TIS/FBI & B & Photosphere & 430 & N/A & img & 40 & 0.02 & N/A & 100 & N/A & 6 \\
 &  & Chromosphere & 396.9 & 0.30 & spec & 40 & 0.02 & 50000 & 500 & 15 & 6 \\
TIS/FBI & V & Chromosphere & 517.3 & 0.30 & spec & 40 & 0.03 & 100000 & 500 & 21 & 10 \\
 &  & Chromosphere & 656.2 & 0.40 & spec & 40 & 0.03 & 100000 & 500 & 21 & 10 \\
TIS/FBI & R & Chromosphere & 854.2 & 0.30 & spec & 40 & 0.05 & 80000 & 500 & 21 & 5 \\
EMBER & IR & Photosphere & 1565 & 1.00 & pol & 8x8 & 0.10 & 100000 & 2000 & N/A & 10 \\
\hline
\textbf{Notes:} & \multicolumn{11}{p{17cm}|}{N/A} \\
\hline
\end{tabular}%
\end{adjustbox}%
\end{table}

\begin{table}[H]
\OPtabletitleCustom{6.5.2}{QPPs and oscillatory reconnection}{OP6.5.2table}{OP6.5.2}
\centering
\vspace{4mm}
\begin{adjustbox}{max width=\textwidth}
\begin{tabular}{|l|c|c|c|c|c|c|c|c|c|c|c|}
\hline
\textbf{Goal:} & \multicolumn{11}{p{17cm}|}{TIS/FBIs: Detect flare emission in photosphere and chromosphere. Search for QPPs correlating with reconnection events. IFS: Determine magnetic field strength and topology of both pre-existing and emerging loops as a function of height. Detect signatures of (oscillating) magnetic reconnection due to the interaction of emerging fields with pre-existing flux in photosphere and chromosphere.} \\
\hline
\makecell[t]{\textbf{Instrument}} & \makecell[t]{\textbf{Optical}\\ \textbf{Arm}} & \makecell[t]{\textbf{Atm}\\ \textbf{Layer}} & \makecell[t]{\textbf{CWL}\\ \textbf{{[nm]}}} & \makecell[t]{\textbf{Spectral}\\ \textbf{Range}\\ \textbf{{[nm]}}} & \makecell[t]{\textbf{Mode}} & \makecell[t]{\textbf{FOV}\\ \textbf{{[\arcsec]}}} & \makecell[t]{\textbf{Sampling}\\ \textbf{{[\arcsec /px]}}} & \makecell[t]{\textbf{R}} & \makecell[t]{\textbf{SNR}} & \makecell[t]{\textbf{\# WL}\\ \textbf{Points}} & \makecell[t]{\textbf{Cadence}\\ \textbf{{[s]}}} \\
\hline
TIS/FBI & B & Photosphere & 430 & N/A & img & 40 & 0.02 & N/A & 100 & N/A & 5 \\
 &  & Chromosphere & 396.9 & 0.30 & spec & 40 & 0.02 & 50000 & 500 & 15 & 5 \\
TIS/FBI & V & Photosphere & 630 & 0.40 & pol & 40 & 0.03 & 100000 & 500 & 15 & 20 \\
 &  & Chromosphere & 656.2 & 0.40 & spec & 40 & 0.03 & 100000 & 500 & 15 & 20 \\
TIS/FBI & R & Chromosphere & 854.2 & 0.30 & pol & 40 & 0.05 & 80000 & 500 & 15 & 15 \\
EMBER & IR & Chromosphere & 1083 & 1.00 & pol & 8x8 & 0.10 & 100000 & 2000 & N/A & 10 \\
\hline
\textbf{Notes:} & \multicolumn{11}{p{17cm}|}{The full active region FOV should be obtained by tiling the IFUs as appropriate.} \\
\hline
\end{tabular}%
\end{adjustbox}%
\end{table}

\begin{table}[H]
\OPtabletitleCustom{6.6.1}{Sunquakes initiated by impulsive heating during flares}{OP6.6.1table}{OP6.6.1}
\centering
\vspace{4mm}
\begin{adjustbox}{max width=\textwidth}
\begin{tabular}{|l|c|c|c|c|c|c|c|c|c|c|c|}
\hline
\textbf{Goal:} & \multicolumn{11}{p{17cm}|}{TIS/FBIs: Detect flare emission in photosphere and chromosphere. Context information. IFS: Verify whether sunquakes are due to impulsive heating by means of simultaneous observations at different wavelengths: search for any time delay between photospheric and chromospheric emission that might shed light on the heating and energy trasnsport mechanisms. Try to discriminate between the possible impulsive heating mechanisms: thick target, backwarming or wave heating (no polarimetry).} \\
\hline
\makecell[t]{\textbf{Instrument}} & \makecell[t]{\textbf{Optical}\\ \textbf{Arm}} & \makecell[t]{\textbf{Atm}\\ \textbf{Layer}} & \makecell[t]{\textbf{CWL}\\ \textbf{{[nm]}}} & \makecell[t]{\textbf{Spectral}\\ \textbf{Range}\\ \textbf{{[nm]}}} & \makecell[t]{\textbf{Mode}} & \makecell[t]{\textbf{FOV}\\ \textbf{{[\arcsec]}}} & \makecell[t]{\textbf{Sampling}\\ \textbf{{[\arcsec /px]}}} & \makecell[t]{\textbf{R}} & \makecell[t]{\textbf{SNR}} & \makecell[t]{\textbf{\# WL}\\ \textbf{Points}} & \makecell[t]{\textbf{Cadence}\\ \textbf{{[s]}}} \\
\hline
TIS/FBI & B & Photosphere & 430 & N/A & img & 40 & 0.02 & N/A & 100 & N/A & 5 \\
 &  & Chromosphere & 396.9 & 0.30 & spec & 40 & 0.02 & 50000 & 500 & 15 & 5 \\
TIS/FBI & V & Chromosphere & 656.2 & 0.40 & spec & 40 & 0.03 & 100000 & 500 & 15 & 4 \\
IFS-M & V & Photosphere & 630 & 1.00 & pol & 8x8 & 0.03 & 100000 & 2000 & N/A & 10 \\
IFS-M & R & Chromosphere & 854.2 & 0.30 & spec & 40 & 0.05 & 80000 & 500 & N/A & 4 \\
EMBER & IR & Photosphere & 1565 & 1.00 & pol & 8x8 & 0.10 & 100000 & 2000 & N/A & 10 \\
\hline
\textbf{Notes:} & \multicolumn{11}{p{17cm}|}{IFS: The full active region FOV should be obtained by tiling the IFS as appropriate.} \\
\hline
\end{tabular}%
\end{adjustbox}%
\end{table}

\begin{table}[H]
\OPtabletitleCustom{6.6.2}{Sunquakes initiated by changes in the Lorentz force during flares}{OP6.6.2table}{OP6.6.2}
\centering
\vspace{4mm}
\begin{adjustbox}{max width=\textwidth}
\begin{tabular}{|l|c|c|c|c|c|c|c|c|c|c|c|}
\hline
\textbf{Goal:} & \multicolumn{11}{p{17cm}|}{TIS/FBIs: Detect flare emission in photosphere and chromosphere during sunquakes. Context information. IFS: Determine magnetic field configuration before and during flares. Detect signatures of sunquakes at photospheric and chromospheric levels.} \\
\hline
\makecell[t]{\textbf{Instrument}} & \makecell[t]{\textbf{Optical}\\ \textbf{Arm}} & \makecell[t]{\textbf{Atm}\\ \textbf{Layer}} & \makecell[t]{\textbf{CWL}\\ \textbf{{[nm]}}} & \makecell[t]{\textbf{Spectral}\\ \textbf{Range}\\ \textbf{{[nm]}}} & \makecell[t]{\textbf{Mode}} & \makecell[t]{\textbf{FOV}\\ \textbf{{[\arcsec]}}} & \makecell[t]{\textbf{Sampling}\\ \textbf{{[\arcsec /px]}}} & \makecell[t]{\textbf{R}} & \makecell[t]{\textbf{SNR}} & \makecell[t]{\textbf{\# WL}\\ \textbf{Points}} & \makecell[t]{\textbf{Cadence}\\ \textbf{{[s]}}} \\
\hline
TIS/FBI & B & Photosphere & 430 & N/A & img & 40 & 0.02 & N/A & 100 & N/A & 5 \\
 &  & Chromosphere & 396.9 & 0.30 & spec & 40 & 0.02 & 50000 & 500 & 15 & 5 \\
TIS/FBI & V & Chromosphere & 656.2 & 0.40 & spec & 40 & 0.03 & 100000 & 500 & 15 & 4 \\
IFS-M & V & Photosphere & 630 & 1.00 & pol & 8x8 & 0.03 & 100000 & 2000 & N/A & 10 \\
IFS-M & R & Chromosphere & 854.2 & 0.30 & spec & 40 & 0.05 & 80000 & 500 & N/A & 4 \\
EMBER & IR & Chromosphere & 1083 & 1.00 & pol & 8x8 & 0.10 & 100000 & 2000 & N/A & 10 \\
\hline
\textbf{Notes:} & \multicolumn{11}{p{17cm}|}{IFS: The full active region FOV should be obtained by tiling the IFUs as appropriate.} \\
\hline
\end{tabular}%
\end{adjustbox}%
\end{table}

\begin{table}[H]
\OPtabletitleCustom{6.7.1}{Changes in the magnetic field configuration during/after flares}{OP6.7.1table}{OP6.7.1}
\centering
\vspace{4mm}
\begin{adjustbox}{max width=\textwidth}
\begin{tabular}{|l|c|c|c|c|c|c|c|c|c|c|c|}
\hline
\textbf{Goal:} & \multicolumn{11}{p{17cm}|}{TIS/FBIs: Detect changes in the sunspot configuration during/after flares. Determine the magnetic field configuration of the flaring site and search for any variation in the vector magnetic field correlated with the flare occurrence. Infer the time sequence leading to the magnetic field sunspot variation. Detect flare signatures at photospheric and chromospheric heights. Context information on sunspots morphology before, during and after the flare. IFS: Context magnetic field at photospheric and chromospheric heights.} \\
\hline
\makecell[t]{\textbf{Instrument}} & \makecell[t]{\textbf{Optical}\\ \textbf{Arm}} & \makecell[t]{\textbf{Atm}\\ \textbf{Layer}} & \makecell[t]{\textbf{CWL}\\ \textbf{{[nm]}}} & \makecell[t]{\textbf{Spectral}\\ \textbf{Range}\\ \textbf{{[nm]}}} & \makecell[t]{\textbf{Mode}} & \makecell[t]{\textbf{FOV}\\ \textbf{{[\arcsec]}}} & \makecell[t]{\textbf{Sampling}\\ \textbf{{[\arcsec /px]}}} & \makecell[t]{\textbf{R}} & \makecell[t]{\textbf{SNR}} & \makecell[t]{\textbf{\# WL}\\ \textbf{Points}} & \makecell[t]{\textbf{Cadence}\\ \textbf{{[s]}}} \\
\hline
TIS/FBI & B & Photosphere & 430 & N/A & img & 40 & 0.02 & N/A & 100 & N/A & 5 \\
 &  & Chromosphere & 396.9 & 0.30 & spec & 40 & 0.02 & 50000 & 500 & 15 & 5 \\
TIS/FBI & V & Chromosphere & 656.2 & 0.40 & spec & 40 & 0.03 & 100000 & 500 & 15 & 4 \\
IFS-M & V & Photosphere & 630 & 1.00 & pol & 8x8 & 0.03 & 100000 & 2000 & N/A & 10 \\
TIS/FBI & R & Chromosphere & 854.2 & 0.30 & pol & 40 & 0.05 & 80000 & 500 & 15 & 16 \\
EMBER & IR & Photosphere & 1565 & 1.00 & pol & 8x8 & 0.10 & 100000 & 2000 & N/A & 10 \\
\hline
\textbf{Notes:} & \multicolumn{11}{p{17cm}|}{IFS: The full active region FOV should be obtained by tiling as appropriate. It may be desirable also to bin e.g. 4x4 pixels to obtain higher SNR and hence increase the cadence.} \\
\hline
\end{tabular}%
\end{adjustbox}%
\end{table}

\begin{table}[H]
\OPtabletitleCustom{6.7.2}{Determine magnetic helicity accumulation, dip and shear angle in flaring active regions}{OP6.7.2table}{OP6.7.2}
\centering
\vspace{4mm}
\begin{adjustbox}{max width=\textwidth}
\begin{tabular}{|l|c|c|c|c|c|c|c|c|c|c|c|}
\hline
\textbf{Goal:} & \multicolumn{11}{p{17cm}|}{TIS/FBIs: Detect flare signatures at photospheric and chromospheric heights. Context information on the active region morphology before, during and after the flare. IFS: Infer the magnetic field configuration and the photospheric flow field to determine the magnetic helicity accumulation in the active region. Determine the dip and shear angle along the main polarity inversion line.} \\
\hline
\makecell[t]{\textbf{Instrument}} & \makecell[t]{\textbf{Optical}\\ \textbf{Arm}} & \makecell[t]{\textbf{Atm}\\ \textbf{Layer}} & \makecell[t]{\textbf{CWL}\\ \textbf{{[nm]}}} & \makecell[t]{\textbf{Spectral}\\ \textbf{Range}\\ \textbf{{[nm]}}} & \makecell[t]{\textbf{Mode}} & \makecell[t]{\textbf{FOV}\\ \textbf{{[\arcsec]}}} & \makecell[t]{\textbf{Sampling}\\ \textbf{{[\arcsec /px]}}} & \makecell[t]{\textbf{R}} & \makecell[t]{\textbf{SNR}} & \makecell[t]{\textbf{\# WL}\\ \textbf{Points}} & \makecell[t]{\textbf{Cadence}\\ \textbf{{[s]}}} \\
\hline
TIS/FBI & B & Photosphere & 430 & N/A & img & 40 & 0.02 & N/A & 100 & N/A & 5 \\
 &  & Chromosphere & 396.9 & 0.30 & spec & 40 & 0.02 & 50000 & 500 & 15 & 5 \\
TIS/FBI & V & Photosphere & 630 & 0.40 & pol & 40 & 0.03 & 100000 & 500 & 15 & 20 \\
 &  & Chromosphere & 656.2 & 0.40 & spec & 40 & 0.03 & 100000 & 500 & 15 & 20 \\
IFS-M & R & Chromosphere & 854.2 & 1.00 & pol & 8x8 & 0.05 & 80000 & 2000 & N/A & 10 \\
EMBER & IR & Photosphere & 1565 & 1.00 & pol & 8x8 & 0.10 & 100000 & 2000 & N/A & 10 \\
\hline
\textbf{Notes:} & \multicolumn{11}{p{17cm}|}{N/A} \\
\hline
\end{tabular}%
\end{adjustbox}%
\end{table}

\begin{table}[H]
\OPtabletitleCustom{6.8.1}{Filaments in flaring active regions}{OP6.8.1table}{OP6.8.1}
\centering
\vspace{4mm}
\begin{adjustbox}{max width=\textwidth}
\begin{tabular}{|l|c|c|c|c|c|c|c|c|c|c|c|}
\hline
\textbf{Goal:} & \multicolumn{11}{p{17cm}|}{IFS: Infer changes in the vector magnetic field and flows during the eruption of the filament. Scientific goal: identify the mechanisms which lead to flaring around filaments and eventually to the eruption of the filament. Predict future filament eruptions.} \\
\hline
\makecell[t]{\textbf{Instrument}} & \makecell[t]{\textbf{Optical}\\ \textbf{Arm}} & \makecell[t]{\textbf{Atm}\\ \textbf{Layer}} & \makecell[t]{\textbf{CWL}\\ \textbf{{[nm]}}} & \makecell[t]{\textbf{Spectral}\\ \textbf{Range}\\ \textbf{{[nm]}}} & \makecell[t]{\textbf{Mode}} & \makecell[t]{\textbf{FOV}\\ \textbf{{[\arcsec]}}} & \makecell[t]{\textbf{Sampling}\\ \textbf{{[\arcsec /px]}}} & \makecell[t]{\textbf{R}} & \makecell[t]{\textbf{SNR}} & \makecell[t]{\textbf{\# WL}\\ \textbf{Points}} & \makecell[t]{\textbf{Cadence}\\ \textbf{{[s]}}} \\
\hline
TIS/FBI & B & Upper Photosphere & 455.4 & 0.30 & pol & 40 & 0.02 & 40000 & 500 & 15 & 15 \\
TIS/FBI & V & Chromosphere & 656.2 & 0.40 & spec & 40 & 0.03 & 100000 & 500 & 15 & 4 \\
IFS-M & V & Photosphere & 630 & 1.00 & pol & 8x8 & 0.03 & 100000 & 2000 & N/A & 10 \\
IFS-M & R & Chromosphere & 854.2 & 1.00 & pol & 8x8 & 0.05 & 80000 & 2000 & N/A & 10 \\
EMBER & IR & Chromosphere & 1083 & 1.00 & pol & 8x8 & 0.10 & 100000 & 2000 & N/A & 10 \\
\hline
\textbf{Notes:} & \multicolumn{11}{p{17cm}|}{IFS: Two observing strategies: (1) Sit and stare with one tile on the filament. (2) Small mosaic with tiles along the filament covering some of the flaring area next to the filament.} \\
\hline
\end{tabular}%
\end{adjustbox}%
\end{table}

\begin{table}[H]
\OPtabletitleCustom{6.9.1}{CME's sources and temporal relation with flares}{OP6.9.1table}{OP6.9.1}
\centering
\vspace{4mm}
\begin{adjustbox}{max width=\textwidth}
\begin{tabular}{|l|c|c|c|c|c|c|c|c|c|c|c|}
\hline
\textbf{Goal:} & \multicolumn{11}{p{17cm}|}{TIS/FBIs: Detect flare signatures at photospheric and chromospheric heights. Context information on the active region morphology before, during and after the flare-CME occurrence. IFS: Search for any brightening or flaring emission which could be spatially and temporally correlated with a CME. Infer the magnetic field configuration of the flux rope involved in the eruption leading to the coronal mass ejection.} \\
\hline
\makecell[t]{\textbf{Instrument}} & \makecell[t]{\textbf{Optical}\\ \textbf{Arm}} & \makecell[t]{\textbf{Atm}\\ \textbf{Layer}} & \makecell[t]{\textbf{CWL}\\ \textbf{{[nm]}}} & \makecell[t]{\textbf{Spectral}\\ \textbf{Range}\\ \textbf{{[nm]}}} & \makecell[t]{\textbf{Mode}} & \makecell[t]{\textbf{FOV}\\ \textbf{{[\arcsec]}}} & \makecell[t]{\textbf{Sampling}\\ \textbf{{[\arcsec /px]}}} & \makecell[t]{\textbf{R}} & \makecell[t]{\textbf{SNR}} & \makecell[t]{\textbf{\# WL}\\ \textbf{Points}} & \makecell[t]{\textbf{Cadence}\\ \textbf{{[s]}}} \\
\hline
TIS/FBI & B & Photosphere & 430 & N/A & img & 40 & 0.02 & N/A & 100 & N/A & 5 \\
 &  & Chromosphere & 396.9 & 0.30 & spec & 40 & 0.02 & 50000 & 500 & 15 & 5 \\
TIS/FBI & V & Chromosphere & 656.2 & 0.40 & spec & 40 & 0.03 & 100000 & 500 & 15 & 4 \\
IFS-M & V & Photosphere & 630 & 1.00 & pol & 8x8 & 0.03 & 100000 & 2000 & N/A & 10 \\
IFS-M & R & Chromosphere & 854.2 & 1.00 & pol & 8x8 & 0.05 & 80000 & 2000 & N/A & 10 \\
EMBER & IR & Photosphere & 1565 & 1.00 & pol & 8x8 & 0.10 & 100000 & 2000 & N/A & 10 \\
\hline
\textbf{Notes:} & \multicolumn{11}{p{17cm}|}{TIS/FBIs: These observations should be complemented with data acquired by coronagraphs. IFS:  Mosaic of a few tiles along magnetic flux rope axis to look for plasma motions and for brightenings in the surroundings.} \\
\hline
\end{tabular}%
\end{adjustbox}%
\end{table}

\begin{table}[H]
\OPtabletitleCustom{7.1.1.1}{Waves in prominences observed in neutral and ionized spectral lines}{OP7.1.1.1table}{OP7.1.1.1}
\centering
\vspace{4mm}
\begin{adjustbox}{max width=\textwidth}
\begin{tabular}{|l|c|c|c|c|c|c|c|c|c|c|c|}
\hline
\textbf{Goal:} & \multicolumn{11}{p{17cm}|}{IFS: in spectroscopic mode with very high-cadence velocity measurements strictly simultaneously, target on prominence. 2D field of view is essential to align spectral lines and correct for differential refraction effects. In polarimetric mode: Context magnetic field in prominence. Possibly, using the same IFS spectropolarimeter for both kinds of measurements and highest cadence possible (see above), then adding up polarimetric data to increase s/n.} \\
\hline
\makecell[t]{\textbf{Instrument}} & \makecell[t]{\textbf{Optical}\\ \textbf{Arm}} & \makecell[t]{\textbf{Atm}\\ \textbf{Layer}} & \makecell[t]{\textbf{CWL}\\ \textbf{{[nm]}}} & \makecell[t]{\textbf{Spectral}\\ \textbf{Range}\\ \textbf{{[nm]}}} & \makecell[t]{\textbf{Mode}} & \makecell[t]{\textbf{FOV}\\ \textbf{{[\arcsec]}}} & \makecell[t]{\textbf{Sampling}\\ \textbf{{[\arcsec /px]}}} & \makecell[t]{\textbf{R}} & \makecell[t]{\textbf{SNR}} & \makecell[t]{\textbf{\# WL}\\ \textbf{Points}} & \makecell[t]{\textbf{Cadence}\\ \textbf{{[s]}}} \\
\hline
IFS-M & B & Photosphere & 396.6-397.1 & 0.50 & spec & 10x20 & 0.05 & 75000 & 500 & N/A & 3 \\
IFS-M & V & Chromosphere & 656.3 & 0.50 & spec & 10x20 & 0.05 & 100000 & 500 & N/A & 3 \\
EMBER & IR & Chromosphere & 1083 & 0.50 & spec & 10x20 & 0.05 & 100000 & 500 & N/A & 3 \\
IFS-M & R & Chromosphere & 854.2 & 0.50 & spec & 10x20 & 0.05 & 80000 & 500 & N/A & 3 \\
\hline
\textbf{Notes:} & \multicolumn{11}{p{17cm}|}{IFS: Extremely high-cadence in velocity is essential. Same field of view as for velocity measurements.} \\
\hline
\end{tabular}%
\end{adjustbox}%
\end{table}

\begin{table}[H]
\OPtabletitleCustom{7.1.1.2}{Waves in prominences observed in neutral and ionized spectral lines}{OP7.1.1.2table}{OP7.1.1.2}
\centering
\vspace{4mm}
\begin{adjustbox}{max width=\textwidth}
\begin{tabular}{|l|c|c|c|c|c|c|c|c|c|c|c|}
\hline
\textbf{Goal:} & \multicolumn{11}{p{17cm}|}{IFS: in spectroscopic mode with very high-cadence velocity measurements strictly simultaneously, target on prominence. 2D field of view is essential to align spectral lines and correct for differential refraction effects. In polarimetric mode: Context magnetic field in prominence. Possibly, using the same IFS spectropolarimeter for both kinds of measurements and highest cadence possible (see above), then adding up polarimetric data to increase s/n.} \\
\hline
\makecell[t]{\textbf{Instrument}} & \makecell[t]{\textbf{Optical}\\ \textbf{Arm}} & \makecell[t]{\textbf{Atm}\\ \textbf{Layer}} & \makecell[t]{\textbf{CWL}\\ \textbf{{[nm]}}} & \makecell[t]{\textbf{Spectral}\\ \textbf{Range}\\ \textbf{{[nm]}}} & \makecell[t]{\textbf{Mode}} & \makecell[t]{\textbf{FOV}\\ \textbf{{[\arcsec]}}} & \makecell[t]{\textbf{Sampling}\\ \textbf{{[\arcsec /px]}}} & \makecell[t]{\textbf{R}} & \makecell[t]{\textbf{SNR}} & \makecell[t]{\textbf{\# WL}\\ \textbf{Points}} & \makecell[t]{\textbf{Cadence}\\ \textbf{{[s]}}} \\
\hline
IFS-M & R & Chromosphere & 854.2 & 0.50 & pol & 10x20 & 0.05 & 80000 & 1000 & N/A & 20 \\
EMBER & IR & Chromosphere & 1083 & 0.50 & pol & 10x20 & 0.05 & 100000 & 1000 & N/A & 20 \\
\hline
\textbf{Notes:} & \multicolumn{11}{p{17cm}|}{IFS: Extremely high-cadence in velocity is essential. Same field of view as for velocity measurements.} \\
\hline
\end{tabular}%
\end{adjustbox}%
\end{table}

\begin{table}[H]
\OPtabletitleCustom{7.1.2}{Prominence-corona transition region (PCTR) dynamics}{OP7.1.2table}{OP7.1.2}
\centering
\vspace{4mm}
\begin{adjustbox}{max width=\textwidth}
\begin{tabular}{|l|c|c|c|c|c|c|c|c|c|c|c|}
\hline
\textbf{Goal:} & \multicolumn{11}{p{17cm}|}{IFS: in spectroscopic mode with very high-cadence velocity measurements strictly simultaneously, target on prominence. 2D field of view is essential to align spectral lines and correct for differential refraction effects. In polarimetric mode: High spatial resolution maps of filaments and prominences at different positions across the disc and at the limb. 2D field of view and polarimetry.} \\
\hline
\makecell[t]{\textbf{Instrument}} & \makecell[t]{\textbf{Optical}\\ \textbf{Arm}} & \makecell[t]{\textbf{Atm}\\ \textbf{Layer}} & \makecell[t]{\textbf{CWL}\\ \textbf{{[nm]}}} & \makecell[t]{\textbf{Spectral}\\ \textbf{Range}\\ \textbf{{[nm]}}} & \makecell[t]{\textbf{Mode}} & \makecell[t]{\textbf{FOV}\\ \textbf{{[\arcsec]}}} & \makecell[t]{\textbf{Sampling}\\ \textbf{{[\arcsec /px]}}} & \makecell[t]{\textbf{R}} & \makecell[t]{\textbf{SNR}} & \makecell[t]{\textbf{\# WL}\\ \textbf{Points}} & \makecell[t]{\textbf{Cadence}\\ \textbf{{[s]}}} \\
\hline
IFS-M & V & Chromosphere & 656.3 & 0.50 & spec & 10x20 & 0.03 & 100000 & 500 & N/A & 5 \\
EMBER & IR & Chromosphere & 1083 & 0.50 & spec & 10x20 & $^{*}$0.03 & 100000 & 500 & N/A & 5 \\
IFS-M & R & Chromosphere & 854.2 & 0.50 & spec & 10x20 & 0.03 & 80000 & 500 & N/A & 5 \\
\hline
\textbf{Notes:} & \multicolumn{11}{p{17cm}|}{IFS: High spatial resolution is the priority.} \\
\hline
\end{tabular}%
\end{adjustbox}%
\end{table}

\begin{table}[H]
\OPtabletitleCustom{7.1.3}{Draining of neutral material from prominences and filaments}{OP7.1.3table}{OP7.1.3}
\centering
\vspace{4mm}
\begin{adjustbox}{max width=\textwidth}
\begin{tabular}{|l|c|c|c|c|c|c|c|c|c|c|c|}
\hline
\textbf{Goal:} & \multicolumn{11}{p{17cm}|}{IFS in polarimetric mode: High spatial resolution maps of filaments and prominences at different positions across the disc and at the limb. 2D field of view and polarimetry.} \\
\hline
\makecell[t]{\textbf{Instrument}} & \makecell[t]{\textbf{Optical}\\ \textbf{Arm}} & \makecell[t]{\textbf{Atm}\\ \textbf{Layer}} & \makecell[t]{\textbf{CWL}\\ \textbf{{[nm]}}} & \makecell[t]{\textbf{Spectral}\\ \textbf{Range}\\ \textbf{{[nm]}}} & \makecell[t]{\textbf{Mode}} & \makecell[t]{\textbf{FOV}\\ \textbf{{[\arcsec]}}} & \makecell[t]{\textbf{Sampling}\\ \textbf{{[\arcsec /px]}}} & \makecell[t]{\textbf{R}} & \makecell[t]{\textbf{SNR}} & \makecell[t]{\textbf{\# WL}\\ \textbf{Points}} & \makecell[t]{\textbf{Cadence}\\ \textbf{{[s]}}} \\
\hline
IFS-M & V & Chromosphere & 656.3 & 0.50 & pol & 40x100 & $^{*}$0.03 & 100000 & 2000 & N/A & 200 \\
EMBER & IR & Chromosphere & 1083 & 0.50 & pol & 40x100 & $^{*}$0.03 & 100000 & 2000 & N/A & 200 \\
IFS-M & R & Chromosphere & 854.2 & 0.50 & pol & 40x100 & 0.03 & 80000 & 2000 & N/A & 200 \\
\hline
\textbf{Notes:} & \multicolumn{11}{p{17cm}|}{IFS: High spatial resolution is the priority.} \\
\hline
\end{tabular}%
\end{adjustbox}%
\end{table}

\begin{table}[H]
\OPtabletitleCustom{7.2.1}{Dynamics of spicules observed in neutral and ionized spectral lines}{OP7.2.1table}{OP7.2.1}
\centering
\vspace{4mm}
\begin{adjustbox}{max width=\textwidth}
\begin{tabular}{|l|c|c|c|c|c|c|c|c|c|c|c|}
\hline
\textbf{Goal:} & \multicolumn{11}{p{17cm}|}{IFS in polarimetric mode: Very high-cadence velocity measurements strictly simultaneously, target on spicules. 2D field of view is essential to align spectral lines and correct for differential refraction effçects. Spectropolarimetry in Ca II 854.2 and He I 1083.} \\
\hline
\makecell[t]{\textbf{Instrument}} & \makecell[t]{\textbf{Optical}\\ \textbf{Arm}} & \makecell[t]{\textbf{Atm}\\ \textbf{Layer}} & \makecell[t]{\textbf{CWL}\\ \textbf{{[nm]}}} & \makecell[t]{\textbf{Spectral}\\ \textbf{Range}\\ \textbf{{[nm]}}} & \makecell[t]{\textbf{Mode}} & \makecell[t]{\textbf{FOV}\\ \textbf{{[\arcsec]}}} & \makecell[t]{\textbf{Sampling}\\ \textbf{{[\arcsec /px]}}} & \makecell[t]{\textbf{R}} & \makecell[t]{\textbf{SNR}} & \makecell[t]{\textbf{\# WL}\\ \textbf{Points}} & \makecell[t]{\textbf{Cadence}\\ \textbf{{[s]}}} \\
\hline
IFS-M & V & Chromosphere & 656.3 & 0.50 & pol & 10x10 & 0.05 & 100000 & 1000 & N/A & 10 \\
EMBER & IR & Chromosphere & 1083 & 0.50 & pol & 10x10 & 0.05 & 100000 & 1000 & N/A & 10 \\
IFS-M & R & Chromosphere & 854.2 & 0.50 & pol & 10x10 & 0.05 & 80000 & 1000 & N/A & 10 \\
\hline
\textbf{Notes:} & \multicolumn{11}{p{17cm}|}{IFS: Extremely high-cadence in velocity is essential.} \\
\hline
\end{tabular}%
\end{adjustbox}%
\end{table}

\begin{table}[H]
\OPtabletitleCustom{7.2.2}{Alignment between the magnetic field in fibrils and disc counterparts of spicules}{OP7.2.2table}{OP7.2.2}
\centering
\vspace{4mm}
\begin{adjustbox}{max width=\textwidth}
\begin{tabular}{|l|c|c|c|c|c|c|c|c|c|c|c|}
\hline
\textbf{Goal:} & \multicolumn{11}{p{17cm}|}{IFS in polarimetric mode: Very deep 2D Spectropolarimetry in Ca II 854.2. IFS to ensure full spectral line simultaneously} \\
\hline
\makecell[t]{\textbf{Instrument}} & \makecell[t]{\textbf{Optical}\\ \textbf{Arm}} & \makecell[t]{\textbf{Atm}\\ \textbf{Layer}} & \makecell[t]{\textbf{CWL}\\ \textbf{{[nm]}}} & \makecell[t]{\textbf{Spectral}\\ \textbf{Range}\\ \textbf{{[nm]}}} & \makecell[t]{\textbf{Mode}} & \makecell[t]{\textbf{FOV}\\ \textbf{{[\arcsec]}}} & \makecell[t]{\textbf{Sampling}\\ \textbf{{[\arcsec /px]}}} & \makecell[t]{\textbf{R}} & \makecell[t]{\textbf{SNR}} & \makecell[t]{\textbf{\# WL}\\ \textbf{Points}} & \makecell[t]{\textbf{Cadence}\\ \textbf{{[s]}}} \\
\hline
IFS-M & R & Chromosphere & 854.2 & 0.50 & pol & 10x40 & 0.05 & 80000 & 5000 & N/A & 15 \\
IFS-M & V & Chromosphere & 656.3 & 0.50 & spec & 10x40 & 0.05 & 100000 & 5000 & N/A & 15 \\
\hline
\textbf{Notes:} & \multicolumn{11}{p{17cm}|}{IFS: Spectral integrity comes at the expense of small FOV. Programs using TIS/FBIs should also be tried.} \\
\hline
\end{tabular}%
\end{adjustbox}%
\end{table}

\begin{table}[H]
\OPtabletitleCustom{7.3.1}{Neutral and ionized material Evershed flow in sunspots}{OP7.3.1table}{OP7.3.1}
\centering
\vspace{4mm}
\begin{adjustbox}{max width=\textwidth}
\begin{tabular}{|l|c|c|c|c|c|c|c|c|c|c|c|}
\hline
\textbf{Goal:} & \multicolumn{11}{p{17cm}|}{IFS in spectroscopic mode: Very high-cadence velocity measurements strictly simultaneously, target on sunspot located at the limb. IFS in polarimetric mode: Sunspot magnetic field vector accompanying velocity measurements.} \\
\hline
\makecell[t]{\textbf{Instrument}} & \makecell[t]{\textbf{Optical}\\ \textbf{Arm}} & \makecell[t]{\textbf{Atm}\\ \textbf{Layer}} & \makecell[t]{\textbf{CWL}\\ \textbf{{[nm]}}} & \makecell[t]{\textbf{Spectral}\\ \textbf{Range}\\ \textbf{{[nm]}}} & \makecell[t]{\textbf{Mode}} & \makecell[t]{\textbf{FOV}\\ \textbf{{[\arcsec]}}} & \makecell[t]{\textbf{Sampling}\\ \textbf{{[\arcsec /px]}}} & \makecell[t]{\textbf{R}} & \makecell[t]{\textbf{SNR}} & \makecell[t]{\textbf{\# WL}\\ \textbf{Points}} & \makecell[t]{\textbf{Cadence}\\ \textbf{{[s]}}} \\
\hline
IFS-M & V & Photosphere & 519.7 & 0.50 & spec & 20x20 & 0.03 & 100000 & 1000 & N/A & 60 \\
IFS-M & B & Photosphere & 457.5 & 0.50 & spec & 20x20 & 0.03 & 75000 & 1000 & N/A & 60 \\
IFS-M & R & Photosphere & 868.8 & 0.50 & spec & 20x20 & 0.03 & 80000 & 1000 & N/A & 60 \\
EMBER & IR & Photosphere & 1564.8 & 1.00 & pol & 20x20 & $^{*}$0.04 & 100000 & 2000 & N/A & 60 \\
\hline
\textbf{Notes:} & \multicolumn{11}{p{17cm}|}{IFS: High-cadence and simultaneous detection of all lines is essential.} \\
\hline
\end{tabular}%
\end{adjustbox}%
\end{table}

\begin{table}[H]
\OPtabletitleCustom{7.4.1}{Observation of magnetic swirls under 2-fluid condition}{OP7.4.1table}{OP7.4.1}
\centering
\vspace{4mm}
\begin{adjustbox}{max width=\textwidth}
\begin{tabular}{|l|c|c|c|c|c|c|c|c|c|c|c|}
\hline
\textbf{Goal:} & \multicolumn{11}{p{17cm}|}{IFS in polarimetric mode: Extremely high cadence measurements of velocities in the photosphere and chromosphere. Spectropolarimetry in Fe I 630 and Ca II 854.2 at the same high cadence of velocity measurements. The polarimetric data will be averaged to increase s/n and determine magnetic field strength.} \\
\hline
\makecell[t]{\textbf{Instrument}} & \makecell[t]{\textbf{Optical}\\ \textbf{Arm}} & \makecell[t]{\textbf{Atm}\\ \textbf{Layer}} & \makecell[t]{\textbf{CWL}\\ \textbf{{[nm]}}} & \makecell[t]{\textbf{Spectral}\\ \textbf{Range}\\ \textbf{{[nm]}}} & \makecell[t]{\textbf{Mode}} & \makecell[t]{\textbf{FOV}\\ \textbf{{[\arcsec]}}} & \makecell[t]{\textbf{Sampling}\\ \textbf{{[\arcsec /px]}}} & \makecell[t]{\textbf{R}} & \makecell[t]{\textbf{SNR}} & \makecell[t]{\textbf{\# WL}\\ \textbf{Points}} & \makecell[t]{\textbf{Cadence}\\ \textbf{{[s]}}} \\
\hline
IFS-M & V & Photosphere & 617.3 & 0.50 & pol & 10x10 & 0.03 & 100000 & 700 & N/A & 2 \\
IFS-M & R & Chromosphere & 854.2 & 0.50 & pol & 10x10 & 0.03 & 80000 & 400 & N/A & 2 \\
TIS/FBI & V & Chromosphere & 656.3 & 2.00 & img & 10 & 0.03 & N/A & 400 & N/A & 2 \\
\hline
\textbf{Notes:} & \multicolumn{11}{p{17cm}|}{IFS: 2D field of view and simultaneous measurements in all three spectral lines.} \\
\hline
\end{tabular}%
\end{adjustbox}%
\end{table}

\begin{table}[H]
\OPtabletitleCustom{7.5.1}{Observations of reconnection and plasmoids in partially ionized plasma}{OP7.5.1table}{OP7.5.1}
\centering
\vspace{4mm}
\begin{adjustbox}{max width=\textwidth}
\begin{tabular}{|l|c|c|c|c|c|c|c|c|c|c|c|}
\hline
\textbf{Goal:} & \multicolumn{11}{p{17cm}|}{TIS/FBIs: Detect fast moving blobs in the chromosphere that are consistent with being plasmoids. IFS: Detect spectroscopic signatures of plasmoids/evolution of tearing mode instability in solar flares, with accompanying magneic field information} \\
\hline
\makecell[t]{\textbf{Instrument}} & \makecell[t]{\textbf{Optical}\\ \textbf{Arm}} & \makecell[t]{\textbf{Atm}\\ \textbf{Layer}} & \makecell[t]{\textbf{CWL}\\ \textbf{{[nm]}}} & \makecell[t]{\textbf{Spectral}\\ \textbf{Range}\\ \textbf{{[nm]}}} & \makecell[t]{\textbf{Mode}} & \makecell[t]{\textbf{FOV}\\ \textbf{{[\arcsec]}}} & \makecell[t]{\textbf{Sampling}\\ \textbf{{[\arcsec /px]}}} & \makecell[t]{\textbf{R}} & \makecell[t]{\textbf{SNR}} & \makecell[t]{\textbf{\# WL}\\ \textbf{Points}} & \makecell[t]{\textbf{Cadence}\\ \textbf{{[s]}}} \\
\hline
IFS-M & V & Photosphere & 617.3 & 0.50 & pol & 10x10 & 0.05 & 100000 & 1000 & N/A & 10 \\
IFS-M & R & Chromosphere & 854.2 & 0.50 & pol & 10x10 & 0.05 & 80000 & 1000 & N/A & 10 \\
EMBER & IR & Chromosphere & 1083 & 0.50 & pol & 10x10 & 0.05 & 100000 & 1000 & N/A & 10 \\
TIS/FBI & V & Chromosphere & 656.3 & 2.00 & img & 60 & 0.03 & N/A & 200 & N/A & 5 \\
TIS/FBI & B & Chromosphere & 396 & 2.00 & img & 60 & 0.01 & N/A & 100 & N/A & 2 \\
\hline
\textbf{Notes:} & \multicolumn{11}{p{17cm}|}{IFS: You ideally want context imaging and spectroscopy in the TR and the corona as well, so coordination with Solar C EUVST is highly desirable.} \\
\hline
\end{tabular}%
\end{adjustbox}%
\end{table}

\begin{table}[H]
\OPtabletitleCustom{7.5.2}{Wave damping by ion-neutral friction as a possible cause of flare chromosphere heating, wight-light flares, and sunquakes}{OP7.5.2table}{OP7.5.2}
\centering
\vspace{4mm}
\begin{adjustbox}{max width=\textwidth}
\begin{tabular}{|l|c|c|c|c|c|c|c|c|c|c|c|}
\hline
\textbf{Goal:} & \multicolumn{11}{p{17cm}|}{TIS/FBIs: Detect flare signatures at photospheric and chromospheric heights. Context information on flare emission (chromospheric heating and evaporation, WL emission, sunquakes signatures). IFS: Measure oscillations of the magnetic field to detect the presence and the damping of Alfv\'en waves propagating downwards during a flare. Detect velocities and abundances of ionized and neutral species in the same region.} \\
\hline
\makecell[t]{\textbf{Instrument}} & \makecell[t]{\textbf{Optical}\\ \textbf{Arm}} & \makecell[t]{\textbf{Atm}\\ \textbf{Layer}} & \makecell[t]{\textbf{CWL}\\ \textbf{{[nm]}}} & \makecell[t]{\textbf{Spectral}\\ \textbf{Range}\\ \textbf{{[nm]}}} & \makecell[t]{\textbf{Mode}} & \makecell[t]{\textbf{FOV}\\ \textbf{{[\arcsec]}}} & \makecell[t]{\textbf{Sampling}\\ \textbf{{[\arcsec /px]}}} & \makecell[t]{\textbf{R}} & \makecell[t]{\textbf{SNR}} & \makecell[t]{\textbf{\# WL}\\ \textbf{Points}} & \makecell[t]{\textbf{Cadence}\\ \textbf{{[s]}}} \\
\hline
IFS-M & V & Photosphere & $^{+}$617.3 & 0.50 & pol & 10x10 & 0.03 & 100000 & 1000 & N/A & 2 \\
IFS-M & R & Chromosphere & $^{+}$854.2 & 0.50 & pol & 10x10 & 0.03 & 80000 & 1000 & N/A & 2 \\
EMBER & IR & Chromosphere & $^{+}$1083 & 0.50 & pol & 10x10 & $^{*}$0.03 & 100000 & 1000 & N/A & 2 \\
TIS/FBI & V & Chromosphere & $^{+}$656.3 & 2.00 & img & 60 & $^{*}$0.03 & N/A & 200 & N/A & 2 \\
TIS/FBI & B & Chromosphere & $^{+}$396 & 2.00 & img & 60 & $^{*}$0.01 & N/A & 100 & N/A & 2 \\
\hline
\textbf{Notes:} & \multicolumn{11}{p{17cm}|}{N/A} \\
\hline
\end{tabular}%
\end{adjustbox}%
\end{table}

\begin{table}[H]
\OPtabletitleCustom{8.1.1}{Spatial fluctuations of scattering polarization in \ion{Sr}{i} 4607\,\AA}{OP8.1.1table}{OP8.1.1}
\centering
\vspace{4mm}
\begin{adjustbox}{max width=\textwidth}
\begin{tabular}{|l|c|c|c|c|c|c|c|c|c|c|c|}
\hline
\textbf{Goal:} & \multicolumn{11}{p{17cm}|}{TIS/FBIs: Detection of spatial fluctuations at sub-granular scales of scattering polarization in Sr i 460.7 nm.} \\
\hline
\makecell[t]{\textbf{Instrument}} & \makecell[t]{\textbf{Optical}\\ \textbf{Arm}} & \makecell[t]{\textbf{Atm}\\ \textbf{Layer}} & \makecell[t]{\textbf{CWL}\\ \textbf{{[nm]}}} & \makecell[t]{\textbf{Spectral}\\ \textbf{Range}\\ \textbf{{[nm]}}} & \makecell[t]{\textbf{Mode}} & \makecell[t]{\textbf{FOV}\\ \textbf{{[\arcsec]}}} & \makecell[t]{\textbf{Sampling}\\ \textbf{{[\arcsec /px]}}} & \makecell[t]{\textbf{R}} & \makecell[t]{\textbf{SNR}} & \makecell[t]{\textbf{\# WL}\\ \textbf{Points}} & \makecell[t]{\textbf{Cadence}\\ \textbf{{[s]}}} \\
\hline
IFS-M & B & Photosphere & 460.75 & 0.10 & pol & 2x2 & 0.05 & 75000 & 3000 & N/A & 5 \\
%\hline
%\textbf{Notes:} & \multicolumn{11}{p{17cm}|}{TIS/FBIs: Stable observations with AO at different limb distances required, between  u = 1.0 and  u = 0.1. The above values for spatial sampling and cadence refer to disk center. For other limb distances, the values have to be scaled according to limb darkening. For the approximate computation of spatial sampling and cadence as in this table, it is assumed that the FWHM of the NB transmission profile is equal to its equivalent width. (2) The values correspond to the optimum spatial sampling and cadence (cf. Appendix A.1), derived from the expected photon flux at line center (cf. EST photon budget), a solar evolution speed of 7 km/s (OP 8.1.1, 8.1.2, 8.1.3), 25 km/s (OP 8.2.1, 8.3.1), and 50 km/s (OP 8.2.2, 8.2.3), respectively, and the required spectral resolution and polarimetric sensitivity. An ideal balanced polarimetric efficiency e = sqrt(3) is assumed for Stokes Q, U, and V. The values have been computed using the photoncount code by J. Leenaarts, which also implements the equations of Appendix A.1.} \\
\hline
\end{tabular}%
\end{adjustbox}%
\end{table}

\begin{table}[H]
\OPtabletitleCustom{8.1.2}{Spatial fluctuations of scattering polarization in C$_2$
molecular lines at 5140\,{\AA}}{OP8.1.2table}{OP8.1.2}
\centering
\vspace{4mm}
\begin{adjustbox}{max width=\textwidth}
\begin{tabular}{|l|c|c|c|c|c|c|c|c|c|c|c|}
\hline
\textbf{Goal:} & \multicolumn{11}{p{17cm}|}{IFS: Detection of spatial fluctuations at granular scales of scattering polarization in C2 molecular lines at 514.0 nm.} \\
\hline
\makecell[t]{\textbf{Instrument}} & \makecell[t]{\textbf{Optical}\\ \textbf{Arm}} & \makecell[t]{\textbf{Atm}\\ \textbf{Layer}} & \makecell[t]{\textbf{CWL}\\ \textbf{{[nm]}}} & \makecell[t]{\textbf{Spectral}\\ \textbf{Range}\\ \textbf{{[nm]}}} & \makecell[t]{\textbf{Mode}} & \makecell[t]{\textbf{FOV}\\ \textbf{{[\arcsec]}}} & \makecell[t]{\textbf{Sampling}\\ \textbf{{[\arcsec /px]}}} & \makecell[t]{\textbf{R}} & \makecell[t]{\textbf{SNR}} & \makecell[t]{\textbf{\# WL}\\ \textbf{Points}} & \makecell[t]{\textbf{Cadence}\\ \textbf{{[s]}}} \\
\hline
IFS-M & V & Photosphere & 514 & 0.30 & pol & 3x3 & 0.19 & 100000 & 30000 & N/A & 19 \\
\hline
\textbf{Notes:} & \multicolumn{11}{p{17cm}|}{IFS: Stable observations with AO at different limb distances required, between  u = 1.0 and  u = 0.1. The above values for spatial sampling and cadence refer to disk center. For other limb distances, the values have to be scaled according to limb darkening. For the approximate computation of spatial sampling and cadence as in this table, it is assumed that the FWHM of the NB transmission profile is equal to its equivalent width. (2) The values correspond to the optimum spatial sampling and cadence (cf. Appendix A.1), derived from the expected photon flux at line center (cf. EST photon budget), a solar evolution speed of 7 km/s (OP 8.1.1, 8.1.2, 8.1.3), 25 km/s (OP 8.2.1, 8.3.1), and 50 km/s (OP 8.2.2, 8.2.3), respectively, and the required spectral resolution and polarimetric sensitivity. An ideal balanced polarimetric efficiency e = sqrt(3) is assumed for Stokes Q, U, and V. The values have been computed using the photoncount code by J. Leenaarts, which also implements the equations of Appendix A.1.} \\
\hline
\end{tabular}%
\end{adjustbox}%
\end{table}

\begin{table}[H]
\OPtabletitleCustom{8.1.3}{Spatial fluctuations of scattering polarization in the \ion{Ti}{i} multiplet around 4530\,\AA }{OP8.1.3table}{OP8.1.3}
\centering
\vspace{4mm}
\begin{adjustbox}{max width=\textwidth}
\begin{tabular}{|l|c|c|c|c|c|c|c|c|c|c|c|}
\hline
\textbf{Goal:} & \multicolumn{11}{p{17cm}|}{IFS: Detection of spatial fluctuations at granular scales of scattering polarization in the Tmultiplet around 453.0 nm.} \\
\hline
\makecell[t]{\textbf{Instrument}} & \makecell[t]{\textbf{Optical}\\ \textbf{Arm}} & \makecell[t]{\textbf{Atm}\\ \textbf{Layer}} & \makecell[t]{\textbf{CWL}\\ \textbf{{[nm]}}} & \makecell[t]{\textbf{Spectral}\\ \textbf{Range}\\ \textbf{{[nm]}}} & \makecell[t]{\textbf{Mode}} & \makecell[t]{\textbf{FOV}\\ \textbf{{[\arcsec]}}} & \makecell[t]{\textbf{Sampling}\\ \textbf{{[\arcsec /px]}}} & \makecell[t]{\textbf{R}} & \makecell[t]{\textbf{SNR}} & \makecell[t]{\textbf{\# WL}\\ \textbf{Points}} & \makecell[t]{\textbf{Cadence}\\ \textbf{{[s]}}} \\
\hline
IFS-M & B & Photosphere & 453 & 1.00 & pol & 2x2 & 0.37 & 75000 & 30000 & N/A & 38 \\
\hline
\textbf{Notes:} & \multicolumn{11}{p{17cm}|}{Same note as 8.1.2} \\
\hline
\end{tabular}%
\end{adjustbox}%
\end{table}

\begin{table}[H]
\OPtabletitleCustom{8.2.1}{The Ca~{\sc i} 4227\,\AA\ resonance line }{OP8.2.1table}{OP8.2.1}
\centering
\vspace{4mm}
\begin{adjustbox}{max width=\textwidth}
\begin{tabular}{|l|c|c|c|c|c|c|c|c|c|c|c|}
\hline
\textbf{Goal:} & \multicolumn{11}{p{17cm}|}{IFS: Detecting spatial and temporal variations of the scattering polarization signal of Ca I 422.7 nm from the core to the wings.} \\
\hline
\makecell[t]{\textbf{Instrument}} & \makecell[t]{\textbf{Optical}\\ \textbf{Arm}} & \makecell[t]{\textbf{Atm}\\ \textbf{Layer}} & \makecell[t]{\textbf{CWL}\\ \textbf{{[nm]}}} & \makecell[t]{\textbf{Spectral}\\ \textbf{Range}\\ \textbf{{[nm]}}} & \makecell[t]{\textbf{Mode}} & \makecell[t]{\textbf{FOV}\\ \textbf{{[\arcsec]}}} & \makecell[t]{\textbf{Sampling}\\ \textbf{{[\arcsec /px]}}} & \makecell[t]{\textbf{R}} & \makecell[t]{\textbf{SNR}} & \makecell[t]{\textbf{\# WL}\\ \textbf{Points}} & \makecell[t]{\textbf{Cadence}\\ \textbf{{[s]}}} \\
\hline
IFS-M & B & Photosphere & 422.7 & 0.60 & pol & 2x2 & 0.23 & 75000 & 3000 & N/A & 7 \\
\hline
\textbf{Notes:} & \multicolumn{11}{p{17cm}|}{Same note as 8.1.2} \\
\hline
\end{tabular}%
\end{adjustbox}%
\end{table}

\begin{table}[H]
\OPtabletitleCustom{8.2.2}{The Ca~{\sc ii} H \& K resonance lines}{OP8.2.2table}{OP8.2.2}
\centering
\vspace{4mm}
\begin{adjustbox}{max width=\textwidth}
\begin{tabular}{|l|c|c|c|c|c|c|c|c|c|c|c|}
\hline
\textbf{Goal:} & \multicolumn{11}{p{17cm}|}{IFS: Detecting spatial fluctuations of scattering polarization in the line core and wings of the Ca ii K line.} \\
\hline
\makecell[t]{\textbf{Instrument}} & \makecell[t]{\textbf{Optical}\\ \textbf{Arm}} & \makecell[t]{\textbf{Atm}\\ \textbf{Layer}} & \makecell[t]{\textbf{CWL}\\ \textbf{{[nm]}}} & \makecell[t]{\textbf{Spectral}\\ \textbf{Range}\\ \textbf{{[nm]}}} & \makecell[t]{\textbf{Mode}} & \makecell[t]{\textbf{FOV}\\ \textbf{{[\arcsec]}}} & \makecell[t]{\textbf{Sampling}\\ \textbf{{[\arcsec /px]}}} & \makecell[t]{\textbf{R}} & \makecell[t]{\textbf{SNR}} & \makecell[t]{\textbf{\# WL}\\ \textbf{Points}} & \makecell[t]{\textbf{Cadence}\\ \textbf{{[s]}}} \\
\hline
IFS-M & B & Chromosphere & 393.37 & 1.00 & pol & 2x2 & 0.29 & 75000 & 3000 & N/A & 5 \\
\hline
\textbf{Notes:} & \multicolumn{11}{p{17cm}|}{Same note as 8.1.2} \\
\hline
\end{tabular}%
\end{adjustbox}%
\end{table}

\begin{table}[H]
\OPtabletitleCustom{8.2.3}{The Ca~{\sc ii} IR triplet}{OP8.2.3table}{OP8.2.3}
\centering
\vspace{4mm}
\begin{adjustbox}{max width=\textwidth}
\begin{tabular}{|l|c|c|c|c|c|c|c|c|c|c|c|}
\hline
\textbf{Goal:} & \multicolumn{11}{p{17cm}|}{IFS: Simultaneous observations of the 3 spectral lines of the Ca ii IR triplet.} \\
\hline
\makecell[t]{\textbf{Instrument}} & \makecell[t]{\textbf{Optical}\\ \textbf{Arm}} & \makecell[t]{\textbf{Atm}\\ \textbf{Layer}} & \makecell[t]{\textbf{CWL}\\ \textbf{{[nm]}}} & \makecell[t]{\textbf{Spectral}\\ \textbf{Range}\\ \textbf{{[nm]}}} & \makecell[t]{\textbf{Mode}} & \makecell[t]{\textbf{FOV}\\ \textbf{{[\arcsec]}}} & \makecell[t]{\textbf{Sampling}\\ \textbf{{[\arcsec /px]}}} & \makecell[t]{\textbf{R}} & \makecell[t]{\textbf{SNR}} & \makecell[t]{\textbf{\# WL}\\ \textbf{Points}} & \makecell[t]{\textbf{Cadence}\\ \textbf{{[s]}}} \\
\hline
IFS-M & R & Chromosphere & 854 & 0.50 & pol & 4x4 & 0.22 & 80000 & 10000 & N/A & 4 \\
\hline
\textbf{Notes:} & \multicolumn{11}{p{17cm}|}{The spectral FOV refers to each line of the Ca ii IR triplet separately and shall be centered on the respective line. Stable observations with AO at different limb distances required, between   u = 1.0 and and  u = 0.1. The above values for spatial sampling and cadence refer to disk center. For other limb distances, the values have to be scaled according to limb darkening.} \\
\hline
\end{tabular}%
\end{adjustbox}%
\end{table}

\begin{table}[H]
\OPtabletitleCustom{8.3.1}{The physics and diagnostic potential of the \ion{Na}{i} D$_1$ and D$_2$ lines}{OP8.3.1table}{OP8.3.1}
\centering
\vspace{4mm}
\begin{adjustbox}{max width=\textwidth}
\begin{tabular}{|l|c|c|c|c|c|c|c|c|c|c|c|}
\hline
\textbf{Goal:} & \multicolumn{11}{p{17cm}|}{IFS: Investigate spatial and temporal variability of the “enigmatic” scattering polarization signal of the Na i D1 line.} \\
\hline
\makecell[t]{\textbf{Instrument}} & \makecell[t]{\textbf{Optical}\\ \textbf{Arm}} & \makecell[t]{\textbf{Atm}\\ \textbf{Layer}} & \makecell[t]{\textbf{CWL}\\ \textbf{{[nm]}}} & \makecell[t]{\textbf{Spectral}\\ \textbf{Range}\\ \textbf{{[nm]}}} & \makecell[t]{\textbf{Mode}} & \makecell[t]{\textbf{FOV}\\ \textbf{{[\arcsec]}}} & \makecell[t]{\textbf{Sampling}\\ \textbf{{[\arcsec /px]}}} & \makecell[t]{\textbf{R}} & \makecell[t]{\textbf{SNR}} & \makecell[t]{\textbf{\# WL}\\ \textbf{Points}} & \makecell[t]{\textbf{Cadence}\\ \textbf{{[s]}}} \\
\hline
IFS-M & V & Upper Photosphere & 589.6 & 1.00 & pol & 3x3 & 0.80 & 100000 & 30000 & N/A & 22 \\
\hline
\textbf{Notes:} & \multicolumn{11}{p{17cm}|}{Stable observations with AO at different limb distances required, between  u = 1.0 and  u = 0.1. The above values for spatial sampling and cadence refer to disk center. For other limb distances, the values have to be scaled according to limb darkening.} \\
\hline
\end{tabular}%
\end{adjustbox}%
\end{table}

 %\newpage
%==================================

%\setcounter{section}{0}
\part{Discussion and Requirements} 

\section{What are the particular strengths of the EST design?} %\phantomsection 

In the years between the Preliminary Design Review and the design that is presented at the Preliminary Design Review in 2025 various aspects have evolved. In the following subsections we summarise the developments. The main drivers for all developments are to achieve a telescope with instrumentation which maximises image quality, polarimetric accuracy, photon flux, and the capability to perform multi-wavelength observations.

\subsection{EST preliminary design study 2008 - 2011}

The conceptual design that was developed during the conceptual design study 2008 - 2011 defines EST to be a 4\,m telescope. To assure highest polarimetric accuracy and compactness, it is carried out as an on-axis telescope mounted on an altitude-azimuth frame, with polarimetrically compensated transfer optics, i.e.,  which is free of introducing net polarisation to the light beam. The rotating transfer optics also compensates for image rotation such that the Coud{\'e} lab does not need to be rotated. This allows for a large room for the suite of EST science instruments. 

\paragraph{On-axis versus off-axis:} Both off-axis and on-axis designs were studied, and both designs have advantages and disadvantages. The off-axis design has the advantage of a clear aperture, such that spiders and obscurations can be avoided. This avoids stray-light and results in an ideal point-spread function. Spiders and a central obscuration are a necessity for on-axis telescopes. They cause the point-spread function of the telescope to have extended `wings'. While this can be corrected in terms of spatial resolution, it  decreases the signal-to-noise ratio. It should be said that this negative effect is small compared to other aberration effects that are typically introduced by the train of telescope mirrors. The spider of an on-axis telescope rotates on the wavefront sensor and the measurement of the distorted wave front becomes complicated.

Off-axis telescopes at a given aperture are larger and heavier than on-axis telescopes. On-axis telescopes being more compact and having less weight can be integrated into higher towers, i.e., further away from ground-layer turbulence dominating the seeing, hereby improving the image quality. In contrast to off-axis telescopes, on-axis telescopes can be constructed such that the optical beam is polarimetrically compensated, thereby facilitating the calibration of polarimetric measurements. Another advantage of an on-axis telescope is that -- at the same aperture -- their cost is considerably smaller than for an off-axis telescope. Based on the these reasons, the choice of an on-axis system was preferred during the conceptual design study.

\paragraph{Open versus closed dome:}  With the open design of an on-axis telescope, EST will have less weight than a telescope with a dome. EST will therefore allow for a high tower which elevates the telescope as high as possible above ground-layer seeing. As this is considered to be a major advantage, a closed dome should be avoided if possible.

\paragraph{Post focus instrumentation and light distribution:} In the EST conceptual design from 2011, the post-focus instrument suite consists of 
\begin{itemize}
\item 5 narrow-band imagers (NBIs) with wavelength ranges (WRs) in nm between 390 -- 500 (WR1), 500 -- 620 (WR2), 620 -- 860(WR3), 800 -- 1100 (WR4), 1500 -- 1800 (WR5). 
\item 4 spectrographs (SPs) with wavelength ranges in nm between 390 -- 560, 560 -- 1100, 700 -- 1600, 1000 -- 2300. These SPs were thought to be configured as long-slit spectrographs, as multi-slit multi-wavelength SPs with an integral field unit (IFU), or as double-pass imaging spectrographs (TUNIS or MSDP type of instruments).
\item 3 Broad Band Imagers (BBIs), two with 380 -- 500, and one with 600 - 900. They are associated to the NBIs of the corresponding wavelength, selecting the desired wavelength range with a suitable filter. Broad Band Imager provide diffraction limited context images with large FOVs at high cadence.
\end{itemize}

The light distribution system foresees exchangeable dichroic and partial beam splitters. This allows either simultaneous observations with all instruments by sharing light of a specific wavelength or observations where a subset of instruments is fed with all light of a specific wavelength.  

\subsection{Design developments  2011 -- 2019}

\paragraph{Instrumentation:} Since the preliminary design phase, some 10 years ago, significant developments on post-focus instrumentation have happened. Prototypes of Integral-Field-Units (IFUs) with solar spectro-polarimetric measurements have successfully been developed and tested. Prototypes of micro-lense arrays and image slicer have proven to be most promising. With such IFUs, the 2D spatial information and the spectral dimension are recorded simultaneously, while traditional spectrographs and narrow band imagers either have long slits that scan the solar surface or take very narrow-band images that sequentially scan through wavelength.
This has fundamental consequences for those IFU systems that record coherent 2D images. {\it Post-factum} image restoration techniques can then be applied to the IFU data. Hereby, highest, diffraction-limited, spatial resolution can be achieved with full spectral integrity.  
%Hitherto, spectral integrity was only guaranteed by long-slit spectrographs, which could not profit from image restoration methods. This also means that IFU systems that do not record coherent images can not take advantage of image restoration techniques and may not be considered as an instrument of choice for EST.

Spatially scanned maps with such long-slit spectrographs ask for a non-rotating solar image, since image rotation squeezes and stretches the solar scene. This effectively shrinks the field-of-view. With IFUs, the requirement of de-rotating the solar image is less stringent, since the larger fields-of-view are build up by mosaics. And since each tile of the mosaic has short exposure times, the image rotation is negligible, also for image restoration techniques. 

\paragraph{Adaptive secondary mirror (ASM):} Another crucial technical development during the last decade is that large-diameter deformable mirrors become available. If such a large-diameter deformable mirror could be placed as a secondary mirror, it appears feasible to integrate all AO and MCAO mirrors within the first seven mirrors of the telescope and its main axes. The transfer optics of the preliminary design that has seven additional mirrors becomes obsolete. Hence, with a deformable secondary mirror, the total number of mirrors could be reduced from fourteen to seven. This will substantially improve the optical quality and the photon flux in the science focus will increase by a factor of about 2 depending on wavelength\footnote{A removal of 7 Ag-mirrors ($R@400{\rm nm} \!\approx\! 0.85, R@1000{\rm nm} \!\approx\! 0.95$) increases the photon flux by a factor of 3 at 400nm and a factor of 1.4 at 1000nm.}. For these reasons, scientists would prefer this solution, but it remains to be seen whether the involved technical risks are acceptable. 

While the removal of the (rotating) transfer optics is of great advantage in terms of photon flux and image quality, it reintroduces image rotation. There are three options how to deal with this: (1) a rotating Coud{\'e} room, (2) a separate image de-rotator, or (3) accept image rotation.  A rotating Coud{\'e} room is the straight-forward solution, but costly and may reduce the space for the instrument platform. Introducing a separate polarimetrically compensated image de-rotator with 5 additional mirrors appears to be a bad solution with respect to optical quality and photon flux. As mentioned above with instrumentation based on IFUs and NBIs,  the problems of image rotation are much less severe and would be acceptable. Image rotation is less acceptable for long-slit spectrographs. They could be equipped with an internal image de-rotator.

Note on image rotation: Depending on the direction of the beam along the elevation axis, the image rotation rate differs between morning and afternoon. Since seeing conditions at Observatorio del Teide (OT, Tenerife) and the Observatorio del Roque de los Muchachos (ORM, La Palma) are typically better in the morning, the telescope design should allow for the smaller image rotation rate in the morning.

\begin{figure}[t]
\includegraphics[width=\textwidth]{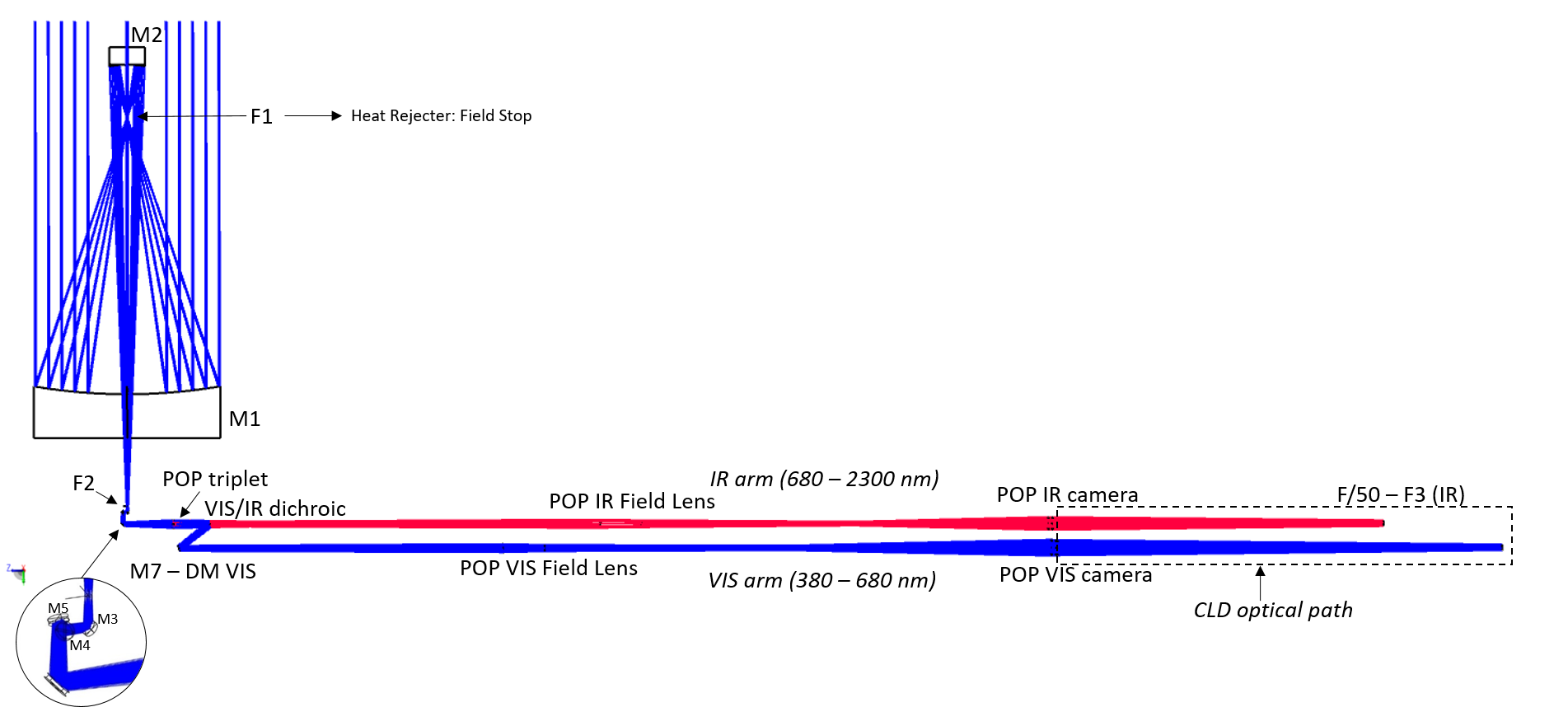}
\caption{This figures sketches the optical layout of the telescope (M1 to M2), the Transfer Optics and Calibration Assembly (TOCA) (M3 to M6), and the Pier Optical Path (POP) bringing the light to the Coudé focus (F3) and the Coudé Light Distribution (CLD).\label{fig_layout}}
\end{figure}

\subsection{Design developments 2019 -- 2025}\label{subsec_2025}

After the submission of the second edition of the SRD in December 2019, the Project Office, a review panel of telescope and instrument experts, and the Science Advisory Group had frequent meetings to transform the science requirements into technical requirements. Those technical requirements led to the design of EST that is presented at the Preliminary Design Reviews in 2024 and 2025. In particular, the PDR design of EST foresees the secondary mirror (M2) to be a deformable mirror, and to use the TOCA mirrors (M3 to M6) as  the MCAO deformable mirrors. This design needs six mirrors and one lense to bring the light to F3 and thereby reduces the numbers of optical surfaces to a minimum. The optical layout is sketched in Figure \ref{fig_layout}.

In the phase between 2019 and 2025 the concept for the post-focus instrumentation developed significantly. These developments were driven and done in close collaborations with the SAG. At the PDR, the suite of scientific instrumentation consists of 
\begin{itemize}
\item three TIS/FBI instruments for three wavelength bands: 680-1000 nm, 500-680 nm, \& 380-500 nm. TIS stands for Tunable Imaging Spectropolarimeters that consist of a narrow-band channel based on Fabry Perot Interferometers, and a broad band channel that can be used as a Fixed Band Imager (FBI).
\item three IFS-M: Integral Field Spectropolarimeters based on micro-lense arrays (-M). Three such instruments are planned for the three wavelength bands: 380-500 nm, 500-680 nm, 680-1000 nm.
\item one IFS-S: Integral Field Spectropolarimeter based on an image slicer (-S). Such an instrument is planned for the near IR for the spectral window around 1083 nm and 1565 nm.
\end{itemize}

\begin{figure}[t]
\includegraphics[width=\textwidth]{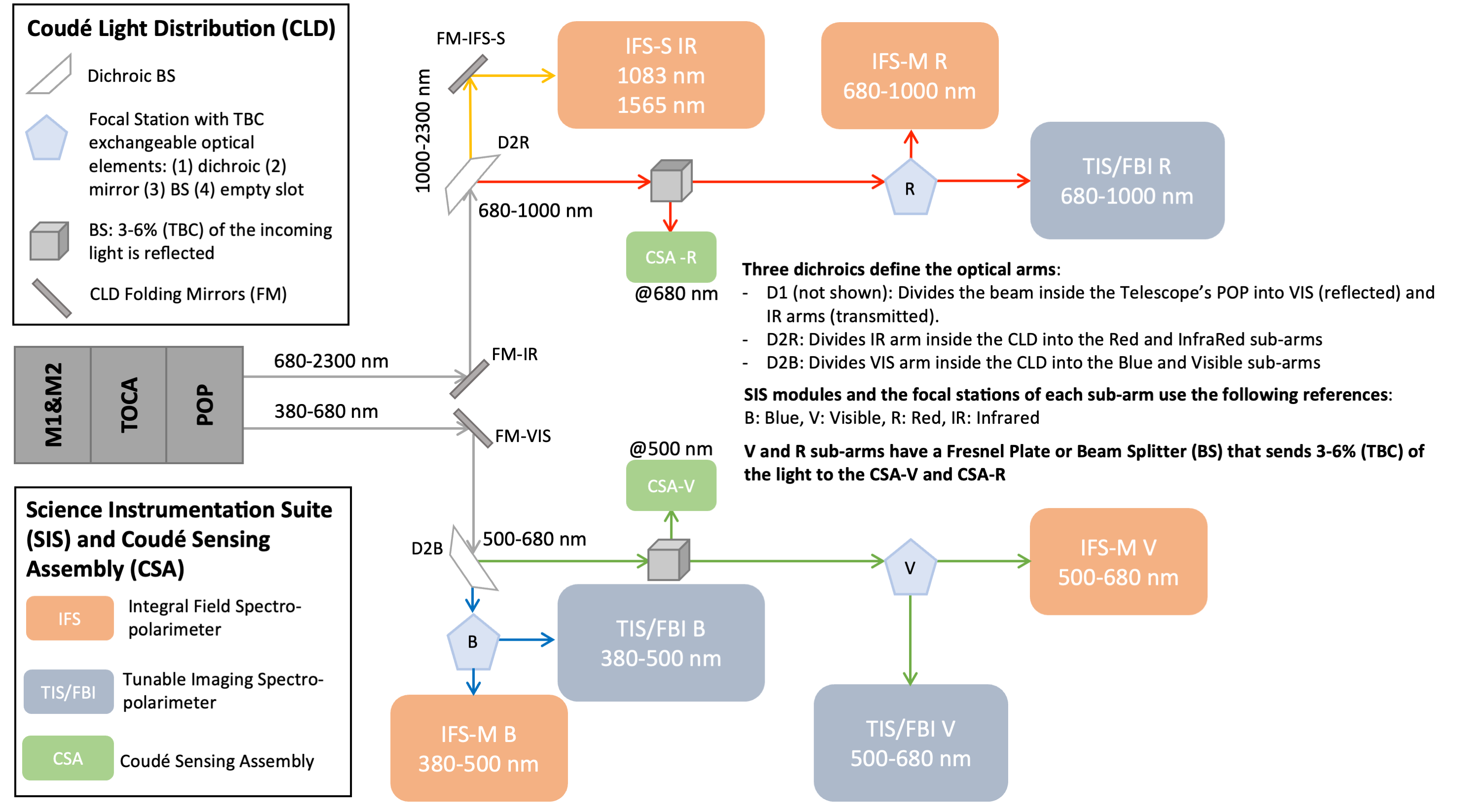}
\caption{Coudé Light Distribution (CLD)\label{fig_CLD} showing how the Science Instrumentation Suite (SIS) is fed with light from the telescope.}
\end{figure}

The Coudé Light Distribution and the Science Instrumentation Suite is sketched in Figure \ref{fig_CLD}. As consequence of this change of concept for the instrumentation the observing programmes were revisited. It was confirmed that the goals of the Observing Programmes (OPs) of the SRD 2019 document can be achieved with the new suite of instruments, and the OP tables were adjusted accordingly. Hence, in this document, we list the updated OP tables in Part II, such that the Coudé Light Distribution is compliant with the SRD OP tables.

\section{Discussion} %\phantomsection

The overall goal of EST is to understand the small-scale processes in the solar atmosphere. The telescope is designed to be a `microscope' providing with best imaging and polarimetric quality and with the highest possible photon throughput. 
To achieve highest imaging quality, the telescope design contains Multi-Conjugate Adaptive Optics (MCAO). MCAO and image reconstruction techniques allow for diffraction limited imaging quality over the entire FOV, including FOVs at and off the limb. An optimal polarimetric performance is assured by assembling the mirrors such that they are polarimetrically compensated. The suite of post-focus instruments will change during its lifetime.

\subsection{Discussion on photon flux} 

Science cases are presented and discussed in Part II (starting on page \pageref{partII}). Modelling the photon flux\footnote{{\it photoncount} app by J. Leenaarts.}, observing programmes were devised for each top-level science case. While doing this, it became obvious that compromises need to be made between spatial resolution, spectral resolution, and SNR. This is based on two fundamental properties: (i) The number of photons collected by pixels that sample the diffraction limit of a given aperture is independent of the aperture. I.e. the advantage of a larger aperture can either be used to increase the spatial resolution {\bf or} to increase the number of photons (SNR). (ii) The solar scene evolves with time. This severely limits the exposure times that are needed to freeze a snapshot of the solar evolution. Hence, the top-level goal for the final design of the telescope is two-folded: 
\begin{enumerate}\item[(1)] EST should be capable to reach the highest possible image quality and spatial resolution, i.e. diffraction limit. With the option to sacrifice spatial resolution in favour of collecting more photons. \item[(2)] The final design must be optimised for the highest possible photon flux, with the premises of securing polarimetric accuracy and sensitivity.\end{enumerate}

\subsection{Discussion on field-of-view (FOV)}

EST is designed to study the small-scale structure of the solar atmosphere. The science cases presented in Part II request to resolve sub-structures from some 0.2 arcsec down to the diffraction limit (0.026 arcsec at 500 nm with a 4 m aperture). While scientists desire to have the FOV as large as possible, the reasons for limitations are manifold: 
\begin{enumerate}
\item[(1)] The size of optical surfaces in the telescope light beam increases with increasing FOV, leading to a substantial increase of costs. This cost increase is particularly high for NBIs. Large FOVs at highest spatial resolution are expensive.

\item[(2)] The performance of the MCAO system degrades with increasing the wavefront-corrected FOV. The expectation is that MCAO will correct a FOV of less than 60 arcsec in diameter.
\end{enumerate}
 
The FOVs of IFUs is expected to be limited to some 10 by 10 arcsec$^2$. These IFUs will be complemented by Narrow Band Imagers (NBIs). For the science cases in part II, a FOV of some 40 by 40 arcsec$^2$ is sufficiently large. The science cases require simultaneous observation in three or more different wavelength regions. I.e., rather than requesting a larger FOV, the science cases require a multitude of NBIs and IFUs that operate in different wavelengths simultaneously. If compromises need to be done, priority will be given to simultaneous multi-wavelength observation over larger FOVs.

If individual FOVs do not need to be larger than 40 by 40 arcsec$^2$, one could, in principle, relax the original requirement for the telescope FOV of 120 by 120 arcsec$^2$. Such large FOVs are traditionally required for context imaging. Instead of designing EST with such a large FOV, one could consider another telescope with a FOV of some 400 arcsec in diameter that delivers spectropolarimetric measurements with a spatial resolution of about 0.5 arcsec. This implies that such a telescope should be equipped with an adaptive optics system. This could also be an extended version of the auxiliary full-disk telescope (AFDT) that was foreseen in the preliminary EST design (cf. EST Conceptual Design Study, 2011).

\subsection{Discussion on pointing requirements}

For the science cases, the absolute pointing accuracy of the telescope on the sky is not relevant. Only the absolute pointing relative to the solar disk is relevant for the solar observer. In the World Coordinate System (Thomson 2005), the position on the solar disk can be determined by the Helioprojective-Cartesian coordinate system. These coordinates are given relative to the center of the solar disk and the separation between two points is given by the corresponding angle separation, i.e. in arcsec. 

Absolute positioning is needed for simultaneous observations with other telescopes. Taking into account that the FOVs of IFUs are expected to be smaller than 10 by 10 arcsec$^2$, an absolute pointing on the solar disk should be on the order of 1 arcsec to guarantee the needed overlap with observations from other telescopes. 

\paragraph{Mosaic with NBIs:} For obtaining a large FOV mosaic with the NBIs, the tiling should be done by re-pointing the telescope. For this, the relative pointing is requested to have an accuracy of 1 arcsec. The re-pointing of the telescope including a closed-loop in the adaptive optics system at the new position should be performed as fast as possible. This will optimise the duty cycle. A time lapse of 2 seconds or less would allow for valuable time cadence of larger FOVs.

\paragraph{Mosaic with IFUs:} In some science cases it is necessary to raster a larger FOV with IFUs. Since their FOV is smaller than those of NBIs, the positioning can not be done with the telescope, but needs an internal mechanism. To avoid redundant overlapping, the pointing accuracy of individual IFU tiles relative to each other should be in the order of 0.1 arcsec. For valuable time series of IFU mosaics the re-pointing time lapse should be as fast as possible. A time lapse of 0.1 sec or less is desirable to be able to acquire time series of mosaics at a reasonable cadence.

\subsection{Discussion on wavelength range}

The telescope should be optimised to cover the spectral range from  \CaIIK 393.4 to \FeI 1564.8 nm. Observations up to 2300 nm are desired. The choice of coatings should favour the transmission in the longer wavelengths between 800 and 1083 nm. This wavelength regime is considered to be more important to measure the magnetic coupling in the chromosphere. The optical train should be optimised to assure highest photon flux at the Ca II IR 854.2 nm as this line is crucial for the most important science cases, and since the time scales of the associated chromospheric signals are shorter than in the photosphere. Requirement for wavelength coverage beyond 1600 nm up to 2300 nm should not compromise performance for wavelengths smaller than 1600\,nm.

\subsection{Discussion on instrumentation and light distribution (prior to 2019)}

The main science driver for EST consists in understanding the magnetic coupling of the solar atmosphere. This is reflected in many of the science cases described in Part II. In the corresponding observing programmes, measurements probing many different layers of the solar atmosphere are required to be taken co-temporally and co-spatially in many different wavelengths, as e.g., \CaIIK 393.4 nm, \MgI 517.3 nm, \FeI 525.0 nm, \FeI 557.6 nm, \HeI 587.6 nm, \FeI 617.3 nm, \FeI 709.0 nm, \Halpha  656.3 nm,  \CaII IR triplet 849.8 nm, 854.2 nm and 866.2 nm,  \SiI 1082.7 nm, \HeI 1083.0 nm, \FeI 1564.8 nm, etc. For the observing programmes in Part II, typically three or more of these lines are required to probe the photosphere and chromosphere. Here again, time and photons are limited, such that it is desirable that multiple instruments operate at different wavelengths simultaneously. Ideally, each instrument would receive all available photons of a specific wavelength.

Efficient NBIs have low spectral resolution (between 30\,000 and 100\,000), meaning that science cases which request high spectral resolution (around 200\,000) need to be served with IFUs or long-slit spectrographs. 

The post-focus instrumentation and light distribution concept of the preliminary design from 2011 is powerful and sophisticated. With this concept, simultaneous observations as desired in the previous paragraph are possible. Yet, the old concept foresees to share light between SPs and NBIs. And IFUs are foreseen to replace classical long-slit spectrographs. As a consequence of the lesson learned during years of developments, one should also consider to replace some of the NBIs by IFUs, or have both types of instruments for some wavelength ranges, depending on whether large FOV or short time cadence and high spectral resolution is needed for the science case. Different types of instruments for the same wavelength may be necessary to achieve the goals of different science cases, but simultaneous observation at a specific wavelength with different types of instruments, i.e. sharing the light at a specific wavelength, is not needed.

In this respect, the consensus of the present EST Science Advisory Group differs from the Preliminary Design Study: the light distribution should be such that each instrument receives all photons of a particular wavelength. Many of the top-level science goals need high SNRs to observe features that have short evolutions time scales. Hence, the exposure time and the observing cadence must be short. Although IFUs have smaller FOVs than NBIs, they have higher spectral resolution and potentially can complete one measurement in a shorter time span. 

Since photon flux and solar evolution time speed are critical, measurements should in general be done simultaneously rather than sequentially in 5 or more spectral lines. I.e., EST should be equipped with a multitude of exchangeable NBIs and IFUs. In the SRD document of December 2019, a concept for the Coudé Light Distribution was not yet developed. In the phase between 2019 and 2025 (see Sect.~\ref{subsec_2025}) such a concept was developed (see Fig.~\ref{fig_CLD}), and all Observing Programme Tables were adjusted to fulfill their goals with the planned suite of science instruments (TIS/FBIs, IFS-Ms, IFS-S, cf. Sect.~\ref{subsec_2025}).

\section{Scientific Requirements}

Which are the specifications that are relevant for the final design of the telescope, and for a preliminary planning of its instrumentation? 

Apart from the specifications listed below, a high signal-to-noise ratio (SNR) is crucial for most of the science goals in Part II of this document. A general consideration of the SNR with respect to spatial and temporal resolution as well as with respect to the diffraction limit is given in Appendix \ref{app_snr}.

\subsection{Telescope and instrument specifications}

\begin{enumerate}
\item FOV:
\begin{enumerate}
\item Telescope FOV (Field stop in F1): Diameter of 125 arcsec corresponding to a square of  90 by 90 arcsec$^2$.
\item Seeing-corrected FOV: 40 by 40 arcsec$^2$.  Seeing at the sites of OT and ORM are dominated by ground-layer turbulence. Therefore a high elevation of the telescope is to be preferred.The MCAO system shall also correct FOVs at and off the solar limb to observe spicules and prominences at the diffraction limit.
\item For context information larger FOVs are essential. Such context information  can be supplied by another telescope that has spectropolarimetric capabilities. 
\item Auxiliary full disk telescope (AFDT): At the site of EST a FDT is needed as a finder telescope for the orientation of the observer on the solar disk. This AFDT should provide full disk images (with FOV of 1 degree$^2$) in \CaIIH or \CaIIK, \Halpha  656.3 nm, possibly \HeI 1083.0 nm, and visible continuum light at a spatial resolution of 1.5 arcsec. The AFDT sketched in the EST Conceptual Design Study (2011) would serve the needs.
\end{enumerate}
\item Pointing:
 \begin{enumerate}
\item Pointing accuracy on solar disk: 1 arcsec. 
\item Relative pointing accuracy for individual tiles of mosaic with TISs: 1 arcsec 
\item Telescope repointing time for tiles in TIS mosaic: 2 sec
\item Relative (internal) pointing accuracy for individual tiles of mosaic with IFUs (small FOVs): 0.1 arcsec
\item Repointing time (internal raster) for tiles in IFS mosaic: 0.1 sec. This is an instrument requirement.
\end{enumerate}
\item Optical quality: Optimised for high photon flux and best optical quality, i.e. optimised for a minimum amount of optical surfaces.
 \begin{enumerate}
\item  Diffraction-limited image quality
\item  The Modulation Transfer Function (MTF) is required to be optimised for spatial scales that correspond to twice the diffraction-limit. At that scale a MTF value of about 70\% relative to an ideal non-obscured aperture telescope is desired.\footnote{The EST baseline design foresees a central obscuration of 1.30\,m, and 4 spiders with a width of 0.05\,m. Relative to an open aperture, this results in a reduction of the MTF to 72\% at a spatial frequency that corresponds to half of the diffraction limit.} The performance of  the multi-conjugate adaptive optics (MCAO) system is required to retain this image quality under excellent Seeing conditions (Fried parameter $R_0 > 10$\,cm and stable high altitude layers).
\end{enumerate}
\item Secondary Mirror under the aspect of maximising photon flux:
\begin{enumerate}
\item Design with Adaptive Secondary Mirror (ASM) is to be preferred if technically feasible.
\item In case of design with ASM: Image rotation is acceptable for TISs and IFUs. Long-slit spectrographs benefit from an internal  image de-rotator. Rotating Coud{\'e} platform to be preferred if it allows for sufficient space for instruments. The image rotation rate should be smallest during the morning\footnote{The choice of direction of the beam in the elevation axis decides on image rotation rate being smaller or larger in the morning relative to the afternoon.}.
\end{enumerate}
\item The polarimetric properties of the solar light need to be measured with high sensitivity and accuracy. Polarimetric sensitivity refers to the ability to detect a signal above the noise and is therefore a requirement of the signal-to-noise ratio. The polarimetric sensitivity is defined as the ratio between the RMS noise value of a given Stokes parameter Q, U or V and the average intensity (Stokes I).  
Polarimetric accuracy quantifies the residual errors in establishing the zero polarisation levels and in removing the crosstalk between Stokes $Q$, $U$, and $V$, i.e., accuracy refers to the residual error after demodulating the measured Stokes parameters.
\begin{enumerate}
\item Measurements shall allow to be sensitive to polarimetric signals at the level of $3\cdot 10^{-5}$. 
\item For the errors, ${\cal E}$, of the demodulation matrix of the measured Stokes vectors, i.e. for the contamination, ${\cal E}_{i,j}$ due to cross talk between Stokes $i$ and $j$, the requirements can be written in the following form\footnote{If $\mathbb{O}$ denotes the modulation matrix, $\mathbb{D}$ the demodulation matrix, and $\mathbb{E}$ the unity matrix, then ${\cal E} = \mathbb{D} \cdot \mathbb{O} - \mathbb{E}$.}: 
\[ {\cal E} = \left(
\begin{array}{*{3}r@{\;\qquad}r}
10^{-2} & 1 & 1 & 0.1 \\
5\cdot10^{-4} & 10^{-2} & 5\cdot10^{-2} & 1\cdot10^{-3} \\
5\cdot10^{-4} & 5\cdot10^{-2} & 10^{-2} &  1\cdot10^{-3} \\
5\cdot10^{-3} & 5\cdot10^{-1} & 5\cdot10^{-1} & 10^{-2} 
\end{array}\right)
\]

After discussions between the Project Office, the review panel of instrument experts, and the SAG in 2020, these requirements were relaxed to the following values:

\[ {\cal E} = \left(
\begin{array}{*{3}r@{\;\qquad}r}
10^{-2} & 1 & 1 & 0.1 \\
{\color{red} 3\cdot10^{-3}} & 10^{-2} & 5\cdot10^{-2} & 1\cdot10^{-3} \\
{\color{red} 3\cdot10^{-3}} & 5\cdot10^{-2} & 10^{-2} &  1\cdot10^{-3} \\
{\color{red} 3\cdot10^{-3}} & 5\cdot10^{-1} & 5\cdot10^{-1} & 10^{-2} 
\end{array}\right)
\]

This implies that 
\begin{enumerate}
\item the scaling accuracy needs to be better than $10^{-2}$, 
\item the residual polarisation measured in an unpolarised region must be smaller than $3\cdot 10^{-3}$ (for $Q$, $U$, and $V$), 
\item the cross talk $V \rightarrow (Q, U)$ is constrained to 0.1\% to allow for scattering polarisation measurements, and 
\item if $Q$ and $U$  are 10\% of $V$ and $V$ is 10\% of $I$, the relative crosstalk contamination is evenly distributed among  $Q$, $U$, and $V$, and amounts to 5\%. 
\end{enumerate}
\end{enumerate}
\item Wavelength coverage: 390 nm - 2300 nm. Transmission is required to be optimised for Ca II IR 854.2 nm. Requirement for wavelength coverage beyond 1600 nm should not compromise performance for wavelengths smaller than 1600nm.
\item Pointing off the solar disk: $< 100$ arcsec. The telescope is required to be able to point to the sun during day time.
\end{enumerate}

\subsection{Considerations for instrumentation and light distribution}

\paragraph{Considerations in the SRD documents of 2019:}\label{considerations_lines}

Many of the top-level science cases request to observe in many different layers of the solar atmosphere simultaneously. Up to three lines with photospheric contributions together with up to three lines forming in the chromosphere are requested. The selected lines for the Zeeman and Doppler diagnostics are one of the following:
\CaIIK \,393.4\,nm, \MgI 517.3\,nm, \FeI 525.0\,nm , \FeI 543.5\,nm (non-magnetic), \FeI 557.6\,nm (non-magnetic), \HeI 587.6\,nm, \FeI 617.3\,nm, \FeI 630.1\,nm, \FeI 630.2\,nm, \FeI 709.0\,nm (non-magnetic), \Halpha  656.3\,nm,  \CaII 854.2\,nm,  \SiI 1082.7\,nm, \HeI 1083.0\,nm, \FeI 1564.8\,nm. 
For the diagnostics using scattering polarization the following lines are proposed: 
\CaIIH, \CaIIK, \CaI 422.7 nm, \TiI 453.0 nm, \SrI 460.7 nm, C$_2$ molecular line at 514.0 nm, \NaI D$_1$ and D$_2$ 589 nm, \CaII IR triplet 849.8 nm, 854.2 nm and 866.2 nm.

Simultaneous observation in up to six different spectral lines are requested by the high-impact science cases. Some of those require larger FOVs, others require spectral integrity and smaller time cadences. I.e., 6 NBIs exchangeable with 6 IFUs for 6 different simultaneous wavelength regimes would suffice to serve science requirements.

The spatial resolution of NBIs and IFUs in some cases should be close to the diffraction limit. However many science cases ask for high SNR that will require to increase the collection area, i.e. reduce the spatial resolution. This could be achieved by a variable image scale or by binning of the recorded pixels.

Broad Band Imagers (BBIs) should be capable to image the solar evolution at the diffraction limit. Therefore a sequential observing procedure is not of advantage. Instead, one should foresee two BBIs for the blue visible and one BBI for the red visible light.

\paragraph{Scientific Instrumentation proposed for the Preliminary Design Review (2025):} 

The upper considerations (Sect.~\ref{considerations_lines}) led into the development of the Coudé Light Distribution and  the Science Instrumentation Suite (SIS) that is presented in Sect.~\ref{subsec_2025}. With this concept, all lines considered in the SRD 2019 can be observed and it has been confirmed that the science goals can be achieved.

%\section{Requirement Summary}
%
%\begin{description}
%\item[Field-of-View (FOV):]
%\item[Polarimetric sensitivity and accuracy:]
%\item[Absolute and relative pointing accuracy:]
%\item[Photonflux and its dependence with wavelength:]
%\item[Spectral resolution]
%\item[Scattered light:]
%\item[Image rotation:]
%\item[Slit orientation:]
%\item[Mosaics:] Timing between tiles? Repositioning AO lock point?
%\item[Multi-line capability:]
%\item[Pointing range:] 200 arcsec off-limb.
%\end{description}

\section{Conclusions}

The overall goal of EST is to understand the small-scale processes in the solar atmosphere. The telescope is designed to be a ‘microscope’ providing with best imaging and polarimetric quality and with the highest possible photon throughput. To achieve highest imaging quality, the telescope design contains Multi-Conjugate Adaptive Optics (MCAO). MCAO and image reconstruction techniques allow for diffraction limited imaging quality over the entire FOV, including FOVs at and off the limb. An optimal polarimetric performance is assured by assembling the mirrors such that they are polarimetrically compensated. The suite of post-focus instruments will change during its lifetime. 

An analysis of the top-level science cases reveals that the corresponding observing programmes demand high polarimetric accuracy and/or high time cadence. This requires that the EST photon throughput should be maximised with highest priority. This implies that an adaptive secondary mirror with corresponding removal of the transfer optics (removal of 7 mirrors) is to be preferred if technically feasible. 

The constraints that results from short effective exposure times to resolve the dynamics of the chromosphere have implications for the EST instrumentation. Although a detailed concept of the post-focus instrumentation is beyond the scope of this document, it is obvious that traditional spectrographs and Fabry-Perot systems alone will not be able to perform the necessary measurements. Instead, integral field units (IFUs) that simultaneously measure spatial and spectral dimensions are required. The latter implies that the individual field-of-views are small. The SAG concludes that the field-of-view is secondary, while being as large as possible.

The document concludes the scientific specifications that are relevant for the final design of the telescope. The free field-of-view should be 125 arcsec in diameter, with a Seeing-corrected field-of-view of 55 arcsec in diameter. As Seeing is dominated by ground-layers, a high elevation of the telescope will be essential. For the required larger field-of-view an auxiliary full disk telescope is specified. The pointing accuracy on the solar disk is specified with 1 arcsec to be achieved in less than 2 sec, and pointings off the solar limb up to 100 arcsec is required to be possible. The optical quality is required to diffraction limited for wavelengths as small as 395 nm. A high priority requirement is that the number of optical surfaces should be as small as possible while the telescope is polarimetrically compensated. The polarimetric sensitivity is specified at a level of $3 \cdot 10^{-5}$, and the errors of the demodulation matrix with 16 elements is specified with individual matrix elements being as small as $5 \cdot10^{-4}$. The wavelength coverage is specified to be 390 nm to 2300 nm. The Ca II IR line around 854 nm has been identified as the key line to probe the magnetic field in the chromosphere, implying that the transmission at 854 nm should have priority.

\part{Appendix, References, \& Abbrevations}
\appendix

\section{Signal-to-noise ratio S/N}\label{app_snr}

\subsection{Optimum spatio-temporal resolution}
\label{sec_optres}

(Alex Feller)

Number of photo-electrons, $Z$, in a resolution element (detector pixel):
\begin{equation}
\label{eq_samp1}
	Z = (S/N)^2 = \Phi \Delta \lambda \, \Delta x^2 \, \Delta t
\end{equation}
$S/N$ is the signal-to-noise ratio. Here we assume that the measurement noise 
is dominated by photon noise which is Poisson distributed. $\Phi$ is the 
photon-electron flux per unit wavelength, time, and resolution element. This quantity 
determines the actual S/N of the measurement and takes into account the overall 
transmission of the beam path from the solar surface through the Earth's 
atmosphere, the telescope optics, as well as the quantum efficiency of the 
detector. 
$\Delta \lambda, \Delta x, \Delta t$ denote the spectral, spatial and 
temporal sampling respectively.

If we assume that spatial and temporal resolution are coupled by a 
characteristic solar evolution speed $v$:
\begin{equation}
\label{eq_samp2}
	\Delta x = v \Delta t,
\end{equation}   
we can determine the optimum temporal sampling (cadence) in order to reach a 
given S/N:
\begin{equation}
\label{eq_samp3}
	\Delta t = \left( \frac{(S/N)^2}{\Phi \, \Delta \lambda \, v^2} 
	\right)^{1/3}
\end{equation}
The optimum spatial sampling is then given by Eq.~(\ref{eq_samp2}). If the 
spatial sampling is lower than this optimum value the decreased photon flux 
per pixel leads to integration times which exceed the timescale of solar 
evolution at the given spatial sampling, leading to a smearing of the observed 
solar scene.  

\subsection{Number of photons at diffraction limit}

(Rolf Schlichenmaier)

If one considers the signal-to-noise ratio at the diffraction limit of the telescope aperture, one needs to take into account that  the size of the aperture increases the photon collection area, while it decreases the size of the diffraction-limited resolution element. The photon flux, $\Phi$, is proportional to the square of the diameter, $D$, of the aperture: $\Phi = \tilde{\Phi} \, D^2$. Hence, Eq.({\ref{eq_samp1}) can be written as
\[ (S/N)^2 = \tilde{\Phi} \, D^2 \, \Delta x^2 \, \Delta \lambda \,  \Delta t
\]
If the size of the resolution element is chosen such that it is given by the diffraction limit of the telescope, $\Delta x = \Delta x_{\rm limit} = \lambda / D$, and if the exposure time is limited by the solar evolution speed (Eq.~\ref{eq_samp2}), $\Delta t = \Delta x_{\rm limit} / v = \lambda / (D \, v)$, then
\[ (S/N)_{\rm limit}^2 =\tilde{\Phi}   \, D^2 \, \Delta x_{\rm limit}^2 \, \Delta \lambda \,    \Delta t_{\rm limit}
                = \tilde{\Phi}   \, D^2 \, \frac{\lambda^2}{D^2} \, \Delta \lambda \,  \frac{\lambda}{D \, v}
                =  \tilde{\Phi}   \, \frac{ \lambda^3 \, \Delta \lambda}{v}  \,  \frac{1}{D}
\]
This consideration demonstrates that bigger telescopes may give better spatial resolution, but at the diffraction limit they do not provide a higher signal-to-noise ratio. Even worse, the solar evolution speed, $v$, limits the allowed exposure time at the diffraction limit, such that the signal-to-noise ratio at the diffraction limit decreases with telescope aperture:
\begin{equation}\label{eq_samp4}
 (S/N)_{\rm limit}^2 \propto \frac{1}{D}
\end{equation}
This somewhat counter-intuitive lesson stresses that high-resolution solar observations are limited by photons, and that solar telescopes must be optimised for photon flux.

The above equations are included in the \emph{photoncount} code by 
Jorrit Leenaarts.

%\section{On-axis versus off-axis (unfinished)}
%
%The on-axis design has the disadvantage that the PSF gets broadened by the spiders and the central obscuration. While image reconstruction techniques can take this into account, it is %unavoidable that this attenuates the signal-to-noise ratio.
%
%The off-axis design has the disadvantage that the mirrors have inclined reflections varying with time. Hereby, one of the polarisation direction gets attenuated more than the other. This asks %for a demanding calibration procedure and time-costly daily calibration measurements, which are needed to assure high polarimetric accuracy. For the attenuated direction of polarisation, a %gain factor has to be applied which also amplifies the polarimetric noise, thereby decreasing the polarimetric sensitvity. 
%
%To be continued...

%\section{Test}
%\subsection{Subtest}
%\subsubsection{Subsubtest}

%\newpage
%\section{References}

\addcontentsline{toc}{section}{B  \hspace{.25em} References}\addtocounter{section}{1}
\bibliographystyle{aa}
\bibliography{SRD}

%\rule{\textwidth}{1pt}

\section{List of Abbreviations}

\begin{tabbing}
ASM  \quad \= $\longrightarrow$ \quad \= Adaptive Secondary Mirror\\[1ex]
AFDT  \> $\longrightarrow$  \> Auxiliary full-disk telescope \\[1ex]
BBI  \> $\longrightarrow$ \> Broad Band Imager (cf. page \pageref{def_bbi}) \\[1ex]
EAST \> $\longrightarrow$ \> European Association for Solar Telescopes: \\[1ex]
 \> \> \verb+http://www.est-east.eu/est/index.php/people/ + \\[1ex]
EST   \> $\longrightarrow$ \> European Solar Telescope:  \verb+http://www.est-east.eu+ \\[1ex]
FoV  \> $\longrightarrow$ \>  Field-of-view \\[1ex]
IFU \> $\longrightarrow$ \> \parbox[t]{14.5cm}{Integral field unit: Instrument that measures 2D-spatial and spectral dimensions simultaneously. The big advantage of such devices is that image restoration techniques can be applied to enable full spectral integrity at highest spatial resolution.
An IFU includes a grating spectrograph, but in contrast to SP which has a long slit, its FOV is 2D. 
(cf. page \pageref{def_ifu})} \\[1ex]
MCAO  \> $\longrightarrow$ \>  \parbox[t]{14.5cm}{Multi-Conjugate Adaptive Optics: EST will be equipped with a ground-layer adaptive optics system (one deformable mirror in the pupil and one TipTilt mirror). In addition, EST will be equipped with a MCAO system that incorporates 4 deformable mirrors conjugated to four different atmospheric heights aiming to correct a FoV of up to 60 arcsec in diameter.}\\[1ex]
NBI \> $\longrightarrow$ \> \parbox[t]{14.5cm}{Narrow Band Imager (Fabry Perot system): 2D images sequentially sample spectral positions (cf. page \pageref{def_nbi}). } \\[1ex]
PSF \> $\longrightarrow$ \> Point-Spread-Function \\[1ex]
$R$ \> $\longrightarrow$ \> Spectral Resolution $R =  \lambda /  \delta\lambda $  \\[1ex]
SAG \> $\longrightarrow$  \> Science Advisory Group \\[1ex]
SNR  \> $\longrightarrow$  \> Signal-to-noise ratio: Definition see page \pageref{app_snr}.\\[1ex]
SP \> $\longrightarrow$ \> \parbox[t]{14.5cm}{Spectropolarimeter based on a long-slit spectrograph. To obtain spatial maps the slit must scan across the image (cf. page \pageref{def_sp}). }\\[1ex]
SRD \> $\longrightarrow$  \> Science Requirement Document 
\end{tabbing}

\end{document}